\documentclass[12pt,a4paper,twoside,openright]{book}

\usepackage{amsmath}
\usepackage{amssymb}
\usepackage[dvips]{graphicx}
\usepackage{epsf}
\usepackage{slashed}
\usepackage{verbatim}
\usepackage{simplewick}
\usepackage{subfigure}
\usepackage{stmaryrd} % \llbracket \rrbracket

\usepackage{color}

\usepackage[czech,english]{babel}
\usepackage[cp1250]{inputenc}

\usepackage[intoc,noprefix]{nomencl}
\makenomenclature

\usepackage{wasysym} % \apprge, \apprle (looks nicer than AMS's \gtrsim, \lesssim)

\usepackage[breaklinks=true]{hyperref}

\newdimen\InnerMargin
\newdimen\OuterMargin

\oddsidemargin=\InnerMargin   % left margin on odd numbered pages (\hoffset inches are added to this value)
\evensidemargin=\OuterMargin  % left margin on even numbered pages (\hoffset inches are added to this value)
\textwidth=\paperwidth
\advance\textwidth by-\InnerMargin
\advance\textwidth by-\OuterMargin

\newdimen\BottomMargin

\textheight=\paperheight
\advance\textheight by-\topmargin
\advance\textheight by-\headheight
\advance\textheight by-\headsep
\advance\textheight by-\topskip
\advance\textheight by-\footskip
\advance\textheight by-\BottomMargin
\newcommand{\bra}[1]{\langle#1\vert}% bra vectors
\newcommand{\ket}[1]{\vert#1\rangle}% ket vectors
\newcommand{\Tr}{\mathop{\rm Tr}\nolimits}% trace operator
\newcommand{\I}{\ensuremath{\mathrm{i}}}% imaginary unit
\newcommand{\e}{\ensuremath{\mathrm{e}}}% Euler number
\renewcommand{\Re}{\mathop{\rm Re}\nolimits}% real part
\renewcommand{\Im}{\mathop{\rm Im}\nolimits}% imaginary part
\renewcommand{\d}{\ensuremath{\mathrm{d}}}% differential
\newcommand{\hc}{\ensuremath{\mathrm{h.c.}}}% hermitean conjugated part
\newcommand{\SU}[1]{\ensuremath{\mathrm{SU}(#1)}}

\newcommand{\OO}[1]{\ensuremath{\mathrm{O}(#1)}}
\newcommand{\U}[1]{\ensuremath{\mathrm{U}(#1)}}

\newcommand{\slesp}[1]{#1 \hspace{-4.3 pt}/}
\renewcommand{\d}{\ensuremath{\mathrm{d}}}% differential
\newcommand{\C}{\ensuremath{c}}% charge conjugation
\newcommand{\T}{\ensuremath{\mathrm{T}}}% matrix transposition
\newcommand{\F}{\ensuremath{\mathrm{F}}}
\newcommand{\EM}{\ensuremath{\mathrm{em}}}
\newcommand{\SM}{\ensuremath{\mathrm{SM}}}

\newcommand{\qm}[1]{``#1''} % quotation marks
\newcommand{\GeV}{\ensuremath{\mathrm{\,GeV}}}
\newcommand{\TeV}{\ensuremath{\mathrm{\,TeV}}}

\DeclareFontFamily{OT1}{cmrx}{}
\DeclareFontShape{OT1}{cmrx}{m}{n}{<->cmr10}{}
\let\saveLongrightarrow\Longrightarrow
\makeatletter
\renewcommand*{\Longrightarrow}{%
    \mathrel{\rlap{\fontfamily{cmrx}\fontencoding{OT1}\selectfont=}%
    \hphantom{\saveLongrightarrow}%
    \llap{$\m@th\Rightarrow$}}}
\makeatother

\usepackage{bbm} %\mathbbm{1}, \mathbbmss{1}, \mathbbmtt{1}
\def\unitmatrix{\mathbbm{1}} %unit matrix (to)
\newcommand{\chapterstar}[1]{
    \chapter*{#1}
    \addcontentsline{toc}{chapter}{#1}
    \markboth{\uppercase{#1}}{\uppercase{#1}}
    }

\begin{document}

%%%%%%%%%%%%%%%%%%%%%%%%%%%%%%%%%%%%%%%%%%%%%%%%%%%%%%%%%%%%%%%%%%%%%%%%
%% TITLE
%%%%%%%%%%%%%%%%%%%%%%%%%%%%%%%%%%%%%%%%%%%%%%%%%%%%%%%%%%%%%%%%%%%%%%%%

\def\be{\begin{equation}}
\def\eet{\,.\end{equation}}
\def\eec{\,,\end{equation}}
\def\ee{\end{equation}}
\def\im{\mathrm{i}}
\def \openone {\leavevmode\hbox{\small1\kern-3.5pt\normalsize1}}
\def \unitmatrix {\leavevmode\hbox{\small1\kern-3.0pt\normalsize1}}
\def\beginm{\left(\begin{array}}
\def\endm{\end{array}\right)}

\pagenumbering{roman}
\pagestyle{empty}

\begin{center}
{\Large \sc Charles University in~Prague\\
Faculty of Mathematics and Physics}

\vspace{0.1cm}

{\large and \\}

\vspace{0.1cm}

{\Large \sc Academy of Sciences of the Czech Republic\\
Nuclear Physics Institute}

\vspace{0.1cm}

\end{center}
\hrule

\vspace*{2cm}
\begin{figure}[h]
\centering \epsfxsize=0.55\hsize \mbox{\epsfbox{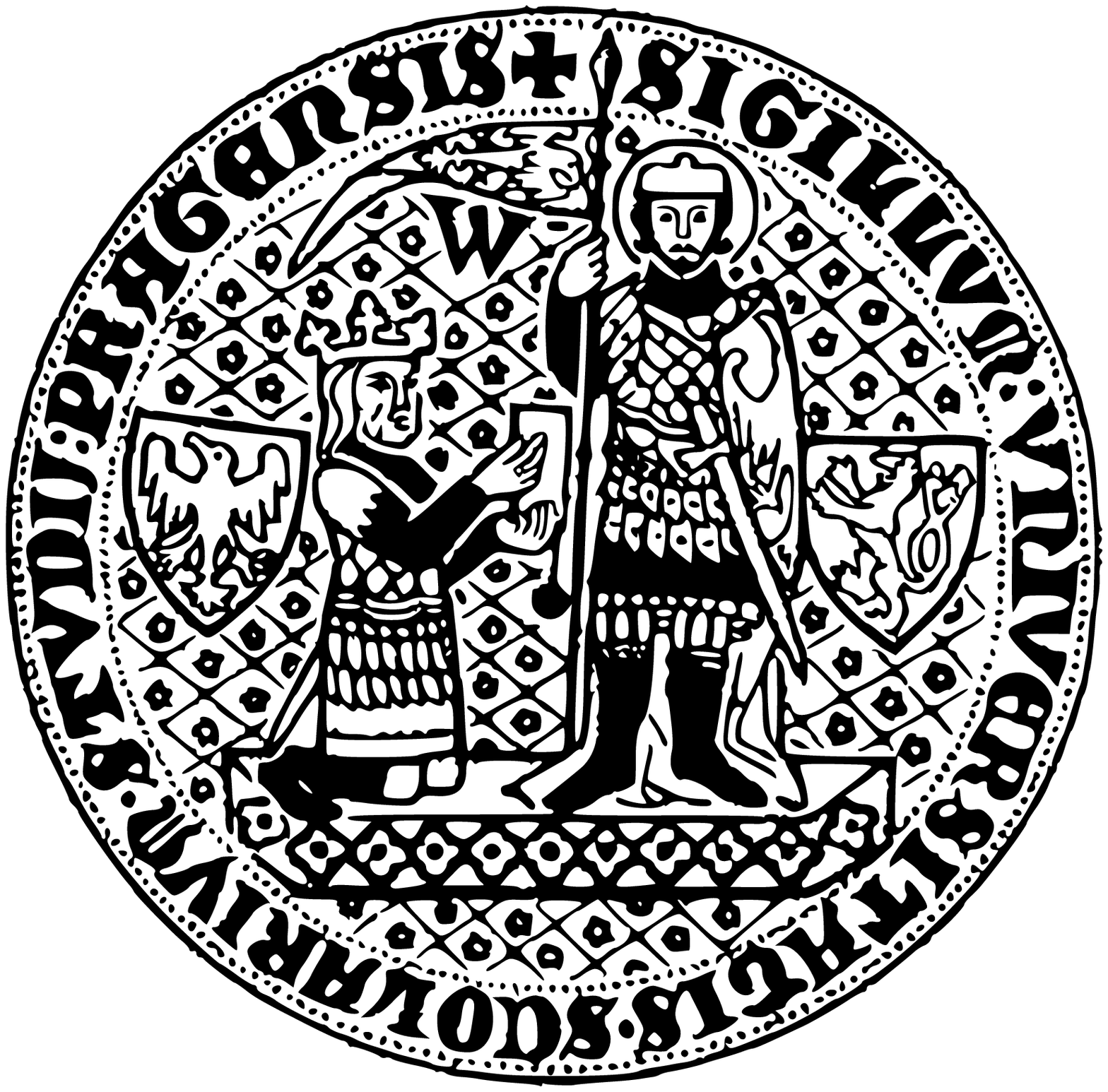}}
\end{figure}
\vspace{2cm}

\begin{center}

{\Large \sc Adam Smetana}\\
\vspace*{1cm}
{\LARGE \bf Electroweak symmetry breaking by dynamically generated masses of quarks and leptons}\\
\vspace*{0.5cm}
Thesis submitted for the degree of Philosophi{\ae} Doctor\\
\vspace*{1cm}
Supervisor of the doctoral thesis:\quad\quad  \textbf{Ing.~Jiří Hošek,~DrSc.},\quad NPI AS CR\\

%Ji\v{r}\'{\i} Ho\v{s}ek

%\vfill
%
%\begin{tabular}{rl}
%
%Study programme, specialization: & Physics, subnuclear physics F9 \\
%%\noalign{\vspace{2mm}}
%%Specialization: & Subnuclear Physics F9 \\
%\end{tabular}

\vspace{1cm}

Prague, May 2013

\end{center}

%%%%%%%%%%%%%%%%%%%%%%%%%%%%%%%%%%%%%%%%%%%%%%%%%%%%%%%%%%%%%%%%%%%%%%%%
%% declaration & acknowledgement
%%%%%%%%%%%%%%%%%%%%%%%%%%%%%%%%%%%%%%%%%%%%%%%%%%%%%%%%%%%%%%%%%%%%%%%%

\newpage
\pagestyle{headings}
\cleardoublepage

%%% Strana s čestným prohlášením k disertační práci

\section*{Acknowledgements}

I would like to thank my supervisor, Ing. Jiří Hošek, DrSc., for guiding me through the years of my PhD studies, for sharing his inspiring ideas and knowledge with me and for insisting on the highest possible quality for this work. I am grateful to Professor RNDr. Jiří Hořejší, DrSc. for his valuable support which enabled me to travel to important scientific events. I would like to express my gratitude to Ing. Stanislav Pospíšil, DrSc., for giving me the opportunity to finish my thesis at the Institute of Experimental and Applied Physics, at the Czech Technical University, Prague. I acknowledge my scientific twin, Petr Beneš, and also other fellows and colleagues Tomáš Brauner, Hynek Bíla, Filip Křížek, Dalibor Nedbal and Karel Smolek for numerous fruitful discussions on the subject of physics. I am indebted to Jiří Hošek, Jiří Adam and Petr Beneš for their thoroughness and for the time they spent reading this manuscript and offering their suggestions.

I acknowledge the IPNP, Charles University in Prague, and the ECT* Trento for the support during the ECT* Doctoral Training Programme 2005. The presented work was supported in part by the Institutional Research Plan AV0Z10480505, by the GACR grant No.~202/06/0734, by the Grant LA08015 of the Ministry of Education of the Czech Republic, by the Research Program MSM6840770029, by the MEIS of Czech Republic LM2011027 and by the project of International Cooperation ATLAS-CERN LA08032. All Feynman diagrams were drawn using the JaxoDraw \cite{Binosi:2003yf,Binosi:2008ig}.

I appreciate the irreplaceable support and encouragement of all my family. I dedicate this thesis to my Grandmother, who always expressed her trust in my potential. Finally and from my heart, I thank Šárka for making the last year of my PhD studies a joyful and meaningful time of development.

May this thesis and all scientific research lead to understanding the way things are.

\vglue 0pt plus 1fill

\noindent
I declare that I carried out this doctoral thesis independently, and only with the cited sources, literature and other professional sources.

\medskip\noindent
I understand that my work relates to the rights and obligations under the Act No.~121/2000 Coll., the Copyright Act, as amended, in particular the fact that the Charles University in Prague has the right to conclude a license agreement on the use of this work as a school work pursuant to Section 60 paragraph 1 of the Copyright Act.

\vspace{10mm}

\hbox{\hbox to 0.5\hsize{%
In ........................ date ........................
\hss}\hbox to 0.5\hsize{%
%Podpis autora
signature
\hss}}

\vspace{20mm}
\newpage

%%%%%%%%%%%%%%%%%%%%%%%%%%%%%%%%%%%%%%%%%%%%%%%%%%%%%%%%%%%%%%%%%%%%%%%%
%% abstract
%%%%%%%%%%%%%%%%%%%%%%%%%%%%%%%%%%%%%%%%%%%%%%%%%%%%%%%%%%%%%%%%%%%%%%%%
%
%
\newpage
\pagestyle{headings}

%% Povinná informační strana disertační práce

% \glqq{flavorů}\grqq

\newdimen\LeftColumn
\newdimen\MiddleColumn
\newdimen\RightColumn
\LeftColumn=3.8cm
\MiddleColumn=0.5cm
\RightColumn=\textwidth
\advance\RightColumn by-\LeftColumn
\advance\RightColumn by-\MiddleColumn

\noindent
\begin{tabular*}{\textwidth}{@{}p{\LeftColumn}@{} @{}p{\MiddleColumn}@{} @{}p{\RightColumn}@{}}
\textbf{Název práce:}              && Narušení elektroslabé symetrie dynamickým generováním hmot kvarků a leptonů
\\ &&\\
\textbf{Autor:}                    && Adam Smetana
\\ &&\\
\textbf{Katedra:}                  && Ústav částicové a jaderné fyziky MFF UK
\\ &&\\
\textbf{Vedoucí disertační práce:} && Ing.~Jiří Hošek,~DrSc., Oddělení teoretické fyziky, Ústav jaderné fyziky, AV ČR, v.~v.~i.
\\ &&\\
\textbf{Abstrakt:}                 && Tématem práce je teoretické studium modelů narušení elektroslabé symetrie způsobené dynamicky generovanými hmotami kvarků a leptonů. (1) Nejprve ukazujeme, že samotná tato základní myšlenka je fenomenologicky akceptovatelná. Proto rozpracováváme zjednodušující model dvou kompozitních Higgsovských dubletů, který popisuje top-kvarkovou a neutrinovou kondenzaci. Z modelu vyplývá, že počet pravotočivých neutrin je vyšší, řádu $\mathcal{O}(100)$. (2) Dále se zabýváme modelem silné Yukawovské dynamiky, ve kterém je dynamického gene\-rování fermionových hmot dosaženo výměnou nových hmotných ele\-mentárních komplexních skalárních dubletních polí. V práci se soustředíme na řešení svázaných Schwingerových--Dysonových rovnic pro fermionové a skalární vlastní energie za použití aproximativních metod. Ukazujeme, že lze dosáhnout silně hierarchického hmotového spektra. (3) Nakonec se zabýváme flavorovým kalibračním modelem, ve kterém je generování fermionových hmot dosaženo pomocí asymptoticky volné, samu sebe narušující flavorové kalibrační dynamiky. Ukazujeme, že kondenzace majoranovského typu pravotočivých neutrin v sextetní flavorové reprezentaci přirozeně způsobuje spontánní narušení flavorové symetrie. Tato kondenzace vede k velkým majoranovským hmotám pravotočivých neutrin.
\\ &&\\
\textbf{Klíčová slova:}            && Dynamické narušení elektroslabé symetrie, top-kvarková kondenzace, neutrinová kondenzace, silná Yukawovská dynamika, flavorová kalibrační dynamika
\end{tabular*}

\newpage

\noindent
\begin{tabular*}{\textwidth}{@{}p{\LeftColumn}@{} @{}p{\MiddleColumn}@{} @{}p{\RightColumn}@{}}
\textbf{Title:}                    && Electroweak symmetry breaking by dynamically generated masses of quarks and leptons
\\ &&\\
\textbf{Author:}                   && Adam Smetana
\\ &&\\
\textbf{Department:}               && Institute of Particle and Nuclear Physics, Faculty of Mathematics and Physics, Charles University
\\ &&\\
\textbf{Supervisor:}               && Ing.~Jiří Hošek,~DrSc., Department of Theoretical Physics, Nuclear Physics Institute, Academy of Sciences of the Czech Republic
\\ &&\\
\textbf{Abstract:}                 && The aim of the thesis is to study models of the electroweak symmetry breaking caused by dynamically generated masses of quarks and leptons. (1) We perform the basic analysis whether the main underlying idea, that the masses of only known fermions can provide the electroweak symmetry breaking, is actually feasible. For that we elaborate a two-composite-Higgs-doublet model of the top-quark and neutrino condensation. The model suggests rather large number, $\mathcal{O}(100)$, of right-handed neutrinos. (2) We analyze the model of strong Yukawa dynamics where the dynamical fermion mass generation is provided by exchanges of new elementary massive complex doublet scalar fields. We focus on solving the coupled Schwinger--Dyson equations for fermion and scalar self-energies by means of approximative methods. We document that strongly hierarchical mass spectra can be reproduced. (3) We elaborate the flavor gauge model where the dynamical fermion mass generation is provided by asymptotically free non-Abelian self-breaking flavor gauge dynamics. We show that the Majorana type condensation of right-handed neutrinos in the flavor sextet representation triggers the complete flavor symmetry breaking. It leads to huge right-handed neutrino Majorana masses.
\\ &&\\
\textbf{Keywords:}                 && Dynamical electroweak symmetry breaking, top-quark condensation, neutrino condensation, strong Yukawa dynamics, flavor gauge dynamics
\end{tabular*}

%%%%%%%%%%%%%%%%%%%%%%%%%%%%%%%%%%%%%%%%%%%%%%%%%%%%%%%%%%%%%%%%%%%%%%%%
%% CONTENTS
%%%%%%%%%%%%%%%%%%%%%%%%%%%%%%%%%%%%%%%%%%%%%%%%%%%%%%%%%%%%%%%%%%%%%%%%

\newpage
\pagestyle{headings}

%%\cleardoublepage
%\addcontentsline{toc}{chapter}{Contents}
\tableofcontents

\cleardoublepage
\phantomsection
\addcontentsline{toc}{chapter}{List of figures}
\renewcommand\listfigurename{List of figures}
\listoffigures

\cleardoublepage
\phantomsection
\addcontentsline{toc}{chapter}{List of tables}
\renewcommand\listtablename{List of tables}
\listoftables

%%%%%%%%%%%%%%%%%%%%%%%%%%%%%%%%%%%%%%%%%%%%%%%%%%%%%%%%%%%%%%%%%%%%%%%%
%% PREFACE
%%%%%%%%%%%%%%%%%%%%%%%%%%%%%%%%%%%%%%%%%%%%%%%%%%%%%%%%%%%%%%%%%%%%%%%%

%\chapterstar{Preface}

\cleardoublepage
\phantomsection
\markboth{\uppercase{List of acronyms}}{\uppercase{List of acronyms}}
\renewcommand{\nomname}{List of acronyms}
\printnomenclature[2cm]

\nomenclature{ETC}{Extended-Technicolor}
\nomenclature{FCNC}{flavor-changing neutral currents}
\nomenclature{GUT}{Grand Unification Theory}
\nomenclature{$\hc$}{Hermitian conjugate}
\nomenclature{LHC}{Large Hadron Collider}
\nomenclature{MAC}{Most Attractive Channel}
\nomenclature{MSSM}{Minimal Supersymmetric Standard Model (of electroweak interactions)}
\nomenclature{NJL}{Nambu--Jona-Lasinio}
\nomenclature{QCD}{Quantum Chromodynamics}
\nomenclature{QED}{Quantum Electrodynamics}
\nomenclature{TC}{Technicolor}
\nomenclature{UV}{ultraviolet}

%%%%%%%%%%%%%%%%%%%%%%%%%%%%%%%%%%%%%%%%%%%%%%%%%%%%%%%%%%%%%%%%%%%%%%%%
%% CONVENTIOS AND NOTATION
%%%%%%%%%%%%%%%%%%%%%%%%%%%%%%%%%%%%%%%%%%%%%%%%%%%%%%%%%%%%%%%%%%%%%%%%

\chapterstar{Conventions and notations}

We list here some of the conventions and notations which are used throughout the text:
\begin{itemize}
\item We use the \qm{natural} units, i.e., we set $c=\hbar=1$.

\item For the Minkowski metric tensor we use the convention
\begin{eqnarray}
\label{symbols:gmunu}
g_{\mu\nu} \ = \  g^{\mu\nu} &=&
\left(\begin{array}{rrrr}
1 & 0 & 0 & 0 \\
0 & -1 & 0 & 0 \\
0 & 0 & -1 & 0 \\
0 & 0 & 0 & -1
\end{array}\right)
\,.
\end{eqnarray}

\item The $\gamma_5$ matrix is defined as $\gamma_5=\I\gamma^0\gamma^1\gamma^2\gamma^3$.

\item We will frequently use the chiral projectors
\begin{equation}
P_L = \frac{1-\gamma_5}{2} \,, \quad\quad P_R = \frac{1+\gamma_5}{2}
\end{equation}
and correspondingly the \emph{left-handed} and \emph{right-handed} fermion fields $\psi_L=P_L\psi$ and $\psi_R=P_R\psi$.

\item For the totally antisymmetric Levi-Civita tensor (symbol) $\varepsilon_{\mu\nu\rho\sigma}$ we adopt the sign convention $\varepsilon_{0123}=+1$.

\item Dirac conjugation $\bar\psi$ of a bispinor $\psi$ is defined as
\begin{eqnarray}
  \bar\psi &\equiv& \psi^\dag\gamma_0 \,,
\end{eqnarray}

\item Charge conjugation $\psi^\C$ of a bispinor $\psi$ is defined as
\begin{eqnarray}
\label{symbols:psiC}
  \psi^\C &\equiv& C \bar\psi^\T \,,
\end{eqnarray}
where $C=\im\gamma_0\gamma_2$.

\item The Pauli matrices are denoted by $\sigma_i$, $i=1,2,3$:
\begin{equation}
\sigma_1 = \left(\begin{array}{rr} 0 &   1  \\  1 &  0 \end{array}\right) \,,\quad
\sigma_2 = \left(\begin{array}{rr} 0 & -\I  \\ \I &  0 \end{array}\right) \,,\quad
\sigma_3 = \left(\begin{array}{rr} 1 &   0  \\  0 & -1 \end{array}\right) \,.\quad
\end{equation}
\item The Gell-Mann matrices are denoted by $\lambda_a$, $a=1,\dots,8$:
\begin{eqnarray}
&&\lambda_1 = \left(\begin{array}{rrr} 0 &   1 &  0 \\  1 &  0 &  0  \\  0 &  0 &  0 \end{array}\right) \,,\quad
\lambda_2 = \left(\begin{array}{rrr} 0 & -\I &  0 \\ \I &  0 &  0  \\  0 &  0 &  0 \end{array}\right) \,,\quad
\lambda_3 = \left(\begin{array}{rrr} 1 &   0 &  0 \\  0 & -1 &  0  \\  0 &  0 &  0 \end{array}\right) \,, \nonumber\\
&&\lambda_4 = \left(\begin{array}{rrr} 0 &   0 &  1 \\  0 &  0 &  0  \\  1 &  0 &  0 \end{array}\right) \,,\quad
\lambda_5 = \left(\begin{array}{rrr} 0 & 0 &  -\im \\ 0 &  0 &  0  \\  \im &  0 &  0 \end{array}\right) \,,\quad
\lambda_6 = \left(\begin{array}{rrr} 0 &   0 &  0 \\  0 & 0 &  1  \\  0 &  1 &  0 \end{array}\right) \,, \nonumber\\
&&\lambda_7 = \left(\begin{array}{rrr} 0 & 0 &  0 \\  0 &  0 &  -\im  \\  0 &  \im &  0 \end{array}\right) \,,\quad
\lambda_8 = \frac{1}{\sqrt{3}}\left(\begin{array}{rrr} 1 &   0 &  0 \\  0 & 1 &  0  \\  0 &  0 &  -2 \end{array}\right) \,.
\end{eqnarray}

\item The Feynman \qm{slash} notation for four-vectors ($\slashed{p}=p_\mu\gamma^\mu$) or partial derivatives ($\slashed{\partial}=\partial_\mu\gamma^\mu$) will be extensively used throughout the text.

\item When denoting the Standard Model gauge coupling constants we use both of two notations
\begin{eqnarray}
g_1 & \equiv & g' \quad \mathrm{for\ weak\ hypercharge},\\
g_2 & \equiv & g \quad \mathrm{for\ weak\ isospin},\\
g_3 & \equiv & g_\mathrm{s} \quad \mathrm{for\ color}.
\end{eqnarray}

%
%%\item The angle brackets $\langle\,\rangle$ will denote the VEV of the $T$-product of the argument, i.e. the Green's function: $\langle \ldots \rangle = \bra{0}T\{\ldots\}\ket{0}$.  In a special case of only one local field this will be simply its VEV, e.g. $\langle \phi \rangle = \bra{0}\phi\ket{0}$.
%\item In analogy with the standard Dirac conjugation for bispinors $\bar\psi=\psi^\dag\gamma_0$

\end{itemize}

%\chapterstar{List of acronyms}
%
%%\section*{Acronyms}
%%\addcontentsline{toc}{section}{Acronyms}
%
%% only for the definitions of acronyms:
%%\newcommand{\acronym}[1]{\textbf{#1}}
%\newcommand{\acronym}[1]{#1}
%
%%\noindent Here we list the acronyms used in the thesis:\pozn{Nevyhodit tuhle větu?}\\
%\begin{tabular}{p{0cm} p{2cm} p{12cm}}
%&\acronym{ETC}   & Extended-Technicolor \\
%&\acronym{FCNC} & flavor-changing neutral currents \\
%&\acronym{GUT}   & Grand Unification Theory \\
%&$\hc$          & Hermitian conjugate \\
%&\acronym{LHC}   & Large Hadron Collider \\
%&\acronym{MAC}   & Most Attractive Channel \\
%&\acronym{MSSM}   & Minimal Supersymmetric Standard Model (of electroweak interactions) \\
%&\acronym{NJL}   & Nambu--Jona-Lasinio \\
%&\acronym{QCD}   & quantum chromodynamics \\
%&\acronym{QED}   & quantum electrodynamics \\
%&\acronym{TC}   & Technicolor \\
%&\acronym{UV}   & ultraviolet \\
%\end{tabular}

%%%%%%%%%%%%%%%%%%%%%%%%%%%%%%%%%%%%%%%%%%%%%%%%%%%%%%%%%%%%%%%%%%%%%%%%
%% BODY
%%%%%%%%%%%%%%%%%%%%%%%%%%%%%%%%%%%%%%%%%%%%%%%%%%%%%%%%%%%%%%%%%%%%%%%%

\chapter{Introduction}
\pagenumbering{arabic}

\label{intro}

\section{Understanding the elementary particle physics}

\subsection{The Standard Model}

Our current knowledge of elementary particle phenomena fits into a single compact Lagrangian $\mathcal{L}_\SM$ defining the Standard Model as a quantum field theory. Three fundamental interactions acting among all known matter fields are introduced in $\mathcal{L}_\SM$ by means of a \emph{gauge principle} applied to three simple subgroups of the symmetry
\begin{equation}\label{GSM}
G_{\mathrm{SM}}\equiv\U{1}_Y\times\SU{2}_L\times\SU{3}_c \,.
\end{equation}
The success of the gauge principle started by formulation of electromagnetic interactions in terms of Quantum Electrodynamics (QED) \cite{Fock:1926fj,Dirac:1927dy} gauging $\U{1}_\EM$. Decades latter it was followed by identification of the $\SU{3}_c$ gauge nature of strong interactions described by Quantum Chromodynamics (QCD) \cite{NambuQCD,Fritzsch:1973pi} and by unification of weak interactions together with QED within a gauge theory based on the electroweak symmetry $\U{1}_Y\times\SU{2}_L\supset\U{1}_\EM$ \cite{Glashow:1961tr,Weinberg:1967tq,Salam:1968rm}.

\subsubsection{The gauge principle}

Certainly the identification of the underlying symmetries \eqref{GSM} gains appeal by itself. It results in a reduction of a number of possible free parameters. It improves predictivity of the description, shapes its field content and expresses our better understanding of the dynamical laws. On top of that there are two conceptual quantum field theoretical reasons to employ the gauge principle. First, it is not possible to consistently construct massless vector boson fields without the gauge principle, as needed for describing photon and gluons. Second, it is not possible to construct a renormalizable quantum field theory of interacting massive vector boson fields without the gauge principle, as needed to describe $W$ and $Z$ bosons. Hard mass terms of $W$ and $Z$ bosons in the Lagrangian would ruin the tree-level unitarity of gauge boson amplitudes, the necessary criterium of perturbative renormalizability of the theory \cite{Cornwall:1974km,LlewellynSmith:1973ey,Joglekar:1973hh,Horejsi:1993hz}. The electroweak gauge symmetry simply forbids them.

The masses of $W$ and $Z$ bosons have to be soft. They are generated from the spontaneous electroweak gauge symmetry breaking,
\begin{equation}\label{EWSB}
\U{1}_Y\times\SU{2}_L\longrightarrow\U{1}_\EM \,,
\end{equation}
via the Anderson--Higgs mechanism \cite{Higgs:1964pj,Guralnik:1964eu,Englert:1964et,Migdal:1965aa,'tHooft:1971rn} in analogy with Meissner effect in superconductivity \cite{Meissner1933aa,Anderson:1963pc}. For that sake, the Standard Model is equipped by an elementary complex Higgs boson doublet. Its electrically neutral component develops an electroweak symmetry breaking vacuum expectation value. The other three components are Nambu--Goldstone modes, consequences of the Goldstone theorem \cite{Goldstone:1961eq,Goldstone:1962es}. They combine with two transverse components of gauge fields providing their longitudinal components and disappear from the particle spectrum of the theory.

The soft gauge boson mass generation is by construction accompanied by a massive scalar field, the Higgs field. Its presence follows from general reasons. Because the model was constructed as the renormalizable gauge theory, the interactions of the Higgs field are properly and automatically adjusted to cancel completely tree-level non-unitarities of all amplitudes. The important step towards final confirmation of the Standard Model picture has been achieved just recently by observing a scalar particle carrying quantum numbers and expected mass of the Higgs field \cite{125Higgs:2012gk,125Higgs:2012gu}.

\subsubsection{Robustness of the Standard Model}

The Standard Model describes much more than just the electroweak symmetry breaking.

The electroweak symmetry is a chiral symmetry. This means that left- and right-handed chiral components of fermion fields transform differently under $\U{1}_Y\times\SU{2}_L$. As a result no hard mass term, which connects left-handed with right-handed electroweakly charged fermion fields, can be written into the Standard Model Lagrangian $\mathcal{L}_\SM$. If we still insisted upon introducing the fermion mass operators in $\mathcal{L}_\SM$, arguing by their super-renormalizability for an innocence of such act, they would induce hard mass terms for electroweak gauge bosons as counter terms for divergent radiative corrections induced by fermion masses. That would destroy the renormalizability of the theory. Therefore the fermion masses have to be soft.

It is a gift that the spontaneous generation of fermion masses can be achieved by using the same Higgs doublet field introduced already for the different purpose of the electroweak gauge boson mass generation. The connection between left- and right-handed fermion fields is provided chiral gauge invariantly via the Yukawa interactions. In fact the Yukawa interactions must be present in the Lagrangian as they are renormalizable and not forbidden by symmetries. All Dirac masses then arise spontaneously when the Higgs field develops the vacuum expectation value.

Another gift is that the Yukawa interactions explicitly break large global inter-family chiral symmetries. Upon the spontaneous fermion mass generation the global chiral symmetries would give rise to unwanted Nambu--Goldstone bosons significantly coupled to fermions. Therefore they would have already been observed, but they were not.

A consistency crosscheck increasing a confidence in the Standard Model gauge construction \eqref{GSM} is provided by a complete gauge axial anomaly cancelation achieved just with that fermion content which is observed in Nature. The gauge anomaly-free balance is a result of delicate interplay of the fermion gauge quantum numbers which would be destroyed by, e.g., other number of colors than $N_\mathrm{C}=3$. It is however not destroyed by complementing the fermion spectrum by right-handed neutrinos. Being the Standard Model singlets, they can be added in an arbitrary number. In fact, their presence is phenomenologically welcome in order to provide masses for observed neutrinos.

The robustness of the Standard Model equipped by just three right-handed neutrinos, one for each fermion generation, is spectacular. It offers everything what
is necessary to satisfy currently most urgent phenomenological needs. Embedding the gauge invariant right-handed neutrino Majorana mass term provides the seesaw mechanism for explanation of tininess of neutrino masses \cite{GellMann:1980vs,Mohapatra:1979ia,Yanagida:1979gs} and guarantees the electric charge quantization \cite{Babu:1989tq}. The model offers dark matter candidates in the form of sterile neutrinos, it has enough neutrino degrees of freedom to fit the neutrino oscillation data and it provides lepton number violation necessary for leptogenetic scenario to generate the baryon asymmetry of the Universe \cite{Canetti:2012vf}.

\subsection{The Standard Model as a phenomenological model}

Despite of the robustness of the Standard Model, it is difficult to accept it as the fundamental theory. Namely the Standard Model does not \emph{explain} fermion masses, it rather \emph{parametrizes} them by one-to-one correspondence with the Yukawa coupling constants. To reproduce vast spectrum of fermion masses and mixing angles, the same number of Yukawa parameters is used. Moreover, because the fermion mass matrices are \emph{directly} proportional to their Yukawa matrices, the wild fermion mass hierarchy is just shared by both.

It is a historical experience that all of the observed spin-0 particles ultimately turned out to be the fermion bound states. In this spirit the Standard Model with its spin-0 Higgs doublet may be easily envisaged as a phenomenological realization of the electroweak gauge symmetry breaking in the same way as the Ginzburg--Landau theory is a phenomenological realization of superconductivity.

Indeed, constructing a phenomenological description of dynamical fermion mass generation based on analogy with superconductivity \cite{Nambu:1961tp}, it is a natural and minimal choice to use a single Higgs doublet field. The electroweakly charged fermions occupy only two types of weak isospin representations. The left-handed fermions are weak isospin doublets and right-handed fermions are weak isospin singlets. The condensate $\langle\bar\psi_R\psi_L\rangle$ responsible for their Dirac masses connects left-handed with right-handed fermion fields. Therefore it is meaningful to assume that the condensate is in fact a vacuum expectation value of a neutral component of a scalar doublet composite operator made of fermion bilinears. The composite operator can be used as an interpolating field for the composite Higgs doublet field indistinguishable from the elementary Standard Model Higgs doublet field.

These considerations are unanimously pointing at the Higgs doublet sector and accusing it of being just a phenomenological description. It is however necessary to stress that it is a phenomenological description of a special kind: it is renormalizable.

\subsubsection{The Standard Model sets its own limits}

The spontaneous breaking of the electroweak symmetry is a non-perturbative act. In the Standard Model the Higgs doublet field $\Phi(x)$ forms energetically favorable electroweak symmetry breaking field configuration $\Phi_{0}\equiv\langle\Phi(x)\rangle$, which is far from being perturbative. It is given by
\begin{equation}\label{vacuum_condition}
\Phi_{0}^\dag\Phi_{0} \equiv \frac{v^2}{2}=\frac{|\mu^2|}{2\lambda} \,,
\end{equation}
as a minimum of appropriately designed Higgs potential
\begin{equation}\label{Higgs_potential}
{\cal V}(\Phi^\dag\Phi)=-|\mu^2|\Phi^\dag\Phi+\lambda(\Phi^\dag\Phi)^2 \,
\end{equation}
where $v$ is the vacuum expectation value representing the electroweak scale.
The operational simplicity of this mechanism is given by the fact that the non-perturbative transition towards better vacuum can be done already at a classical level prior to the quantization. After the quantization however it is necessary to check stability of the mechanism with respect to quantum corrections.

It turns out that the parameters $\mu$ and $\lambda$ are not as stable against quantum corrections as necessary to claim that the Standard Model is the fundamental theory of elementary particles. The parameter $\lambda$ runs according to its renormalization group equation coupled with other parameters, mainly with the one for top-quark Yukawa coupling parameter $y_t$. Depending on its initial value at the electroweak scale $\lambda(v)$ with respect to $y_t$, the $\lambda$ runs either to the Landau pole \cite{Maiani:1977cg,Lindner:1985uk} or to zero \cite{Cabibbo:1979ay,Sher:1988mj} at some energy scale $\Lambda$. The most recent results based on the measured value of Higgs mass suggest that the latter possibility may happen already at $\Lambda\sim10^{11}\,\mathrm{GeV}$ well below the Planck scale \cite{Degrassi:2012ry,Bednyakov:2013eba,Chetyrkin:2013wya}. These three-loop calculations are however still weighted by large systematic errors so that the positivity of $\lambda$ up to the Planck scale cannot be ruled out.

The parameter $\mu$, and so the electroweak scale $v$ \eqref{vacuum_condition}, acquires quadratically divergent quantum corrections in contrast to the rest of Standard Model parameters which acquire only logarithmically divergent quantum corrections. This causes a questionable tension commonly referred to as the \emph{gauge hierarchy problem},
\begin{equation}\label{GHP}
v\ll\Lambda \,,
\end{equation}
where $\Lambda$ is some scale up to which we keep encountering the quantum corrections. Even though the electroweak scale $v$ plays a central role in the mass generation, as all masses are proportional to it, the Standard Model does not at all explain its origin let alone its small value. Through the non-dynamical parameter $\mu$ the electroweak scale $v$ is just put into the Lagrangian and its value is to be fixed by a single measurement.

The quantum instabilities are however not a reason for despair. The Higgs potential \eqref{Higgs_potential} was introduced into the Lagrangian in a rather ad hoc way without any vigilance or leading principle in contrast to the rest of the Lagrangian. The only purpose of the Higgs potential was to make the elementary scalar condense. It makes more sense to consider the Higgs potential rather as a \emph{phenomenological parametrization} of the electroweak symmetry breaking. In this light there remains only one question about the range of validity of this phenomenological description. The appearance of the quantum instability at the energy scale $\Lambda$ should be understood as an indication that the Standard Model is a good phenomenological description in the range of energies $(0,\Lambda)$ and breaks down for energies above $\Lambda$. %It is needed to stress that current experimental data suggest that in this respect the Standard model works up to surprisingly high energies.

\subsubsection{The Standard Model calls for beyond}

We have seen that there exists a good motivation \emph{not} to consider the Standard Model to be the fundamental theory. Therefore some new more fundamental theory should replace it above some scale $\Lambda$. From the perspective of the new theory, the Standard Model is then considered as an infrared \emph{effective theory}. To be fundamental, the new theory should \emph{explain} Higgs sector parameters, including Yukawa parameters, and stabilize the electroweak scale $v$ with respect to the scale $\Lambda$.

From the point of view of the Standard Model as an effective theory, the stabilization of the electroweak scale $v$ is just a matter of renormalization. We need one experimental measurement to fix $v$ and no reference to the cutoff $\Lambda$ is needed anymore \cite{Malinsky:2012tp}. It is not the Standard Model which suffers from the gauge hierarchy problem. However, we are led to think in terms of some underlying theory relevant at energies above $\Lambda$. In that moment, it is this underlying theory which has to deal with the gauge hierarchy problem and explain the smallness \eqref{GHP} of its effective parameter $v$ in terms of its fundamental parameters.

\subsection{QCD as a prototype of a fundamental theory}

The QCD is defined by its Lagrangian as a non-Abelian gauge theory \cite{Yang:1954ek} of quarks. In the chiral limit (assuming only two flavors, $u$ and $d$, for the sake of simplicity) when $m_u=m_d=0$, the Lagrangian is free of mass parameters. At the classical level the theory is scale invariant. The only parameter in the Lagrangian is a dimensionless strong gauge coupling constant $g_\mathrm{s}$ which can be viewed as a unit of color charge carried by quarks and gluons. So far this does not seem to be helpful in generating masses.

The QCD is however of quantum nature and quantum fluctuations of QCD vacuum provide screening of the color dependent on a distance from which the color is measured. Actual calculation reveals that unlike the electric charge in QED, the color charge effectively vanishes at asymptotically short distances. This is expressed by the result of one-loop calculation of running color charge at asymptotically large momenta $q^2$
\begin{equation}\label{running_QCD}
\alpha_\mathrm{s}(q^2)\equiv\frac{g_{\mathrm{s}}^2(q^2)}{4\pi}\simeq\frac{4\pi}{(11-\tfrac{2}{3}n_f)\ln{q^2/\Lambda_{\mathrm{QCD}}^2}} \,,
\end{equation}
where $n_f$ is a number of QCD flavors. The dimensionful and theoretically arbitrary quantity $\Lambda_{\mathrm{QCD}}$ appears as a result of renormalization enforced by the presence of divergences within the perturbation calculation. It is fixed by experiment to be roughly $300\,\mathrm{MeV}$.

The formula \eqref{running_QCD} expresses the \emph{asymptotic freedom} \cite{Vanyashin:1965aa,Khriplovich:1969aa,'tHooft:1971fh,'tHooft:1971rn,Gross:1973id,Politzer:1973fx}, shared in general by non-Abelian gauge theories with not too many flavors. Within this type of theories, quantum fluctuations contributing to all various observables are damped at large momenta by the asymptotically vanishing coupling constant. Therefore any necessity of some higher cutoff scale $\Lambda$ disappears and the theory is said to be ultraviolet (UV) complete and it is a true candidate for being a fundamental theory.

Through the formula \eqref{running_QCD} the dimensionful quantity $\Lambda_{\mathrm{QCD}}$ fully determines the strength of the color coupling at short distances. It is the phenomenon called \emph{dimensional transmutation}. It tells us that at the quantum level the theory is \emph{not} scale invariant any more. In other words, it determines how does the attractive force between two color charges grow when one pulls them apart from each other. The scale $\Lambda_{\mathrm{QCD}}$ then refers to a distance where the force is getting large, $\alpha_\mathrm{s}\sim1$. Although we do not know how does the color charge behave at momenta smaller than $\Lambda_{\mathrm{QCD}}$, we are experiencing that it does not allow to isolate individual quarks and gluons, it rather keeps them confined in small bags of a characteristic radius given by $\Lambda_{\mathrm{QCD}}$. The low-energy spectrum of QCD consists of colorless bound states of quarks and gluons, the spectrum of hadrons.

The enhanced attraction among colored fields at low momenta $q^2\lesssim\Lambda_{\mathrm{QCD}}^2$ changes the structure of the vacuum. It triggers the formation of the chiral condensate
\begin{equation}\label{chiral_condensate}
\langle\bar u_R u_L+d_R d_L\rangle\ne0\,,
\end{equation}
which breaks the global chiral $\SU{2}_L\times\SU{2}_R$ symmetry of the Lagrangian down to its vector subgroup $\SU{2}_{L+R}$. Purely on the basis of this symmetry group pattern, the spectrum of hadrons is determined. Vast spectrum of hadrons occupies in principle all available multiplets of the unbroken $\SU{2}_{L+R}$. Out of them, three pions are massless (still considering the chiral limit) because they are the Nambu--Goldstone bosons of the broken chiral symmetry. Other hadrons are massive. Their masses are directly proportional to the only dimensionful quantity in the theory, $\Lambda_{\mathrm{QCD}}$. Various ratios of their masses are fully determined just as group theoretical factors.

To appreciate the spectacular power of QCD we can recapitulate: Just by assuming a symmetry structure of the Lagrangian and of the vacuum and by fixing a single mass scale $\Lambda_{\mathrm{QCD}}$ experimentally, the QCD reproduces rich low-energy spectrum of states, their masses and couplings. This is how a fundamental theory should be.

It is instrumental to notice that when coupled to the electroweak gauge dynamics, the QCD via its chiral condensate \eqref{chiral_condensate} spontaneously breaks the electroweak symmetry. There are however at least two evidences that the QCD cannot be the only source of the electroweak symmetry breaking. First, the magnitude of the QCD chiral condensate being of order of $\Lambda_{\mathrm{QCD}}$ is so small that it would generate roughly $2000$ times lighter electroweak bosons \cite{Farhi:1980xs}. Second the pions are observed as a part of particle spectrum instead of having been combined with the electroweak bosons.

\section{New physics beyond the Standard Model}

\subsection{Beyond-Standard models from naturalness}

The gauge hierarchy problem has stimulated an extensive use of the \emph{naturalness principle} when developing the more fundamental theory beyond the Standard Model. It claims that the most natural value of the scale of new physics is just one order of magnitude above the electroweak scale. Most of the beyond-Standard models deal with the problem by bringing the scale $\Lambda$ near above the electroweak scale $v$. This is an appealing solution as it predicts a variety of new phenomena within the scope of our current experimental abilities.

Many of the beyond-Standard models postulate some new symmetry which provides a \emph{cancelation} of quadratic divergences of the mass parameter $\mu$. The new symmetry connects the Higgs field with some other field whose mass enjoys protection of its own symmetry from acquiring the quadratically divergent corrections. Thereby the Higgs field mass parameter $\mu$ can enjoy the protection as well. There are three protective symmetries at the disposal in the Standard Model: the chiral symmetry for fermions, the gauge invariance for gauge bosons and the shift invariance for Nambu--Goldstone bosons. Correspondingly, it is the supersymmetry \cite{Ramond:1971gb,Neveu:1971rx,Gervais:1971ji,Golfand:1971iw,Volkov:1973ix,Wess:1974tw} connecting Higgs and chiral fermion fields resulting in the Minimal Supersymmetric Model (MSSM) \cite{Haber:1984rc,Nilles:1983ge,Chung:2003fi} and its extensions. Next, it is the five-dimensional gauge invariance connecting four-dimensional gauge boson fields with the Higgs field being the fifth gauge field component giving rise to the Gauge-Higgs unification models \cite{Pomarol:1998sd,Burdman:2002se,Csaki:2002ur}. Finally, it is the shift invariance which counts the Higgs field among pseudo-Nambu--Goldstone fields and it is used in the Little Higgs models \cite{ArkaniHamed:2002qx,Gregoire:2002ra,ArkaniHamed:2002qy,Low:2002ws,Kaplan:2003uc}.

The other big class of beyond-Standard models uses a \emph{suppression} of the scale of new physics. The models of Extra Dimensions \cite{ArkaniHamed:1998rs,Randall:1999ee,Randall:1999vf} explain the suppression factor geometrically from a nontrivial either curvature or topology of higher dimensional space-time. The Technicolor models \cite{Weinberg:1979bn,Susskind:1978ms} and their Extended-Technicolor descendants \cite{Holdom:1984sk,Akiba:1985rr,Yamawaki:1985zg,Appelquist:1986an,Appelquist:1987fc} imitate the excellent qualities of QCD and explain the suppression by means of asymptotic freedom. Because the dynamical electroweak symmetry breaking scenario of the (Extended-)Technicolor models is closely related to our approach, we will expose it in more details later in section \ref{Technicolor}.

Despite the extensive and promising effort dedicated to developing natural and \emph{realistic} beyond-Standard models, none of them really represents a systematic solution of the gauge hierarchy problem. They rather represent a postponement of the problem by one-step-higher scale which again demands to be stabilized with respect to the Planck scale for instance. Moreover, some introduce a plethora of new fields and corresponding number of parameters in order to reproduce what already Standard Model has done by means of much less effort.

It is easy to make a conclusion that there is need for new physics for both theoretical and phenomenological reasons. But it is difficult to say that it will be discovered soon as the Standard Model persists to serve us as an excellent description of high-energy physics. By agreement of the Standard Model predictions with collider experimental data at the electroweak scale we are in fact probing the field content of our description already up to higher energies than $2\,\mathrm{TeV}$ through quantum corrections. This may simply mean that the naturalness principle is not the way the Nature follows. Equally well, the Standard Model may be valid up to a very high scale, like the seesaw scale, the scale of the Grand Unified Theory (GUT), or even the Planck scale. In that case we should find a systematic solution how to not only parametrize but really how to explain the extremely large hierarchies within the quantum field theory. This may require completely new concepts upon which the model of the new physics should be built.

\subsection{Beyond-Standard models from analogy with\\ superconductivity}

The Standard Model and all of the beyond-Standard models mentioned above do introduce elementary fields primarily designed to be responsible for the electroweak symmetry breaking and for the electroweak gauge boson mass generation. The presence of ordinary fermions is irrelevant for achieving the electroweak symmetry breaking in these models.

In contrast, there are approaches, including ours, directly following the analogy with superconductors \cite{Nambu:1961tp,Bardeen:1957mv,Freundlich:1970kn}, which identify or introduce some \emph{dynamics} acting among usual fermions, quarks and leptons, whose primary purpose is to generate their masses. The electroweak symmetry breaking then comes automatically. The value of the electroweak scale $v$ is then roughly given by the values of the heaviest fermion masses $m_f$. Apparently the top-quark mass contributes significantly.

We present here a non-exhaustive list of the models (we present the author, the year and the used dynamics):
\begin{itemize}
    \item \cite{Hosek:1982cz} Ho\v{s}ek, (1982), $\U{1}$ new gauge dynamics
    \item \cite{Kimura:1984zv} Kimura and Munakata, (1984), four-fermion interaction
    \item \cite{Nambu:1988mr} Nambu (1988), not specified
    \item \cite{Miransky:1988xi,Bardeen:1989ds} Miransky, Tanabashi, Yamawaki (1989); Bardeen, Hill, Lindner (1989), \\ \vphantom{\cite{Miransky:1988xi,Bardeen:1989ds} }four-fermion interaction\footnote{ These models are formulated only for a top-quark mass generation. To date it was practically the only mass relevant for the electroweak symmetry breaking. These models can be understood as simplifications relevant for the models of dynamical generation of all quark and lepton masses. }
    \item \cite{Nagoshi:1990wk} Nagoshi, Nakanishi and Tanaka (1990), $\SU{3}$ flavor gauge dynamics
    \item \cite{Cvetic:1992xn} Cveti\v{c} (1992), flavor gauge dynamics
    \item \cite{Gribov:1994jy} Gribov (1994), U(1) weak hypercharge gauge dynamics
    \item \cite{Bashford:2003yg} Bashford (2003)
    \item \cite{Brauner:2004kg,Benes:2008ir} Ho\v{s}ek, Brauner (2004); Bene\v{s}, Brauner, Smetana (2008), Yukawa dynamics
    \item \cite{Wetterich:2006ii,Schwindt:2008gj} Wetterich (2006);  Schwindt and Wetterich (2008), Chiral tensor dynamics
    \item \cite{Hosek:2009ys} Ho\v{s}ek (2009), $\SU{3}$ flavor gauge dynamics
\end{itemize}
Although these models stand rather solitary over the history of the electroweak symmetry breaking and often without referring one to the other, they apparently still gain some attention till today.

These models are characterized by the coupling constant $h$ and by the scale $\Lambda$. Typically $\Lambda$ is very big in order to provide sufficiently large contributions of fermion masses to the electroweak scale, and constrained not to exceed the Planck scale $\Lambda<\Lambda_\mathrm{Planck}$. Therefore these models have to deal with enormous hierarchy $m_f\sim v\ll\Lambda$. In principle it is conceivable that the gauge hierarchy is achieved by some critical scaling \cite{Miransky:1996pd,Braun:2010qs}. The critical scaling however represents mere reparametrization of the gauge hierarchy problem in terms of an extreme vicinity of the coupling parameter $h$ to its critical value $h_{\mathrm{crit}}$,
\begin{equation}
\frac{v}{\Lambda}\ll1\quad\longrightarrow\quad|h-h_{\mathrm{crit}}|\ll1 \,.
\end{equation}
A true explanation of the gauge hierarchy problem should rely on a principle which can provide tiny but not infinitesimal super-criticality of the coupling parameter.

\section{This thesis}

Our ultimate ambition is to construct a fundamental theory of elementary fermion masses in a similar sense as the QCD is the fundamental theory of hadrons and their masses. We postulate a dynamics acting among usual fermions whose primary purpose is to generate their masses. The fermion masses then break the electroweak symmetry.

Operationally, the underlying dynamics generates electroweak symmetry breaking fermion self-energies $\Sigma(p^2)$, in general, complex momentum-dependent matrices. They form the electroweak symmetry breaking parts of the full fermion propagators
\begin{equation}\label{inverse_fermion_propagator}
S^{-1}(p)=\slashed{p}-\Sigma(p^2)P_L-\Sigma^\dag(p^2)P_R \,,
\end{equation}
where $P_{L,R}\equiv\tfrac{1}{2}(1\mp\gamma_5)$ are the chiral projectors. Of course in \eqref{inverse_fermion_propagator}, the renormalization of wave function should be taken into account as well. However, because our main point is to study the spontaneous chiral symmetry breaking triggered by finite chirality-changing self-energies $\Sigma(p^2)$, we do not consider the wave function renormalization in our work in order to keep our point clear.
The poles in the propagators define the fermion masses as solutions of the equation
\begin{equation}\label{m_from_Sigma}
\det\big[p^2-\Sigma^\dag(p^2)\Sigma(p^2)\big]=0 \,.
\end{equation}
In the models which we deal with in our work, the knowledge of the fermion self-energies is completely essential, not only for reproducing the fermion mass spectrum. The self-energies of quarks and leptons determine masses of electroweak gauge bosons. They also determine the Yukawa couplings of various composite (pseudo-)Nambu--Goldstone bosons to their constituent fermions.

The self-energies are the solutions of Schwinger--Dyson equations which are however often beyond our ability to solve. Therefore in practical calculations we often resort to various approximations. The simplest and the most feasible approach consists of approximating the underlying dynamics by a four-fermion interaction. Then the momentum-dependent self-energies are approximated by a momentum-independent fermion condensates being solution of much simpler gap equations.

Our program starts by obligatory investigation of the capability of the fermion self-energies to reproduce correct values of $W$ and $Z$ boson masses. Because the heaviest known fermion is the top-quark it is natural to formulate a simplified model where the contributions of lighter fermions are simply neglected. This leads to formulation of the top-quark condensation model \cite{Miransky:1988xi,Bardeen:1989ds,Cvetic:1997eb}. In section \ref{top_quark_condensation_model} we summarize important results of the top-quark condensation approach. We will remind that the top-quark alone is simply too light to saturate the electroweak scale completely. Furthermore it predicts too heavy Higgs boson.

The models of dynamical fermion mass generation which take into account a generation of neutrino masses via the seesaw mechanism have potential to correctly reproduce the value of the electroweak scale. If the neutrino Dirac mass is large enough, then the neutrino condensate is strong enough to complement the electroweak scale \cite{Martin:1991xw,Cvetic:1992ps,Antusch:2002xh}. In that case the electroweak scale is saturated by both the mass of top-quark $m_t$ and the Dirac mass of neutrinos $m_D$
\begin{equation}
v\sim m_t,\,m_D \,.
\end{equation}
We call this scenario the \emph{top-quark and neutrino condensation scenario} and it is subject of chapter \ref{top_and_nu_condensation} based on our original work \cite{Smetana:2013hm}. We reformulate the top-quark and neutrino condensation scenario in terms of a two composite Higgs doublet model and confront it with the most recent experimental data. Mainly, we reproduce the $125\,\mathrm{GeV}$ particle observed at the Large Hadron Collider (LHC) \cite{125Higgs:2012gk,125Higgs:2012gu}.

The following two chapters are fully dedicated to presentation of our models of dynamics underlying the mass generation of quarks and leptons.

In chapter \ref{part_strong_Yukawa_dynamics} we present our early attempt based on the strong Yukawa dynamics mediated by two complex scalar doublets \cite{Benes:2008ir}. We are able to formulate the interconnected Schwinger--Dyson equations for both fermion and scalar self-energies and to solve them simultaneously. For that we resort to reasonable approximations and use numerical methods.

In the last chapter~\ref{gfd} we formulate a model of strong $\SU{3}_\F$ flavor gauge dynamics relying mainly on our achievements published in \cite{Smetana:2011tj}. The flavor gauge model has an ambition of being the fundamental theory of fermion masses and of the electroweak symmetry breaking: It is asymptotically free and it is characterized by a single gauge coupling parameter. Like in the QCD, within the flavor gauge model all masses are, at least in principle, calculable factors of the fundamental scale of the flavor gauge dynamics $\Lambda_\F$. Unlike the QCD, the flavor gauge dynamics does not confine otherwise it could not describe the quarks and leptons as the flavored asymptotic states. For that purpose the flavor gauge dynamics is formulated as a strongly coupled \emph{chiral gauge theory} which we believe can self-break rather than confine.

The idea of replacing the Higgs sector of the Standard Model by asymptotically free flavor gauge dynamics offers several versions of the model. They are distinguished by a right-handed neutrino content. We will analyze them and bring arguments why we favor just one of them. The preferred version of the model is non-minimal in the sense that it is defined by richer flavor structure of right-handed neutrinos, one sextet and four anti-triplets. The presence of the right-handed neutrino flavor sextet is crucial for two reasons. First, its condensation at the seesaw scale provides a huge right-handed neutrino Majorana mass, thus it naturally and dynamically forms a basis for the seesaw mechanism. Second, at this high scale the flavor sextet condensation breaks completely the flavor gauge symmetry. The resulting massive flavor gauge bosons mediate an attraction among the electroweakly charged fermions. At some lower scale, the attraction results in a dynamical generation of their masses and the electroweak symmetry breaking.

The fermion self-energies, which are the basic elements of the models studied in this thesis, are the consequences of the strong coupling. Reliable methods to calculate them are not available. Therefore we must rely on approximative methods or just to assume their existence. Then the phenomenological outcomes of the models can be made just on the level of qualitative estimates and conjectures. That is why we left these partial results to appendices. In the first appendix~\ref{approx_methods_for_SDE} we summarize the approximate methods of solving the Schwinger--Dyson equations for the fermion self-energies which we have used in the course of elaborating the two models. In the second appendix~\ref{NG_vertices} we discuss the composite Nambu--Goldstone boson coupling to their constituent fermions, which are determined in terms of the fermion self-energies. If the spontaneously broken symmetry is gauged then these couplings determine the gauge boson masses. In the last appendix~\ref{Fierz} we present the derivation of the Fierz identities which are helpful when modeling the underlying dynamics by four-fermion interactions.

\chapter{Dynamical electroweak symmetry breaking}
%\pagenumbering{arabic}
%\input{M.tex}

\label{dynamical_models}

Our work belongs to the category of beyond-Standard models where the electroweak symmetry is broken dynamically. It is closely related to the main representatives of this category, to the (Extended-)Technicolor models and to the Top-quark Condensation models. That is why we dedicate the whole chapter to present main ideas upon which they are built.

\section{Technicolor}
\label{Technicolor}

The most natural and the oldest solution to the gauge hierarchy problem follows the scenario already realized in Nature: The scale of QCD and consequent masses of hadrons are arbitrarily small relative to the Planck scale. The clue lies in the asymptotic freedom of a non-Abelian gauge theory, whose example is the QCD. The scale $\Lambda$ of an asymptotically free theory, above which the gauge coupling parameter $g$ tends to the zero UV fixed point as \eqref{running_QCD}
\begin{equation}\label{running}
g^2(q^2>\Lambda^2)\propto\frac{1}{\ln {\frac{q^2}{\Lambda^2}}} \,,
\end{equation}
represents a natural cutoff for quantum corrections. The quantum corrections are kept under control and their effect decreases above the cutoff $\Lambda$ without any reference to the huge Planck scale. Turning the logic around and keeping the reference to the Planck scale, because of the tiny value of the coupling constant at the Planck scale $g(\Lambda_{\mathrm{Pl}}^2)\ll1$ the huge exponential suppression of scales is in work
\begin{equation}\label{QCD_critical_scaling}
\Lambda^2\sim\ \e^{-1/g^2(\Lambda_{\mathrm{Pl}}^2)}\Lambda_{\mathrm{Pl}}^2 \,.
\end{equation}

\subsubsection{The technicolor}

Technicolor (TC) models \cite{Weinberg:1979bn,Susskind:1978ms} assume the existence of a new strong asymptotically free gauge technicolor dynamics $\SU{N_\mathrm{TC}}$ characterized by its scale $\Lambda_{\mathrm{TC}}$ and the coupling constant $g_\mathrm{TC}$. The technicolor dynamics is acting among new fermions, the techniquark fields $Q(x)$. The techniquarks are electroweakly charged similarly to the usual quarks so that their hard mass terms are forbidden. Instead, the Lagrangian possesses even larger chiral symmetry $G_{\chi \mathrm{TC}}$ out of which $\SU{2}_L\times\U{1}_Y$ electroweak subgroup is gauged. Due to the electroweak dynamics the chiral symmetry $G_{\chi \mathrm{TC}}$ is only approximate except for the gauged subgroup. The electroweak symmetry breaking is achieved by a generation of techniquark self-energy $\Sigma_Q(p^2)$ spontaneously breaking the chiral symmetry
\begin{equation}
G_{\chi \mathrm{TC}}\ \longrightarrow\ H_{\chi \mathrm{TC}}\ \supset\ \U{1}_{\mathrm{em}}\big(\times\SU{2}_{\mathrm{custodial}}\big)
\,.
\end{equation}
It gives rise to a set of composite pseudo-Nambu--Goldstone fields and three true composite Nambu--Goldstone fields, the technipions $\pi_{\mathrm{TC}}^a(x)$, corresponding to the electroweak symmetry breaking. The three technipions couple to the corresponding broken electroweak currents
\begin{subequations}
\begin{eqnarray}
\bra{0}j^{\mu}_\pm(x)\ket{\pi_{\mathrm{TC}}^\pm(q)} & = & -\im q^\mu F_\pm\e^{-\im x\cdot q} \,, \\
\bra{0}j^{\mu}_0(x)\ket{\pi_{\mathrm{TC}}^0(q)} & = & -\im q^\mu F_0\e^{-\im x\cdot q}\,.
\end{eqnarray}
\end{subequations}
The conservation of the electromagnetic charge $\U{1}_{\mathrm{em}}$ implies $F_+=F_-$. The custodial symmetry $\SU{2}_{\mathrm{custodial}}$ protects the relation $F_\pm=F_0\equiv F_\mathrm{TC}$.

The technipions $\pi_{\mathrm{TC}}^a(x)$ have the same quantum numbers as the QCD pions $\pi_{\mathrm{QCD}}^a(x)$, thus they mix. The mixing results in the fact that the observed pion states have an admixture of technicolor states, roughly given by \cite{Farhi:1980xs,Chanowitz:1985hj}
\begin{eqnarray}\label{pion_state}
|\pi^a\rangle & = & \cos{\theta_\pi}|\pi_{\mathrm{QCD}}^a\rangle +\sin{\theta_\pi}|\pi_{\mathrm{TC}}^a\rangle \,,
\end{eqnarray}
where $\tan{\theta_\pi}\equiv\frac{f_\pi}{F_\mathrm{TC}}$, where $f_\pi$ is the QCD pion decay constant. The orthogonal states are then the `would-be' Nambu--Goldstone states
\begin{eqnarray}\label{EWpion_state}
|\pi_{W,Z}^a\rangle & = & -\sin{\theta_\pi}|\pi_{\mathrm{QCD}}^a\rangle +\cos{\theta_\pi}|\pi_{\mathrm{TC}}^a\rangle \,,
\end{eqnarray}
which are combined with the electroweak gauge bosons to give them masses
\begin{eqnarray}
M_W & = & \frac{1}{2}g \sqrt{F_\mathrm{TC}^2+f_{\pi}^2} \,, \\
M_Z & = & \frac{1}{2}\sqrt{g^2+{g'}^2}\sqrt{F_\mathrm{TC}^2+f_{\pi}^2} \,.
\end{eqnarray}
The QCD pion decay constant $f_\pi\simeq93\,\mathrm{MeV}$ is very small compared to the electroweak scale $v$, thus it contributes negligibly to the $W$ and $Z$ boson masses. In order to saturate their values, the technipion decay constant has to be
\begin{equation}
F_{\mathrm{TC}}\simeq v\doteq 246\,\mathrm{GeV} \,.
\end{equation}
Consequently, as $F_{\mathrm{TC}}\gg f_\pi$, the mixing \eqref{pion_state} and \eqref{EWpion_state} of QCD pions and technipions is truly negligible.

The technicolor dynamics can be defined simply by scaling-up of the QCD dynamics. It defines the technicolor scale in terms of the QCD scale $\Lambda_\mathrm{QCD}\sim300\,\mathrm{MeV}$ as
\begin{equation}
\Lambda_{\mathrm{TC}}\sim
\sqrt{\frac{3}{N_{\mathrm{TC}}}}\frac{F_{\mathrm{TC}}}{f_\pi}\Lambda_{\mathrm{QCD}}\simeq
\sqrt{\frac{3}{N_{\mathrm{TC}}}}794\,\mathrm{GeV} \,.
\end{equation}
The scale $\Lambda_{\mathrm{TC}}$ sets the typical magnitude of masses of various bound-states, technimesons and technibaryons. These states saturate the unitarity of scattering amplitudes of the longitudinally polarized electroweak gauge bosons. In the formula for $F_{\mathrm{TC}}$, \eqref{technipoin_F} below, the integral is dominated by low momenta, therefore
\begin{equation}\label{Sigma0_sim_LambdaTC}
\Sigma_Q(0)\sim\Lambda_{\mathrm{TC}} \,.
\end{equation}

The technicolor is a beautiful idea and, like QCD, it has the potential to be the fundamental theory. In a natural way, it stabilizes the electroweak scale with respect to the Planck scale. However, it turns on its own inner scale tensions when applied to the fermion mass generation. In the rest of this section, we will briefly demonstrate that the technicolor dynamics cannot be simply QCD-like otherwise a conflict between magnitude of fermion masses and sufficient suppression of flavor-changing neutral currents (FCNC) pops up.

\subsubsection{The extended technicolor}

\begin{figure*}[t]
\begin{center}
\includegraphics[width=0.7\textwidth]{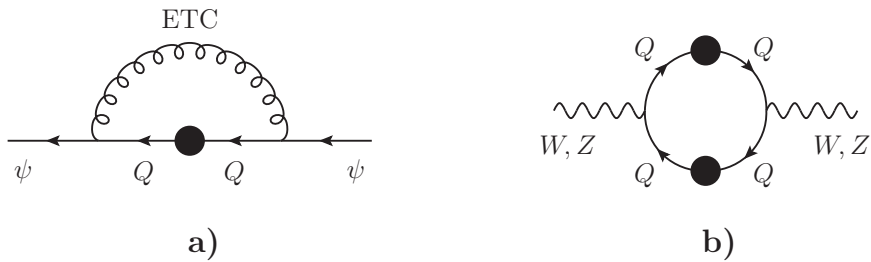}
\caption[Usual fermion masses and $W$ and $Z$ boson masses within the ETC models]{ a) The loop expression for usual fermion mass \eqref{ETC_fermion_mass_eq}. The formation of the techniquark condensate is communicated to the usual fermion sector via the exchange of massive ETC gauge bosons. b) The loop expression for masses of $W$ and $Z$. This is the pictorial representation of the Pagels--Stokar formula \eqref{technipoin_F}.  }
\label{ETC_fermion_mass}
\end{center}
\end{figure*}

Even though the technicolor breaks the electroweak symmetry, it does not provide the mass generation for usual fermions, because it is not coupled to them. An additional interaction which connects usual fermions with the techniquarks is needed. This is provided by embedding $N_\F=3$ flavor copies (families) of ordinary fermions together with corresponding techniquarks in a representation of the extended-technicolor (ETC) gauge dynamics $\SU{N_{\mathrm{ETC}}}$ \cite{Dimopoulos:1979es,Eichten:1979ah}, where $N_{\mathrm{ETC}}\equiv N_\F+N_{\mathrm{TC}}$.

Masses of flavors differ, hence the ETC gauge symmetry is not a property of the fermion mass spectrum. It has to be spontaneously broken down to its vector-like confining technicolor subgroup according to
\begin{equation}\label{ETCtoTC}
\SU{N_{\mathrm{ETC}}}\rightarrow\SU{N_{\mathrm{TC}}} \,.
\end{equation}
The ETC gauge symmetry has to be \emph{chiral} otherwise it could not be broken \cite{Eichten:1979ah}. At this point the technicolor construction looses its beauty as it revives the issue of chiral gauge symmetry breaking. Some mechanism of the spontaneous ETC symmetry breaking has to be added. Clearly, invoking condensing scalars is a possibility but it would negate the main original idea of dynamical symmetry breaking.

In a better way, it is sometimes assumed that the chiral gauge ETC dynamics is strongly coupled and that it \emph{self-breaks}. That is why our model of flavor gauge dynamics described in chapter~\ref{gfd} is relevant for ETC models. The flavor gauge model can be rephrased as ``the ETC model without TC'' as its symmetry breaking pattern of completely broken flavor symmetry can be understood in terms of \eqref{ETCtoTC} as $\SU{N_{\mathrm{ETC}}=N_\F}\rightarrow\emptyset$.

After the spontaneous symmetry breaking at some scale $\Lambda_{\mathrm{ETC}}>\Lambda_{\mathrm{TC}}$, corresponding ETC gauge bosons acquire masses of order $\Lambda_{\mathrm{ETC}}$. Because the exchange of the massive ETC gauge bosons provides the FCNC \cite{Eichten:1979ah} among ordinary fermions, their masses have to be adequately large, setting $\Lambda_{\mathrm{ETC}}>10^3\,\mathrm{TeV}$. In order to keep the idea of suppression of hierarchy problem, the ETC dynamics is canonically assumed to be again asymptotically free above $\Lambda_{\mathrm{ETC}}$.

Ordinary fermion masses are generated via the diagram Fig.~\ref{ETC_fermion_mass}a) being the pictorial representation of the formula in the Landau gauge
\begin{equation}\label{ETC_fermion_mass_eq}
\Sigma_\psi(q^2) = -3\im g_{\mathrm{ETC}}^2\int\frac{\d^4k}{(2\pi)^4}\frac{1}{(k-q)^2-\Lambda_{\mathrm{ETC}}^2}\frac{\Sigma_Q(k^2)}{k^2-\Sigma_{Q}^2(k^2)} \,.
\end{equation}
From this equation the fermion mass can be approximated after the Wick rotation as
\begin{equation}\label{ordinary_fermion_mass}
m_\psi\approx\Sigma_\psi(0) = \frac{g_{\mathrm{ETC}}^2}{\Lambda_{\mathrm{ETC}}^2}\frac{3}{8\pi^2} \int_{0}^{\Lambda_{\mathrm{ETC}}^2}\d x\frac{x\Sigma_Q(x)}{x+\Sigma_{Q}^2(x)} \,.
\end{equation}
The technipion decay constant $F_{\mathrm{TC}}$ is given by the Pagels--Stokar formula \cite{Pagels:1979hd}
%\footnote{According to more recent studies \cite{Benes:2012hz} we use the corrected formula with a factor $\tfrac{1}{2}$ rather than original $\tfrac{1}{4}$ in the derivative term.}
pictorially represented by the diagram in Fig.~\ref{ETC_fermion_mass}b) for which we derived the approximate formula in appendix \eqref{simplified_Wick_PS},
\begin{equation}\label{technipoin_F}
F_{\mathrm{TC}}^2 = \frac{N_\mathrm{TC}}{8\pi^2}\int_{0}^{\Lambda_{\mathrm{ETC}}^2}\d x\,x\frac{\Sigma_Q(x)^{2}}{\big(x+\Sigma_{Q}^2(x)\big)^2} \,.
\end{equation}
The scale $\Lambda_{\mathrm{ETC}}$ is used to cutoff the integrals referring to the asymptotic freedom of the ETC dynamics.
It is clear that the techniquark self-energy $\Sigma_Q(p^2)$ is instrumental for the electroweak symmetry breaking and mass generation.
At first sight the two integrals differ by their sensitivity to the high-momentum details of $\Sigma_Q(p^2)$. Naively the integral for $m_\psi$ in \eqref{ordinary_fermion_mass} is quadratically sensitive, while the integral for $F_\mathrm{TC}$ in \eqref{technipoin_F} is only logarithmically sensitive. That is why $\Sigma_Q(0)\sim\Lambda_\mathrm{TC}$ \eqref{Sigma0_sim_LambdaTC} in order to get the value of $F_\mathrm{TC}$ appropriate for the electroweak boson masses.

\subsubsection{Technicolor governed by infrared fixed point}

It is therefore of key importance to know the behavior of $\Sigma_Q(p^2)$ in the region of momenta $\Lambda_{\mathrm{TC}}^2<p^2<\Lambda_{\mathrm{ETC}}^2$. If dynamically generated, the fermion self-energy has a general Euclidean high-momentum dependence (to be found, e.g., in \cite{Yamawaki:1996vr})
\begin{equation}\label{asymptotics}
\Sigma_Q(p^2\gg\Lambda_{\mathrm{TC}}^2)\ \sim\ \frac{\Lambda_{\mathrm{TC}}^3}{p^2}\exp\Big[\int_{0}^t\gamma_m(t')\d t'\Big]\,,\ \ \ t\equiv\tfrac{1}{2}\ln\tfrac{p^2}{\Lambda_{\mathrm{TC}}^2} \,,
\end{equation}
where $\gamma_m(t)$ is the anomalous dimension of the mass operator. The dependence of the anomalous dimension on $t$ is provided merely through the $t$ dependence of the coupling constant $g_\mathrm{TC}(t)$
\begin{equation}
\gamma_m(t)=\gamma_m\big(g_\mathrm{TC}(t)\big) \,.
\end{equation}
Thus, through the anomalous dimension $\gamma_m(t)$, the evolution of the coupling constant determines the high-momentum dependence of the $\Sigma_Q(p^2)$.

\begin{figure*}[t]
\begin{center}
\includegraphics[width=1.0\textwidth]{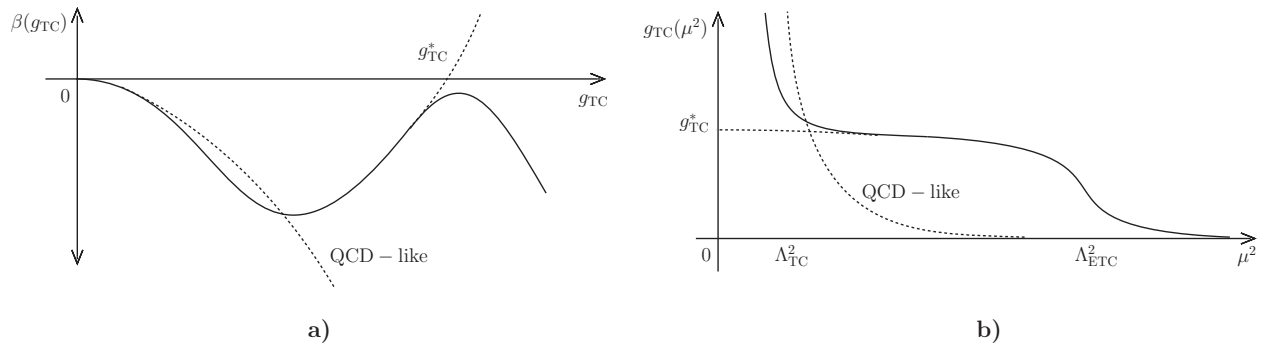}
\caption[Cartoon of walking TC dynamics]{  Cartoon a) of $\beta$-function $\beta(g_\mathrm{TC})$ and b) of running coupling constant $g_\mathrm{TC}$ in the ETC theory governed by an infrared fixed point $g_{\mathrm{TC}}^*$ (solid line) in comparison with a QCD-like theory (dotted line). }
\label{walking}
\end{center}
\end{figure*}

If the technicolor were QCD-like, i.e., its coupling constant were running according to \eqref{running} all the way above $\Lambda_{\mathrm{TC}}$, the $\Sigma_Q(p^2)$ would be damping quickly above $\Lambda_{\mathrm{TC}}$, for illustration see Fig.~\ref{walking} (dotted line). While this would not affect too much the value of $F_\mathrm{TC}$, it would, together with FCNC constrain $\Lambda_{\mathrm{ETC}}>10^3\,\mathrm{TeV}$, allow only unacceptably small ordinary fermion masses. Therefore the high-momentum enhancement of the $\Sigma_Q(p^2)$ is needed in order to enhance the ordinary fermion masses while keeping the proper value of $F_\mathrm{TC}$. This can be achieved if over the range $\Lambda_{\mathrm{TC}}^2<p^2<\Lambda_{\mathrm{ETC}}^2$ the technicolor dynamics is governed by a non-trivial fixed point $g^{*}_\mathrm{TC}$. It is the idea pioneered in \cite{Holdom:1981rm} and further elaborated in \cite{Holdom:1984sk,Akiba:1985rr,Yamawaki:1985zg,Appelquist:1986an,Appelquist:1987fc}.

Within the range $\Lambda_{\mathrm{TC}}^2<p^2<\Lambda_{\mathrm{ETC}}^2$, the QCD-like technicolor is compared to the fixed-point-governed technicolor using the formula \eqref{asymptotics}:
\begin{center}
\begin{tabular}{rlll}
QCD-like: & \quad\hspace{-0.02\textwidth} $g_{\mathrm{TC}}^2(t)\sim t^{-1}$ \,, & \quad\hspace{-0.02\textwidth} $\gamma_m(t)\sim g_{\mathrm{TC}}^2(t)\sim t^{-1}$ \,, & \quad\hspace{-0.02\textwidth} $\Sigma_Q(p^2)\sim\frac{\Lambda_{\mathrm{TC}}^3}{p^2}\Big(\ln\tfrac{p^2}{\Lambda_{\mathrm{TC}}^2}\Big)^a$ \ , \\
fixed\ point: & \quad\hspace{-0.02\textwidth} $g_\mathrm{TC}(t)\sim g^{*}_\mathrm{TC}$ \,, & \quad \hspace{-0.02\textwidth} $\gamma_m(t)\sim\gamma_m\big(g^{*}_\mathrm{TC}\big)=\mathrm{const.}$ \,,     & \quad\hspace{-0.02\textwidth} $\Sigma_Q(p^2)\sim\frac{\Lambda_{\mathrm{TC}}^3}{p^2}\Big(\frac{p^2}{\Lambda_{\mathrm{TC}}^2}\Big)^{\gamma_m/2}$ \ ,
\end{tabular}
\end{center}
where the exponent $a$ is given by details of the technicolor dynamics and its order of magnitude is $|a|\sim1$.

The most popular improved ETC models are based on the walking technicolor dynamics \cite{Holdom:1981rm}. The walking technicolor dynamics is asymptotically free above $\Lambda_{\mathrm{ETC}}$ but, contrary to the QCD, below $\Lambda_{\mathrm{ETC}}$ it is attracted by an approximate infrared fixed point. That slows down the evolution of the coupling constant. Around the scale $\Lambda_{\mathrm{TC}}$ the dynamics finally confines. Over the relevant range of momenta $\Lambda_{\mathrm{TC}}^2<p^2<\Lambda_{\mathrm{ETC}}^2$ the anomalous dimension is approximately constant, $\gamma_m(t)\sim\gamma_m\big(g^{*}_\mathrm{TC}\big)$. Such gauge theories do exist \cite{Dietrich:2006cm,Catterall:2007yx}, for illustration see Fig.\ref{walking} (solid line). The comfortable splitting of $\Lambda_{\mathrm{TC}}$ and $\Lambda_{\mathrm{ETC}}$ is achieved, e.g., within the Minimal Walking Technicolor model \cite{Foadi:2007ue}. For completeness notice that the same enhancement lifts up masses of pseudo-Nambu--Goldstone bosons, which would be otherwise unacceptably light as well.

%
%\section{Reparametrization of gauge hierarchy problem}
%
%There is a class of beyond-Standard models which do not belong to neither category because they simply do not aim to solve the gauge hierarchy problem. They rather reparametrize it into a language of dynamical electroweak symmetry breaking. The most known example is the top-quark condensation model exposed in the following section.

\section{Top-quark condensation models}\label{top_quark_condensation_model}

The mechanism of the electroweak symmetry breaking manifests itself by the fact that both the top-quark mass $m_t$ and the Higgs boson mass $M_h$ are of the same order as the electroweak scale $v$, which is the scale for masses of the weak bosons, i.e., $M_W=g v/2$ and $M_{Z}=\sqrt{g^2+{g'}^2}v/2$. A simple contemplation leads to the suspicion that both the longitudinal components of the weak vector bosons and the Higgs boson are in fact bound states containing top-quark. This is the idea of Top-quark Condensation models introduced in \cite{Hosek:1985jr,Miransky:1988xi,Bardeen:1989ds}.

The top-quark condensation is underlain by some new dynamics characterized by some scale $\Lambda_t$ and coupling constant $g$. It can be for instance some asymptotically free gauge dynamics whose symmetry gets broken below $\Lambda_t$ providing masses of the order $\sim\Lambda_t$ for its gauge bosons. Then the infrared effects of the new dynamics can be parametrized by four-fermion interactions induced by the exchange of the massive gauge bosons, thus weighted by the scale $\Lambda_t$. For the top-quark, the four-fermion interaction upon the Fierz identity \eqref{Dirac_Fierz_a} can be written as
\begin{equation}\label{4f_induced}
  (\bar{t}_R\gamma^\mu t_R)\frac{g^2}{\Lambda_{t}^2}(\bar{t}_L\gamma_\mu t_L)\equiv(\bar{t}_L t_R)\frac{g^2}{\Lambda_{t}^2}(\bar{t}_R t_L) \,.
\end{equation}
This interaction is part of the $\SU{2}_L\times\U{1}_Y$ invariant Lagrangian ${\cal L}_t$, which defines the Top-quark Condensation model. In a suggestive form, the four-top-quark interaction \eqref{4f_induced} can be introduced as
\begin{equation}\label{tttt_interaction}
{\cal L}_t\supset \frac{\kappa_t}{\Lambda_{t}^2}(\bar{t}_Lt_R)(\bar{t}_Rt_L) \equiv \frac{\kappa_t}{4\Lambda_{t}^2}\Big[(\bar{t}t)^2-(\bar{t}\gamma_5t)^2\Big] \,,
\end{equation}
where we have introduced the coupling parameter $\kappa_t\propto g^2$. This is in the analogy with the old proposal of Nambu and Jona-Lasinio (NJL) \cite{Nambu:1961tp} formulated as the model of a spontaneous chiral $\U{1}$ global symmetry breaking.

\subsubsection{The top-quark mass}

The four-top-quark interaction generates a top-quark mass $m_t$ which can be approximated by a solution of the gap equation
\begin{eqnarray}
m_t & = & \im \frac{\kappa_t}{\Lambda_{t}^2}\int_{\Lambda_t}\frac{\d^4k}{(2\pi)^4}\frac{m_t}{k^2-m_{t}^2} \label{top_quark_gap_eq} \\
& \stackrel{\mathrm{Wick}}{=} & \frac{\kappa_t}{16\pi^2\Lambda_{t}^2} \int_{0}^{\Lambda_{t}^2}\d x\frac{xm_t}{x+m_{t}^2} \,.
\end{eqnarray}
The cutoff $\Lambda_t$ expresses the fact that the underlying dynamics is assumed to become weaker above that scale.

The gap equation \eqref{top_quark_gap_eq} should be compared with the ETC formula \eqref{ordinary_fermion_mass}. The ETC formula is mere \emph{expression} for the top-quark mass $m_f\sim\tfrac{g_{\mathrm{ETC}}^2}{\Lambda_{\mathrm{ETC}}^2}\langle\bar QQ\rangle$ in terms of the techniquark condensate $\langle\bar QQ\rangle$ given by the integral. On the other hand, \eqref{top_quark_gap_eq} is the algebraic transcendent \emph{equation} for the top-quark mass
\begin{equation}\label{gap_eq}
m_t = \frac{\kappa_t m_t}{16\pi^2}\left(1-\frac{m_{t}^2}{\Lambda_{t}^2}\ln\frac{\Lambda_{t}^2+m_{t}^2}{m_{t}^2}\right) \,.
\end{equation}
Within this approach, the top-quark mass generation is a non-perturbative phenomenon which exhibits several characteristic features common to various dynamical mass generation models:
\begin{itemize}
\item The trivial solution $m_t=0$ exists.
\item There is a critical value of the coupling parameter $\kappa_t\equiv\kappa_{t,\mathrm{crit.}}=16\pi^2$ below which only the trivial solution exists and above which also the non-trivial solution $m_t\ne0$ exists.
\item The non-trivial solution for the top-quark mass $m_t\ne0$ is proportional to the only mass parameter in the model $\Lambda_t$ through the numerical scaling factor $f(\kappa_t)$ as
    \begin{equation}\label{tqc_critical_scaling}
    m_t=f(\kappa_t)\Lambda_t\,.
    \end{equation}
\item At the critical value of the coupling constant the scaling factor is a non-analytic function $f(\kappa_t)\simeq\sqrt{1-\frac{\kappa_{t,\mathrm{crit.}}}{\kappa_t}}$ providing arbitrarily strong amplification $m_t\ll\Lambda_t$ when $\kappa_t\rightarrow\kappa_{t,\mathrm{crit.}}$.
\end{itemize}

\subsubsection{The electroweak scale}

The original NJL model was immediately applied by authors as a model of pions \cite{Nambu:1961fr} arising as pseudo-Nambu--Goldstone bosons from spontaneous breaking of chiral isospin symmetry explicitly violated by nucleon bare masses. These were found as massless poles in fermion scattering amplitudes in the pseudo-scalar channel. In the scalar channel a massive pole was found. It corresponds to a composite $\sigma$-particle with mass twice as big as the fermion mass. In the top-quark condensation model, the situation is analogous, but the spontaneously broken symmetry is the gauge electroweak symmetry and the Nambu--Goldstone modes are eaten by the $W$ and $Z$ bosons. The NJL treatment of the model leads to the composite Higgs boson with the mass $M_h=2m_t$.

Once the top-quark mass is generated the electroweak symmetry is broken and the electroweak bosons acquire masses, $M_{W}^2=g^2F_{W}^2/4$ and $M_{Z}^2=(g^2+{g'}^2)F_{Z}^2/4$. The dimensionful factor $F_{Z}$ is given by the Pagels--Stokar formula, for which we derived approximate formula in appendix \eqref{simplified_Wick_PS},
%\begin{equation}\label{topquark_F}
%F_{\mathrm{Z}}^2 = \frac{N_\mathrm{C}}{8\pi^2}\int_{0}^{\Lambda_{t}^2}\d x\,x\frac{\Sigma_t(x)^{2}-\tfrac{x}{2}\tfrac{\d}{\d x}\Sigma_t(x)^{2}}{\big(x+\Sigma_{t}^2(x)\big)^2} \,,
%\end{equation}
\begin{equation}\label{topquark_F}
F_{\mathrm{Z}}^2 = \frac{N_\mathrm{C}}{8\pi^2}\int_{0}^{\Lambda_{t}^2}\d x\,x\frac{\Sigma^{2}_t(x)}{\big(x+\Sigma_{t}^2(x)\big)^2} \,,
\end{equation}
which is analogous to \eqref{technipoin_F}. For $F_{W}$ there is a similar formula. Contrary to the TC models, here the loop integral is formed by top-quark propagators. It expresses the assumption that it is the top-quark condensation which stands behind the electroweak symmetry breaking. Following the same level of approximation as in \eqref{top_quark_gap_eq} we use the cutoff $\Lambda_t$ and $\Sigma_t(p^2)=m_t$ and get the formulae
\begin{subequations}\label{mu_W_mu_Z}
\begin{eqnarray}
F_{Z}^2 & = & \frac{N_c}{8\pi^2}m_{t}^2\left[\ln\frac{\Lambda_{t}^2}{m_{t}^2}-\frac{1}{2}\right] \label{mu_Z} \,, \\
F_{W}^2 & = & \frac{N_c}{8\pi^2}m_{t}^2\left[\ln\frac{\Lambda_{t}^2}{m_{t}^2}+1\right] \label{mu_W}\,.
\end{eqnarray}
\end{subequations}
notice that the bigger the $\Lambda_t$ is, the closer to unity is the $\rho$-parameter, $\rho\approx1$, which is defined as usual
\begin{equation}\label{rho}
\rho\equiv\frac{M_{W}^2}{M_{Z}^2\cos^2\theta_W}=\frac{F_{W}^2}{F_{Z}^2} \,.
\end{equation}
This actually simulates the effect of the custodial symmetry which is not the property of the Lagrangian $\mathcal{L}_t$ \eqref{tttt_interaction}.
The equations \eqref{gap_eq} and \eqref{mu_Z} (neglecting the difference between $F_W$ and $F_Z$) have to be satisfied simultaneously. By setting $F_Z=v\doteq246\,\mathrm{GeV}$ and $m_t\doteq172\,\mathrm{GeV}$ we fix the condensation scale $\Lambda_t$ from the equation \eqref{mu_Z},
\begin{equation}
\Lambda_t\sim10^{14}\,\mathrm{GeV} \,.
\end{equation}
The equation \eqref{gap_eq} then fixes the coupling constant $\kappa_t$
\begin{equation}\label{fine_tun_kappat}
\kappa_t\approx(1+m_{t}^2/\Lambda_{t}^2)\kappa_{t,\mathrm{crit.}} \,.
\end{equation}
This result means enormous fine-tuning of the coupling constant.

The lesson is the following. By assuming some dynamics characterized by the scale $\Lambda_t$ leading effectively to the four-fermion interaction \eqref{tttt_interaction}, the electroweak symmetry breaking can in principle take place. It turns out however that \emph{the hierarchy problem is not solved}, because numerically $\Lambda_t$ is pushed very high in order to get the correct values of $M_{W,Z}$ and $m_t$. The extreme fine tuning now reappears in the value of the coupling parameter $\kappa_t$ \eqref{fine_tun_kappat}.

\subsubsection{Including the QCD effect}

The situation becomes even much more inconvenient when more sophisticated approximations are used. In particular, the effect of QCD dynamics to the top-quark masses is significant and it should be taken into account. It leads to the momentum dependent self-energy $\Sigma_t(p^2)$ which is monotonically decreasing and given by the well-known formula \cite{Lane:1974us,Politzer:1976tv}
\begin{equation}\label{t_sigma_p}
\Sigma_t(p^2)\simeq m_t\left[\frac{\alpha_\mathrm{s}(p^2)}{\alpha_\mathrm{s}(m_{t}^2)}\right]^{4/7}\,,
\end{equation}
where $\alpha_\mathrm{s}(q^2)=g_{\mathrm{s}}^2(q^2)/4\pi$ is the QCD effective charge
\begin{equation}\label{qcd_charge}
\alpha_\mathrm{s}^{-1}(q^2)=\alpha_\mathrm{s}^{-1}(m_{t}^2)+\frac{7}{4\pi}\ln\frac{q^2}{m_{t}^2}\,.
\end{equation}
Using the improved top-quark self-energy \eqref{t_sigma_p} within the safely simplified Pagels--Stokar formula \eqref{topquark_F}
\begin{equation}\label{simplified_PS_top}
F_{Z}^2=\frac{3}{8\pi^2}\int^{\Lambda_{t}^2}_{m_{t}^2}\frac{\d k^2}{k^2}\Sigma_{t}^2(k^2)
\end{equation}
we get the improvement of the relation \eqref{mu_Z}
\begin{equation}\label{muZ_in_top}
F_{Z}^2=\frac{3m_{t}^2}{2\pi\alpha_\mathrm{s}(m_{t}^2)}\left(1-\left[\frac{\alpha_\mathrm{s}(\Lambda_{t}^2)}{\alpha_\mathrm{s}(m_{t}^2)}\right]^{1/7}\right) \,.
\end{equation}

\subsubsection{The Higgs boson mass}

The electroweak symmetry breaking driven by the top-quark condensation is accompanied by a composite Higgs boson. %According to \cite{Gribov:1994jy} the Higgs boson follows from the existence of the degenerate scalar states with opposite parity at short distance.
The Higgs boson mass can be estimated by the formula \cite{Gribov:1994jy,Blumhofer:1995zw} similar to \eqref{simplified_PS_top}
\begin{equation}\label{simplified_MH}
M_{h}^2=\frac{3}{2\pi^2F_{Z}^2}\int^{\Lambda_{t}^2}_{m_{t}^2}\frac{\d k^2}{k^2}\Sigma_{t}^4(k^2) \,.
\end{equation}
It reflects the momentum dependence of the $\Sigma_t(p^2)$ and thus it is an improvement of the NJL relation $M_h=2m_t$. Using \eqref{t_sigma_p}, it leads to the expression for the Higgs boson mass
\begin{equation}\label{MH_in_top}
M_{h}^2=\frac{2m_{t}^4}{3\pi\alpha_\mathrm{s}(m_{t}^2)F_{Z}^2}\left(1-\left[\frac{\alpha_\mathrm{s}(\Lambda_{t}^2)}{\alpha_\mathrm{s}(m_{t}^2)}\right]^{9/7}\right) \,.
\end{equation}

The two equations \eqref{muZ_in_top} and \eqref{MH_in_top} relate masses $M_{W,Z}$, $M_h$ and $m_t$. The single free parameter is the scale of the new dynamics, $\Lambda_t$. As Tab.~\ref{Tab_Higgs_top_mass} shows such a simple top-quark condensation scenario is strictly incompatible with the experimental values of $M_{W,Z}$, $m_t$, and $M_h$. There are two main problems. First, the top-quark does not weight enough to saturate the electroweak scale. Second, even if it were heavy enough, the mass of the Higgs boson as the top-quark bound-state would come out larger than the top-quark mass, which is in contradiction with the observation of the Higgs mass $M_h\simeq125\,\mathrm{GeV}$.

\begin{table}[t]
\begin{center}
\begin{tabular}{ll|ccc}
 & $\Lambda_t\ [\mathrm{GeV}]$ & $\ \ 10^{10}\ $ & $\ \ 10^{19}\ $ & $\ \ 10^{42}\ $  \\
\hline
\hline
a)\ \ & $m_t\ [\mathrm{GeV}]$ & $305$ & $252$ & $215$  \\
 & $M_h\ [\mathrm{GeV}]$ & $450$ & $338$ & $257$  \\
\hline
\hline
b)\ \ & $F_Z\ [\mathrm{GeV}]$ & $140$ & $168$ & $196$  \\
 & $M_h\ [\mathrm{GeV}]$ & $258$ & $231$ & $205$  \\
\hline
\hline
\end{tabular}
\end{center}
\caption[Top-quark and Higgs masses and the electroweak scale within the Top-quark Condensation model]{\small For different values of the top-quark condensation scale $\Lambda_t=10^{10}\,\mathrm{GeV}$ (axion), $\Lambda_t=10^{19}\,\mathrm{GeV}$ (Planck), $\Lambda_t=10^{42}\,\mathrm{GeV}$ (Landau), and taking $\alpha_\mathrm{s}(m_{t}^2)\simeq0.11$, using the equations \eqref{muZ_in_top} and \eqref{MH_in_top}, we evaluate a) the $M_h$ and $m_t$ masses while keeping the electroweak gauge boson masses at their experimental values, i.e., $F_Z\doteq246\,\mathrm{GeV}$, b) the $M_h$ and $F_Z$ while keeping the top-quark mass at its experimental value, i.e., $m_t\doteq172\,\mathrm{GeV}$. }
\label{Tab_Higgs_top_mass}
\end{table}

\chapter{Top-quark and neutrino condensation}
%\pagenumbering{arabic}
%\input{3x.tex}

\label{top_and_nu_condensation}

\subsubsection{Motivation}

The key idea of our work is to achieve the electroweak symmetry breaking by generating masses of the known fermions, quarks and leptons.
Prior to any attempts to build a fundamental theory of fermion masses, it is necessary to check whether the fermion masses can actually saturate the electroweak scale.

In general, the dynamical lepton and quark mass generation and the electroweak gauge boson mass generation are accompanied by a set of composite particles. They saturate the unitarity of all amplitudes. These particles are bound-states, whose fields are originally not present in the Lagrangian. They are composites of elementary fields. The minimal scenario is that a single parity-even composite scalar is formed. This composite state then acts in direct analogy with the Standard Model Higgs boson. Further on, we will use the adjective ``Higgs'' in a broader sense for referring to any such boson connected with the electroweak symmetry breaking phenomenon.

The electroweakly charged fermions occupy only two types of weak isospin representations. The left-handed fermions $\psi_L$ are the weak isospin doublets and the right-handed fermions $\psi_R$ are the weak isospin singlets. Their condensates $\langle\bar\psi_R\psi_L\rangle$ responsible for their Dirac masses connect left-handed with right-handed fermion fields. Therefore it is meaningful to assume that the condensates are in fact the vacuum expectation values of neutral components of weak isospin doublet structures of fermion bilinears. The fermion bilinears then can be used as interpolating fields for the composite Higgs doublet fields. Therefore for each Dirac condensate, i.e., for each fermion Dirac mass, there is effectively one composite Higgs doublet below the condensation scale. Therefore our work relies on the assumption that for any model of dynamical fermion mass generation, the appropriate description below a condensation scale deals with a \emph{multitude of composite Higgs doublets}. %This is in correspondence with the fact that the custodial symmetry is operative in this type of models as it is well documented within the top-quark condensation model by \eqref{mu_W_mu_Z}.

Within this context, out of the electroweakly charged fermions, the left-handed neutrinos are special as they can in principle form the condensate of the Majorana type $\langle\bar\nu_{L}\nu_{L}^\C\rangle$. According to the same argumentation, this condensate belongs to the weak isospin triplet structure of lepton bilinears. Therefore the composite Higgs triplet would be the appropriate description. This condensate however corresponds to tiny neutrino masses, and thus we envisage its negligible role in the electroweak symmetry breaking mechanism. Therefore we do not treat it in any more detail further in our work.

All condensates, i.e., vacuum expectation values of all Higgs doublets, contribute to the value of the electroweak scale. The magnitude of individual contributions is proportional to the mass of the corresponding fermion. Out of charged fermions, only the top-quark is heavy enough to contribute significantly to the electroweak scale by its condensate $\langle\bar t_Rt_L\rangle$. Therefore in order to assess the suitability of the models of dynamical fermion mass generation, it should be sufficient to resort to the top-quark condensation model. The effective description then deals only with a single composite Higgs doublet.

However as it was summarized in section~\ref{top_quark_condensation_model}, although the actual calculation gives a correct order of magnitude of masses, there are two essential failures of the top-quark-alone condensation model when confronted with experiment. First, the top-quark is observed to be too light to saturate the electroweak scale $v$. Keeping the condensation scale below the Planck scale, the top-quark condensation can provide only at most 68\,\% of the $W$ and $Z$ boson masses as follows from Tab.~\ref{Tab_Higgs_top_mass}. Second, the composite Higgs boson is predicted to be too heavy, in all available calculations $M_h>m_t$. This prediction was ruled out already before the actual measurement of $125\,\mathrm{GeV}$ particle at the LHC \cite{125Higgs:2012gk,125Higgs:2012gu}. For a review see \cite{Cvetic:1997eb}.

%In this work we study the way relevant for the all matter scenario which significantly improve the situation.

%First, we need to identify yet another source potent to saturate the value of the electroweak scale beside the top-quark condensate $\langle\bar t_Rt_L\rangle$. The remaining observed quarks and charged leptons bring no improvement in this respect as they are way too light and contribute by truly negligible amount to the electroweak scale. At this point one could be easily seduced to invoke some new yet unobserved fermions with high enough mass, like, for instance, fourth-generation fermions or techniquarks. This however may not be necessary.

However, among the known fermions, there is potentially yet another source of the electroweak symmetry breaking naturally present in the form of the neutrino Dirac mass $m_D$ provided that the seesaw mechanism is at work. This of course amounts to assuming the existence of right-handed neutrinos, which are almost mandatory these days. If the neutrino Dirac mass is of the order of the electroweak scale, $m_D\sim v$, then the neutrino condensate $\langle\bar\nu_R\nu_L\rangle$ is strong enough to complement the electroweak scale \cite{Martin:1991xw,Cvetic:1992ps,Antusch:2002xh}. The electroweak scale is therefore linked to the mass of top-quark $m_t$ and to the Dirac mass of neutrinos $m_D$
\begin{equation}
v\sim m_t,\,m_D \,.
\end{equation}
We call this scenario as the \emph{top-quark and neutrino condensation scenario}.

Once we have identified two main fermion sources of the electroweak symmetry breaking we can resort to the effective description using correspondingly two composite Higgs doublets. In this chapter we will check the suitability of this improved scenario.

\subsubsection{The model}

The idea of the top-quark and neutrino condensation was addressed already in the past. First, Martin \cite{Martin:1991xw} investigated the model in which the idea was implemented in the simplest possible way. He invoked a factorization assumption on four-fermion interactions which resulted in the low-energy description with only single Higgs doublet. He reached the correct value of the top-quark mass, but from present day perspective, the model suffers from exhibiting too heavy Higgs boson particle, in the same way as the original top-quark-alone condensation models \cite{Miransky:1988xi,Bardeen:1989ds}. Ten years later, the issue was addressed again by Antusch et al.~\cite{Antusch:2002xh}. They confirmed the usefulness of the incorporation of the neutrino condensation for obtaining the correct value of the top-quark mass. Further, they suggested that the two-Higgs doublet low-energy description is worthwhile to study in more detail. Another ten years later we are addressing the idea once again \cite{Smetana:2013hm} and confront it with the new experimental evidence of the $125\,\mathrm{GeV}$ boson excitation.

The existence of right-handed neutrinos is extremely well motivated. An addition of already three of them to the Standard Model \cite{Canetti:2012vf} can simultaneously explain all three puzzles of dark matter, neutrino oscillations and baryon asymmetry of the Universe. Generally, the number of right-handed neutrino types participating in the seesaw mechanism is not constrained by any upper limit, see references \cite{Ellis:2007wz,Heeck:2012fw}. As claimed there, higher number of the order ${\cal O}(100)$ is even well motivated within some string constructions. Large number of right-handed neutrinos ${\cal O}(100)$ has also an improving effect on the standard thermal leptogenesis \cite{Eisele:2007ws}. Being of order ${\cal O}(10-100)$ it can serve as the reason for large lepton mixing angles \cite{Feldstein:2011ck}. Our motivation is to simulate the low-energy effects to the electroweak symmetry breaking of the flavor gauge dynamics studied in detail in chapter~\ref{gfd}, which is assumed to underlie both the top-quark and neutrino condensations. The consistence of the flavor gauge model requires the existence of right-handed neutrinos in flavor \emph{triplets}. Therefore we study the dependence of our results on the number of right-handed neutrino flavor triplets $N$. We denote them as
\begin{equation}\label{nuRs}
  \nu_{Rs}\,, \quad s=1,\dots,N \,.
\end{equation}

The neutrino mass spectrum is not known. For our analysis of the electroweak symmetry breaking the precise form of the neutrino mass spectrum does not play an essential role. Therefore we just simulate it by the most simple choice for the neutrino mass matrix. It is characterized by flavor diagonal Dirac masses $m_{Ds}$, by a common right-handed Majorana mass $M_R$ and by the number $N$ of right-handed neutrino flavor triplets. By this simplification we can control the order of magnitude of active neutrino masses but do not reproduce any details of the neutrino physics. This would require specifying the underlying dynamics.

\section{Saturation of the electroweak scale}\label{Saturation_of_the_electroweak_scale}

First of all let us show how we get rid of the necessity of having the top-quark condensation scale $\Lambda_t$ enormously big. For comparison see section~\ref{top_quark_condensation_model}. Thanks to the neutrino condensation, the condensation scale $\Lambda_t$ can be considerably lower than the Planck scale. Up to that scale the top-quark self-energy $\Sigma_t(p^2)$ is described by the Ansatz obtained from combining \eqref{t_sigma_p} and \eqref{qcd_charge}
\begin{equation}\label{tt_sigma_p}
\Sigma_t(p^2)\simeq m_t\left[1+\frac{7}{4\pi}\alpha_\mathrm{s}(m_{t}^2)\ln\frac{p^2}{m_{t}^2}\right]^{-4/7}\theta\big(\Lambda_{t}^2-p^2\big)\,.
\end{equation}

Because neutrinos do not feel the QCD dynamics we adopt the Ansatz for their Dirac self-energy to be constant up to the condensation scale $\Lambda_\nu$ contrary to the top-quark case. At the condensation scale we cut-off the Dirac self-energy. Our Ansatz for neutrino Dirac self-energy is
\begin{equation}\label{nuD_sigma_p}
\Sigma_{Ds}(p^2)\simeq m_{Ds}\vartheta\big(\Lambda_{\nu}^2-p^2\big) \,,
\end{equation}
where $s=1,\dots,N$ labels $N$ right-handed neutrino triplets. In order to have a seesaw mechanism we assume an Ansatz for the complete neutrino self-energy in the Nambu--Gorkov formalism in the form
\begin{equation}\label{nu_sigma_p}
\Sigma_\nu(p^2)=\beginm{cccc} 0 & \Sigma_{D1}(p^2) & \cdots & \Sigma_{DN}(p^2) \\ \Sigma_{D1}(p^2) & {\cal M}_R &  &  \\ \vdots &  & \ddots &  \\ \Sigma_{DN}(p^2) &  &  & {\cal M}_R \endm \,.
\end{equation}
Here we adopt the simplest seesaw pattern of the neutrino self-energy in order to simplify the calculation significantly, as we announced in advance. Therefore in \eqref{nuD_sigma_p} and \eqref{nu_sigma_p} the $3\times3$ blocks are assumed to be diagonal
\begin{subequations}\label{numM_sigma_p}
\begin{eqnarray}
m_{Ds} & = & \beginm{ccc} m_{Ds1} &  &  \\  & \hspace{-0.2cm}m_{Ds2} &  \\  &  & \hspace{-0.2cm}m_{Ds3} \endm \,, \label{Dirac_nu_mass}\\
{\cal M}_R & = & \beginm{ccc} M_R &  &  \\  & M_R &  \\  &  & M_R \endm \,,
\end{eqnarray}
\end{subequations}
where all mass parameters are the real and positive numbers.

Important restriction on the neutrino condensation scale $\Lambda_\nu$ comes from the decoupling theorem \cite{Appelquist:1974tg}. If the Majorana mass $M_R$ were bigger than $\Lambda_\nu$, then the correspondingly heavy right-handed neutrinos would decouple from the dynamics before they would manage to condense with the left-handed neutrinos. Therefore, like in \cite{Martin:1991xw}, we assume
\begin{equation}\label{non_decoupling_condition}
\Lambda_\nu>M_R\,,
\end{equation}
and call it the \emph{non-decoupling condition}.

Explanation of values of the neutrino masses
\begin{equation}
m_\nu\approx0.2\,\mathrm{eV}
\end{equation}
by the seesaw mechanism with the neutrino Dirac masses $m_D\sim v$ forces us to assume the value of the right-handed neutrino Majorana mass to be
\begin{equation}
M_R\sim10^{14}\,\mathrm{GeV}\,.
\end{equation}

Our goal is to saturate the electroweak mass scale $v$
\begin{equation}\label{ews_scale_t_nu}
v^2=v_{t}^2+v_{\nu}^2 \,
\end{equation}
by contributions from the top-quark $v_t$ and from the neutrinos $v_\nu$.
In order to evaluate $v_t$ and $v_\nu$ we use the formulae analogous to the Pagels--Stokar formula \eqref{topquark_F}
\begin{subequations}
\begin{eqnarray}
v_{t}^2 & = & \frac{N_c}{8\pi^2}\int^{\Lambda_{t}^2}_{m_{t}^2}k^2\d k^2\frac{\Sigma_{t}^2(k^2)}{k^2+\Sigma_{t}^2(k^2)}\,, \label{t_to_muZ} \\
v_{\nu}^2 & = & \frac{1}{8\pi^2}\int^{\Lambda_{\nu}^2}_{0}k^2\d k^2 \Tr\Big(\Sigma_\nu(k^2)\big\{\Sigma_\nu(k^2),{\cal P}_{\nu_L}\big\} \big[k^2+\Sigma_{\nu}^2(k^2)\big]^{-1}{\cal P}_{\nu_L}\big[k^2+\Sigma_{\nu}^2(k^2)\big]^{-1}\Big) \,. \hspace{1.2cm}\label{nu_to_muZ} %\nonumber
\end{eqnarray}
\end{subequations}
The formulae are derived under the assumption that the momentum dependence of the self-energies is mild and thus the derivative terms, typical for the original Pagels--Stokar formula \cite{Pagels:1979hd,Benes:2012hz}, are negligible. The derivation of the formula for $v_t$ \eqref{t_to_muZ} is performed in appendix \eqref{simplified_Wick_PS}. The formula for $v_{\nu}$ \eqref{nu_to_muZ} was derived in \cite{Benes:2012hz} and it is written in the Nambu--Gorkov formalism.
%\begin{equation}
%\Sigma_\nu(p^2)=\beginm{cc}
%    \Sigma_{L}(p^2) & \Sigma_{D}(p^2) \\ \Sigma_{D}^\T(p^2) & \Sigma_{R}(p^2) \endm \,.
%\end{equation}
The projector
\begin{equation}
{\cal P}_{\nu_L}=\beginm{cc} \openone & 0 \\ 0 & 0 \endm
\end{equation}
reflects the fact that only the left-handed neutrinos are electroweakly charged, for instance, the anti-commutator misses $M_R$ completely,
\begin{equation}
\big\{\Sigma_\nu,{\cal P}_{\nu_L}\big\}=\beginm{cc} 0 & \Sigma_{D} \\ \Sigma_{D}^\T & 0 \endm \,.
\end{equation}
%Because of the enormously mild momentum dependence of self-energies and because of the very high cut-off given by the condensation scales, the derivative terms in the Pagels--Stokar formula \eqref{topquark_F} contribute by negligible amount.

\begin{figure*}[t]
\begin{center}
\includegraphics[width=0.7\textwidth]{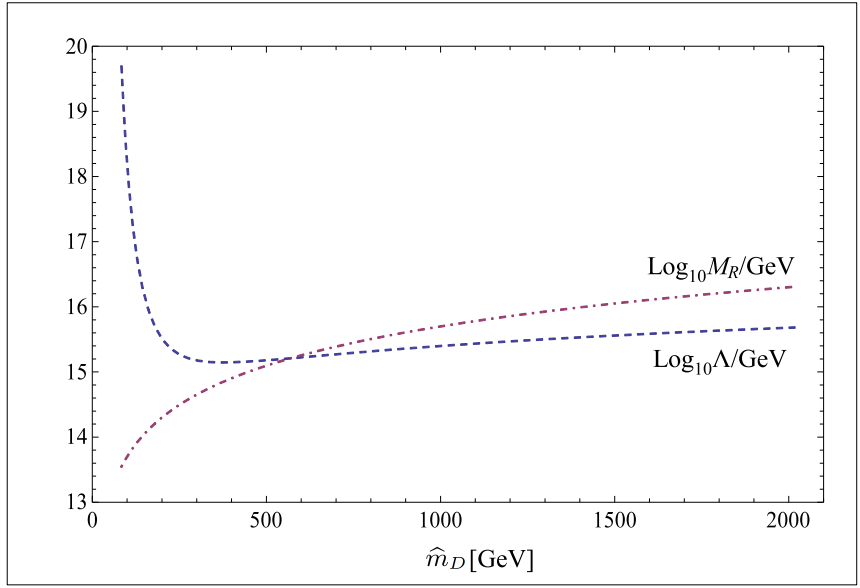}
\caption[Top-quark and neutrino condensation scale $\Lambda$ and right-handed neutrino Majorana mass $M_R$ depending on the neutrino Dirac mass $\widehat{m}_D$]{ \small We plot the dependence of the right-handed neutrino Majorana mass $M_R$ and the common condensation scale $\Lambda\equiv\Lambda_t=\Lambda_\nu$ on the parameter $\widehat{m}_{D}$ in the logarithmic scale. The parameter $\widehat{m}_{D}$ is the length of the vector of neutrino Dirac masses, $(m_{D1},\dots,m_{DN})$, in the family degenerate case \eqref{Snu}. }
\label{plot_Lambda_Sm}
\end{center}
\end{figure*}

While the top-quark integral in \eqref{t_to_muZ} is calculated numerically due to the complicated Ansatz \eqref{tt_sigma_p}, the neutrino integral \eqref{nu_to_muZ} can be calculated analytically because of the simple Ansatz \eqref{nuD_sigma_p}, \eqref{nu_sigma_p} and \eqref{numM_sigma_p}. The result in the limit of $m_{Dsi}\ll M_{R}$ is
\begin{equation}
v_{\nu}^2=\frac{1}{8\pi^2}\sum_{i}\sum_{s}m_{Dsi}^2\ln\frac{M_{R}^2+\Lambda_{\nu}^2}{M_{R}^2} \,,
\end{equation}
where $s=1,\dots,N$ labels the right-handed neutrino generational triplets and $i=1,2,3$ labels the three generations.
In fact if we want to describe the neutrino masses $m_{\nu i}$, the mass parameters $m_{Dsi}$ and $M_R$ are not independent and are related by the seesaw formula
\begin{equation}
m_{\nu i}=\frac{\sum_{s}m_{Dsi}^2}{M_{R}} \,.
\end{equation}
Therefore the only free parameters are actually the condensation scale $\Lambda_\nu$ and three sums of squares of neutrino Dirac masses $\sum_s m_{si}^2$. If we further assume, for the sake of simplicity, that the three light neutrinos are degenerate with a common mass $m_{\nu}$ then we end up with only two free parameters
\begin{equation}
v_{\nu}^2=\frac{3}{8\pi^2}\widehat{m}_{D}^2\ln\frac{\big(\widehat{m}_{D}^2/m_{\nu}\big)^2+\Lambda_{\nu}^2}{\big(\widehat{m}_{D}^2/m_{\nu}\big)^2} \,,
\end{equation}
where
\begin{equation}\label{Snu}
\widehat{m}_{D}^2=\sum_{s=1}^N m_{Dsi}^2 \,.
\end{equation}

Under the assumption that behind the condensation of both top-quark and neutrinos is the same dynamics, e.g., the flavor gauge dynamics described in chapter \ref{gfd}, we can set $\Lambda\equiv\Lambda_t=\Lambda_\nu$ for the sake of simplicity. In Fig.~\ref{plot_Lambda_Sm} we plot our result in the form of the dependence of $\Lambda$ and $M_R$ on the neutrino Dirac mass parameter $\widehat{m}_{D}$. Important observations are the following:
\begin{itemize}
\item The solution for $\Lambda$ has the minimum at $\widehat{m}_{D}^{\mathrm{min}}\doteq375\,\mathrm{GeV}$, see Fig.~\ref{plot_Lambda_Sm}. It means that, under assumption of common condensation scale, below certain value $\Lambda^{\mathrm{min}}\doteq1.4\times10^{15}\,\mathrm{GeV}$ there is no way how to saturate the electroweak scale by the neutrino together with the top-quark contribution.
\item At $\widehat{m}_{D}=569\,\mathrm{GeV}$ the right-handed neutrino Majorana mass becomes greater than the condensation scale $\Lambda$. This breaks the non-decoupling condition \eqref{non_decoupling_condition} and invalidates the participation of right-handed neutrinos on the Dirac type condensation with left-handed neutrinos.
\item Phenomenologically acceptable values are obtained within rather narrow window of $\widehat{m}_{D}\in(103\,\mathrm{GeV},569\,\mathrm{GeV})$. The upper value corresponds to the breaking of the non-decoupling condition \eqref{non_decoupling_condition} and the lower value corresponds to $\Lambda=10^{18}\,\mathrm{GeV}$ which already approaches the Planck scale.
\end{itemize}

This analysis shows the potential of the scenario of the electroweak symmetry breaking by the dynamical fermion mass generation, provided that the seesaw mechanism is at work. The analysis suggests that the condensation scale lies somewhere between the GUT scale and the Planck scale.

\section{Model of top-quark and neutrino condensation}

We have not addressed the issue of bound states and their mass spectrum yet. This we will make in this section within a particular semi-realistic model based on a four-fermion interaction. We will resort to the effective description using composite Higgs boson fields and calculate the mass spectrum with the help of the renormalization group apparatus. The effective description is valid below the condensation scale $\Lambda$, common to both the top-quark and the neutrino condensations.

The goal of the model is to describe the two electroweak scale contributions from top-quark and from neutrinos \eqref{ews_scale_t_nu} by means of just two composite Higgs doublets. We parametrize the condensation of 3 left-handed neutrinos with $3N$ right-handed neutrinos by just a single Higgs doublet. This is far from being a general case where $3N$ neutrino composite Higgs doublets are expected to arise, one for each individual Dirac mass $m_{Dsi}$, which we have introduced in \eqref{nuD_sigma_p}, \eqref{nu_sigma_p} and \eqref{Dirac_nu_mass}. Nevertheless, this simplification is sufficient to make conclusions about the electroweak symmetry breaking. On the other hand, it is too rough to make conclusions about the neutrino spectrum other than just an overall order of magnitude estimate.

The simplification of a single neutrino Higgs doublet is achieved by taking a factorization assumption about the four-neutrino interaction. Instead of working with a general four-neutrino interaction ${\cal L}_{4\nu}^\mathrm{general}$ induced by some underlying gauge dynamics like in \eqref{4f_induced}, e.g., by the flavor gauge dynamics, we work with its special case ${\cal L}_{4\nu}^\mathrm{factor}$ following from the factorization assumption
\begin{equation}
{\cal L}_{4\nu}^\mathrm{general}=- G_{ss'}(\bar \nu_{Rs}\nu_L)(\bar \nu_L\nu_{Rs'})
\quad\longrightarrow\quad
{\cal L}_{4\nu}^\mathrm{factor}=- G_\nu(\sum_s\bar \nu_{Rs}\nu_L)(\sum_{s'}\bar \nu_L\nu_{Rs'}) \,.
\end{equation}
Similar type of a factorization assumption is just what Martin made in \cite{Martin:1991xw} in order to describe both the top-quark and neutrino condensation by a single composite Higgs doublet. In our model it allows us to evaluate $v_t$ and $v_\nu$ by means of the condensates
\begin{subequations}\label{tnucondensation}
\begin{eqnarray}
  v_t   &\propto& \langle\bar{t}_R t_L\rangle+\hc \,,\\
  v_\nu &\propto& \langle\sum_s\bar{\nu}_{Rs} \nu_L\rangle+\hc\,.
\end{eqnarray}
\end{subequations}

Contrary to the previous section, the simplified four-neutrino dynamics generates all neutrino Dirac masses degenerate, commonly denoted just by $m_D$. The equation \eqref{Snu} then becomes
\begin{equation}
\widehat{m}_{D}^2=N m_{D}^2 \,.
\end{equation}
The dynamics of the model will determine the value of the neutrino Dirac mass $m_D$ via the corresponding effective Yukawa coupling constant. Therefore instead of $\widehat{m}_{D}$ there will be the number of right-handed neutrinos $N$ as a free parameter.

\subsection{Underlying Lagrangian}

For the purpose of our analysis we define our simplified model by the four-fermion interaction
\begin{equation}\label{4f}
{\cal L}_{4f} = - G_t(\bar t_Rq_L)(\bar q_Lt_R) - G_\nu(\sum_s\bar \nu_{Rs}\ell_L)(\sum_{s'}\bar \ell_L\nu_{Rs'}) \,,
\end{equation}
where $q_L=\beginm{c}t_L \\ b_L\endm$ and $\ell_L=\beginm{c}\nu_L \\ e_L\endm$. The Lagrangian is designed to provide us just with the condensation \eqref{tnucondensation}.

Only the third generation of quarks participates in the interaction. The fields $t_R$ and $q_L$ are color triplets. On the other hand, because we suppose that all neutrino Dirac masses are of the order of the electroweak scale, then within the simplified model we are letting all three generations of leptons to participate in the interaction. Therefore the fields $\nu_R$ and $\ell_L$ are flavor triplets. Additionally, the left fields are weak isospin doublets. All three types of indices are suppressed. The explicitly written index $s=1,\dots,N$ labels $N$ right-handed neutrino flavor triplets. By this simplified dynamics we are going to generate only the top-quark and neutrino masses.

If the underlying dynamics is such that the four-fermion interactions follow from an exchange of neutral and colorless gauge bosons, then there are only these two types of effective terms relevant for the top-quark and neutrino condensation. No mixing terms like $\propto(\bar t_Rq_L)(\bar \ell_L\nu_{R})$ appear. There could appear also various four-fermion interactions of other leptons and quarks, but we neglect them here as they play the negligible role in the electroweak symmetry breaking.

In order to have the seesaw mechanism in the model, we introduce the right-handed neutrino Majorana mass term. We take it degenerate and diagonal for the same sake of simplicity
\begin{equation}\label{L_M_R}
{\cal L}_{M_R} = -\frac{1}{2}M_R\bar{\nu}_{Rs}\nu_{Rs}^\C+\hc \,,
\end{equation}
where $\nu^{\C}_{R}=C\bar{\nu}^{\T}_R$ and $C=\im\gamma_0\gamma_2$.
%We dare to accept these simplifications as our aim is not to reproduce the neutrino phenomenology exactly.

The model is defined by the Lagrangian
\begin{eqnarray}
{\cal L} & = & {\cal L}_{\mathrm{usual}}+{\cal L}_{\mathrm{model}} \,, \\
{\cal L}_{\mathrm{model}} & = & {\cal L}_{4f}+{\cal L}_{M_R} \,,
\end{eqnarray}
where ${\cal L}_{\mathrm{usual}}$ contains kinetic terms of all known fermions, their Standard Model gauge interactions and pure gauge boson terms.

Let us add a short comment here. Notice that in fact only a single linear combination of right-handed neutrino triplets couples to the left-handed lepton fields. Therefore we could have reformulated the whole program in terms of a new field $n_R=\sum_s\nu_{Rs}$ which would be the only field participating in the seesaw mechanism, see the eigenvalues of the resulting neutrino mass matrix \eqref{nu_mass_matrix}. This transition to another field basis would however shuffle the right-handed neutrino Majorana mass term \eqref{L_M_R}. Of course the physical results are independent of the chosen basis. By working in the basis of $\nu_{Rs}$ we keep a simple form of the  mass term \eqref{L_M_R} and refer directly to a field content of some underlying dynamics.

\subsection{Symmetries}

The Lagrangian ${\cal L}_{\mathrm{model}}$ has the well separated quark and lepton sectors. On the classical level, it is invariant under the global symmetry\footnote{The rest of standard fermions and their corresponding symmetries are of course present in the model in order to provide proper anomaly cancelation, but we do not treat them explicitly here as they do not participate in the symmetry breaking in our simplified analysis. Due to the factorization assumption the three generations of leptons exhibit a single common symmetry group.}
\begin{eqnarray}\label{Gmodel}
G_{\mathrm{model}} &=& \big[\SU{2}\times\U{1}^2\big]_q\times\big[\SU{2}\times\U{1}\big]_\ell \,. \label{doubled_symm}
\end{eqnarray}
Let us shortly comment the symmetry pattern \eqref{doubled_symm}. The quark sector of ${\cal L}_{\mathrm{model}}$ consists of two chiral quark multiplets $q_L$ and $t_R$. Each of them carries one $\U{1}$ phase while only $q_L$ is the $\SU{2}$ doublet. In the lepton sector, the situation is more complicated due to the number of $3N$ right-handed neutrinos. Instead of the $\U{3N}$ symmetry, however, the right-handed neutrino sector would possess a single common $\U{1}$ symmetry due to the factorization assumption, if it were not broken by the Majorana mass term \eqref{L_M_R}. Because of the presence of the Majorana mass term, the right-handed neutrino sector does not carry any symmetry. Therefore, in the lepton sector, there is only one $\U{1}$ subgroup carried by $\ell_L$ which is at the same time the $\SU{2}$ doublet.

One subgroup of $G_{\mathrm{model}}$ is the electroweak $\SU{2}_L\times\U{1}_Y$ gauge symmetry group. The electroweak interactions explicitly break the symmetry $G_{\mathrm{model}}$, so the symmetry of the full Lagrangian ${\cal L}$ is
\begin{equation}\label{EW_breaking}
G\ =\ \SU{2}_L\times\U{1}_Y\times\U{1}_{B}\times\U{1}_{X} \,,
\end{equation}
%The isospin-off-diagonal interactions break the un-gauged charged currents and tight the $\U{1}$ quantum numbers of left-handed doublet components together. The isospin-diagonal interactions leave the $\U{1}$ subgroups unaffected. That is why we end up with only three $\U{1}$ subgroups.
among which $\SU{2}_L\times\U{1}_Y$ are the weak isospin and weak hypercharge gauge symmetries, $\U{1}_B$ is the baryon number and $\U{1}_X$ is the axial symmetry
\begin{equation}\label{U1X}
\U{1}_X:\ \ (q_L,t_R,\ell_L,\nu_R)=(-1,0,1,0) \,.
\end{equation}
We are making this choice of $X$ charges in order to have $X(\bar t_Rq_L)=-1$ and $X(\bar \nu_R\ell_L)=+1$. It is this symmetry which prevents the top-quark and neutrino sectors from mixing.

On the quantum level, the group $\U{1}_{X}$ has an axial anomaly due to both the electroweak and the QCD dynamics. Additionally, the anomaly can be given by some new not specified dynamics underlying the four-fermion interaction \eqref{4f} like, e.g., the gauge flavor dynamics described in chapter \ref{gfd}. %Our ignorance about the anomaly stems from not specifying the complete structure of the model in order to be flexible in our analysis.
In the following, we will simply parameterize the strength of the anomaly by the mass parameter $\mu_{t\nu}$ introduced by hand into the effective Lagrangian.

The dynamically generated Dirac masses for top-quark and neutrinos break spontaneously the $G_{\mathrm{model}}$ symmetry \eqref{Gmodel} (the symmetry of the classical Lagrangian with the electroweak dynamics turned off) down to
\begin{equation}\label{Gmodel_H}
G_{\mathrm{model}}\ \stackrel{m_t,m_\nu}{\longrightarrow}\ \big[\U{1}^2\big]_q\times\big[\U{1}\big]_\ell \,.
\end{equation}
It would give rise to 6 massless Nambu--Goldstone bosons.

There are however effects of both the electroweak dynamics and of the axial anomaly which eliminate the 6 massless states completely.
The electroweak interactions change the spontaneous symmetry breaking pattern to
\begin{equation}\label{G_H}
G\ \stackrel{m_t,m_\nu}{\longrightarrow}\ \U{1}_\mathrm{em}\times\U{1}_{B} \,.
\end{equation}
Out of the 6 states, now only three are the true Nambu--Goldstone states, but they are eaten by the electroweak gauge bosons. The other two form a single charged pseudo-Nambu--Goldstone particle whose mass results from the explicit breaking by the electroweak dynamics and it is therefore proportional to the electroweak gauge coupling constants. The remaining single state stays massless if we neglect the effect of the $\U{1}_{X}$ axial anomaly, otherwise it is the pseudo-Nambu--Goldstone boson with the mass proportional to the parameter $\mu_{t\nu}$.

\subsection{Two Higgs doublet description}\label{SecIII}

Effectively, the top-quark and neutrino condensation can be described by the condensation of two composite Higgs doublets
\begin{eqnarray}
H_t & \sim & (\bar t_Rq_L)\,, \\
H_\nu & \sim & (\sum_s\bar \nu_{Rs}\ell_L) \,.
\end{eqnarray}
Using them we can rewrite the four-fermion interaction \eqref{4f} via the Hubbard--Stratonovich transformation \cite{Stratonovich:1957aa,Hubbard:1959aa} as
\begin{eqnarray}\label{boson_lagr}
{\cal L}_{4f} & = & - y_{t0}(\bar q_Lt_R)H_t - y_{\nu0}(\sum_s\bar \ell_L\nu_{Rs})H_\nu+\hc +\mu_{t0}^{2}H_{t}^\dag H_{t}+\mu_{\nu0}^{2}H_{\nu}^\dag H_{\nu} \,.
\end{eqnarray}
It is completely equivalent Lagrangian to \eqref{4f}. As far as $H_t$ and $H_\nu$ are non-propagating auxiliary fields one can use their trivial equations of motion to arrive at \eqref{4f}.

Below the condensation scale $\Lambda$, the interactions from \eqref{boson_lagr} generate by radiative corrections all operators allowed by the symmetries. Among the operators, there are kinetic terms for the composite Higgs doublets $H_{t}$ and $H_{\nu}$ and their quartic self-interactions. Keeping only the renormalizable operators we effectively obtain a two-Higgs-doublet model
%\begin{eqnarray}\label{eff_lagrangian}
%{\cal L}_\mathrm{eff} & = & Z_{H_{t}}|D H_{t}|^2 + Z_{H_{\nu}}|D H_{\nu}|^2 - {\cal V}(Z_{H_{t}}^{1/2}H_t,Z_{H_{\nu}}^{1/2}H_\nu) \nonumber\\
%& & - Z_{H_{t}}^{1/2}y_t(\bar q_Lt_R)H_t - Z_{H_{\nu}}^{1/2}y_\nu(\sum_s\bar \ell_L\nu_{Rs})H_\nu + \hc  \nonumber\\
%& & + {\cal L}_\mathrm{usual} + {\cal L}_{M_R}  \,.
%\end{eqnarray}
%
\begin{eqnarray}\label{eff_lagrangian}
{\cal L}_\mathrm{eff} & = & |D H_{t}|^2 + |D H_{\nu}|^2 - {\cal V}(H_t,H_\nu)  - y_t(\bar q_Lt_R)H_t - y_\nu(\sum_s\bar \ell_L\nu_{Rs})H_\nu + \hc  \nonumber\\
& & + {\cal L}_\mathrm{usual} + {\cal L}_{M_R}  \,.
\end{eqnarray}
The potential for the two Higgs doublets invariant with respect to the $G$ symmetry \eqref{EW_breaking} is
\begin{eqnarray}\label{2Higgs_potential}
{\cal V} & = & {\cal V}_{0}+{\cal V}_{\mathrm{EW}}+{\cal V}_{\mathrm{soft}} \\
{\cal V}_{0} & = & -\mu_{t}^2H_{t}^\dag H_{t}-\mu_{\nu}^2H_{\nu}^\dag H_{\nu} +\tfrac{1}{2}\lambda_t(H_{t}^\dag H_{t})^2+\tfrac{1}{2}\lambda_\nu(H_{\nu}^\dag H_{\nu})^2  \\
{\cal V}_{\mathrm{EW}} & = & \lambda_{t\nu}(H_{t}^\dag H_{t})(H_{\nu}^\dag H_{\nu}) +\lambda_{t\nu}'(H_{t}^\dag H_{\nu})(H_{\nu}^\dag H_{t})  \,.
\end{eqnarray}
We sort the terms in the potential ${\cal V}$ according to their primary origin. Those terms denoted by ${\cal V}_{0}$ are generated due to the four-fermion interaction irrespectively of the presence of the electroweak dynamics which provides only corrections to their magnitude. The terms denoted by ${\cal V}_{\mathrm{EW}}$, on the other hand, are generated only because of the presence of the electroweak dynamics. They vanish in the limit of vanishing electroweak coupling constants. They provide a bridge between the top-quark and neutrino sectors.

In order to take into account the axial anomaly of $\U{1}_X$ of the original theory we introduce additional term which mixes the two Higgs doublets and breaks explicitly the $\U{1}_X$ symmetry
\begin{eqnarray}
{\cal V}_{\mathrm{soft}} & = & -\mu_{t\nu}^2H_{t}^\dag H_{\nu}+\hc \,.
\end{eqnarray}
This term cannot be generated at any loop order either by the four-fermion interaction or by the electroweak dynamics. In this work we use $\mu_{t\nu}$ as a free parameter.

Apart from $\mu_{t\nu}$ all the parameters of the Lagrangian ${\cal L}_\mathrm{eff}$, i.e., $y$'s, $\mu$'s, and $\lambda$'s, run with the renormalization scale $\mu$ according to the renormalization group equations towards the condensation scale $\mu=\Lambda$. At the condensation scale they are linked to the values of the underlying Lagrangian ${\cal L}$, i.e., $y_0$'s, $\mu_0$'s, and $\lambda_0$'s, through the field renormalization factors. Because the mixing parameter $\mu_{t\nu}$ cannot be obtained by radiative corrections, it is not the subject of the renormalization group equations and as such it acts as a free parameter. The effect of such a mixing parameter was studied in \cite{Luty:1990bg} in the context of top-quark and bottom-quark two-Higgs-doublet model.

The quartic stability of the potential is given by the conditions \cite{Luty:1990bg}
\begin{subequations}\label{stability_condition}
\begin{eqnarray}
&&\lambda_t,\ \lambda_\nu>0\,, \\
&&\sqrt{\lambda_t\lambda_\nu}>-\lambda_{t\nu}-\lambda_{t\nu}' \,.
\end{eqnarray}
\end{subequations}
The parameter setting in the range
\begin{equation}
\lambda_{t\nu}'<0\ \ \mathrm{and}\ \ \mu_{t\nu}^2>0
\end{equation}
leads to the minimum of the potential which conserves the electric charge.

\subsection{Electroweak scale and fermion masses}

The electroweak symmetry is broken once the composite Higgs doublets $H_t$ and $H_\nu$ develop their nonzero vacuum expectation value
\begin{equation}\label{Htnu_cond}
\langle H_t\rangle=\frac{1}{\sqrt{2}}\beginm{c} 0 \\ v_t \endm \quad \mathrm{and} \quad \langle H_\nu\rangle=\frac{1}{\sqrt{2}}\beginm{c} 0 \\ v_\nu \endm \,,
\end{equation}
where $v_t$ and $v_\nu$ are the top-quark and neutrino condensates \eqref{tnucondensation}. We define the $\beta$-angle by
\begin{equation}\label{beta}
\tan\beta\equiv\frac{v_{t}}{v_{\nu}} \,
\end{equation}
and choose the convention that $\beta\in\langle0,\tfrac{\pi}{2}\rangle$.

The condensation generates Dirac masses for the top-quark and neutrinos obtained just by plugging \eqref{Htnu_cond} into the effective Lagrangian \eqref{eff_lagrangian}. We get
\begin{eqnarray}
m_{t} & = & y_{t}(\mu=m_t)v_{t}/\sqrt{2} \,, \label{top_mass} \\
m_{D}(\mu) & = & y_{\nu}(\mu)v_{\nu}/\sqrt{2} \,,  \label{nuDirac_mass}
\end{eqnarray}
where $y_{t,\nu}(\mu)$ are the running Yukawa coupling constants.

Contrary to the top-quark mass, for the neutrino Dirac mass $m_D$ we do not specify the scale $\mu$ because it is not a mass eigenvalue. The neutrino mass eigenvalues are obtained from the complete neutrino mass matrix $\mathbf{M}_\nu$ after encountering hard right-handed Majorana masses from the Lagrangian \eqref{L_M_R}. It has the same form as the neutrino self-energy \eqref{nu_sigma_p}
\begin{equation}\label{nu_mass_matrix}
\mathbf{M}_\nu = \beginm{cccc}0 & \mathbf{m}_D & \cdots & \mathbf{m}_D \\ \mathbf{m}_D & \mathcal{M}_R & & \\ \vdots & & \ddots & \\  \mathbf{m}_D & & & \mathcal{M}_R  \endm \,,
\end{equation}
where $\mathbf{m}_D\equiv m_D\openone$ from \eqref{nuDirac_mass} and $\mathcal{M}_R\equiv M_R\openone$, where $\openone$ is the $3\times3$ unit matrix in the flavor space. This matrix is non-singular and it has $3(N+1)$ massive Majorana eigenstates. Three of them have the degenerate masses
\begin{equation}\label{smaller_mnu}
\frac{1}{2}\left(M_R-\sqrt{4Nm_{D}^2+M_{R}^2}\right)\approx-\frac{N m_{D}^2}{M_R} \,,
\end{equation}
other three mass eigenstates have the degenerate masses
\begin{equation}
\frac{1}{2}\left(M_R+\sqrt{4Nm_{D}^2+M_{R}^2}\right)\approx M_R
\end{equation}
and finally $3(N-1)$ mass eigenstates have the degenerate masses equal to $M_R$. Clearly, the light neutrino masses are identified with the smaller eigenvalues \eqref{smaller_mnu} so we can write the seesaw formula
\begin{equation}\label{nu_mass}
m_{\nu}=\frac{N y_{\nu}^2(\mu=m_\nu)v_{\nu}^2}{2M_R} \,.
\end{equation}
The number of right-handed neutrino triplets $N$ enters the calculation through the formula \eqref{nu_mass}, and also through the renormalizaton group equations \eqref{Yukawa_RGE} and \eqref{quartic_RGE} introduced later.

Within the two-composite-Higgs-doublet model the values of Yukawa couplings around the electroweak scale are calculable using their renormalization group evolution down from the condensation scale $\Lambda$. Because $\Lambda$ is very large the couplings are only weakly sensitive to their initial values $y_{t,\nu}(\Lambda)$ as they have enough ``time'' to approach an infrared fixed point \cite{Hill:1985tg}.

The equation \eqref{top_mass} for the top-quark mass, $m_t\doteq172\,\mathrm{GeV}$, fixes $v_t$. From the relation \eqref{ews_scale_t_nu} we determine $v_\nu$, a portion of the electroweak scale left for neutrinos. Finally from \eqref{nu_mass} we determine the right-handed neutrino Majorana mass $M_R$ based on the assumption that $m_\nu\lesssim0.2\,\mathrm{eV}$. Having $m_D\sim v$ implies roughly $M_R\gtrsim10^{14}\,\mathrm{GeV}$. Here the construction closes by the non-decoupling condition \eqref{non_decoupling_condition}
\begin{equation}\label{non_decoupling_cond}
\Lambda_\mathrm{Planck}>\Lambda>M_R\,.
\end{equation}

\subsection{Higgs boson masses}
\label{Sec_Higgs_boson_Masses}
After the electroweak symmetry breaking given by \eqref{Htnu_cond}, we rewrite the effective Lagrangian \eqref{eff_lagrangian} in the unitary gauge in terms of two neutral scalars $\phi^{(0)}_t$ and $\phi^{(0)}_\nu$, one neutral pseudo-scalar $A$ and one charged scalar $H^\pm$
\begin{equation}\label{Htnu_redef}
H_t=\beginm{c} \im\cos\beta H^+ \\ \tfrac{1}{\sqrt{2}}(v_t+\phi^{(0)}_t-\im\cos\beta A) \endm \quad \mathrm{and} \quad H_\nu=\beginm{c} -\im\sin\beta H^+ \\ \tfrac{1}{\sqrt{2}}(v_\nu+\phi^{(0)}_\nu+\im\sin\beta A) \endm \,.
\end{equation}
The quadratic Lagrangian contains a mixing of the neutral scalars $\phi^{(0)}_t$ and $\phi^{(0)}_\nu$. Its diagonalization results in the mass eigenstates $h$ and $H$ characterized by the mixing angle $\alpha$ according to
\begin{eqnarray}
H & = & \sqrt{2}\big[\phi^{(0)}_t\sin\alpha+\phi^{(0)}_\nu\cos\alpha\big] \,,\\
h & = & \sqrt{2}\big[\phi^{(0)}_t\cos\alpha-\phi^{(0)}_\nu\sin\alpha\big] \,,\\
\tan2\alpha & = & \frac{(\lambda_{t\nu}+\lambda_{t\nu}')v_t v_\nu - \mu_{t\nu}^2}{-\tfrac{1}{2}(v_{t}^2\lambda_t-v_{\nu}^2\lambda_\nu)-\mu_{t\nu}^2\cot2\beta} \,.
\end{eqnarray}
We choose the convention that $\alpha\in\langle-\tfrac{\pi}{2},0)$. In this case the lighter Higgs scalar is always $h$.
The Higgs boson mass spectrum follows as
\begin{subequations}\label{Higgs_masses}
\begin{eqnarray}
M_{H,h}^2 & = & \frac{1}{2}f_{\pm}(t=\ln M_{H,h}) \,, \\
M_{A}^2 & = & \frac{2\mu_{t\nu}^2}{\sin2\beta} \,, \\
M_{H^\pm}^2 & = & \frac{2\mu_{t\nu}^2}{\sin2\beta}-\frac{1}{2}\lambda_{t\nu}'(t=\ln M_{H^\pm})v^2 \,,
\end{eqnarray}
\end{subequations}
where
%\begin{widetext}
\begin{eqnarray}
f_{\pm}(t) & = &  v_{t}^2\lambda_t(t)+v_{\nu}^2\lambda_\nu(t)+\frac{2\mu_{t\nu}^2}{\sin2\beta}\pm\sqrt{A(t)} \,, \nonumber\\
A(t)       & = &  \big(v_{t}^2\lambda_t(t)-v_{\nu}^2\lambda_\nu(t)\big)^2+4v_{t}^2v_{\nu}^2\big(\lambda_{t\nu}(t)+\lambda_{t\nu}'(t)\big)^2
                    +\frac{4\mu_{t\nu}^4}{\sin^22\beta} \nonumber\\
               & & -2\mu_{t\nu}^2\big[v_{t}^2\lambda_t(t)\tan\beta+v_{\nu}^2\lambda_\nu(t)\cot\beta
                   -v_tv_\nu\big(\lambda_t(t)+\lambda_\nu(t)+4\lambda_{t\nu}(t)+4\lambda_{t\nu}'(t)\big)\big] \,.\nonumber
\end{eqnarray}
%\end{widetext}
We have introduced the logarithmic renormalization scale $t\equiv\ln\mu$.

\begin{figure}[t]
\begin{center}
%\begin{tabular}{cc}
\includegraphics[width=0.7\textwidth]{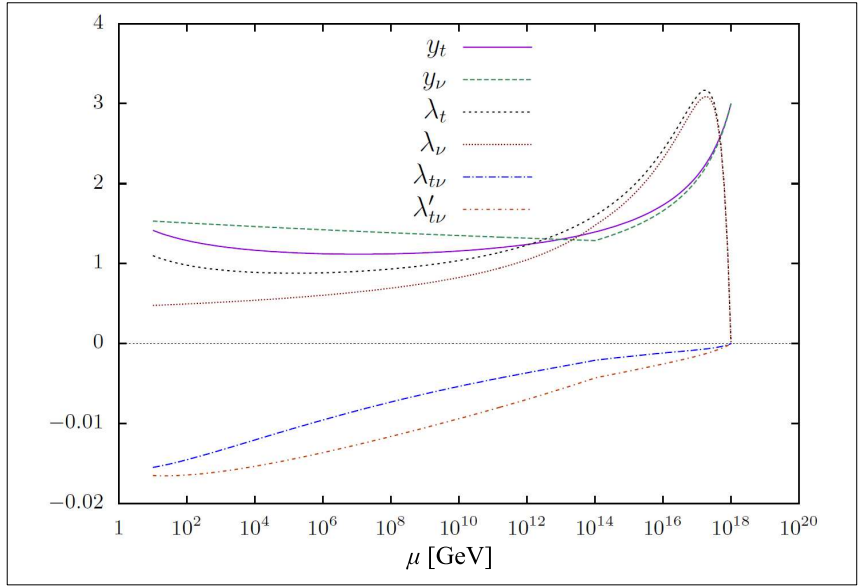}
%\framebox{\scalebox{0.67}{\input{RGE}}}
%\end{tabular}
\end{center}
\caption[Renormalization group evolution of parameters of the top-quark and neutrino condensation model ]{ \small The renormalization group evolution of $y_t$, $y_\nu$, $\lambda_t$, $\lambda_\nu$, $\lambda_{t\nu}$ and $\lambda_{t\nu}'$ for the parameter setting \eqref{param_set}. The role of $M_R=10^{14}\,\mathrm{GeV}$ is visible as a threshold in the evolution. }
\label{plot_RGE}
\end{figure}

In the following we will show that our model indeed meets with the stability condition \eqref{stability_condition} as it leads to
\begin{eqnarray}\label{cond_1}
\lambda_{t\nu},\ \lambda_{t\nu}'<0
\end{eqnarray}
and
\begin{equation}\label{cond_2}
\lambda_t,\ \lambda_\nu\gg|\lambda_{t\nu}|,\ |\lambda_{t\nu}'| \,.
\end{equation}
For illustration, see Fig.~\ref{plot_RGE}.
In order to conserve the electric charge we will consider only the values
\begin{equation}
\mu_{t\nu}^2\geq0\,.
\end{equation}
Let us now analyze two important limits:
\begin{enumerate}
\item First, by setting $\lambda_{t\nu},\lambda_{t\nu}'\rightarrow0$, we switch the effect of the electroweak interactions off, and by setting $\mu_{t\nu}^2=0$, we preserve the $\U{1}_X$ symmetry. The top-quark and neutrino condensation causes the symmetry breaking \eqref{G_H}. In this limit the spectrum of bosons changes to
    \begin{eqnarray}
    & M_{A}^2=M_{H^\pm}^2 = 0\,, & \\
    & M_{H}^2=\mathrm{Max}\big\{v_{t}^2\lambda_t,v_{\nu}^2\lambda_\nu\big\}\,,& \\
    & M_{h}^2=\mathrm{Min}\big\{v_{t}^2\lambda_t,v_{\nu}^2\lambda_\nu\big\} \,, & \label{MH2_mu_0}
    \end{eqnarray}
    where we identify three uneaten Nambu--Goldstone bosons. The two Higgs scalars do not mix as $\alpha=-\tfrac{\pi}{2}$. The lighter Higgs scalar is
    \begin{eqnarray}
    \mathrm{for}\ \ \ v_{t}^2\lambda_t>v_{\nu}^2\lambda_\nu & : & \ \ \ h=\sqrt{2}\Re H^{0}_{\nu}\,, \\
    \mathrm{for}\ \ \ v_{t}^2\lambda_t<v_{\nu}^2\lambda_\nu & : & \ \ \ h=\sqrt{2}\Re H^{0}_{t}\,.
    \end{eqnarray}
    %Typically we will obtain $\lambda_t\sim 1$ and $\lambda_\mu\sim 0.5$
\item Second, we set $\lambda_{t\nu},\lambda_{t\nu}'\rightarrow0$ again, but we let $\mu_{t\nu}\gg v$.
    In this limit, the spectrum of bosons changes to
    \begin{eqnarray}
    & M_{H}^2 = M_{A}^2=M_{H^\pm}^2 = \frac{2\mu_{t\nu}^2}{\sin2\beta} \,, & \label{degenerate_MH}\\
    & M_{h}^2 = \tfrac{1}{8}v^2\big[4\lambda_t(\sin^2\beta+\sin^4\beta)+4\lambda_\nu(\cos^2\beta+\cos^4\beta)  & \label{MH2_mu_inf}
    \end{eqnarray}
    In this limit four degrees of freedom $H$, $A$ and $H^\pm$ get degenerate masses proportional to $\mu_{t\nu}$ and decouple from the low-energy physics. One degree of freedom $h$ stays light. It is the mixture of top-quark and neutrino neutral composite scalars which is characterized by $\tan2\alpha=\tan2\beta$, hence the mixing angle $\alpha=\beta-\tfrac{\pi}{2}$.
\end{enumerate}
To obtain actual values of masses we need to evolve the running parameters from their initial values at the condensation scale down to the electroweak scale. This we will do in section \ref{SecV}.

\subsection{Interactions of mass eigenstates}

We list here several interactions of the lighter Higgs scalar $h$ and of the charged Higgs boson $H^\pm$, which are particularly important for present phenomenological analyses:
\begin{subequations}
\begin{eqnarray}
{\cal L}_{\mathrm{Yukawa}}^{h} & = & -\frac{m_t}{v}C_th\bar tt \,, \label{Higgs_Yukawa}\\
{\cal L}_{\mathrm{gauge}}^{h} & = & -C_Vh\left(\frac{2M_{W}^2}{v}W^+W^- + \frac{M_{Z}^2}{v}Z^2\right) \,, \label{Higgs_gauge}\\
{\cal L}_{\mathrm{H^\pm}}^{h} & = & -C_{H^\pm}vhH^+H^-\,, \label{charged_Higgs_Higgs}
%{\cal L}_{\mathrm{Yukawa}}^{H^\pm} & = & \frac{m_t}{v}C_{tb}\bar tbH^+ + \frac{m_D}{v}C_{\nu\ell}\bar\nu\ell H^+ \,, \label{charged_Higgs_Yukawa}
\end{eqnarray}
\end{subequations}
where the scaling coupling factors are
\begin{subequations}\label{coupling_parameters}
\begin{eqnarray}
C_t & = & \frac{\cos\alpha}{\sin\beta} \,, \\
C_V & = & \sin(\beta-\alpha) \,, \\
C_{H^\pm} & = & \sin\beta\cos\alpha\big(\sin^2\beta\lambda_{t\nu}+\cos^2\beta(\lambda_t-\lambda_{t\nu}')  \big) \\
            & & -\cos\beta\sin\alpha\big(\cos^2\beta\lambda_{t\nu}+\sin^2\beta(\lambda_\nu-\lambda_{t\nu}')\big) \nonumber\,.
\end{eqnarray}
\end{subequations}
They measure the departure from the Standard Model, which is characterized by $C_t=1$, $C_V=1$ and $C_{H^\pm}=0$.

Knowledge of these three scaling factors is necessary to determine the one-loop induced decay width of $h\rightarrow\gamma\gamma$ using the well known analytic expression to be found, e.g., in \cite{Djouadi:2005gj}:
\begin{equation}\label{Gamma_h_gammagamma}
\Gamma(h\rightarrow\gamma\gamma)=\frac{\alpha^2M_{h}^3}{256\pi^3v^2}\left|\frac{4}{9}N_\mathrm{C}C_tA_f(\tau_t)+C_VA_V(\tau_W)+\frac{v^2}{2M_{H^\pm}^2}C_{H^\pm}A_S(\tau_{H^\pm})\right|
\end{equation}
where $\alpha\doteq1/137$ is the fine-structure constant and
\begin{equation}
\tau_t=\frac{M_{h}^2}{4m_{t}^2} \,,\ \ \tau_W=\frac{M_{h}^2}{4M_{W}^2}\,,\ \ \tau_{H^\pm}=\frac{M_{h}^2}{4M_{H^\pm}^2}\,.
\end{equation}
The loop functions for scalars, fermions and vector bosons are
\begin{equation}
\begin{array}{c}
\begin{array}{rcl}
A_S(x) & = & -\big(x-f(x)\big)x^{-2} \,, \\
A_f(x) & = & 2\big(x+(x-1)f(x)\big)x^{-2} \,, \\
A_V(x) & = & -\big(2x^2+3x+3(2x-1)f(x)\big)x^{-2} \,,
\end{array} \\
\\
\mathrm{for}\ \
f(x)=\left\{
\begin{array}{l}
x\le1:\ \ \arcsin^2{\sqrt{x}} \,, \\
x>1:\ \ -\frac{1}{4}\left(\log\frac{1+\sqrt{1-1/x}}{1-\sqrt{1-1/x}}-\im\pi\right) \,.
\end{array}
\right.
\end{array}
\end{equation}

\subsection{Renormalization group equations}\label{SecV}

The parameters of the low-energy Lagrangian ${\cal L}_\mathrm{eff}$ run with the renormalization scale $\mu$ according to equations of the renormalization group. The exhaustive analysis of renormalization group equations for two-Higgs-doublet models is to be found in \cite{Hill:1985tg}. The compositeness of the Higgs doublets is expressed by the fact that the Lagrangian \eqref{eff_lagrangian} is equivalent to the Higgs-less Lagrangian \eqref{4f} or \eqref{boson_lagr} at the condensation scale $\Lambda$. From sewing two Lagrangians together at $\Lambda$, a set of boundary conditions for $\mu\rightarrow\Lambda$ follows:
\begin{eqnarray}
& y_t \rightarrow \infty\,,\ \ y_\nu \rightarrow \infty\,, \ \ y_t/y_\nu \rightarrow y_{t0}/y_{\nu0} \,, & \nonumber\\
& \lambda_t/y_{t}^4 \rightarrow 0\,,\ \ \lambda_\nu/y_{\nu}^4 \rightarrow 0\,, & \nonumber\\
& \lambda_{t\nu}/y_{t}^2y_{\nu}^2 \rightarrow 0\,,\ \ \lambda_{t\nu}'/y_{t}^2y_{\nu}^2 \rightarrow 0\,, & \nonumber\\
& \mu_{t}^2/y_{t}^2 \rightarrow \mu_{t0}^{2}/y_{t0}^{2}\,,\ \ \mu_{\nu}^2/y_{\nu}^2 \rightarrow \mu_{\nu0}^{2}/y_{\nu0}^{2}\,. &
\end{eqnarray}
In practice, for actual numerical calculation, we will use the boundary conditions
\begin{eqnarray}\label{boundary_conditions}
y_t(\ln\Lambda)=Y_t \,, \ \ y_\nu(\ln\Lambda)=Y_\nu \,, \nonumber\\
\lambda_t(\ln\Lambda)=0 \,, \ \
\lambda_\nu(\ln\Lambda)=0 \,, \nonumber\\
\lambda_{t\nu}(\ln\Lambda)=0 \,, \ \
\lambda_{t\nu}'(\ln\Lambda)=0 \,,
\end{eqnarray}
where $Y_t$ and $Y_\nu$ are finite numbers on which the low-energy result depends only very weakly.

\noindent Further, we will restrict our analysis only to one-loop order.

The presence of the second Higgs doublet affects the $t=\ln\mu$ evolution of the gauge coupling constants governed by the one-loop renormalization group equations in comparison with the Standard Model case. Here we use the notation that $g_1$ refers to the weak hypercharge, $g_2$ refers to the weak isospin and $g_3$ refers to the color gauge couplings. As boundary conditions we use their experimental values at $\mu=M_Z\doteq91.1\GeV$ used also in \cite{Luty:1990bg},
\begin{subequations}\label{g123}
\begin{eqnarray}
16\pi^2\tfrac{\d}{\d t}g_1 = \hphantom{-}7g_{1}^3 \,,&\ \ &g_{1}^2(\ln M_Z)\doteq0.127 \,, \\
16\pi^2\tfrac{\d}{\d t}g_2 = -3g_{2}^3 \,,&\ \ &g_{2}^2(\ln M_Z)\doteq0.425 \,, \\
16\pi^2\tfrac{\d}{\d t}g_3 = -7g_{3}^3 \,,&\ \ &g_{3}^2(\ln M_Z)\doteq1.440 \,.
\end{eqnarray}
\end{subequations}
The renormalization group equations for Yukawa coupling constants are (from \cite{Hill:1985tg})
\begin{eqnarray}
16\pi^2\tfrac{\d}{\d t}y_t   & = & y_t\big[\tfrac{9}{2}y_{t}^2-\tfrac{17}{12}g_{1}^2-\tfrac{9}{4}g_{2}^2-8g_{3}^2\big] \,, \label{Yukawa_RGE} \\
16\pi^2\tfrac{\d}{\d t}y_\nu & = & y_\nu\big[3(N+\tfrac{1}{2})\theta(\mu-M_R)y_{t}^2-\tfrac{3}{4}g_{1}^2-\tfrac{9}{4}g_{2}^2\big] \,. \nonumber
\end{eqnarray}
The $\theta$-function stands for the threshold at $\mu=M_R$ below which the heavy right-handed neutrinos decouple from the system. The renormalization group equations for the quartic coupling constants are (from \cite{Hill:1985tg})
%\begin{widetext}
%\begin{eqnarray}
%16\pi^2\tfrac{\d}{\d t}\lambda_t   & = & 12\lambda_{t}^2+4\lambda_{t\nu}^2+4\lambda_{t\nu}^2\lambda_{t\nu}'+2{\lambda_{t\nu}'}^2
%                                   +\lambda_{t}\big[12y_{t}^2-3g_{1}^2-9g_{2}^2\big]-12y_{t}^4+\tfrac{3}{4}(g_{1}^4-2g_{1}^2g_{2}^2+3g_{2}^4) \,, \nonumber\\
%16\pi^2\tfrac{\d}{\d t}\lambda_\nu & = & 12\lambda_{\nu}^2+4\lambda_{t\nu}^2+4\lambda_{t\nu}^2\lambda_{t\nu}'+2{\lambda_{t\nu}'}^2
%                                   +\lambda_{\nu}\big[12N\theta(t-M_R)y_{\nu}^2-3g_{1}^2-9g_{2}^2\big]-12N y_{\nu}^4+\tfrac{3}{4}(g_{1}^4-2g_{1}^2g_{2}^2+3g_{2}^4) \,, \nonumber\\
%16\pi^2\tfrac{\d}{\d t}\lambda_{t\nu} & = & 2(\lambda_{\nu}+\lambda_{t})(3\lambda_{t\nu}+\lambda_{t\nu}')+4\lambda_{t\nu}^2+2{\lambda_{t\nu}'}^2
%                                   +\lambda_{t\nu}\big[6y_{t}^2+6N\theta(t-M_R)y_{\nu}^2-3g_{1}^2-9g_{2}^2\big]+\tfrac{3}{4}(g_{1}^4-2g_{1}^2g_{2}^2+3g_{2}^4) \,, \nonumber\\
%16\pi^2\tfrac{\d}{\d t}\lambda_{t\nu}' & = & \lambda_{t\nu}'\big[\lambda_{\nu}+\lambda_{t}+8\lambda_{t\nu}^2+4{\lambda_{t\nu}'}^2
%                                   +6y_{t}^2+6N\theta(t-M_R)y_{\nu}^2-3g_{1}^2-9g_{2}^2\big]+3g_{1}^2g_{2}^2 \,. \label{quartic_RGE}
%\end{eqnarray}
%\end{widetext}
\begin{eqnarray}
16\pi^2\tfrac{\d}{\d t}\lambda_t   & = & 12\lambda_{t}^2+4\lambda_{t\nu}^2+4\lambda_{t\nu}^2\lambda_{t\nu}'+2{\lambda_{t\nu}'}^2  \label{quartic_RGE}\\
                                   & & +\lambda_{t}\big[12y_{t}^2-3g_{1}^2-9g_{2}^2\big]-12y_{t}^4+\tfrac{3}{4}(g_{1}^4-2g_{1}^2g_{2}^2+3g_{2}^4) \,, \nonumber\\
16\pi^2\tfrac{\d}{\d t}\lambda_\nu & = & 12\lambda_{\nu}^2+4\lambda_{t\nu}^2+4\lambda_{t\nu}^2\lambda_{t\nu}'+2{\lambda_{t\nu}'}^2 \nonumber\\
                                   & & +\lambda_{\nu}\big[12N\theta(\mu-M_R)y_{\nu}^2-3g_{1}^2-9g_{2}^2\big]-12N y_{\nu}^4+\tfrac{3}{4}(g_{1}^4-2g_{1}^2g_{2}^2+3g_{2}^4) \,, \nonumber\\
16\pi^2\tfrac{\d}{\d t}\lambda_{t\nu} & = & 2(\lambda_{\nu}+\lambda_{t})(3\lambda_{t\nu}+\lambda_{t\nu}')+4\lambda_{t\nu}^2+2{\lambda_{t\nu}'}^2 \nonumber\\
                                   & &+\lambda_{t\nu}\big[6y_{t}^2+6N\theta(\mu-M_R)y_{\nu}^2-3g_{1}^2-9g_{2}^2\big]+\tfrac{3}{4}(g_{1}^4-2g_{1}^2g_{2}^2+3g_{2}^4) \,, \nonumber\\
16\pi^2\tfrac{\d}{\d t}\lambda_{t\nu}' & = & \lambda_{t\nu}'\big[\lambda_{\nu}+\lambda_{t}+8\lambda_{t\nu}^2+4{\lambda_{t\nu}'}^2+6y_{t}^2+6N\theta(\mu-M_R)y_{\nu}^2-3g_{1}^2-9g_{2}^2\big]+3g_{1}^2g_{2}^2 \,. \nonumber
\end{eqnarray}
According to Luty \cite{Luty:1990bg} the one loop renormalization evolution of the dimensionless parameters does not depend on the presence of the mixing parameter $\mu_{t\nu}$.

\begin{figure}[t]
%\begin{center}
\begin{tabular}{cc}
\includegraphics[width=0.5\textwidth]{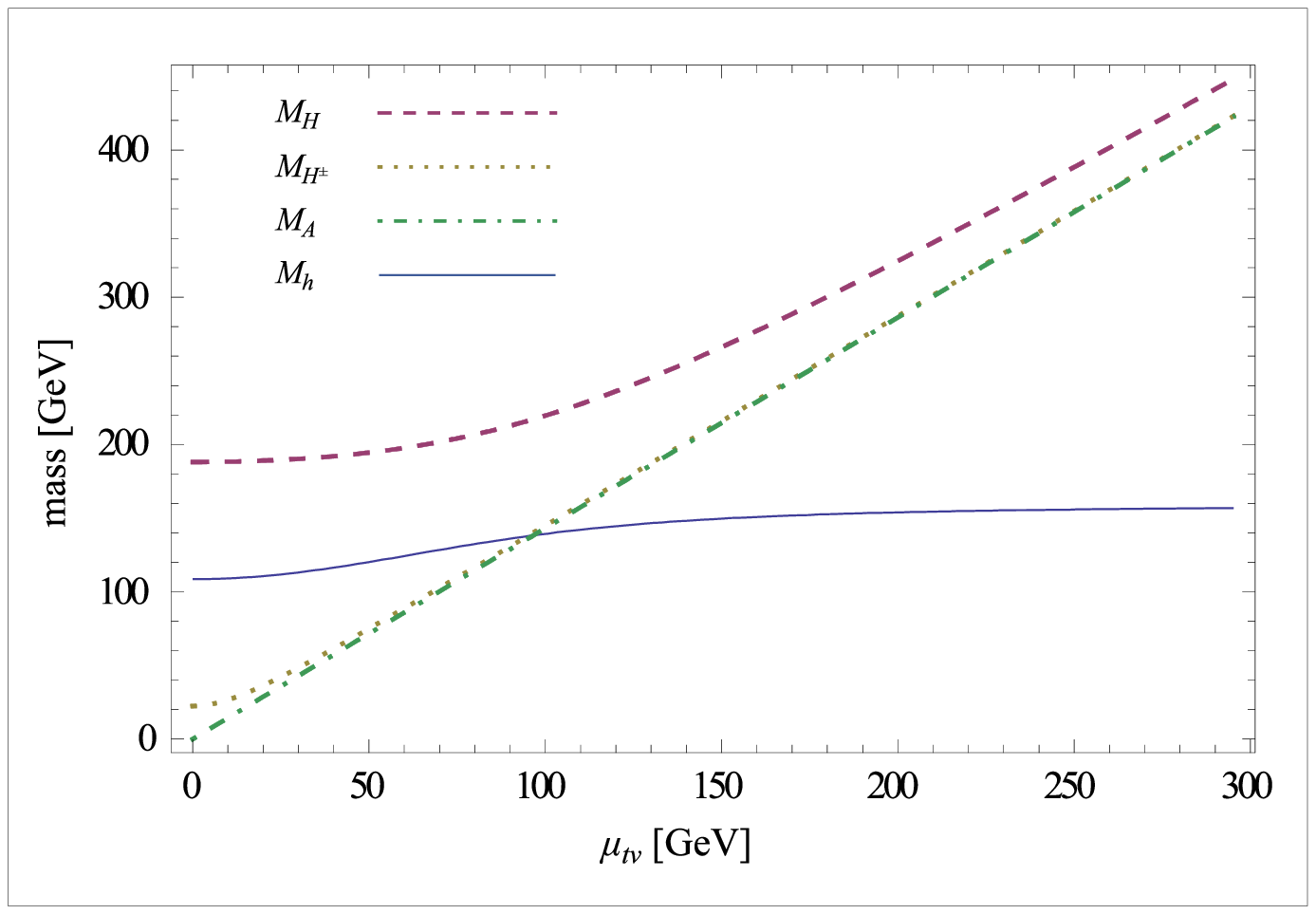} &
%\framebox{\scalebox{0.67}{\input{HiggsMasses}}} \\
\includegraphics[width=0.5\textwidth]{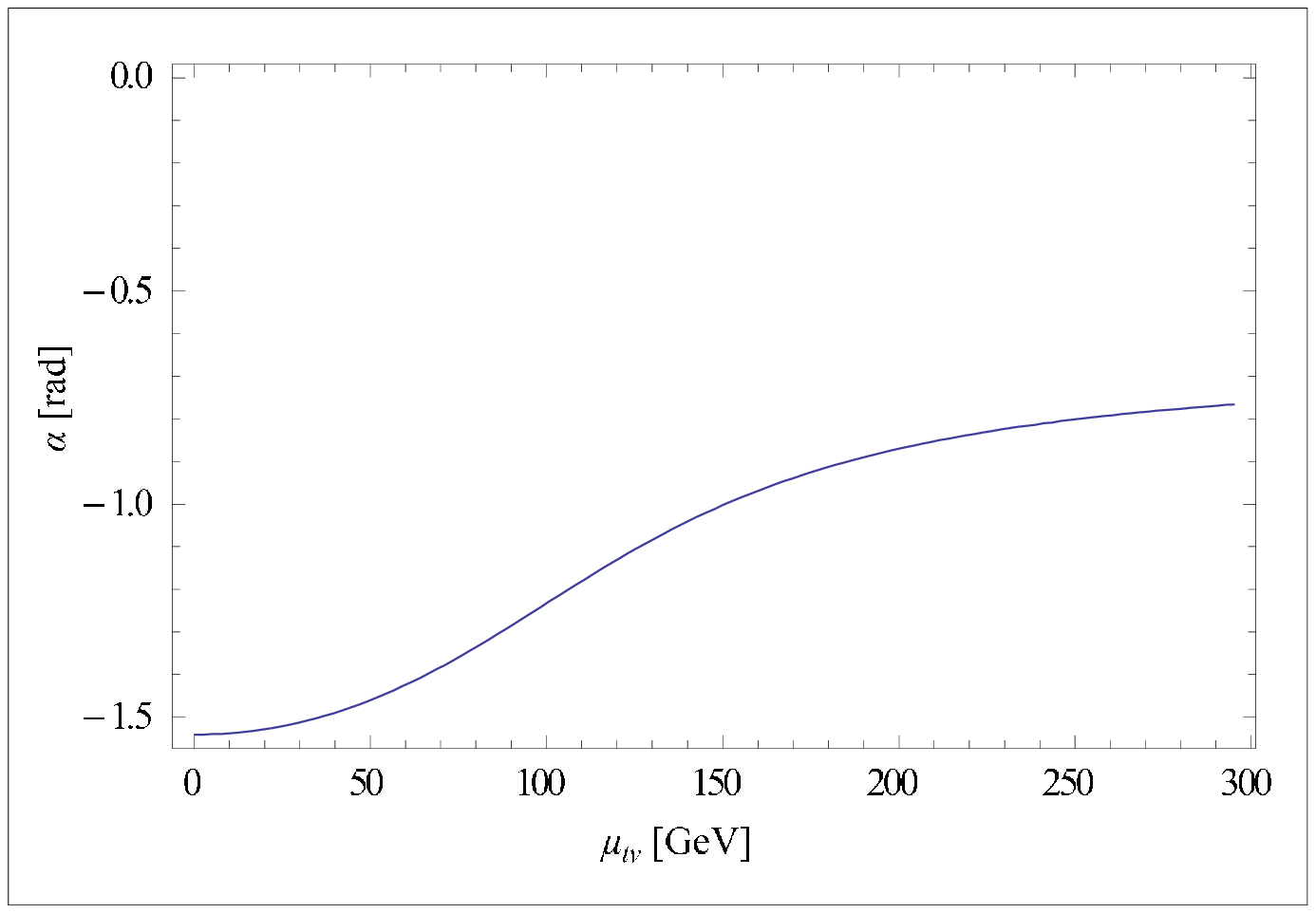} \\
%\framebox{\scalebox{0.67}{\input{gamma}}} \\
a) & b)
\end{tabular}
\caption[Higgs masses $M_{H,h}$, $M_{A}$ and $M_{H^\pm}$, and mixing angle $\gamma$ as functions of $\mu_{t\nu}$ ]{ \small a) Higgs masses $M_{H,h}$, $M_{A}$ and $M_{H^\pm}$ as functions of $\mu_{t\nu}$ for $\Lambda=10^{18}\,\mathrm{GeV}$, $N=1$ and fixed neutrino mass to be $m_\nu=0.2\,\mathrm{eV}$ (i.e. $M_R\sim10^{14}\,\mathrm{GeV}$). b) Mixing of two scalar Higgs bosons $H$ and $h$ as a function of $\mu_{t\nu}$. At the condensation scale, the Yukawa couplings $y_{t,\nu}$ start at the value $Y_{t,\nu}=3$, and the quartic couplings $\lambda$ start at zero value. }
\label{plot_HiggsMasses}
%\end{center}
\end{figure}

\subsection{Results}\label{SecVI}

Before we present our results we briefly describe the strategy of the analysis. Our input parameters are $\Lambda$, $M_R$ and $\mu_{t\nu}$, out of which $M_R$ is fixed by reproducing the neutrino mass $m_\nu=0.2\,\mathrm{eV}$. Strictly speaking, there are additional two parameters $Y_t$ and $Y_\nu$ which however have only very mild effect on the result and we take them quite arbitrarily to be
\begin{equation}\label{Yuk_set}
Y_t=Y_\nu=3 \,.
\end{equation}
On top of that we have the freedom to choose the number $N$ of right-handed neutrino triplets.

The steps of calculation are the following:
\begin{enumerate}
\item We solve analytically the equations \eqref{g123} for gauge coupling constants.
\item We numerically evolve the Yukawa and quartic coupling constants according to the equations \eqref{Yukawa_RGE} and \eqref{quartic_RGE} with the boundary conditions \eqref{boundary_conditions} and \eqref{Yuk_set}.
\item From the experimental value of the top quark mass we determine $v_t$ using the equation \eqref{top_mass}.
\item We calculate $v_\nu$ and $\beta$ from equations \eqref{ews_scale_t_nu} and \eqref{beta}, respectively.
\item The equation \eqref{nu_mass} gives us the value for the neutrino mass $m_\nu$. Changing the value of $M_R$ we repeat the calculation from the point 2.~and iterate the value of neutrino mass to get $m_\nu=0.2\,\mathrm{eV}$.
\item Using the equations \eqref{Higgs_masses} we calculate the Higss boson mass spectrum.
\end{enumerate}

\subsubsection{Renormalization group evolution}

The renormalization group evolution of the dimensionless parameters is plotted in Fig.~\ref{plot_RGE} for
\begin{equation}\label{param_set}
N=1,\ \Lambda=10^{18}\,\mathrm{GeV},\ M_R=10^{14}\,\mathrm{GeV}\,.
\end{equation}
As we mentioned before, at one-loop order the result does not depend on the parameter $\mu_{t\nu}$. It gives us typical result which does not change qualitatively much with changing $\Lambda$ and $M_R$. We can see that at the electroweak scale $\Lambda_\mathrm{EW}$
\begin{eqnarray}
&\lambda_{t\nu},\lambda_{t\nu}'<0 \,,& \\
&100\times\big(|\lambda_{t\nu}|,|\lambda_{t\nu}'|\big)\sim\big(\lambda_{t},\lambda_{\nu}\big) \,, \label{lambda_mag} &
\end{eqnarray}
so the stability conditions \eqref{cond_1} and \eqref{cond_2} are fulfilled. On top of that, taking $\mu_{t\nu}^2>0$ we assure that the condensate will be electrically neutral.

\subsubsection{Mass spectrum of Higgs bosons}

The typical result \eqref{lambda_mag} allows us to neglect $\lambda_{t\nu}$, $\lambda_{t\nu}'$ in favor of $\lambda_t$, $\lambda_\nu$. That is why the limits analyzed at the end of section~\ref{Sec_Higgs_boson_Masses} are useful for us. We can roughly estimate the mass of the lighter Higgs scalar $h$ and the mixing angle $\alpha$ to lie in the interval
\begin{equation}
\left.\begin{array}{rcl}M_{h}&\simeq&\langle113,160)\mathrm{GeV} \\
                \alpha&\simeq&\langle-\tfrac{\pi}{2},-0.7)\end{array}\right\}\ \ \mathrm{for}\ \ \mu_{t\nu}=\langle0,\infty)\mathrm{GeV}
\end{equation}
calculated from \eqref{MH2_mu_0} and \eqref{MH2_mu_inf} with input parameters \eqref{param_set} and using estimated values from Fig.~\ref{plot_RGE}, \eqref{ews_scale_t_nu} and \eqref{top_mass}
\begin{eqnarray}
&v_t\simeq187\,\mathrm{GeV}\,,\ v_\nu\simeq160\,\mathrm{GeV} \,,& \\
&\lambda_t\simeq1.0\,,\ \lambda_\nu\simeq0.5 \,. &
\end{eqnarray}
It represents a promising improvement with respect to the previous results of the single Higgs doublet top-quark condensation models, compared with the Higgs boson mass in Tab.~\ref{Tab_Higgs_top_mass}.

Now, let us investigate the solutions of the Higgs boson mass spectrum without approximations. In Fig.~\ref{plot_HiggsMasses}a) we plot the dependence of Higgs boson masses on the mixing parameter $\mu_{t\nu}$. We use $\Lambda=10^{18}\,\mathrm{GeV}$ while the right-handed neutrino Majorana mass we fix from demanding $m_\nu=0.2\,\mathrm{eV}$. It turns out to be roughly $M_R\sim10^{14}\,\mathrm{GeV}$.

For lower values of $\mu_{t\nu}$ the bosons $A$ and $H^\pm$ are very light, the mass $M_{A}$ even vanishes for vanishing $\mu_{t\nu}$ reflecting the spontaneous breaking of the exact $\U{1}_X$ symmetry \eqref{U1X}. In Fig.~\ref{plot_HiggsMasses}b) where we plot the dependence of the $h$-$H$ mixing angle $\alpha$ on $\mu_{t\nu}$, it can be seen that the lighter scalar $h$ is composed mainly of neutrinos, $\alpha\sim-\tfrac{\pi}{2}$, and therefore its coupling to top-quark is suppressed, see \eqref{Higgs_Yukawa}.

Increasing $\mu_{t\nu}$ translates into the lifting of masses of the Higgs bosons, $H$, $A$ and $H^\pm$. They soon become growing linearly and nearly degenerate. On the other hand, the mass of the lighter scalar $h$ is only mildly sensitive to the increase of $\mu_{t\nu}$ and quite soon saturates just below $160\,\mathrm{GeV}$. On top of that it is acquiring gradually larger admixture from a top-quark composite state, reaching the value over $\alpha\sim-0.8$. $M_{A}$ minimizes the spectrum of $H$, $A$ and $H^\pm$ for all positive values of $\mu_{t\nu}$ in our model.

Setting $\Lambda=10^{18}\,\mathrm{GeV}$, $m_\nu=0.2\,\mathrm{eV}$ and $N=1$ we can reach the lighter Higgs boson mass of the desired value
\begin{equation}
M_{h}=125\,\mathrm{GeV}\ \ \mathrm{for}\ \ \mu_{t\nu}\doteq62\,\mathrm{GeV}\,.
\end{equation}
The value $\mu_{t\nu}\doteq62\,\mathrm{GeV}$ translates into the Higgs boson mass spectrum
\begin{equation}
M_{H}\doteq198\,\mathrm{GeV}\,,\ \ M_{A}\doteq88\,\mathrm{GeV}\,,\ \ M_{H^\pm}\doteq91\,\mathrm{GeV}\,,
\end{equation}
which apparently contradicts the data \cite{Chatrchyan:2012vca,Aad:2012tj}. These values can be altered by changing the model parameters, the condensation scale $\Lambda$ and the number of right-handed neutrino triplets $N$.

By an order of magnitude decrease of $\Lambda$ for fixed number $N$ and $M_h=125\,\mathrm{GeV}$ we decrease the parameter $\mu_{t\nu}$ and also change the value of $\tan\beta$ according to Tab.~\ref{mutn_tanbeta_lambda}.
The values for $\Lambda$ above the Planck scale are shown only for curiosity, otherwise we avoid them further in our analysis.

By increasing the number of the right-handed neutrino triplets $N$ for a given $\Lambda$ we increase the value of $\mu_{t\nu}$ as seen in Tab.~\ref{mutnmin_mutnmax_lambda}. On the other hand the value of $\tan\beta$ is completely insensitive to the change of $N$.

Surprisingly, the number $N$ has an upper limit given by either of two conditions: the non-decoupling condition $\Lambda>M_R$ \eqref{non_decoupling_cond}, or the Higgs potential stability condition \eqref{cond_1} and \eqref{cond_2}. In the former case, increasing $N$ requires increasing $M_R$ in order to keep $m_\nu=0.2\,\mathrm{eV}$ according to the equation \eqref{nu_mass}. So for sufficiently high $N$ the mass $M_R$ runs over the condensation scale $\Lambda$. In the latter case, the increase of $N$ decreases the $\lambda_\nu(\mu)$ in infrared region so that it eventually runs negative around the electroweak scale.

\begin{figure}[h]
%\begin{center}
\begin{tabular}{c}
\begin{tabular}{cc}
\includegraphics[width=0.5\textwidth]{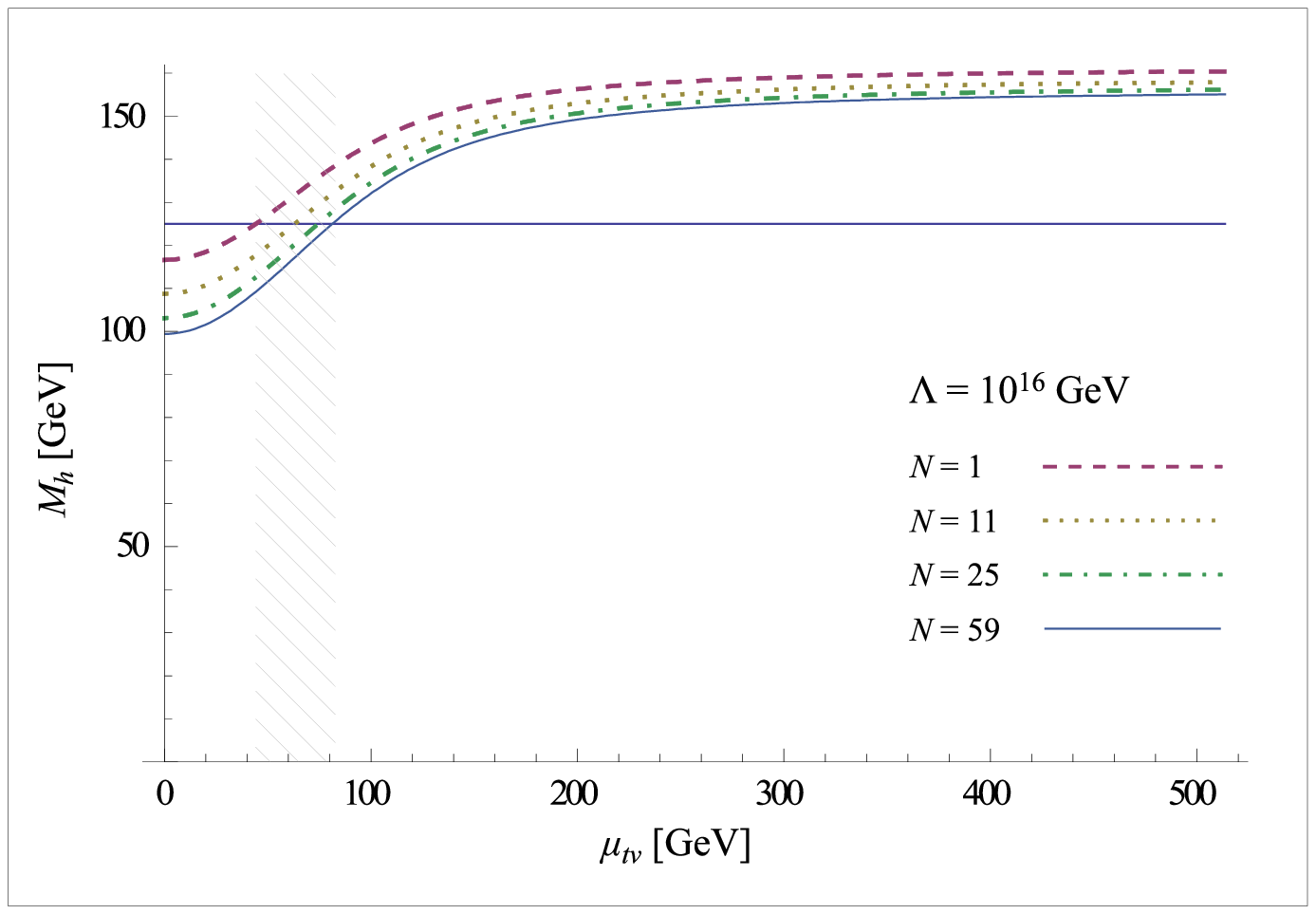} & \includegraphics[width=0.5\textwidth]{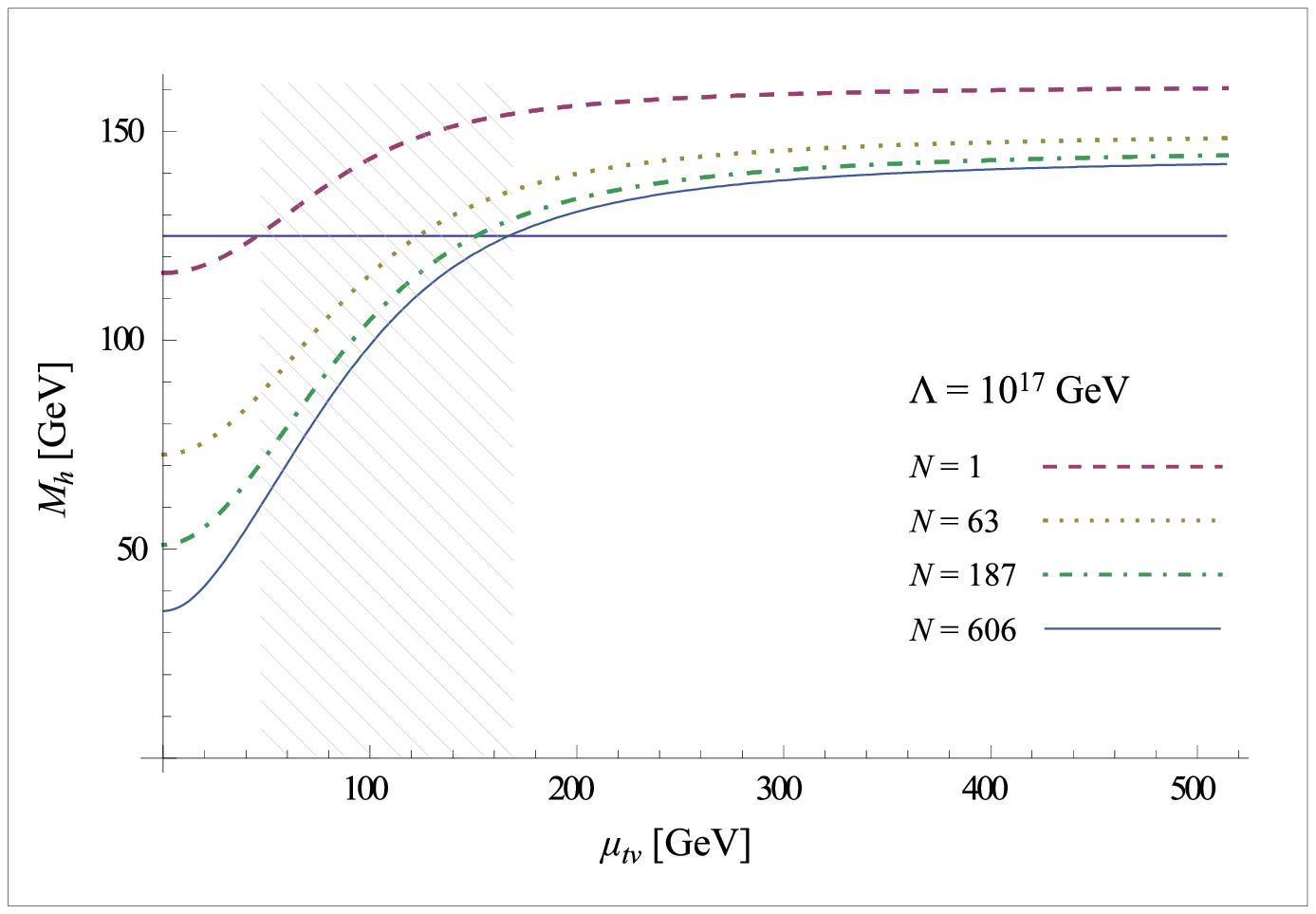} \\
%\vphantom{$lq$}
a) & b)
\end{tabular}
 \\
\includegraphics[width=0.5\textwidth]{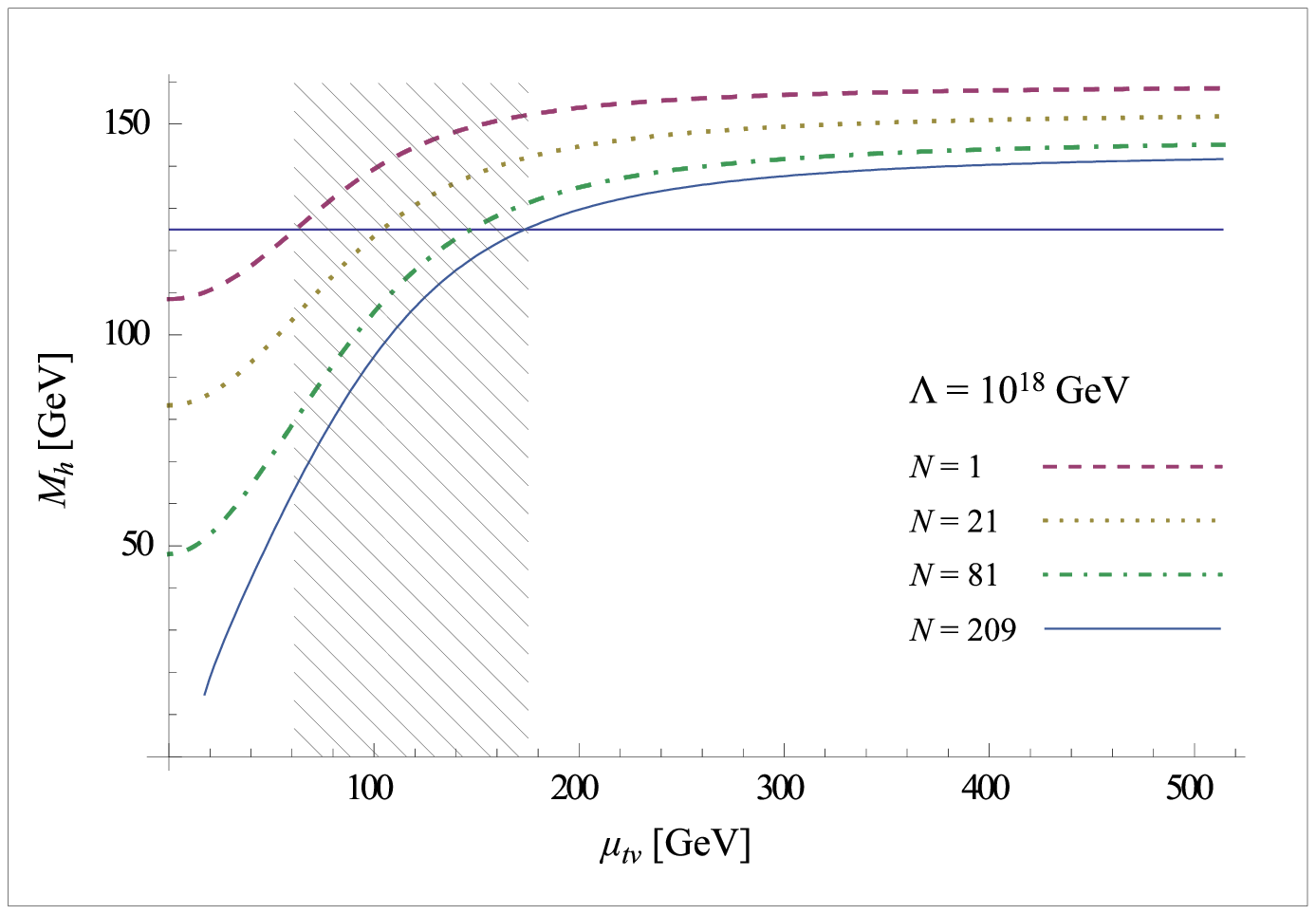} \\
%\vphantom{$lq$}
c)
\end{tabular}
\caption[ Higgs mass $M_{h}$ for various numbers of right-handed neutrino triplets as a function of $\mu_{t\nu}$ ]{ \small The lightest scalar Higgs mass $M_{h}$ for various numbers of right-handed neutrino triplets as a function of $\mu_{t\nu}$ for a) $\Lambda=10^{16}\,\mathrm{GeV}$, b) $\Lambda=10^{17}\,\mathrm{GeV}$, c) $\Lambda=10^{18}\,\mathrm{GeV}$, and for fixed neutrino mass to be $m_\nu=0.2\,\mathrm{eV}$. The solid horizontal line visualizes the $125\,\mathrm{GeV}$ value. The hatched area visualizes the interval of $\mu_{t\nu}$ values corresponding to $M_{h}=125\,\mathrm{GeV}$ shown in Tab.~\ref{mutnmin_mutnmax_lambda} as well. }
\label{plot_LightestHiggsMass}
%\end{center}
\end{figure}

In Figs.~\ref{plot_LightestHiggsMass} we plot $M_{h}$ for various $N$ from $1$ to $N^\mathrm{max}$ for three cases $\Lambda=10^{16},\,10^{17},\,10^{18}\,\mathrm{GeV}$ and we read out the intervals $(\mu_{t\nu}^\mathrm{min},\mu_{t\nu}^\mathrm{max})$ of only possible values for $\mu_{t\nu}$ that correspond to $M_{h}=125\,\mathrm{GeV}$. We show it in Tab.~\ref{mutnmin_mutnmax_lambda} together with the corresponding minimal and maximal masses for $H^\pm$, $(M_{H^\pm}^\mathrm{min},M_{H^\pm}^\mathrm{max})$.

In Tab.~\ref{mutnmin_mutnmax_lambda} we show the maximum number $N^\mathrm{max}$ as well. For the case $\Lambda=10^{16}\,\mathrm{GeV}$ and $\Lambda=10^{17}\,\mathrm{GeV}$ the number $N^\mathrm{max}$ is actually not the maximal value allowed by either of conditions. It rather corresponds to maximizing the parameter $\mu_{t\nu}^\mathrm{max}$. Increasing $N$ above $N^\mathrm{max}$ causes the backward decrease of $\mu_{t\nu}$. The maximal number of right-handed neutrino triplets, above which the non-decoupling condition is broken, is $N=158$ ($N\simeq1500$) for $\Lambda=10^{16}\,\mathrm{GeV}$ ($\Lambda=10^{17}\,\mathrm{GeV}$). In the case $\Lambda=10^{18}\,\mathrm{GeV}$, the maximum number $N^\mathrm{max}=209$ is given by the Higgs potential stability.

\subsubsection{Light Higgs scalar coupling strengths}

\begin{figure}[t]
%\begin{center}
\begin{tabular}{cc}
\includegraphics[width=0.5\textwidth]{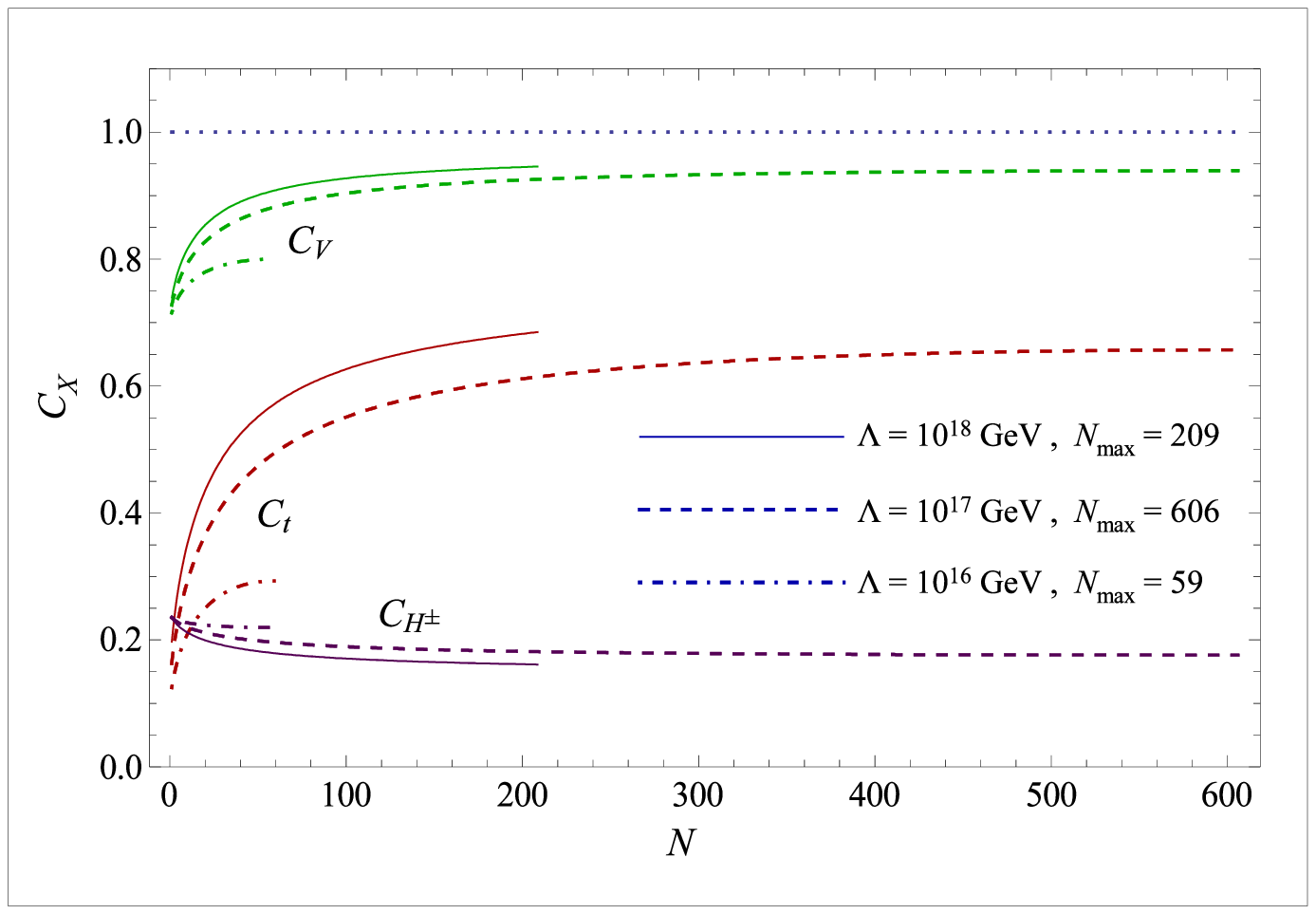} &
%\framebox{\scalebox{0.67}{\input{HiggsMasses}}} \\
\includegraphics[width=0.5\textwidth]{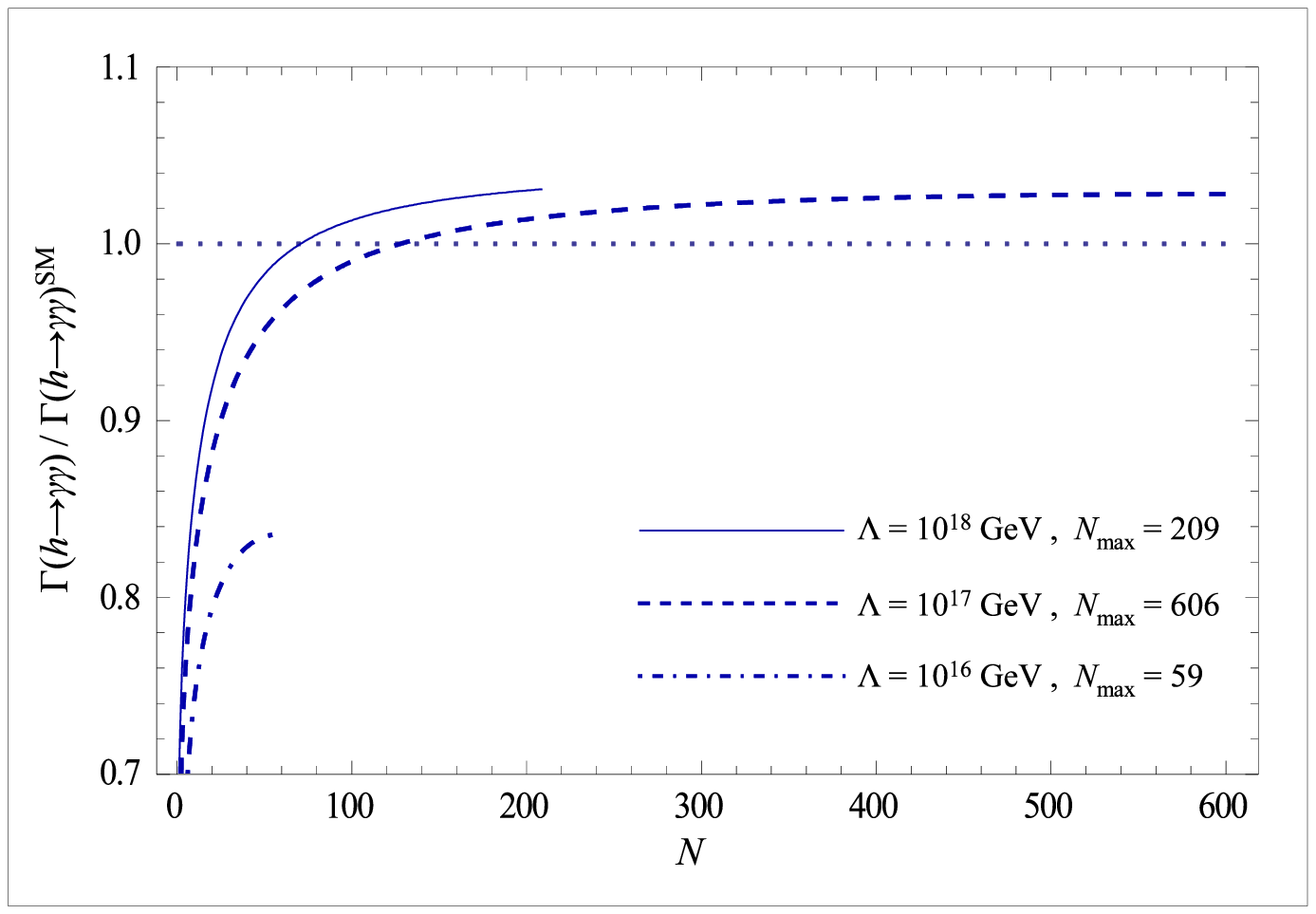} \\
%\framebox{\scalebox{0.67}{\input{gamma}}} \\
a) & b)
\end{tabular}
\caption[Relative coupling parameters $C_t$, $C_V$, and $C_{H^\pm}$, and decay width $\Gamma(h\rightarrow\gamma\gamma)$ depending on number of right-handed neutrino triplets $N$ ]{ \small For three cases $\Lambda=10^{16},\,10^{17},\,10^{18}\,\mathrm{GeV}$ we plot a) the $N$-dependence of relative coupling parameters $C_t$, $C_V$, and $C_{H^\pm}$ defined in \eqref{coupling_parameters};
b) the dependence of the decay width $\Gamma(h\rightarrow\gamma\gamma)$ relative to the Standard Model value $\Gamma(h\rightarrow\gamma\gamma)^\mathrm{SM}$ on $N$. }
\label{plot_couplings}
%\end{center}
\end{figure}

It is mandatory to ask how strongly the candidate for the $125\,\mathrm{GeV}$ resonance, the lighter Higgs scalar $h$, couples to the fermions and the gauge bosons. We study the couplings relative to the Standard Model case of $h$ to $W$ and $Z$ bosons given by $C_V$, to the top-quark given by $C_t$, and to the charged Higgs bosons given by $C_{H^\pm}$, defined in \eqref{coupling_parameters}. We plot their dependence on the number of right-handed neutrino triplets $N$ in Fig.~\ref{plot_couplings}a) for three cases $\Lambda=10^{16},\,10^{17},\,10^{18}\,\mathrm{GeV}$. The scaling coupling factor $C_V$ approaches the Standard Model value for larger $N$ in the cases $\Lambda=10^{17}\,\mathrm{GeV}$ and $\Lambda=10^{18}\,\mathrm{GeV}$. On the other hand, the coupling to the top-quark given by $C_t$ stays rather suppressed in comparison with the Standard Model in all three cases.

All three coupling parameters are relevant for the loop-induced decay of $h$ to two photons. The dependence of the decay width $\Gamma(h\rightarrow\gamma\gamma)$ relative to the Standard Model value \mbox{$\Gamma(h\rightarrow\gamma\gamma)^\mathrm{SM}$} on the number of right-handed neutrino triplets $N$ is plotted in Fig.~\ref{plot_couplings}b). A slight enhancement occurs only for higher values of $N$ for the cases $\Lambda=10^{17}\,\mathrm{GeV}$ and $\Lambda=10^{18}\,\mathrm{GeV}$. The decay widths are calculated using the well known analytic expression \eqref{Gamma_h_gammagamma}.

\begin{table}[t]
\begin{center}
\begin{tabular}{c|cc}
$\Lambda\,[\mathrm{GeV}]$ & $\mu_{t\nu}\,[\mathrm{GeV}]$ & $\tan{\beta}$  \\
\hline
\hline
$10^{16}$ & $44$ & $1.183$  \\
$10^{17}$ & $54$ & $1.215$  \\
$10^{18}$ & $62$ & $1.245$  \\
\hline
\hline
$10^{24}$ & $86$ & $1.401$  \\
$10^{40}$ & $101$ & $1.624$ \\
\hline
\end{tabular}
\end{center}
\caption[Values of $\mu_{t\nu}$ and $\tan{\beta}$ depending on the condensation scale $\Lambda$]{\small Values of $\mu_{t\nu}$ and $\tan{\beta}$ depending on $\Lambda$ while keeping $m_\nu=0.2\,\mathrm{eV}$ and $M_{h}=125\,\mathrm{GeV}$ for $N=1$. }
\label{mutn_tanbeta_lambda}
\end{table}

\begin{table}[t]
\begin{center}
\begin{tabular}{rc|cc|cc}
 &  & $\mu_{t\nu}^\mathrm{min}$  & $\mu_{t\nu}^\mathrm{max}$ & $M_{H^\pm}^\mathrm{min}$ & $M_{H^\pm}^\mathrm{max}$  \\
$\Lambda\,[\mathrm{GeV}]$ \vline& $N^\mathrm{max}$ & $[\mathrm{GeV}]$  & $[\mathrm{GeV}]$ & $[\mathrm{GeV}]$ & $[\mathrm{GeV}]$  \\
\hline
\hline
$10^{16}$\ \ \vline& $59$  & $44$ & $81$  & $66$ & $117$ \\
$10^{17}$\ \ \vline& $606$ & $54$ & $164$ & $80$ & $234$ \\
$10^{18}$\ \ \vline& $209$ & $62$ & $173$ & $92$ & $249$ \\
\hline
\hline
\end{tabular}
\end{center}
\caption[ Number of right-handed neutrino triplets and charged Higgs boson mass depending on the condensation scale $\Lambda$ ]{\small We show the maximum number of right-handed neutrino triplets $N^\mathrm{max}$ for three cases $\Lambda=10^{16},\,10^{17},\,10^{18}\,\mathrm{GeV}$. Next, we show the only intervals for $\mu_{t\nu}$,  $(\mu_{t\nu}^\mathrm{min},\mu_{t\nu}^\mathrm{max})$ corresponding to the interval $(1,N^\mathrm{max})$, allowed by $M_{h}=125\,\mathrm{GeV}$. Finally, we show the corresponding intervals for charged Higgs boson masses $(M_{H^\pm}^\mathrm{min},M_{H^\pm}^\mathrm{max})$. }
\label{mutnmin_mutnmax_lambda}
\end{table}

\subsection{Discussion}\label{SecVII}

In this work we have chosen the two-Higgs doublet model as the next-to-simplest model accommodating the leading idea of our interest -- the top-quark and neutrino condensation as a sufficient source of the electroweak symmetry breaking -- as it was proposed in \cite{Antusch:2002xh}.

Our analysis shows that it is possible to simultaneously reproduce correct values for top-quark mass $m_t$, the electroweak scale $v$, the $125\,\mathrm{GeV}$ boson mass, and the neutrino mass $m_\nu$ below observational upper limit, despite rather limited manoeuvring space for participating parameters \eqref{non_decoupling_cond}. Notice that the system is quite constrained and it could have easily happened that not all of the conditions were satisfied.

The number of right-handed neutrino types participating in the seesaw mechanism is not constrained phenomenologically by any upper limit. The model however exhibits an interesting feature of providing the upper limit on that number.

Next, we present two aspects of the model which are relevant for present phenomenology: the mass spectrum of additional Higgs bosons $H$, $A$ and $H^\pm$, and the coupling strengths of the $125\,\mathrm{GeV}$ Higgs boson to the top-quark and gauge bosons. We study their dependence on $N$ and on the condensation scale $\Lambda$. Generally speaking, the higher values of $N$ and $\Lambda$ are preferred, because they lead to higher values of additional Higgs boson masses, and to the coupling strengths closer to the Standard Model values. For example, for $\Lambda=10^{18}$ and $N=100$ we obtain the charged Higgs boson mass $M_{H^\pm}\doteq223\,\mathrm{GeV}$ and the coupling constant of $h$ to $W$ and $Z$ at $93\%$ level of the Standard Model value.

The confrontation of these two aspects with the experimental constraints in the following two subsections however should be taken with a grain of salt for three reasons. First, the model analyzed in this work is only a semi-realistic model: it ignores the mass generation of fermions other than the top-quark and neutrinos. Second, it is subject to simplification of the neutrino sector. Third, it is not possible to directly link our model to one of the standard types of two-Higgs-doublet models, for which the data analyses are available. They usually deal with full arsenal of Yukawa interactions with charged fermions. In our model, we avoid Yukawa interactions with lighter fermions in favour of the Yukawa coupling of one of the Higgs doublets to right-handed neutrinos.

\subsubsection{Mass of the charged Higgs boson}

The values of $M_{H^\pm}$ and $\tan\beta$ accessible in the model for the higher number of right-handed neutrino triplets $N$ lie in the ranges
\begin{equation}
M_{H^\pm}\simeq(200-250)\mathrm{GeV}\,,\ \ \tan\beta\simeq(1.2-1.25)\,,
\end{equation}
see Tab.~\ref{mutn_tanbeta_lambda} and Tab.~\ref{mutnmin_mutnmax_lambda}.

Direct searches for the charged Higgs boson at LHC give a lower limit for its mass $M_{H^\pm}>160\,\mathrm{GeV}$ \cite{Chatrchyan:2012vca,Aad:2012tj}. The excluded mass interval corresponds to the below-threshold production $t\rightarrow H^+b$. The analyses are made under the assumption of $100\%$ decay ratio via $H^+\rightarrow\tau\nu$ and within one of the MSSM scenario. The lower limit $M_{H^\pm}>160\,\mathrm{GeV}$ translates in our model into the lower limit for $N$
\begin{subequations}\label{Lambda}
\begin{eqnarray}
N>38  \ \ & \mathrm{for} & \ \ \Lambda=10^{17}\,\mathrm{GeV} \,, \\
N>20  \ \ & \mathrm{for} & \ \ \Lambda=10^{18}\,\mathrm{GeV} \,,
\end{eqnarray}
\end{subequations}
while the case $\Lambda=10^{16}\,\mathrm{GeV}$ is excluded.

Indirect searches in $B$-physics are more stringent in setting the lower limits, but they are more model-dependent at the same time. For review see \cite{Branco:2011iw}. For example, the limit $M_{H^\pm}>300-400\,\mathrm{GeV}$ from the $B\rightarrow X_s\gamma$ decay is set for type-II\footnote{The nomenclature of Type I and Type II of two-Higgs-doublet models was introduced in \cite{Hall:1981bc}. The two types differ by the Higgs couplings to fermions. } two-Higgs-doublet model \cite{Donoghue:1978cj}, and the limit $M_{H^\pm}>160\,\mathrm{GeV}$($M_{H^\pm}>500\,\mathrm{GeV}$)\footnote{The limit in parenthesis follows from $B\rightarrow X_s\gamma$ but it is very sensitive to assumptions and to input parameters.} from $Z\rightarrow b\bar{b}$ and from $B$- and $K$-meson mixing is set for type-I two-Higgs-doublet model \cite{Haber:1978jt}. Even though it is rather speculative to apply those constraints to our model, they indicate that the model appears to be at the edge of viability.

\subsubsection{Production and decays of lighter Higgs scalar}

In order to successfully identify the lighter Higgs scalar $h$ with the observed $125\,\mathrm{GeV}$ boson, it should exhibit coupling properties to other particles which lead to the observed phenomena.

Provided the higher number of right-handed neutrino triplets $N>100$ and $\Lambda=10^{18}\,\mathrm{GeV}$, the $h$ scaling coupling factors \eqref{coupling_parameters} characteristic for the model are at the level
\begin{equation}\label{CtCV}
C_t\simeq(0.63-0.69) \,,\ \ C_V\simeq(0.93-0.95) \,,\ \
\end{equation}
see Fig.~\ref{plot_couplings}a).

This result is to be compared with the ATLAS \cite{ATLAS:2012wma} and CMS \cite{CMS:2012wwa} results for corresponding quantities.\footnote{In the ATLAS and CMS analyses the overall fermion scaling factor $C_F$ is used, instead of $C_t$, which scales only the top-quark Yukawa interaction in our model. In any case $C_t=C_F$.} For example, the gluon fusion cross-section $gg\rightarrow h$ scales with a factor $C_{t}^2$ being induced by top-quark loop, or the partial decay width for $h\rightarrow WW$ scales with a factor $C_{V}^2$. The best-fit values over all observed production-decay modes are
\begin{eqnarray}
(C_t,C_V) & \simeq & (1.0,1.2) \ \ \ \mathrm{ATLAS} \,, \\
(C_t,C_V) & \simeq & (0.5,1.0) \ \ \ \mathrm{CMS} \,.
\end{eqnarray}
The point \eqref{CtCV} lies within the ATLAS $95\,\%$ confidence level range and within the CMS less than $68\,\%$ confidence level range. The analyses were made under the assumption that no non-Standard Model particles contribute to the total decay width, what is reasonable assumption for our model, as they all are heavier than the lighter Higgs scalar $h$.

Out of the individual decay channels, we discuss $h\rightarrow\gamma\gamma$. Provided $N>100$ and $\Lambda=10^{18}\,\mathrm{GeV}$, the enhancement of the partial decay width $\Gamma(h\rightarrow\gamma\gamma)$ with respect to the Standard Model can be achieved at the level of $(1-3)\%$, see Fig.~\ref{plot_couplings}b),
\begin{equation}
\frac{\Gamma(h\rightarrow\gamma\gamma)}{\Gamma(h\rightarrow\gamma\gamma)^\mathrm{SM}}\simeq(1.01-1.03) \,.
\end{equation}
The signal strength $\mu$ for $h\rightarrow\gamma\gamma$ channel is measured with a $2\sigma$ excess with respect to the Standard Model \cite{ATLAS:2012wma,CMS:2012wwa}
\begin{equation}
\mu(h\rightarrow\gamma\gamma) = \frac{\sigma_h}{\sigma_{h}^\mathrm{SM}}\frac{\Gamma(h\rightarrow\gamma\gamma)/\Gamma_\mathrm{tot.}}{\Gamma(h\rightarrow\gamma\gamma)^\mathrm{SM}/\Gamma_{\mathrm{tot.}}^\mathrm{SM}}\sim(1.5-2.0)\,.
\end{equation}
The main part of the $h$ production cross section $\sigma_h$ is given by gluon fusion cross-section $\sigma(gg\rightarrow h)$ which scales as $C_{t}^2\simeq(0.39-0.47)$ in our model. This suppression can be however compensated by the suppression of the total decay width $\Gamma_\mathrm{tot.}$, which scales as a linear combination of the $C_{t}^2$ and $C_{V}^2$ factors, both presenting a suppression. According to its parameter setting the model does not profit from the presence of charged Higgs boson in order to enhance significantly the $h\rightarrow\gamma\gamma$ signal strength with respect to the Standard Model prediction.

\section{Viability of the fermion mass generation scenario}\label{SecVIII}

The analysis of the top-quark and neutrino condensation model is relevant for checking suitability of the scenario of the electroweak symmetry breaking by dynamically generated masses of quarks and leptons. The seesaw mechanism characterized by the seesaw scale $M_R\gtrsim10^{14}\,\mathrm{GeV}$ has to be the integral part of such scenario. Then it is possible to simultaneously reproduce a mass spectrum of top-quark, neutrinos, $W$ and $Z$ bosons and observed $125\,\mathrm{GeV}$ boson without any unnatural parameter tension.

The presence of more than a single Higgs doublet is instrumental in achieving lighter neutral scalar $h$ as the candidate for the $125\,\mathrm{GeV}$ particle due to the appropriate mixing among components of the doublets. The coupling strengths of $h$ differ from the Standard Model values, but the experimental data have not a decisive power yet to discriminate among them.

It is the same presence of more than a single Higgs doublet which brings a phenomenological danger of unacceptably light additional Higgs mass eigenstates, like charged and pseudoscalar Higgs bosons. Some indirect constraints on their masses indicate that the model appears to be at the edge of viability.

For lifting their masses in a general two-Higgs doublet model the soft mixing among Higgs doublets $\mu_{12}H_{1}^\dag H_{2}+\hc$ is instrumental \cite{Luty:1990bg,Smetana:2013hm}. Due to the mixing all mass eigenstates but one, the analogue of the Standard Model Higgs boson, acquire the contribution into their masses in the form $\tfrac{2\mu_{12}^2}{\sin2\beta}$, where the mixing angle $\beta$ is defined via the ratio of two Higgs doublet vacuum expectation values $v_1$ and $v_2$ as $\tan\beta\equiv\tfrac{v_2}{v_1}$, compare with \eqref{Higgs_masses} and \eqref{degenerate_MH}. The lifting of the additional Higgs boson masses can proceed by both increasing the mixing parameter $\mu_{12}$ and/or splitting the vacuum expectation values so that $\sin2\beta\ll1$. Via the mixing the additional Higgs bosons can be made arbitrarily well decoupled from a single Standard Model-like Higgs scalar, which stays with its small mass in contact to the electroweak scale.

For example, in the top-quark and neutrino condensation model \cite{Smetana:2013hm} where $\tan\beta=\tfrac{v_t}{v_\nu}\sim1$, see Tab.~\ref{mutn_tanbeta_lambda}, the mixing parameter has to be $\mu_{t\nu}={\cal O}(100)\mathrm{GeV}$ in order to have charged Higgs boson mass above experimental lower limits of several hundreds $\mathrm{GeV}$. In the type-II two composite Higgs doublet model \cite{Luty:1990bg} where the second condensate is formed by bottom-quark, the lifting of the masses of the additional Higgs bosons is provided by $\tan\beta=\tfrac{v_t}{v_b}\sim40$. With the mixing parameter $\mu_{tb}$ of the same magnitude as $\mu_{t\nu}$ the charged Higgs boson mass is larger by factor 5, (in the case of electron instead of bottom-quark, the factor would be even 400). An analogous mechanism in principle could apply in the multi-Higgs doublet model in order to decouple the multitude of Higgs boson mass eigenstates. Therefore it is conceivable that the low-energy effective theory of models of dynamical fermion mass generation is close to the Standard Model.

If we keep only well established and model independent constraints then there remains some room for the top-quark and neutrino condensation scenario, provided that the condensation scale is $\Lambda\sim10^{17-18}\,\mathrm{GeV}$ and the number of right-handed neutrinos participating in the seesaw mechanism is ${\cal O}(100-1000)$.

There are two detail-independent predictions of the scenario. First, the light Higgs boson $h$ has rather big admixture of the neutrinos given by $\alpha\sim-0.8$. The mixing factor suppresses its Yukawa coupling with the top-quark and eventually with other charged fermions at the level of $\sim60\,\%$ in comparison with the Standard Model. Second, the scenario provides an upper limit on the additional Higgs boson masses which is rather low, $<250\,\mathrm{GeV}$. Through both predictions the scenario should be easily and definitely falsifiable by delivering more data from LHC in the near future.

\chapter{Model of strong Yukawa dynamics}
%\pagenumbering{arabic}
%\input{1x.tex}

\label{part_strong_Yukawa_dynamics}

In this chapter we will study our early attempt \cite{Benes:2008ir} to formulate a model of dynamical quark and lepton mass generation. For the purpose of this thesis, it will serve as a playground for investigating directly the dynamical formation of fermion masses from the underlying dynamics and for studying possibilities of achieving the fermion mass hierarchy.

The dynamics which provides the fermion mass generation in this case is the strong Yukawa dynamics -- the attractive force among fermions caused by the exchange of scalar boson fields. We will show that if the Yukawa coupling constants are strong enough then there exists a non-trivial and electroweak symmetry breaking solution of gap equations. However rather than using the fermion condensates and their gap equations we will formulate the mass generation in terms of chirality-changing momentum-dependent fermion self-energies being subjects of the Schwinger--Dyson equations.

We will demonstrate that in principle the electroweak symmetry breaking can be achieved without developing the vacuum expectation value of the elementary scalar fields. Instead the symmetry can be broken by a formation of the symmetry breaking parts of two-point Green's functions, i.e., the fermion and also the scalar propagators. In order to study this particular phenomenon, we will completely avoid the eventuality that the elementary scalar fields develop their vacuum expectation values, not even dynamically. %We are however aware of the fact that in the full treatment of the model, once the electroweak symmetry is spontaneously broken, all Green's functions allowed by the remaining unbroken symmetries will manifest, including the one-point Green's functions, i.e., including the vacuum expectation values of neutral scalar fields.

There is an indisputable advantage in invoking Yukawa interactions as the underlying dynamics. The Lagrangian containing only the gauge bosons and usual fermions would possess large global chiral symmetry. Fermion masses would break this large symmetry spontaneously giving rise to a number of unwanted Nambu--Goldstone bosons. To avoid this, the Yukawa interactions are very useful as they have the power to break completely the chiral symmetry. The price to be paid is however as high as in the Standard Model, the number of parameters participating in the fermion mass generation is at least the same as the number of fermion masses themselves.

However the benefit of this approach in comparison with the Standard Model lies in the possibility to achieve large fermion mass hierarchy while having moderate hierarchy among the Yukawa parameters. This is clearly not an over-idealistic aim. The fact that the Yukawa dynamics is strongly coupled promises the critical scaling of fermion masses. With respect to some characteristic scale of the theory, $M$, the fermion masses $m$ can be made in principle arbitrarily small, $m=f(y)M$, by some scaling function $f(y)$ depending on the Yukawa parameters $y$. This is in analogy with the scaling within the QCD or TC \eqref{QCD_critical_scaling}, or with the critical scaling of top-quark mass in the Top-quark Condensation model \eqref{tqc_critical_scaling}. The same scaling is conceivable to apply for the fermion mass ratios, because the ratio $m_1/m_2=f_1(y_1)/f_2(y_2)$ typically exhibits a critical behavior as well.

The mass scale $M$ in this model however does not occur dynamically. It is introduced in the form of bare masses of the two complex scalar doublets via their hard mass terms. In this respect the model is not better than the Standard Model with its $\mu$ parameter in the Higgs potential \eqref{Higgs_potential}. The model of strong Yukawa dynamics can underlie the mass generation but being not UV complete it should not be understood as the fundamental theory.

\section{The model}

Defining the model, we keep the same field content as in the Standard Model including usual fermions and gauge bosons, but we replace the Higgs doublet field by two complex massive scalar doublet fields $N$ and $S$. Also for the sake of aesthetics and in order to address neutrino masses, we introduce three right-handed neutrino fields $\nu_R$.

The scalar fields $N$ and $S$ couple to the gauge bosons of $\SU{3}_c\times\SU{2}_L\times\U{1}_Y$ according to their representation setting:
\begin{equation}
N:\ (\mathbf{1},\mathbf{2},+1)\,,\ \ \ S:\ (\mathbf{1},\mathbf{2},-1)\,.
\end{equation}
This means that they can be written in electric charge eigenstate components as
\begin{eqnarray}
N = \left( \begin{array}{c} N^{0} \\ N^{-} \end{array} \right)
\,,\quad\quad S = \left( \begin{array}{c} S^{+} \\ S^{0}
\end{array} \right) \,.
\end{eqnarray}
The scalar Lagrangian including their gauge interactions is written as
\begin{equation}
{\cal L}_\mathrm{scalar} = (D^\mu N)^\dag D_\mu N - M_{N}^2N^\dag N + (D^\mu S)^\dag D_\mu S - M_{S}^2S^\dag S \,,
\end{equation}
where $D_\mu$ is the covariant derivative
\begin{equation}
D_\mu=\partial_\mu-\im g' \frac{Y}{2}B_\mu-\im g T_{L}^a A^{a}_\mu \,.
\end{equation}
%providing the scalar interactions with the electroweak gauge bosons of hypercharge, $B_\mu(x)$, and left isospin, $A^{a}_\mu(x)$.
The parameters $M_{N}$ and $M_{S}$ are the bare masses of the scalar doublet fields $N$ and $S$.

The tree-level scalar potential is very simple:
\begin{equation}
{\cal V}_0 = M_{N}^2N^\dag N +  M_{S}^2S^\dag S \,.
\end{equation}
It is designed according to our intention to avoid the scalar vacuum expectation value at least at the tree-level. Therefore, contrary to the Higgs boson doublet in the Standard Model, the mass terms stand with the ``proper'' sign, provided that $M_{N,S}^2>0$. As the mass terms themselves form the potential which is already stable, there is in fact no need to stabilize it by quartic coupling terms. Because our purpose is to study the effect of the Yukawa interactions, we deliberately do not introduce any of the quartic terms for the sake of simplicity. We are aware of the fact that in the full treatment of the model, they would be generated as counter terms.

For our purpose the Yukawa interactions of the scalar doublet fields with fermions are instrumental. Here the weak hypercharge assignment determines that the `northern' scalar $N$ couples to the up-type right-handed fermions $u_R$ and $\nu_R$, while the `southern' scalar $S$ couples to the down-type right-handed fermions $d_R$ and $e_R$. The Yukawa Lagrangian is written as
\begin{equation}
{\cal L}_\mathrm{Yukawa} = Y^{ij}_u\bar q^{i}_Lu^{j}_RN + Y^{ij}_d\bar q^{i}_Ld^{j}_RS + Y^{ij}_\nu\bar\ell^{i}_L\nu^{j}_RN + Y^{ij}_e\bar\ell^{i}_Le^{j}_RS + \hc \,,
\end{equation}
where the indices $i,j=1,\dots,3$ label three generations of fermions, i.e., $u^{j}\in\{u,c,t\}$, $d^{j}\in\{d,s,b\}$, $e^{j}\in\{e,\mu,\tau\}$ and $\nu^{j}\in\{\nu_e,\nu_\mu,\nu_\tau\}$. Additionally we use the notation $q^{i}_L=\beginm{c}u^{i}_L\\d^{i}_L\endm$ and $\ell^{i}_L=\beginm{c}\nu^{i}_L \\ e^{i}_L\endm$.

Actually, ${\cal L}_\mathrm{Yukawa}$ is analogous to the Yukawa interactions in the MSSM. Also like in the MSSM we do not consider the interactions of charge conjugated scalar fields. In our model they could have been obtained from ${\cal L}_\mathrm{Yukawa}$ after substitution $N\rightarrow\im\sigma_2S^*$, $S\rightarrow\im\sigma_2N^*$ and $Y\rightarrow Y'$. These terms, otherwise allowed by the gauge symmetries, can be forbidden by imposing an invariance under the special parity transformation $(d_R,e_R,S)\rightarrow-(d_R,e_R,S)$, for instance.

In general, the Yukawa coupling constants are complex matrices. For our purpose however we will content ourselves with diagonal matrices with real entries, $Y_u=\mathrm{Diag}\{y_u,y_c,y_t\}$, $Y_d=\mathrm{Diag}\{y_d,y_s,y_b\}$, $Y_e=\mathrm{Diag}\{y_e,y_\mu,y_\tau\}$ and $Y_\nu=\mathrm{Diag}\{y_{\nu_e},y_{\nu_\mu},y_{\nu_\tau}\}$.

The model is defined by the Lagrangian
\begin{equation}
{\cal L} = {\cal L}_\mathrm{usual} + {\cal L}_\mathrm{scalar} + {\cal L}_\mathrm{Yukawa} \,.
\end{equation}
The Lagrangian ${\cal L}_\mathrm{usual}$ contains pure terms of gauge bosons, kinetic terms of fermions and the interaction terms of both. The Lagrangians ${\cal L}_\mathrm{scalar}$ and ${\cal L}_\mathrm{Yukawa}$ substitute the Standard Model part of Lagrangian containing the Higgs doublet field.

\subsection{Spectrum in the weak coupling regime}

In the regime of weak Yukawa couplings both the generation of fermion masses and the electroweak symmetry breaking do not occur. The particle spectrum of the model in this regime consists of massless fermions, massless gauge bosons, and massive complex scalar bosons with tree-level masses
\begin{eqnarray}\label{bare_mass_spectrum_X}
&& M_{N^{\pm}}=M_{N^{0}}=M_N\,, \\
&& M_{S^{\pm}}=M_{S^{0}}=M_S\,.
\end{eqnarray}
This is encoded in the bare propagators of the corresponding fields.
%For our purpose of studying fermion mass spectrum, we will not consider the electroweak boson mass generation having in mind that it comes out automatically.
The fermion inverse bare propagators are massless
\begin{equation}
\big[S^{0}_\psi(p)\big]^{-1}\equiv\langle\psi\bar\psi\rangle_{0}^{-1}=\slesp{p} \,,
\end{equation}
where $\psi=u^i,d^i,e^i,\nu^i$ denotes all types of fermions in the model.
For later convenience it is better to define the bare scalar boson inverse propagators for neutral and for charged components separately.
\begin{subequations}\label{scalar_propagator_bare}
\begin{eqnarray}
\big[D^{0}_N(p^2)\big]^{-1}&\equiv&\langle\Phi_N\Phi_{N}^\dag\rangle_{0}^{-1} = \left( \begin{array}{cc}
p^2 - M_{N}^2 & 0 \\
0 & p^2 - M_{N}^2
\end{array} \right)\,,
\\
\big[D^{0}_S(p^2)\big]^{-1}&\equiv&\langle\Phi_S\Phi_{S}^\dag\rangle_{0}^{-1} = \left( \begin{array}{cc}
p^2 - M_{S}^2 & 0 \\
0 & p^2 - M_{S}^2
\end{array} \right)\,,
\\
\big[D^{0}_{NS}(p^2)\big]^{-1}&\equiv&\langle\Phi_{NS}\Phi_{NS}^\dag\rangle_{0}^{-1} = \left( \begin{array}{cc}
p^2 - M_{S}^2 & 0 \\
0 & p^2 - M_{N}^2
\end{array} \right)\,,
\end{eqnarray}
\end{subequations}
written in the Nambu--Gorkov-like basis
\begin{equation}
\Phi_{N} = \left( \begin{array}{c} N^{0} \\ N^{0\dag}
\end{array} \right) \,,\quad \Phi_{S} = \left( \begin{array}{c}
S^{0} \\ S^{0\dag} \end{array} \right) \,,\quad \Phi_{NS} =
\left( \begin{array}{c} S^{+} \\ N^{-\dag} \end{array} \right) \,.
\end{equation}
We use the notation reflecting the electric charge conservation, where $S^+\equiv S^{-\dag}$ and $N^+\equiv N^{-\dag}$.

Now, the fermion mass generation can be formulated in the spirit of the Nambu's and Jona-Lasinio's pioneering work \cite{Nambu:1961tp}. First the masses are assumed and then their existence is approved by finding corresponding non-trivial solutions of dynamical equations.

\subsection{Mass assumptions}

We assume that the electroweak symmetry breaking self-energies $\mathbf{\Sigma}_\psi(p^2)$ for all fermions $\psi=u^i,d^i,e^i,\nu^i$ are actually generated. We do not consider any flavor mixing of fermions and for every fermion including neutrinos we assume only Dirac type self-energy for simplicity. Thus the self-energy for a given fermion $\psi$ has the form of \eqref{Sigma_obecne}
\begin{equation}\label{Sigma_Yukawa}
\mathbf{\Sigma}_\psi(p^2)\equiv\langle\psi_R\bar\psi_L\rangle_{1\mathrm{PI}}+\hc=\Sigma_\psi(p^2)P_L+\Sigma_{\psi}^*(p^2)P_R \,,
\end{equation}
where $\Sigma_\psi(p^2)$ are complex non-matrix functions. Under the assumption of existence of $\mathbf{\Sigma}_\psi(p^2)$ and under the simplification of neglecting the fermion wave function renormalization we can write for the fermion inverse full propagator
\begin{equation}\label{inverse_fermion_propagator_Yukawa}
\big[S_\psi(p)\big]^{-1}\ \equiv\ \langle\psi\bar\psi\rangle^{-1}\ =\ \big[S^{0}_\psi(p)\big]^{-1}-\big(\langle\psi_R\bar\psi_L\rangle_{1\mathrm{PI}}+\hc\big)
\ =\ \slesp{p}-\mathbf{\Sigma}_{\psi}(p^2) \,.
\end{equation}
Inverting the formula we get \eqref{fermion_propagator}
\begin{equation}
S_\psi(p) = \frac{\slesp{p}+\mathbf{\Sigma}_{\psi}^*(p^2)}{p^2-|\Sigma_{\psi}(p^2)|^2} \,.
\end{equation}
The Dirac mass $m_\psi$ of a fermion $\psi$ is given as a pole of the propagator, thus its square solves the equation
\begin{equation}\label{secular_psi}
p^2-|\Sigma_{\psi}(p^2)|^2=0
\end{equation}
with respect to $p^2$.
Clearly the full propagator can be split into the chiral symmetry preserving, $\propto\slesp{p}$, and chiral symmetry breaking, $\propto\mathbf{\Sigma}_{\psi}^*(p^2)$, parts. For latter convenience of writting the Schwinger--Dyson equation in a more compact form, we denote the chiral symmetry breaking part of the fermion propagator as\footnote{Sometimes we deliberately suppress the square power in momentum argument for the sake of saving a place in the following Schwinger--Dyson equations. The actual dependence on either $p^2$ or $p^\mu$ should be clear from the corresponding definition of a given function. }
\begin{equation}\label{SB_fermion_propagators}
\big[S^{\Sigma}_\psi(p)\big]^* \equiv \frac{\mathbf{\Sigma}_{\psi}^*(p^2)}{p^2-|\Sigma_{\psi}(p^2)|^2} \,.
\end{equation}

\begin{figure}
  \centering
  \includegraphics[width=0.5\textwidth]{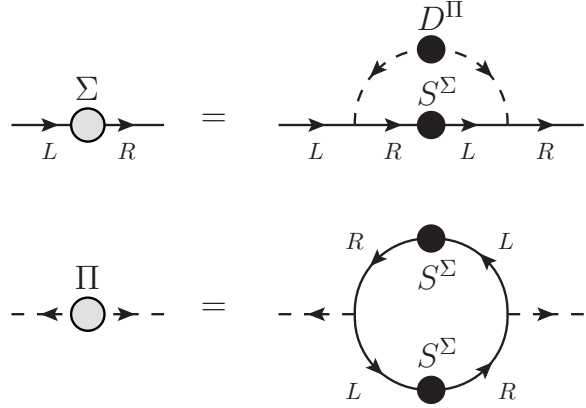}
  \caption[Schwinger--Dyson equations within the model of strong Yukawa dynamics]{The figure shows the structure of the Schwinger--Dyson equations \eqref{SDE_Yukawa_Minkowski}. The letters $L$ and $R$ denote the left- and right-handed fermions and $S^\Sigma$ and $D^\Pi$ denote the chiral symmetry breaking parts of the fermion \eqref{SB_fermion_propagators} and scalar \eqref{SB_scalar_propagators} propagators. On the right-hand sides of the equations there are loop expressions for $\Sigma(p^2)$ and $\Pi(p^2)$ self-energies. We can see how the existence of either of the self-energies brings about the existence of the other. }\label{SDE_Yukawa}
\end{figure}

So far we have only assumed the actual existence of the electroweak symmetry breaking self-energy $\mathbf{\Sigma}_{\psi}(p^2)$.
%and the symmetry breaking part of the propagator $S^{\Sigma}_\psi(p)$.
To justify its existence dynamically we should be able to construct loop expressions for $\mathbf{\Sigma}_{\psi}(p^2)$. This is possible only under assumption that the symmetry breaking parts of the complex boson propagators are generated as well. In the pictorial representation of the loop expressions in Fig.~\ref{SDE_Yukawa}, it is exhibited by boson propagators with unusual arrows providing an anomalous boson mixing $\langle N^{0}N^{0}\rangle$, $\langle S^{0}S^{0}\rangle$ and $\langle S^{+}N^{-}\rangle$. The formation of the anomalous scalar two-point Green's functions lies in the heart of the model.

We define the symmetry breaking parts $\mathbf{\Pi}_N(p^2)$, $\mathbf{\Pi}_S(p^2)$ and $\mathbf{\Pi}_{NS}(p^2)$ of the full scalar propagators as
\begin{eqnarray}
\mathbf{\Pi}_N(p^2)&\equiv&
\left( \begin{array}{cc} 0 & \langle N^{0}N^{0}\rangle_{1\mathrm{PI}} \\ \langle N^{0\dag}N^{0\dag}\rangle_{1\mathrm{PI}} & 0 \end{array} \right) =
\left( \begin{array}{cc} 0 & \Pi_N(p^2) \\ \Pi_{N}^{*}(p^2) & 0 \end{array} \right) \,, \\
\mathbf{\Pi}_S(p^2)&\equiv&
\left( \begin{array}{cc} 0 & \langle S^{0}S^{0}\rangle_{1\mathrm{PI}} \\ \langle S^{0\dag}S^{0\dag}\rangle_{1\mathrm{PI}} & 0 \end{array} \right) =
\left( \begin{array}{cc} 0 & \Pi_S(p^2) \\ \Pi_{S}^{*}(p^2) & 0 \end{array} \right) \,, \\
\mathbf{\Pi}_{NS}(p^2)&\equiv&
\left( \begin{array}{cc} 0 & \langle S^{+}N^{-}\rangle_{1\mathrm{PI}} \\ \langle N^{+}S^{-}\rangle_{1\mathrm{PI}} & 0 \end{array} \right) =
\left( \begin{array}{cc} 0 & \Pi_{NS}(p^2) \\ \Pi_{NS}^{*}(p^2) & 0 \end{array} \right) \,.
\end{eqnarray}
Under the assumption of the existence of $\mathbf{\Pi}_{X}(p^2)$, $X=N,S,NS$,\footnote{In the following we will use the index $X$ to label all three types of anomalous scalar self-energies $N,S,NS$. When it is necessary to denote just the scalars $N,S$ we will use the label $Y$. } and again under the simplification of neglecting the scalar wave function renormalization we can write for the scalar inverse full propagators
\begin{equation}
\big[D_X(p)\big]^{-1}\equiv\langle\Phi_X\Phi_{X}^\dag\rangle^{-1}=\big[D^{0}_X(p)\big]^{-1}-\mathbf{\Pi}_{X}(p^2) \,.
\end{equation}
Writing it in more detail we get
\begin{eqnarray}
\big[D_N(p^2)\big]^{-1} &\equiv&
\left( \begin{array}{cc}
\langle N^{0}N^{0\dag}\rangle_{0}^{-1} & \langle N^{0}N^{0}\rangle_{1\mathrm{PI}} \\
\langle N^{0\dag}N^{0\dag}\rangle_{1\mathrm{PI}} & \langle N^{0\dag}N^{0}\rangle_{0}^{-1}
\end{array} \right) =
\left( \begin{array}{cc} p^2 - M_{N}^2 & -\Pi_N(p^2) \\ -\Pi_{N}^{*}(p^2) & p^2 - M_{N}^2 \end{array} \right) \,, \\
\big[D_S(p^2)\big]^{-1} &\equiv&
\left( \begin{array}{cc}
\langle S^{0}S^{0\dag}\rangle_{0}^{-1} & \langle S^{0}S^{0}\rangle_{1\mathrm{PI}} \\
\langle S^{0\dag}S^{0\dag}\rangle_{1\mathrm{PI}} & \langle S^{0\dag}S^{0}\rangle_{0}^{-1}
\end{array} \right) =
\left( \begin{array}{cc} p^2 - M_{S}^2 & -\Pi_S(p^2) \\ -\Pi_{S}^{*}(p^2) & p^2 - M_{S}^2 \end{array} \right) \,, \\
\big[D_{NS}(p^2)\big]^{-1} &\equiv&
\left( \begin{array}{cc} \langle S^{+}S^{-}\rangle_{0}^{-1} & \langle S^{+}N^{-}\rangle_{1\mathrm{PI}} \\ \langle N^{+}N^{-}\rangle_{1\mathrm{PI}} & \langle N^{+}S^{-}\rangle_{0}^{-1} \end{array} \right) =
\left( \begin{array}{cc} p^2 - M_{S}^2 & -\Pi_{NS}(p^2) \\ -\Pi_{NS}^{*}(p^2) & p^2 - M_{N}^2 \end{array} \right) \,.\quad\quad
\end{eqnarray}
Inverting the formulae we get
\begin{eqnarray}
D_N(p^2) &=&
\left( \begin{array}{cc}
p^2 - M_{N}^2 & \Pi_N(p^2) \\ \Pi_{N}^{*}(p^2) & p^2 - M_{N}^2
\end{array} \right)\frac{1}{(p^2-M_{N}^2)^2-|\Pi_N|^2} \,, \\
D_S(p^2) &=&
\left( \begin{array}{cc}
p^2 - M_{S}^2 & \Pi_S(p^2) \\ \Pi_{S}^{*}(p^2) & p^2 - M_{S}^2
\end{array} \right)\frac{1}{(p^2-M_{S}^2)^2-|\Pi_S|^2} \,, \\
D_{NS}(p^2) &=&
\left( \begin{array}{cc}
p^2 - M_{N}^2 & \Pi_{NS}(p^2) \\ \Pi_{NS}^{*}(p^2) & p^2 - M_{S}^2
\end{array} \right)\frac{1}{(p^2-M_S^2)(p^2-M_N^2)-|\Pi_{NS}|^2} \,.
\end{eqnarray}
The mass spectrum of scalars after the symmetry breaking gets substantially changed. The squared masses are derived as a set of poles of the scalar propagators, thus as solutions of the equations
\begin{subequations}\label{secular_equation_X}
\begin{eqnarray}
(p^2-M_{N}^2)^2-|\Pi_N(p^2)|^2 &=& 0 \,, \label{secular_equation_N} \\
(p^2-M_{S}^2)^2-|\Pi_S(p^2)|^2 &=& 0 \,, \label{secular_equation_S} \\
(p^2-M_S^2)(p^2-M_N^2)-|\Pi_{NS}(p^2)|^2 &=& 0 \, \label{secular_equation_SN}
\end{eqnarray}
\end{subequations}
with respect to $p^2$.
Clearly the full propagators can be split into the symmetry preserving, diagonal, and symmetry breaking, off-diagonal, matrix elements. For latter convenience we denote the symmetry breaking propagators as
\begin{subequations}\label{SB_scalar_propagators}
\begin{eqnarray}
D_N^{\Pi}(p) &=& \frac{\Pi_N(p^2)}{(p^2-M_{N}^2)^2-|\Pi_N(p^2)|^2} \,, \label{propagator_N} \\
D_S^{\Pi}(p) &=& \frac{\Pi_S(p^2)}{(p^2-M_{S}^2)^2-|\Pi_S(p^2)|^2} \,, \label{propagator_S} \\
D_{NS}^{\Pi}(p) &=& \frac{\Pi_{NS}(p^2)}{(p^2-M_S^2)(p^2-M_N^2)-|\Pi_{NS}(p^2)|^2} \,. \label{propagator_SN}
\end{eqnarray}
\end{subequations}

So far we have only assumed the existence of the symmetry breaking self-energies $\mathbf{\Pi}_{X}(p^2)$ and the symmetry breaking parts of the propagators $D_{X}^{\Pi}(p)$. To justify them dynamically we should be able to construct the loop expression for $\mathbf{\Pi}_{X}(p^2)$. But for that we already have the ingredients. As can be seen in Fig.~\ref{SDE_Yukawa}, what is needed is just the symmetry breaking fermion self-energies $\mathbf{\Sigma}_\psi(p^2)$.

Notice that, assuming that $\Pi_X(p^2)\rightarrow0$, the symmetry breaking parts of the propagators $D_{X}^{\Pi}(p)$ defined in \eqref{SB_scalar_propagators} behave like
\begin{equation}
D_{X}^{\Pi}(p)\ \stackrel{p^2\rightarrow\infty}{\longrightarrow}\ \frac{\Pi_X(p^2)}{p^4} \,
\end{equation}
for asymptotically large momenta. This is extremely important as it improves the convergence properties of integrals in the Schwinger--Dyson equations. The resulting self-energies $\Sigma_\psi(p^2)$ and $\Pi_X(p^2)$ then come out perfectly finite as they should, because there are no counter terms allowed by symmetries of the Lagrangian which would renormalize them. The convergent behavior can be better appreciated by realizing that the anomalous propagators \eqref{SB_scalar_propagators} can be rewritten as a difference of two ``standard'' propagators, e.g.,
\begin{equation}\label{difference_propagator}
  D_N^{\Pi}(p) = \frac{1}{2}\left(\frac{1}{p^2-M_{N}^2-|\Pi_N(p^2)|}-\frac{1}{p^2-M_{N}^2+|\Pi_N(p^2)|}\right) \,.
\end{equation}

Under the simplifying assumption of constant and real-valued self-energies $\Sigma_\psi(p^2)=m_\psi$ and $\Pi_X(p^2)=\mu_{X}^2$, we can see their effect on the mass spectrum. The equations \eqref{secular_psi} and \eqref{secular_equation_X} are then easily solved. The fermion mass is simply given directly by the magnitude of the self-energy $m_\psi$. The scalar mass spectrum \eqref{bare_mass_spectrum_X} is split due to the presence of the self-energy $\Pi_X(p^2)$. For $Y=N,S$
\begin{subequations}\label{scalar_mass_spectrum}
\begin{eqnarray}
M_{Y^{0}_{1,2}}^2 & = & M_{Y}^2\mp\mu_{Y}^2 \,, \\
M_{C^{+}_{1,2}}^2 & = & \frac{1}{2}\left[M_{N}^2+M_{S}^2\mp\sqrt{\big(M_{N}^2-M_{S}^2\big)^2+4\mu_{NS}^4}\,\right] \,,
\end{eqnarray}
\end{subequations}
the neutral mass eigenstates are
\begin{eqnarray}
Y^{0}_{1} = \sqrt{2}\Im Y^{0} \,, &\ & Y^{0}_{2} = \sqrt{2}\Re Y^{0} \,
\end{eqnarray}
and the charge mass eigenstates are
\begin{eqnarray}
C^{\pm}_{1} & = & \cos\alpha_C S^{\pm}-\sin\alpha_C N^{\pm} \,, \\
C^{\pm}_{2} & = & \sin\alpha_C S^{\pm}+\cos\alpha_C N^{\pm} \,,
\end{eqnarray}
where
\begin{equation}
\tan\alpha_C=\frac{-(M_{N}^2-M_{S}^2)+\sqrt{\big(M_{N}^2-M_{S}^2\big)^2+4\mu_{NS}^4}}{2\mu_{NS}^2} \,.
\end{equation}

\subsection{Mass generation}

Now we employ the dynamics in order to approve the generation of the self-energies which have been merely assumed in the previous subsection. For this purpose we use the Schwinger--Dyson equations depicted in Fig.~\ref{SDE_Yukawa}.
\begin{subequations}\label{SDE_Yukawa_Minkowski}
\begin{eqnarray}
\Sigma_{U^{i}}(p) & = & \im y_{U^{i}}^2\int\frac{\mathrm{d}^4k}{(2\pi)^4}\,\big[S_{U^{i}}^\Sigma(k)\big]^*\,D_{N}^\Pi(k-p)
                             +\im y_{U^{i}} y_{D^{i}}\int\frac{\mathrm{d}^4k}{(2\pi)^4}\,\big[S_{D^{i}}^\Sigma(k)\big]^*\,D_{NS}^\Pi(k-p) \,,\hspace{1.3cm} \label{1SDE_Minkowski} \\
\Sigma_{D^{i}}(p) & = & \im y_{D^{i}}^2\int\frac{\mathrm{d}^4k}{(2\pi)^4}\,\big[S_{D^{i}}^\Sigma(k)\big]^*\,D_{S}^\Pi(k-p)
                             +\im y_{D^{i}} y_{U^{i}}\int\frac{\mathrm{d}^4k}{(2\pi)^4}\,\big[S_{U^{i}}^\Sigma(k)\big]^*\,D_{NS}^\Pi(k-p) \,, \\
\Pi_N(p) & = & -2\im\sum_{i}N_\mathrm{C}y_{U^{i}}^2\int\frac{\mathrm{d}^4k}{(2\pi)^4}\,S_{U^{i}}^{\Sigma}(k)\,S_{U^{i}}^{\Sigma}(k-p) \,, \\
\Pi_S(p) & = & -2\im\sum_{i}N_\mathrm{C}y_{D^{i}}^2\int\frac{\mathrm{d}^4k}{(2\pi)^4}\,S_{D^{i}}^{\Sigma}(k)\,S_{D^{i}}^{\Sigma}(k-p) \,, \\
\Pi_{NS}(p) & = & -2\im\sum_{i}N_\mathrm{C}y_{D^{i}} y_{U^{i}}\int\frac{\mathrm{d}^4k}{(2\pi)^4}\,S_{U^{i}}^{\Sigma}(k)\,S_{D^{i}}^{\Sigma}(k-p) \,,
\end{eqnarray}
\end{subequations}
where $U^i$ denotes six up-type fermions, $u^i$ and $\nu^i$, $D^i$ denotes six down-type fermions, $d^i$ and $e^i$, and $N_\mathrm{C}$ is the number of colors, $N_\mathrm{C}=3$ for quarks and $N_\mathrm{C}=1$ for leptons.
These are the Schwinger--Dyson equations which can be obtained more rigorously by means of the Cornwall--Jackiw--Tomboulis formalism \cite{Cornwall:1974vz}. The details of derivation can be found in \cite{Benes:2012hz}.

Several important notes are in order:
\begin{itemize}
  \item The model leads to the system of coupled integral equations which should be solved simultaneously.
  \item Free parameters of the system of equations are the Yukawa coupling constants $y_{U^{i}}$ and $y_{D^{i}}$, and the scalar bare masses $M_N$ and $M_S$. Whole mass spectrum is in principle calculable in terms of these parameters.
  \item The exchange of massive scalars induces the amplitudes of flavor-changing neutral currents. They can be suppressed by setting the scalar bare masses to be large, presumably $M_{N,S}>10^{6}\,\mathrm{GeV}$. In that case, to reproduce the fermion mass spectrum we need a huge hierarchy $\Sigma_\psi(p^2)\ll M_{N,S}$ which should be provided by a critical scaling as a result of the equations. The scalar self-energies are given by the integrals of expressions composed by fermion self-energies. Therefore it is reasonable to expect their similar order of magnitude, $\Sigma_\psi^2(p^2)\sim\Pi_X(p^2)$, and thus also
      \begin{equation}\label{scalar_splitting}
      \Pi_X(p^2)\ll M_{N,S}^2 \,.
      \end{equation}
      According to \eqref{scalar_mass_spectrum}, we expect that all scalars come out similarly heavy with splitting of the order of the electroweak scale.
  \item To solve the equations is an extremely complex issue. We will therefore resort to approximations described in the next section and in appendix~\ref{approx_methods_for_SDE}. The first approximation is that we will study the Wick-rotated Schwinger--Dyson equations and assume that the self-energies $\Sigma_\psi(p^2)$ and $\Pi_X(p^2)$ are real-valued:
      \begin{subequations}\label{SDE_Yukawa_Euclid}
      \begin{eqnarray}
      && \hspace{-0.8cm} \Sigma_{U^{i}}(p) = y_{U^{i}}^2\int\frac{\mathrm{d}^4k}{(2\pi)^4}
                    \frac{\Sigma_{U^{i}}(k)}{k^2+\Sigma_{U^{i}}^2(k)}\frac{\Pi_N(k-p)}{[(k-p)^2+M_{N}^2]^2-\Pi_N^2(k-p)} \label{1SDE_Euclidean}\\
              &  &\hspace{-0.4cm} +y_{U^{i}} y_{D^{i}}\int\frac{\mathrm{d}^4k}{(2\pi)^4}
                    \frac{\Sigma_{D^{i}}(k)}{k^2+\Sigma_{D^{i}}^2(k)}\frac{\Pi_{NS}(k-p)}{[(k-p)^2+M_{N}^2][(k-p)^2+M_{S}^2]-\Pi_{NS}^2(k-p)} \,,\nonumber\\
      && \hspace{-0.8cm} \Sigma_{D^{i}}(p) = y_{D^{i}}^2\int\frac{\mathrm{d}^4k}{(2\pi)^4}
                    \frac{\Sigma_{D^{i}}(k)}{k^2+\Sigma_{D^{i}}^2(k)}\frac{\Pi_S(k-p)}{[(k-p)^2+M_{S}^2]^2-\Pi_S^2(k-p)} \\
              &  &\hspace{-0.4cm} +y_{D^{i}} y_{U^{i}}\int\frac{\mathrm{d}^4k}{(2\pi)^4}
                    \frac{\Sigma_{U^{i}}(k)}{k^2+\Sigma_{U^{i}}^2(k)}\frac{\Pi_{NS}(k-p)}{[(k-p)^2+M_{N}^2][(k-p)^2+M_{S}^2]-\Pi_{NS}^2(k-p)} \,,\nonumber\\
      && \hspace{-0.8cm} \Pi_N(p) = 2\sum_{i}N_\mathrm{C}y_{U^{i}}^2\int\frac{\mathrm{d}^4k}{(2\pi)^4}
                    \frac{\Sigma_{U^{i}}(k)}{k^2+\Sigma_{U^{i}}^2(k)}\frac{\Sigma_{U^{i}}(k-p)}{(k-p)^2+\Sigma_{U^{i}}^2(k-p)} \,,\\
      && \hspace{-0.8cm} \Pi_S(p) = 2\sum_{i}N_\mathrm{C}y_{D^{i}}^2\int\frac{\mathrm{d}^4k}{(2\pi)^4}
                    \frac{\Sigma_{D^{i}}(k)}{k^2+\Sigma_{D^{i}}^2(k)}\frac{\Sigma_{D^{i}}(k-p)}{(k-p)^2+\Sigma_{D^{i}}^2(k-p)} \,,\\
      && \hspace{-0.8cm} \Pi_{NS}(p)= 2\sum_{i}N_\mathrm{C}y_{U^{i}}y_{D^{i}}\int\frac{\mathrm{d}^4k}{(2\pi)^4}
                    \frac{\Sigma_{U^{i}}(k)}{k^2+\Sigma_{U^{i}}^2(k)}\frac{\Sigma_{D^{i}}(k-p)}{(k-p)^2+\Sigma_{D^{i}}^2(k-p)} \,.
      \end{eqnarray}
      \end{subequations}
  \item By the Wick rotation we have got rid of the otherwise unavoidable mass poles present in the Minkowski propagators. Another type of potential poles however still survives. These are the poles in the scalar propagators eventually appearing due to the minus signs in the denominators. Invoking the physical assumption \eqref{scalar_splitting} we keep the equations safely far from that case.
  \item Assuming that the self-energies $\Sigma_\psi(p^2)$ and $\Pi_X(p^2)$ fall off with momentum, we can see that the loop integrals are finite as they should be because there are no counter terms allowed by symmetries of the Lagrangian which would renormalize them. The assumption of the decrease of self-energies is reasonable, as it can be seen from the Wick rotated equations: Starting with $\Sigma_\psi(p^2)=\mathrm{const.}$ and $\Pi_X(p^2)=\mathrm{const.}$, the integrals in the fermion equations are finite (the order of integrand is $\sim k^{-6}$) and they imply that $\Sigma_\psi(p^2)\approx p^{-2}$ for $p^2\rightarrow\infty$. Plugging such $\Sigma_\psi(p^2)$ into the scalar equations leads to the finite integrals there and it implies that $\Pi_X(p^2)\approx p^{-4}$ for $p^2\rightarrow\infty$. Plugging this $\Pi_X(p^2)$ into the fermion equations back again, it improves the convergent properties of the integral even more and leads to even faster decrease of the $\Sigma_\psi(p^2)$. This process can be iterated and one could expect that the decrease of the exact solution is more rapid than any power-law function. This is indeed confirmed by the numerical analysis within the simplified Abelian version of the model \cite{Benes:2006ny}.
  \item The interconnection of the equations is spectacular. The up-type fermions from a given doublet contribute to the equations of their down-type partners, and vice versa, due to the exchange of charged scalars. Further, the up-type and down-type fermions influence each other self-energies through the equation for $\Pi_{NS}(p^2)$. Even if the Yukawa coupling matrices are taken flavor diagonal, the self-energies from different generations influence each other via the equations for $\Pi_{X}(p^2)$. All up-type fermions contribute to $\Pi_{N}(p^2)$, all down-type fermions contribute to $\Pi_{S}(p^2)$ and all fermion doublets contribute to  $\Pi_{NS}(p^2)$. This is the feature following from the non-Abelian nature of the electroweak symmetry. Further we will study also special case which leads to the, so called, Abelian form.
  \item Finding a non-trivial solution of Schwinger--Dyson equations is not enough to claim that the theory develops a symmetry breaking vacuum. It is necessary to check whether the non-trivial solution corresponds to the field configuration of the lowest energy, i.e., whether the non-trivial solution defines a true minimum of the effective potential \cite{Cornwall:1973ts}.
\end{itemize}

\subsection{Electroweak symmetry breaking}

Both $\Sigma_\psi(p^2)$ and $\Pi_X(p^2)$ break spontaneously the electroweak symmetry $\SU{2}_L\times\U{1}_Y$ down to $\U{1}_\mathrm{em}$. This gives rise to three Nambu--Goldstone modes which combine with the electroweak gauge boson fields. They together excite massive $W$ and $Z$ bosons. Their masses $M_{W}^2$ and $M_{Z}^2$ are the quantities calculable in terms of all functions $\Sigma_\psi(p^2)$ and $\Pi_X(p^2)$.

\begin{figure}
  \centering
  \includegraphics[width=0.7\textwidth]{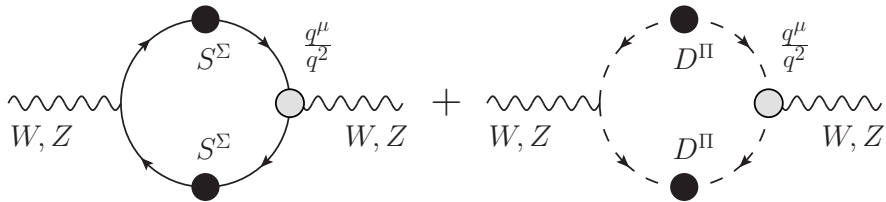}
  \caption[Fermion and scalar contributions to the polarization tensor of $W$ and $Z$ bosons]{The figure shows that both fermions and scalars contribute to the polarization tensor of $W$ and $Z$. The symbols $S^\Sigma$ and $D^\Pi$ denote the symmetry breaking parts of the fermion \eqref{SB_fermion_propagators} and scalar \eqref{SB_scalar_propagators} propagators. }
  \label{fig_MWMZ}
\end{figure}

The calculation follows the strategy presented in appendix~\ref{pole_vertices}. The anomalous electroweak symmetry breaking propagators of both fermions \eqref{SB_fermion_propagators} and scalars \eqref{SB_scalar_propagators} constitute the contributions to the dressed polarization tensors of the electroweak gauge bosons depicted in Fig.~\ref{fig_MWMZ}. The proper vertices marked by grey blobs exhibit the Nambu--Goldstone poles visualised explicitly in \eqref{Gamma_NG} as massless Nambu--Goldstone propagators. The Nambu--Goldstone propagators are connected to the $W$ and $Z$ boson fields by a bilinear coupling function $\Lambda_{ab}(q^2)$, which can be approximated by fermion and scalar loops according to \eqref{Lambda_bilinear} and \eqref{I_def}. This construction actually reveals a composition of the Nambu--Goldstone modes and therefore a composition of the longitudinal components of $W$ and $Z$. They are the fermion and scalar bound states.

For the masses $M_{W}^2$ and $M_{Z}^2$ there are therefore the sum rules \cite{Benes:2008ir}
\begin{eqnarray}\label{MW_MZ_sumrule_Yukawa}
M_Z^2 & = & \frac{\sqrt{g^2+{g'}^2}}{4} \Big( I_{N}+I_{S}+\sum_i I_{U^i}+\sum_i I_{D^i} \Big) \,, \\
M_W^2 & = & \frac{g^2}{4} \Big( I_{S,NS}+I_{N,NS}+\sum_{i} I_{\mathcal{D}^i}\Big) \,,
\end{eqnarray}
where $\mathcal{D}^i\equiv\beginm{c}U^i \\ D^i\endm$. The contributing terms $I$ are given by integrals for $Y=N,S$
\begin{subequations}\label{MW_MZ_Yukawa}
\begin{eqnarray}
I_{Y} & = & 2\im\int\frac{\d^4 k}{(2\pi)^4} k^2 |D^{\Pi}_Y(k)|^2 \,, \\
I_{Y,NS} & = & \im\int\frac{\d^4 k}{(2\pi)^4}
k^2 D^{\Pi}_Y(k)D^{\Pi}_{NS}(k) \frac{|\Pi_Y(k)|^2+|\Pi_{NS}(k)|^2}{\Pi_Y(k)\Pi_{NS}(k)} \,, \\
&& \nonumber \\
I_{\psi} & = & -2\im N_\mathrm{C} \int\frac{\d^4 k}{(2\pi)^4} |S^{\Sigma}_\psi(k)|^2 \,, \\
I_{\mathcal{D}^i} & = & -2\im N_\mathrm{C} \int\frac{\d^4 k}{(2\pi)^4}
S^{\Sigma}_{U^i}(k)S^{\Sigma}_{D^i}(k) \frac{|\Sigma_{U^i}(k)|^2+|\Sigma_{D^i}(k)|^2}{\Sigma_{U^i}(k)\Sigma_{D^i}(k)} \,.
\end{eqnarray}
\end{subequations}
The expression for $I_\psi$ formally coincides with that we have derived in appendix \eqref{simplified_Wick_PS}. Here number of colors $N_\mathrm{C}$ was taken into account. The formulae \eqref{MW_MZ_Yukawa} represent a good approximation assuming that the self-energies have mild momentum dependence, thus they can be approximated by a constant over the relevant range of momenta.

\section{Numerical analysis}

In this section we will solve numerically the set of coupled Schwinger--Dyson equations. We will consider the equations in the Euclidean form \eqref{SDE_Yukawa_Euclid} obtained from \eqref{SDE_Yukawa_Minkowski} by the Wick rotation. In order to study the dependence of the solution on the model parameters we will use the approximate \emph{trial method} described in appendix \ref{trial_method}.

Before studying the feasibility to obtain a mass spectrum of all fermions, we will explore the equations with only two fermions. Within this simpler case we will search for a fermion mass solution $m_1$ and $m_2$, which exhibits the hierarchy
\begin{equation}\label{amplification_of_scales}
m_1\ll m_2\ll M_{N,S}\ \ \mathrm{while}\ \ y_1\sim y_2 \,.
\end{equation}
Later we will try to reproduce the realistic mass spectrum of both fermions and electroweak gauge bosons. It will turn out that a good approximation is to restrict the main calculation to a system of equations for scalars with top and bottom quarks only. The other fermion self-energies can be calculated subsequently. The result of this numerical analysis was published in the work \cite{Benes:2008ir}.

\subsection{Abelian case}

\begin{figure*}[t]
\begin{center}
\includegraphics[width=1\textwidth]{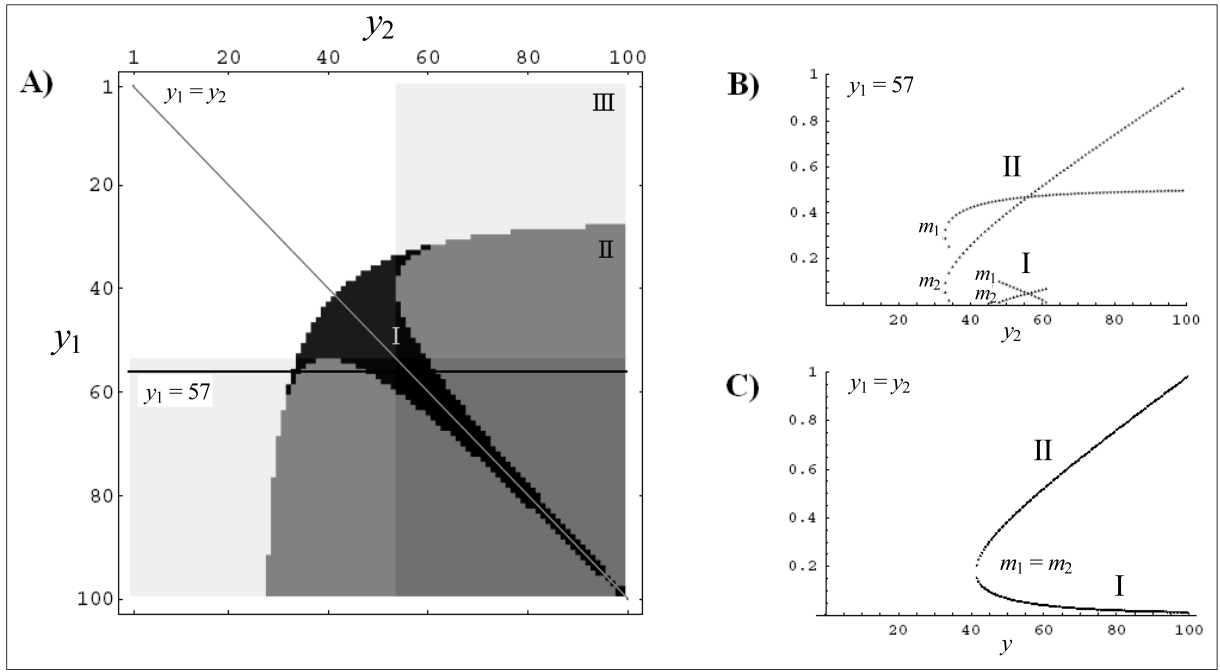}\vspace{-1cm}
\end{center}
\caption[($y_1$,$y_2$)-map of fermion masses $m_1$ and $m_2$]{ At the left figure A) we show the ($y_1$,$y_2$)-map of the solutions for $M_S=1$, and $\Lambda=0.3$. There are three types of solutions denoted as (I), (II) and (III). Find their definition in the text. Next two figures on the right hand side are two cuts of the ($y_1$,$y_2$)-map showing only the solutions (I) and (II): B) the cut
$y_1=57$, and C) the cut along $y\equiv y_1=y_2$. } \label{fig_ye_yd_map}
\end{figure*}

The Abelian form of the Schwinger--Dyson equations can be obtained by setting the charged scalar self-energy to zero, i.e., $\Pi_{NS}(p)=0$. Then the equations \eqref{SDE_Yukawa_Euclid} split into two separate sets of equations of the same form, one for up-type fermions and the other for down-type fermions. We write here the set of equations for down-type fermions
\begin{subequations}\label{Aequation}
\begin{eqnarray}
\Sigma_{D^i}(0) & = & y_{D^i}^2\int_{0}^{\infty}\frac{k^2\mathrm{d}k^2}{16\pi^2}
                    \frac{\Sigma_{D^i}(k)}{k^2+\Sigma_{D^i}^2(k)}\frac{\Pi_{S}(k)}{(k^2+M_{S}^2)^2-\Pi_{S}^2(k)} \,,\\
\Pi_{S}(0) & = & 2
\sum_{i}N_\mathrm{C}y_{D^i}^2\int_{0}^{\infty}\frac{k^2\mathrm{d}k^2}{16\pi^2}
                    \left(\frac{\Sigma_{D^i}(k)}{k^2+\Sigma_{D^i}^2(k)}\right)^2 \,, \label{Pi_equation}
\end{eqnarray}
\end{subequations}
where we have set $p=0$ and performed the angular integration. We assume that the self-energies are real-valued functions. The set of equations for up-type fermions can be obtained by a simple substitution $D\rightarrow U$ and $S\rightarrow N$. Let us add a note that the same set of equations as \eqref{Aequation} is obtained under dynamical assumption that the self-energies of up-type fermions are not generated, e.g., for the reason that $y_{U^i}$ stay subcritical. This is equivalent to setting $y_{U^i}=0$.

Using the simplest step-function Ansatz with a common cutoff $\Lambda$,
\begin{subequations}\label{Ansatz}
\begin{eqnarray}
%\Sigma_{u^i}(p) & \rightarrow & m_{u^i}\theta(\Lambda^2-p^2) \,,\\
\Sigma_{D^i}(p^2) & = & m_{D^i}\theta(\Lambda^2-p^2) \,,\\
%\Pi_N(p) & \rightarrow & \mu_{N}^2\theta(\Lambda^2-p^2) \,,\\
\Pi_S(p^2) & = & \mu_{S}^2\theta(\Lambda^2-p^2) \,,%\\
%\Pi_{NS}(p) & \rightarrow & \mu_{NS}^2\theta(\Lambda^2-p^2) \,,
\end{eqnarray}
\end{subequations}
the integrals are easily calculable. We get the set of algebraic equations which amounts to seven unknown variables $m_{D^i}$ and $\mu_S$, and seven dimensionless parameters $y_{D^i}$ and $M_S/\Lambda$:
\begin{subequations}\label{equations_abelian}
\begin{eqnarray}
m_{D^i} & = & \frac{m_{D^i}y_{D^i}^2}{32\pi^2}\left[f(m_{D^i}^2,M_{S}^2+\mu_{S}^2)-f(m_{D^i}^2,M_{S}^2-\mu_{S}^2)\right] \,,\\
\mu_{S}^2 & = & -\sum_{i}N_\mathrm{C}\frac{m_{D^i}^2y_{D^i}^2}{8\pi^2}\left(\frac{\Lambda^2}{m_{D^i}^2+\Lambda^2}+\log{\frac{m_{D^i}^2}{m_{D^i}^2+\Lambda^2}}\right) \,,
\end{eqnarray}
\end{subequations}
where
\begin{equation}\label{funkce_f_mM}
f(a,b)=\frac{1}{a-b}\left(a\log{\frac{a}{a+\Lambda^2}}-b\log{\frac{b}{b+\Lambda^2}}\right) \,.
\end{equation}
We identify the parameters $m_{D^i}$ from the Ansatz \eqref{Ansatz} directly with fermion masses.

Let us first consider the case with two fermions only, e.g., $D^1$ and $D^2$ with masses $m_1$ and $m_2$, setting $N_\mathrm{C}=1$, and scan numerically the parameter space $(y_1,y_2)$ for some fixed $\Lambda$ and $M_S$. We are obtaining the ($y_1$,$y_2$)-map plotted in Fig.~\ref{fig_ye_yd_map}, which exhibits qualitative features characteristic for the model.

In Fig.~\ref{fig_ye_yd_map}A) we show the ($y_1$,$y_2$)-map of the solutions of the equations \eqref{equations_abelian} for $M_S=1$, and $\Lambda=0.3$. The grey level indicates the number of solutions.
\begin{itemize}
    \item One type of solution is such that one of the fermion self-energies is vanishing, $m_{1}=0$. The self-energy of the other fermion is non-trivial (non-vanishing), $m_{2}\ne0$, for the values of its coupling constant $y_{2}>54$, and it does not depend on the value of the other coupling constant, $m_{2}\ne f(y_{1})$. Analogously, there is the same type of solution with interchanged indices $1\leftrightarrow2$ These solutions are represented in the figure by two transparent rectangle areas which we denote as the solution (III).

    \item More interesting solutions are those which are non-trivial for both fermions, $m_1\ne0$ and $m_2\ne0$. These solutions lie in the darker rounded area for $y_1\gtrsim26$ and $y_2\gtrsim26$. There are two types of non-trivial solutions, one is smaller, denoted by (I), and the other is larger, denoted by (II). These solutions should be better understood from the next two figures on the right hand side. These are two cuts of the ($y_1$,$y_2$)-map showing the non-trivial solutions (I) and (II): B) is the cut $y_1=57$, and C) is the cut $y_1=y_2$.
\end{itemize}
The larger solution (II), $m_{1,2}\sim M_S$, increases with $y_{1,2}$. The smaller solution (I) in the thinner area, $m_{1,2}\ll M_S$, decreases with increasing $y_{1,2}$. This is the interesting solution. See two cuts of the map B) and C). Along the diagonal axis, $y\equiv y_1=y_2$, which defines the axis of a mirror symmetry $D^1\leftrightarrow D^2$, the self-energies are equal $\Sigma_1=\Sigma_2$ and behave according to the plot C). Following the diagonal axis starting in the origin one first intersects the critical line in the critical value $y=y_\mathrm{crit}\approx41$. In the point $y_\mathrm{crit}$ a single nontrivial solution $\Sigma_1=\Sigma_2\neq0$ appears. From that critical point, for $y>y_\mathrm{crit}$, two nontrivial solutions split apart.

We can make conclusions:
\begin{itemize}
    \item We are able to explore the area of smaller Yukawa couplings hardly accessible to the numerical-iterative method. Here, we recover the critical line under which only trivial solution exists and no electroweak symmetry breaking appears.
%\textbf{ii}) In the case that one of the fermion self-energies vanishes, e.g. $\Sigma_e\equiv0$, the system looses its sensitivity to the corresponding coupling, $y_e$. There is one critical value of the second coupling, $y_{d,\mathrm{crit}}$. Above the critical value two nonzero solutions of $\Sigma_d$ arise (see the lighter rectangle areas III on the ($y_e$,$y_d$)-map).
    \item For larger values of Yukawa couplings non-trivial solutions of the equations \eqref{equations_abelian} and electroweak symmetry breaking appear.
    \item The solutions (I) exhibit critical scaling necessary for achieving the amplification of scales \eqref{amplification_of_scales}. First, the solutions are by order of magnitude smaller than $M_S$ and even decrease with increasing $y_{1,2}$. Second, they form a characteristic `x'-shape, see Fig.~\ref{fig_ye_yd_map}B), what means that at the edge of the area (I) the ratio $m_1/m_2$ can be made arbitrarily small while keeping $y_1/y_2\approx1$.
\end{itemize}

The approximate trial method is a suitable tool not only for scanning the parameter space, but it also allows us to invert the procedure. We can search for all six Yukawa coupling constants $y_{D^i}$ in order to reproduce the realistic down-type fermion mass spectrum
\begin{eqnarray}\label{d_spectrum}
m_{e}=0.5\,\mathrm{MeV}\,, & m_{\mu}=106\,\mathrm{MeV}\,, & m_{\tau}=1.8\,\mathrm{GeV} \,, \nonumber\\
m_{d}=6\,\mathrm{MeV}\,, & m_{s}=100\,\mathrm{MeV}\,, & m_{b}=4.2\,\mathrm{GeV} \,. \label{down_mass_spectrum}
\end{eqnarray}

\begin{figure*}[t]
\begin{center}
\includegraphics[width=0.6\textwidth]{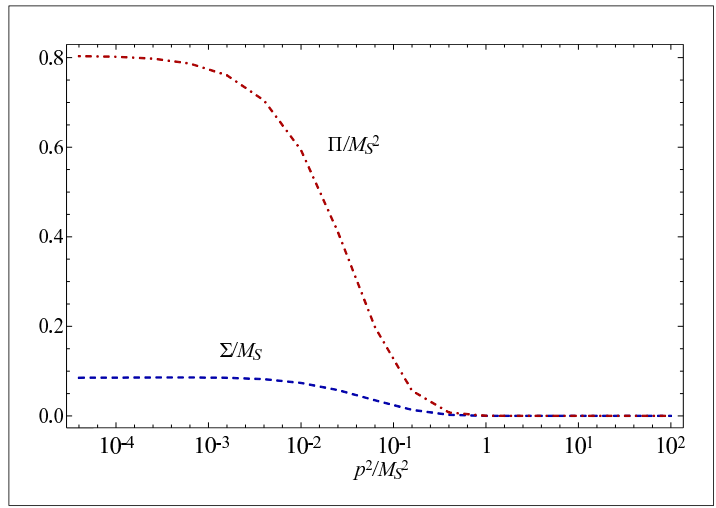}\vspace{-1cm}
\end{center}
\caption[Numerical solution of Abelian Schwinger--Dyson equations]{ Solutions $\Sigma_{\mathrm{NI}}(p^2)$ and $\Pi_{\mathrm{NI}}(p^2)$ of the Abelian Schwinger--Dyson equations \eqref{Aequation} obtained by a numerical-iterative method. The author is grateful to Petr Bene\v{s} for doing the numerical iterative computation and providing data for this figure. } \label{AnsatzNI}
\end{figure*}

For this purpose, instead of using the step-function with a cutoff as an Ansatz, we take the best trial functions we have, i.e., the results of the
numerical iterative method $\Sigma_{\mathrm{NI}}(p^2)$ and $\Pi_{\mathrm{NI}}(p^2)$ depicted in Fig.~\ref{AnsatzNI}. We are adopting a non-trivial assumption that solutions for various parameter settings can be converted from one to the other simply by multiplying by a constant scaling factor, see \eqref{Ansatz_SigmaNI} and the text therein. We take the Ansatz
\begin{subequations}\label{Ansatz_NI}
\begin{eqnarray}
\Sigma_{D^i}(p^2) & = & \sigma_{D^i}f^{\Sigma}(p^2)=\sigma_{D^i}\Sigma_{\mathrm{NI}}(p^2)/\Sigma_{\mathrm{NI}}(0) \,, \\
\Pi_S(p^2) & = & \mu_{S}^2f^{\Pi}(p^2)=\mu_{S}^2\Pi_{\mathrm{NI}}(p^2)/\Pi_{\mathrm{NI}}(0) \,.
\end{eqnarray}
\end{subequations}
Let us briefly describe the computational procedure. First, we fix the scaling factor $\sigma_{D^i}$ in order to reproduce known values of fermion masses \eqref{down_mass_spectrum} as a solution of the equation
\begin{equation}
m_{D^i}=\sigma_{D^i}f^{\Sigma}(m_{D^i}^2) \,.
\end{equation}
Therefore we know completely the Ans\"{a}tze for all six fermion self-energies. This we use to express the scaling factor $\mu_S$ from the equation \eqref{Pi_equation} as a polynomial of the unknown Yukawa coupling constants. We get
\begin{equation}
\mu_{S}^2(y_{D^1},\dots,y_{D^6})=\sum_{i}y_{D^i}^2\mu_{D^i}^2 \,,
\end{equation}
where $\mu_{D^i}^2$ are numbers, the contributions from individual fermions calculated from the equation \eqref{Pi_equation} by numerical integration. Now we have all necessary ingredients to write six equations for the only unknown variables, for the six Yukawa coupling constants $y_{D^i}^2$,
\begin{equation}
1 = y_{D^i}^2\int_{0}^{\infty}\frac{k^2\mathrm{d}k^2}{16\pi^2}
                    \frac{f^\Sigma(k)}{k^2+\big[\sigma_{D^i}f^{\Sigma}(k)\big]^2}\frac{\mu_{S}^2(y_{D^1},\dots,y_{D^6})f^{\Pi}(k)}{(k^2+M_{S}^2)^2-\big[\mu_{S}^2(y_{D^1},\dots,y_{D^6})f^{\Pi}(k)\big]^2} \,.
\end{equation}

We show the results of the numerical computation in Tab.~\ref{tabulka}.
The solutions clearly exhibit features characteristic for the solutions (I) in Fig.~\ref{fig_ye_yd_map}. Despite of the wide range of the fermion mass ratios, the biggest being $m_{b}/m_{e}\approx10^4$, all of the Yukawa couplings are very close to each other. The couplings approach each other with increasing $M_S$ what is equivalent to the decreasing and narrowing behavior of the solution (I). The overall magnitude of the Yukawa couplings is driven by the heaviest fermion. The reason is that the heaviest fermion contributes by dominant portion into the scalar self-energy $\Pi_S(p^2)$. Therefore in order to determine the overall magnitude of Yukawa couplings, it is sufficient to solve only the simplified set of equations with the single fermion. This determines how far in the area (I) does the solution lie. Approximately, the couplings of other fermions can be obtained from their individual equations using $\Pi_S(p^2)$ already determined by the heaviest fermion.
In order to achieve the fermion mass hierarchy, the other Yukawa couplings lie nearer to the edge of the area (I) where they are subjects of a critical scaling by the `x'-shape.

The self-energies given by the Ansatz \eqref{Ansatz_NI} depicted in Fig.~\ref{AnsatzNI} drop down very quickly. In particular, at the scales $\sim M_S$ they are negligible. Therefore also the splitting of the scalar masses \eqref{scalar_mass_spectrum} turns out to be negligible. We believe that this feature of negligible splitting is just a remnant of the chosen method based on the scaling assumption \eqref{Ansatz_NI}.

%\begin{widetext}
\begin{table*}[t]
%\tbl{Down-Yukawa couplings}
\begin{center}
\begin{tabular}{c|cccccc|c}
$M_S\,[\mathrm{GeV}]$ & $y_e$ & $y_\mu$ & $y_\tau$ & $y_d$ & $y_s$ & $y_b$ & $\mu_S\,[\mathrm{GeV}]$ \\
\hline
10   & 60.96 & 60.92   & 58.30   & 60.96   & 60.92   & 80.15   & 37 %& 10 & 4.5
\\
100  & 65.52    & 65.54    & 68.02    & 65.52    & 65.54    & 73.25    & 91  \\
1000 & 179.4 & 179.4 & 179.5 & 179.4 & 179.4 & 179.8 & 407 \\
\hline
\hline
\end{tabular}
\end{center}
\caption[Yukawa coupling constants reproducing realistic mass spectrum of down-type fermions]{The Yukawa couplings $y_{D^i}$ and the scaling parameter $\mu_S$ for three values of bare scalar mass, $M_S=10\GeV$, $M_S=100\GeV$, $M_S=1\TeV$. These are the solutions which reproduce the realistic mass spectrum of down-type fermions \eqref{d_spectrum}. Where the same values are shown, it means that the Yukawa coupling parameters differ at higher decimal place. } \label{tabulka}
\end{table*}
%\end{widetext}

\subsection{Non-Abelian case}

Equipped by the understanding obtained in the Abelian case analysis, we will consider now the set of equations \eqref{SDE_Yukawa_Euclid} in its complete non-Abelian form. The essential part of the scalar self-energies $\Pi_X$ comes from the heaviest fermions, top- and bottom-quarks. Therefore contributions from lighter fermion doublets can be neglected. As the scalar self-energies are the only bridges between different fermion doublets we can pick up and separately solve the set of merely five equations for $\Sigma_t$, $\Sigma_b$, and $\Pi_N$, $\Pi_S$, $\Pi_{NS}$. The rest of the fermion self-energies can be calculated using already known scalar self-energies.

Under the simple step-function Ansatz
\begin{subequations}\label{NA_Ansatz}
\begin{eqnarray}
\Sigma_t(p^2) & = & m_t\theta(\Lambda^2-p^2) \,,\\
\Sigma_b(p^2) & = & m_b\theta(\Lambda^2-p^2) \,,\\
\Pi_N(p^2) & = & \mu_{N}^2\theta(\Lambda^2-p^2) \,,\\
\Pi_S(p^2) & = & \mu_{S}^2\theta(\Lambda^2-p^2) \,,\\
\Pi_{NS}(p^2) & = & \mu_{NS}^2\theta(\Lambda^2-p^2) \,,
\end{eqnarray}
\end{subequations}
the integrals in the equations \eqref{SDE_Yukawa_Euclid} for $p=0$ are easily calculable
\begin{subequations}\label{NA_equations_tb}
\begin{eqnarray}
m_t & = & \frac{m_ty_{t}^2}{32\pi^2}\left(f(m_{t}^2,M_{N^{0}_{2}}^2)-f(m_{t}^2,M_{N^{0}_{1}}^2)\right) \nonumber\\
    & & + \frac{m_b y_ty_b}{32\pi^2}\frac{2\mu_{NS}^2}{M_{N}^2+M_{S}^2-2M_{C^{+}_{1}}^2}\left(f(m_{b}^2,M_{C^{+}_{2}}^2)-f(m_{b}^2,M_{C^{+}_{1}}^2)\right) \,,\\
m_b & = & \frac{m_by_{b}^2}{32\pi^2}\left(f(m_{b}^2,M_{S^{0}_{2}}^2)-f(m_{b}^2,M_{S^{0}_{1}}^2)\right) \nonumber\\
    & & + \frac{m_t y_ty_b}{32\pi^2}\frac{2\mu_{NS}^2}{M_{N}^2+M_{S}^2-2M_{C^{+}_{1}}^2}\left(f(m_{t}^2,M_{C^{+}_{2}}^2)-f(m_{t}^2,M_{C^{+}_{1}}^2)\right) \,,\\
\mu_{N}^2 & = & -N_\mathrm{C}\frac{m_{t}^2y_{t}^2}{8\pi^2}\left(\frac{\Lambda^2}{m_{t}^2+\Lambda^2}+\log{\frac{m_{t}^2}{m_{t}^2+\Lambda^2}}\right) \,,\\
\mu_{S}^2 & = & -N_\mathrm{C}\frac{m_{b}^2y_{b}^2}{8\pi^2}\left(\frac{\Lambda^2}{m_{b}^2+\Lambda^2}+\log{\frac{m_{b}^2}{m_{b}^2+\Lambda^2}}\right) \,,\\
\mu_{NS}^2 & = & -N_\mathrm{C}\frac{m_tm_by_ty_b}{8\pi^2}f(m_{t}^2,m_{b}^2) \,,
\end{eqnarray}
\end{subequations}
where the mass parameters $M_{N^{0}_{1,2}}^2$, $M_{S^{0}_{1,2}}^2$ and $M_{C^{+}_{1,2}}^2$ are defined in \eqref{scalar_mass_spectrum} and the function $f(m_1,m_2)$ is defined in \eqref{funkce_f_mM}. We numerically search for solutions of this set of equations.

\begin{figure*}[t]
%\begin{center}
\begin{tabular}{cc}
\includegraphics[width=0.42\textwidth]{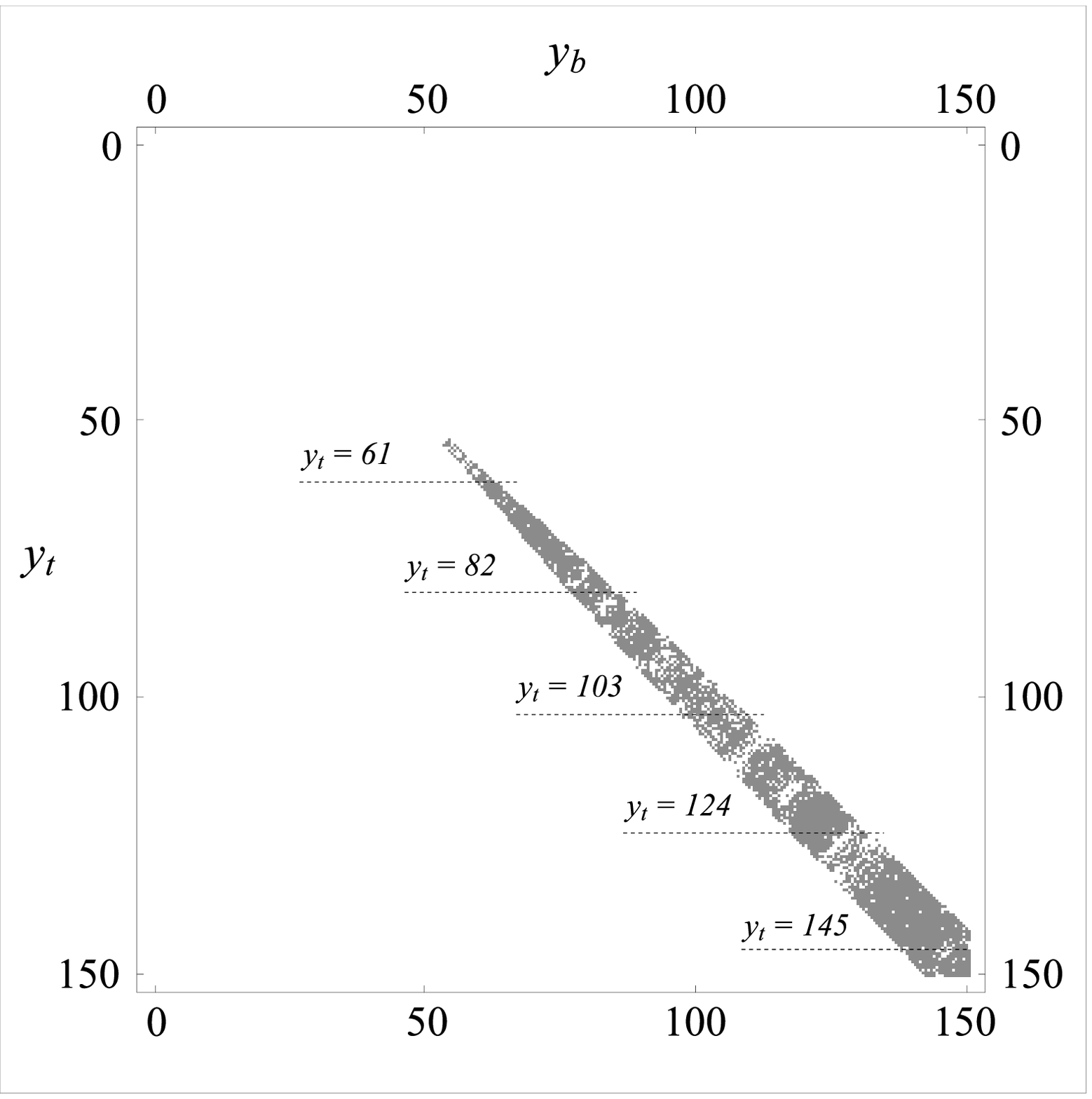} & \hspace{-0.8cm}
\includegraphics[width=0.60\textwidth]{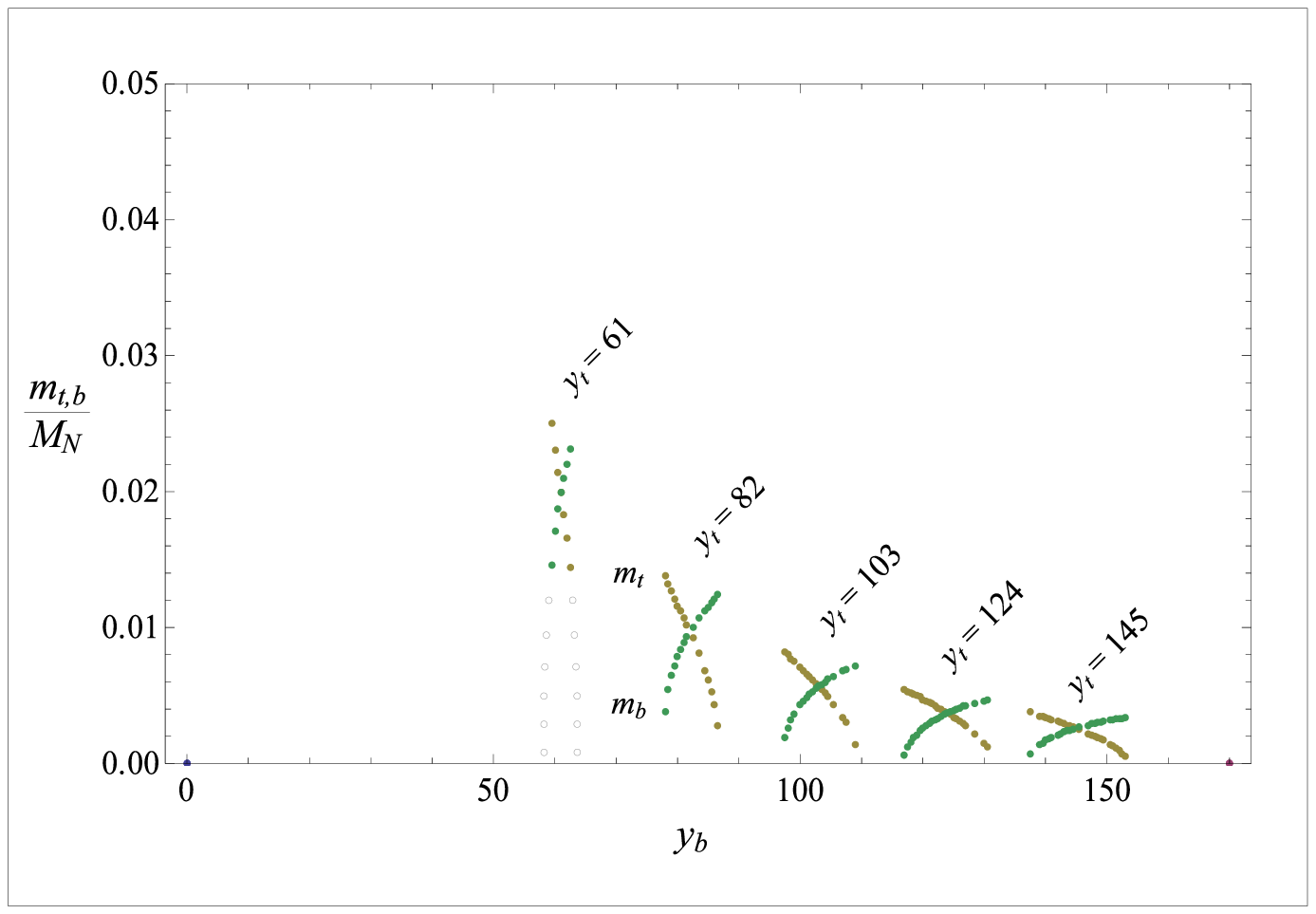}
\end{tabular}
\caption[$(y_t,y_b)$-map of fermion masses $m_t$ and $m_d$ for $M_S=M_N$ ]{ At the left figure, the grey points represent Yukawa couplings $(y_t,y_b)$ for which we found numerically a non-trivial solution of the equations \eqref{NA_equations_tb} for both fermions for $M_S=M_N$ and $\Lambda=0.3M_N$. At the right figure, we show cuts of the $(y_t,y_b)$-map of solutions. The solutions exhibit typical `x'-shape which is instrumental for achieving the fermion mass hierarchy. } \label{scan_NA_MN1_L03}
%\end{center}
\end{figure*}

First, we set $M_S=M_N$ and $\Lambda=0.3M_N$, and we scan over the Yukawa coupling parameter space, $y_t=1,\dots,150$ and $y_b=1,\dots,150$, searching for variables $m_t/M_N$ and $m_b/M_N$ solving the equations \eqref{NA_equations_tb}. The result is shown in Fig.~\ref{scan_NA_MN1_L03}. Out of the multiple solution the figure shows only the interesting part of solutions, which is the analogy of the solutions (I) in the Abelian case in Fig.~\ref{fig_ye_yd_map}. It exhibits qualitatively the same features, mainly the `x'-shape, with single exception that now the area is getting wider with increasing Yukawa coupling parameters instead of getting narrower.

\begin{figure*}[t]
\begin{center}
\begin{tabular}{c}
\includegraphics[width=0.7\textwidth]{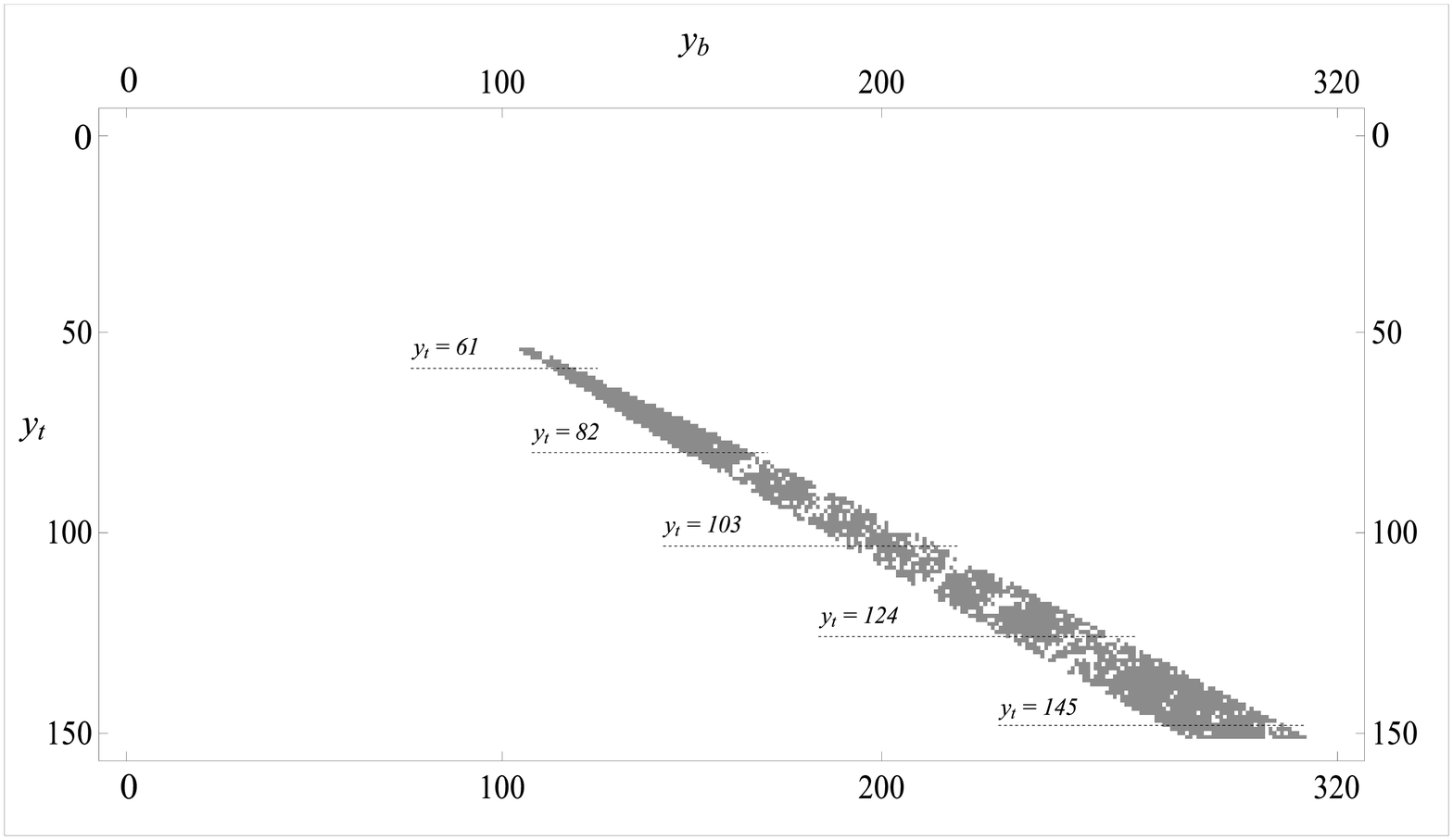} \vspace{-0.6cm} \\
\hspace{0.1mm}\includegraphics[width=0.715\textwidth]{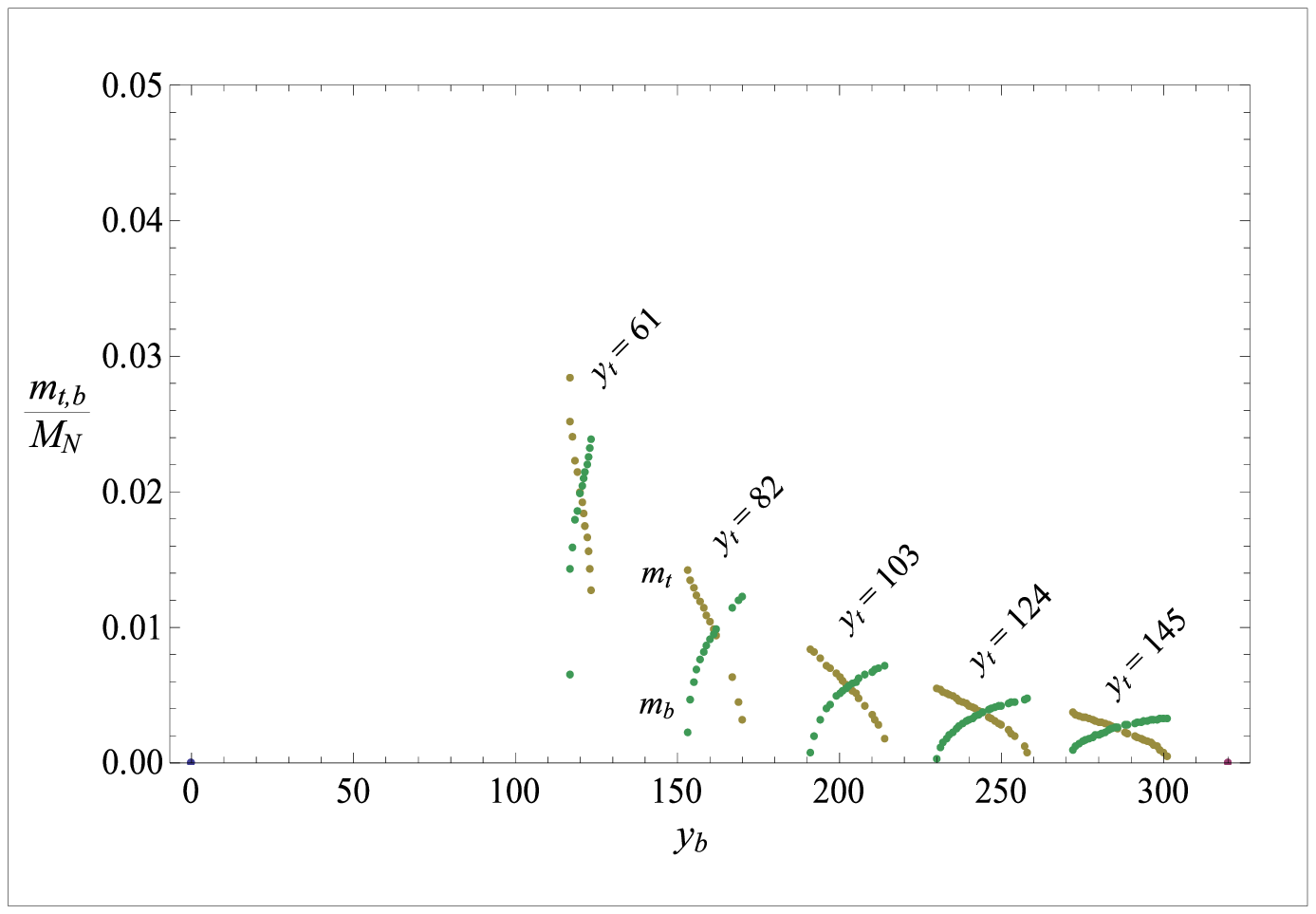}
\vspace{-0.6cm}
\end{tabular}
\end{center}
\caption[$(y_t,y_b)$-map of fermion masses $m_t$ and $m_d$ for $M_S=2M_N$]{ At the upper figure, the grey points represent Yukawa coupling setting $(y_t,y_b)$ for which we found numerically a non-trivial solution of the equations \eqref{NA_equations_tb} for both fermions for $M_S=2M_N$ and $\Lambda=0.3M_N$. At the lower figure, we show cuts of the $(y_t,y_b)$ map of solutions. } \label{scan_NA_M1N2_L03}
\end{figure*}

Second, we set $M_S=2M_N$ and $\Lambda=0.3M_N$, and we scan over the Yukawa coupling parameter space, $y_t=1,\dots,150$ and $y_b=1,\dots,320$. The result is shown in Fig.~\ref{scan_NA_M1N2_L03}. We observe that, at least for moderate ratio $M_S/M_N$, the effect of changing $M_S$ relatively to $M_N$ follows a simple conjectured approximate rule
\begin{equation}
\tan\alpha\equiv\left.\frac{y_b}{y_t}\right|_{\Sigma_t=\Sigma_b}\simeq\frac{M_S}{M_N} \,,
\end{equation}
where $\alpha$ gives a direction along which the solutions stretch.

For obtaining more realistic results we will again invert the procedure and search for the values of the Yukawa coupling parameters which reproduce the realistic mass spectrum
\begin{equation}
m_{t}=172.4\,\mathrm{GeV}\,,\ \ \ m_{b}=4.2\,\mathrm{GeV} \,.
\end{equation}
We can use the two free scalar mass parameters $M_N$ and $M_S$ for trying to reproduce also the electroweak gauge boson masses $M_W$ and $M_Z$. For that the mass sum rules \eqref{MW_MZ_sumrule_Yukawa} represent two conditions by which we can fix $M_N$ and $M_S$. The only free parameter is the cutoff $\Lambda$.

With the step-function Ansatz \eqref{NA_Ansatz} the integrals \eqref{MW_MZ_Yukawa} from the sum rule are calculable yielding the algebraic expressions, for $Y=N,S$,
\begin{eqnarray}
I_{Y} & = & \frac{1}{128\pi^2}\left(\frac{2\Lambda^2(\Lambda^2M_{Y}^2+M_{Y^{0}_1}^2M_{Y^{0}_2}^2)}{(\Lambda^2+M_{Y}^2)^2-\mu_{Y}^4}
     -\frac{M_{Y^{0}_1}^2M_{Y^{0}_2}^2}{\mu_{Y}^2}\log\frac{M_{Y^{0}_2}^2(\Lambda^2+M_{Y^{0}_1}^2)}{M_{Y^{0}_1}^2(\Lambda^2+M_{Y^{0}_2}^2)}\right) \,, \\
I_{Y,NS} & = & \frac{1}{128\pi^2}\frac{\mu_{Y}^4+\mu_{NS}^4}{\mu_{Y}^2(M_{N}^2+M_{S}^2-2M_{C^{+}_1}^2)} \\
     & & \times\left(g(M_{Y^{0}_1}^2,M_{C^{+}_1}^2)+g(M_{Y^{0}_2}^2,M_{C^{+}_2}^2)-g(M_{Y^{0}_1}^2,M_{C^{+}_2}^2)-g(M_{Y^{0}_2}^2,M_{C^{+}_1}^2)\right) \,, \nonumber \\
&& \nonumber \\
I_{\psi} & = & -N_\mathrm{C}\frac{m_{\psi}^2}{32\pi^2}\left(\frac{\Lambda^2}{m_{\psi}^2+\Lambda^2}+\log{\frac{m_{\psi}^2}{m_{\psi}^2+\Lambda^2}}\right) \,, \\
I_{\mathcal{D}_{t,b}} & = & -N_\mathrm{C}\frac{1}{32\pi^2}(m_{t}^2+m_{b}^2)f(m_{t}^2,m_{b}^2) \,,
\end{eqnarray}
where
\begin{equation}
g(a,b)=\frac{1}{a-b}\left(a^2\log{\frac{a}{a+\Lambda^2}}-b^2\log{\frac{b}{b+\Lambda^2}}\right)
\end{equation}
and the function $f(m_1,m_2)$ is defined in \eqref{funkce_f_mM}.

We were able to find several reliable solutions. In Tab.~\ref{tab_NA_solutions} we show how successful are the solutions in reproducing the correct $W$ and $Z$ boson mass spectrum and the $\rho$-parameter, see \eqref{MW_MZ_Yukawa},
\begin{equation}
\rho=\frac{M_{W}^2}{\cos^2\theta_\mathrm{W}M_{Z}^2}=\frac{I_{N}+I_{S}+I_{t}+I_{b}}{I_{S,NS}+I_{N,NS}+I_{\mathcal{D}_{t,b}}} \,.
\end{equation}
For the Weinberg angle $\theta_\mathrm{W}$ we take the experimental value at the scale $M_Z$, $\cos^2\theta_\mathrm{W}(M_Z)\doteq0.768$. We can see that for lower values of the cutoff $\Lambda$ we can achieve correct value of the $\rho$-parameter, but the electroweak gauge bosons turn out to be too light (the first three rows in Tab.~\ref{tab_NA_solutions}). Clearly to make them heavier the cutoff $\Lambda$ should be increased. In that case the scalar mass parameters $M_N$ and $M_S$ do scale with $\Lambda$. Here we were limited by the precision capacity of our numerical procedure to increase simultaneously the cutoff $\Lambda$ and the scalar mass parameters $M_N$ and $M_S$ to the level necessary for obtaining correct values of the gauge boson masses $M_W$ and $M_Z$. Thus we relaxed our constraint on the $\rho$-parameter in order to find the magnitude of the cutoff which reproduces correct value at least of $M_W$ (the fourth row in Tab.~\ref{tab_NA_solutions}).

To reach the solution completely consistent with the mass spectrum $m_t$, $m_b$, $M_W$ and $M_Z$, further numerical investigations are needed. In order to find more reliable solutions the numerical iterative method should be used. The numerical iterative method was used within the context of the Abelian model \cite{Brauner:2005hw,Benes:2006ny}. The numerical results can be found in a condensed form in \cite{Benes:2012hz}. It is however extremely demanding on numerical capacities to scan the whole parameter space. Therefore here in the context of realistic model, we have used the trial method which allowed us to get qualitative idea about the behavior of the solution. Qualitatively similar behavior of the solution was presented in \cite{Benes:2006ny}. Also there, the ($y_1$,$y_2$)-map, similar to Fig.~\ref{scan_NA_MN1_L03} and Fig.~\ref{fig_ye_yd_map}, was calculated with the characteristic `x'-shape of the non-trivial solutions appearing along the diagonal axis.

%\begin{widetext}
\begin{table*}[t]
%\tbl{Down-Yukawa couplings}
\begin{center}
\begin{tabular}{ccc|cc|ccc}
$\Lambda\,[\mathrm{GeV}]$ & $M_N\,[\mathrm{GeV}]$ & $M_S\,[\mathrm{GeV}]$ & $y_t$ & $y_b$ & $M_W\,[\mathrm{GeV}]$ & $M_Z\,[\mathrm{GeV}]$ & $\rho$ \\
\hline
\hline
$30$   & $19.5$ & $12.0$   & $53.4$ & $38.4$ & $38.6$ & $43.8$ & $1.01$ \\
$300$  & $219$ & $121$ & $165$ & $112$ & $46.3$ & $52.6$ & $1.01$ \\
$3000$ & $2335$ & $1210$ & $508$ & $335$ & $52.9$ & $60.1$ & $1.01$ \\
$39\times10^{6}$ & $2335$ & $1210$ & $378$ & $278$ & $80.4$ & $98.4$ & $0.87$ \\
\hline
\hline
\end{tabular}
\end{center}
\caption[Sample of solutions $m_t$, $m_b$, $M_W$ and $M_Z$ ]{ We show here a sample of solutions for a complete model which aims to reproduce the mass spectrum of $m_t$, $m_b$, $M_W$ and $M_Z$ without any big hierarchy among the parameters of the model. } \label{tab_NA_solutions}
\end{table*}
%\end{widetext}

%\vspace{0.5cm}
%\begin{center}*\end{center}
%\vspace{0.5cm}

\section{Conclusions}

The key mechanism of the presented model is the formation of the anomalous two-point Green's function $\langle\phi\phi\rangle$ of a complex scalar field $\phi(x)$. It represents a direct analogy of microscopic theory of superconductivity \cite{Bardeen:1957mv} applied to a scalar field instead of to the electron field. The mechanism is interesting by itself and we believe that it can be used for modeling the superfluid behavior of non-relativistic many-boson systems \cite{Imry1968:aa,Pashitskii:2002aa,Alexandrov:2004aa} interacting with fermions.

In our work we have applied the mechanism of the anomalous scalar propagator formation in order to underlie the electroweak symmetry breaking within the model of strong Yukawa dynamics. We have documented that the quark and lepton masses can be generated dynamically.
%Therefore the model is the representative of the models studied in this thesis, the models of electroweak symmetry breaking induced by dynamical generation of quark and lepton masses.

The dynamically generated fermion self-energies exhibit a behavior typical for this type of models. It has a power to reproduce hierarchies of mass spectra. In particular the ratio of two fermion masses $m_1/m_2$ can be made arbitrarily small while keeping $y_1/y_2\approx1$. This is however achieved for the price of the extremely well fine-tuned Yukawa couplings with respect to their critical values. We should also mention that we have neglected or omitted some conceptual aspects when constructing the model. It will require further care in order to develop a completely consistent and realistic model.

\chapter{Model of flavor gauge dynamics}
%\pagenumbering{arabic}
%\input{2x.tex}

\label{gfd}

\subsubsection{The flavor gauge model}

In this chapter we present the flavor gauge model \cite{Hosek:2009ys,Hosek:NagoyaProceeding,Benes:2011gi} which pretends to be a fundamental theory of both fermion masses and the consequent electroweak symmetry breaking. It is formulated by substituting the Higgs sector of the Standard Model by a new \emph{flavor gauge dynamics}. The threefold replication of the fermion generations is taken as an advantage. The new dynamics is introduced by gauging the flavor or family index of quarks and leptons as an index of a fundamental representation of $\SU{3}_\F$ group. This gives rise to an octet of flavor gauge bosons. Their exchange between the chiral fields of quarks and leptons provides an attraction necessary to generate their masses dynamically.

For the dynamical fermion mass generation to be possible, the flavor gauge dynamics must not be in the domain of attraction of an infrared stable origin \cite{Stern:1976jg}. That would not allow for a cancelation of massless poles in the full propagators and for the generation of the massive ones. Therefore the flavor gauge dynamics is defined as an asymptotically free non-Abelian gauge theory in the manner of the QCD, the real example of the dynamical chiral symmetry breaking.  %Such flavor gauge dynamics is a UV complete theory.

The flavor gauge symmetry protects the masslessness of eight \emph{flavor gauge fields}. However, we do not observe any massless flavor `photons' which could be identified with excitations of the octet of flavor gauge fields. They even should not excite an octet of confining flavor `gluons' because we do observe the flavor as a quantum number of asymptotic fermion states. In our model, the flavor gauge fields excite an octet of massive flavor vector bosons. This has a field-theoretic interpretation that the gauge flavor symmetry is spontaneously broken and the general Anderson--Higgs mechanism is realized \cite{Anderson:1963pc,Higgs:1964pj,Guralnik:1964eu,Englert:1964et,'tHooft:1971rn}. It also corresponds to the fact that the flavor is not a symmetry of the ground state as it is clearly exhibited by the wild spectrum of fermion masses.

The field-theoretically consistent implementation of the spontaneous gauge flavor symmetry breaking is possible only if the flavor dynamics is formulated as a \emph{chiral} gauge theory \cite{Vafa:1983tf}. Although, conceptually, the chiral gauge theories are not fully understood \cite{Nielsen:1981hk,Poppitz:2010at}, their irreplaceable role in particle physics is manifested by the chiral electroweak dynamics. The chiral nature of the flavor symmetry does not allow for any hard mass scale to be present at the level of the Lagrangian. This is another feature attributed to fundamental theories.

The chiral nature of the flavor symmetry is achieved by the specific \emph{non-vector-like} setting of fermion flavor representations. The $\SU{3}_\F$ gauging of flavor is not new \cite{Wilczek:1978xi,Ong:1978tq,Davidson:1979wt,Chakrabarti:1979vy,Yanagida:1979gs,Yanagida:1980xy,Berezhiani:1990wn,Nagoshi:1990wk}. What is new is the ingenious embedding of quark and lepton chiral components into either \emph{triplets} or \emph{anti-triplets}. The purpose of such setting is \emph{not only} to make the flavor symmetry non-vector-like, but also it allows to distinguish mass matrices of up-quarks, down-quarks and charged leptons. In this light, the choice of the dimension of the flavor group $\SU{3}_\F$ is not a random act. It possesses three-dimensional complex representations, which are instrumental in defining a non-vector-like theory.

The non-vector-like setting of the flavor symmetry has yet another welcome consequence. The mutual alignment of both the Standard Model gauge symmetries and the flavor gauge symmetry completely avoids the presence of all non-Abelian global symmetries of the electroweakly charged fermion sector of the Lagrangian. All of the few remaining Abelian symmetries (including the baryon and lepton number, $B$ and $L$) but one, the analogue of $B-L$ symmetry, are at the quantum level broken by anomalies. This is common to models with Yukawa interactions (including the Standard Model) and not worse than there.

The consistency of gauging the flavor of quarks and leptons requires the existence of additional fermion fields. They make the flavor gauge dynamics free of gauge anomalies. We postulate a number of \emph{right-handed neutrinos}. The fact that the right-handed neutrinos are almost mandatory to reproduce the observed phenomena brings just another supporting evidence in favor of the flavor gauge model.

\subsubsection{The self-breaking flavor gauge dynamics}

We must emphasize that there is no need of additional dynamics other than the flavor gauge dynamics in order to spontaneously break the flavor symmetry. We do not invoke any elementary scalar field to develop its flavored vacuum expectation value. We rather let the flavor gauge dynamics to \emph{self-break} its own symmetry through the generation of the fermion masses via the formation of fermion self-energies $\Sigma_\psi(p^2)$ \eqref{m_from_Sigma}. The self-energies breaking the flavor symmetries are the consequences of the non-vector-like flavor setting of fermions. The flavor symmetry breaking does not need to be put in by hand as, for example, in the Standard Model, where the Yukawa couplings are \emph{set} in the way to reproduce the wildly flavor variant fermion mass spectrum.

The self-breaking of the flavor gauge dynamics is accompanied by the formation of flavor symmetry breaking massless poles in the flavor gauge field polarization function.
\begin{equation}
\Pi_{ab}(q^2)\stackrel{q^2\rightarrow0}{=}-\frac{M_{ab}^2}{q^2} \,.
\end{equation}
%defined from the transverse polarization tensor
%\begin{equation}\label{polarization_tensor}
%\Pi_{ab}^{\mu\nu}(q)\equiv-(g^{\mu\nu}q^2-q^\mu q^\nu)\Pi_{ab}(q^2) \,.
%\end{equation}
These massless poles cancel the massless pole in the full flavor gauge boson propagator and produce a set of massive poles instead. That corresponds to the generation of the flavor gauge boson masses $M_{ab}^2$. The massless poles do not have their origin in the axial anomaly, like in the Schwinger's 2d QED model \cite{Schwinger:1962tp}, rather they have genuinely dynamical origin firmly justified by the existence of massless Nambu--Goldstone modes of the spontaneously broken chiral flavor symmetry.

For the Nambu--Goldstone modes there are no corresponding elementary scalars in the Lagrangian. Therefore they are composites of the elementary fermions and flavor gauge bosons. The formation of these bound states can happen as a consequence of the strong coupling. No methods are available to reliably calculate the self-breaking of the strongly coupled chiral gauge theory. In the flavor gauge model it amounts to find the non-perturbative flavor symmetry breaking solutions for $\Sigma_\psi(p^2)$ and $\Pi_{ab}(q^2)$ of the model equations.

It is however completely essential to know the fermion self-energies $\Sigma_\psi(p^2)$. It is not only for reproducing the fermion mass spectrum. The self-energies of the heaviest electroweakly charged fermions determine the magnitude of the electroweak scale $v$. The self-energies $\Sigma_\psi(p^2)$ also determine the interactions of various composite (pseudo-)Nambu--Goldstone bosons to their constituent fermions. The right-handed neutrino self-energy provides a huge Majorana mass $M_R$, the necessary prerequisite for the seesaw mechanism for the tiny active neutrino masses. The right-handed neutrino Majorana masses, being the greatest of all fermion masses, determine the magnitude of the flavor gauge symmetry breaking scale $\Lambda_\F$. Presumably they also determine the magnitude of the flavor gauge boson masses, $M_{ab}^2\sim M_{R}^2$. Therefore, instead of calculating the fermion self-energies $\Sigma_\psi(p^2)$ we often assume their existence.

Our analysis of the top-quark and neutrino condensation model described in chapter~\ref{top_and_nu_condensation} suggests that $M_{R}\gtrsim10^{14}\GeV$. Hence in order to have phenomenologically successful model we have to deal with huge hierarchy between the scale of the flavor gauge dynamics $\Lambda_\F$ and the electroweak scale $v$. We argue that this hierarchy is due to the proximity of the effective coupling parameters to their critical values.

\subsubsection{The structure of the chapter}

When presenting the flavor gauge model in this chapter we almost exclusively follow the author's original work \cite{Smetana:2011tj}.

We start by defining the model and establishing its \emph{perturbative} properties. We exhaustively identify all possible ways how to complement the fermion content of the model by the right-handed neutrinos in order to complete the flavor gauge anomaly cancelation on one hand, and to keep the asymptotic freedom on the other hand. Various but not many flavor settings of the right-handed neutrinos define various but not many versions of the flavor gauge model. We argue that there is only one right-handed neutrino setting that defines viable and preferred version of the model. The preferred version is non-minimal in the sense that it contains the right-handed neutrinos not only in the flavor triplet but also in flavor sextet representations. This is common to the various former proposals, e.g., \cite{Martin:1992aq}.

Later in this chapter we discuss the \emph{non-perturbative} flavor symmetry breaking. In the original version of the flavor gauge model \cite{Benes:2011gi}, the flavor symmetry breaking happens via the condensation of \emph{all fermions} and their mass generation happens at once. Here we deal with the modification that only \emph{right-handed neutrino} condensation of Majorana type $\langle\overline{\nu_{R}^\C}\nu_{R}\rangle$ triggers the flavor symmetry breaking at very high energies. Therefore only the right-handed neutrinos acquire their Majorana masses at these energies. Both the mass generation of other fermions and the electroweak symmetry breaking are postponed to lower energies.

The preferred version provides appealing features:
(i) It is chiral, i.e., the only mass scale $\Lambda_\F$ comes from the dimensional transmutation of the running flavor coupling constant.
(ii) Due to the presence of the flavor sextet, the flavor dynamics is non-vector-like by itself, thus it can self-break.
(iii) The sextet right-handed neutrino Majorana pairing leads to a huge right-handed neutrino Majorana mass. This promises to dynamically generate the seesaw pattern of the neutrino mass matrix in a natural way.

The resulting dynamics at the energies below the scale of the flavor symmetry breaking can be described by four-fermion interactions for still massless electroweakly charged fermions, induced by the exchange of already massive flavor gauge bosons. At even lower scales the four-fermion interactions trigger the fermion condensation of the Dirac type $\langle\bar\psi_R\psi_L\rangle$, which breaks the electroweak symmetry.

The dynamically generated fermion mass matrices break spontaneously all (true or anomalouos) global chiral symmetries of the model. As a result, numerous (pseudo-)Nambu--Goldstone fermion-composites appear in the particle spectrum of the model. Namely, among those bosons there are numerous majorons \cite{Chikashige:1980ui,Schechter:1981cv}, the Weinberg--Wilczek axion \cite{Weinberg:1977ma,Wilczek:1977pj} and the Anselm--Uraltsev arion \cite{Anselm:1981aw}. In \cite{Smetana:2011tj} we have elaborated the analysis of the neutrino-composite bosons while neglecting the others. It is however not a priori clear how to proceed with the complete analysis of all these bosons together, thus we present here just conjectures of their properties.

%they do \emph{not} present a phenomenological danger in form of long-range force \cite{Gelmini:1982zz}.

\section{Lagrangian and parameters of the model}

The \emph{flavor gauge model} is defined by gauging the flavor index of usual fermions by means of the $\SU{3}_\F$ gauge symmetry. The new flavor gauge dynamics completely substitutes the Higgs sector of the Standard Model and does not make use of any other elementary scalar. The relevant part of the Lagrangian reads as
\begin{eqnarray}\label{L_model}
{\cal L}_\F & = & -\frac{1}{4}F^{a}_{\mu\nu}F^{\mu\nu a}
%-\frac{1}{4}G^{i}_{\mu\nu}G^{\mu\nu i}-\frac{1}{4}W^{m}_{\mu\nu}W^{\mu\nu m}-\frac{1}{4}Y_{\mu\nu}Y^{\mu\nu} \nonumber\\
%-\frac{1}{4}F_{\mu\nu}^aF^{\mu\nu a}-\frac{1}{4}G_{\mu\nu}^iG^{\mu\nu i}-\frac{1}{4}W_{\mu\nu}^aW^{\mu\nu a}-\frac{1}{4}Y_{\mu\nu}Y^{\mu\nu} \\
               +\sum_{\psi_{Rr}}\bar\psi_{Rr}\im\slashed{D}_r\psi_{Rr}
               +\sum_{\psi_{Lr}}\bar\psi_{Lr}\im\slashed{D}_r\psi_{Lr} \,,
\end{eqnarray}
where $r$ denotes various flavor representations of fermion chiral fields $\psi_{Rr}$ and $\psi_{Lr}$. The gauge field strength tensors and the covariant derivative are given as
\begin{eqnarray}
%Y_{\mu\nu} & = & \partial_\mu B_{\nu}-\partial_\nu B_{\mu} \,,\\
%W^{m}_{\mu\nu} & = & \partial_\mu A^{m}_{\nu}-\partial_\nu A^{m}_{\mu}+g\epsilon^{mno}A^{n}_{\mu}A^{o}_{\nu} \,,\\
%G^{i}_{\mu\nu} & = & \partial_\mu A^{i}_{\nu}-\partial_\nu A^{i}_{\mu}+g_sf^{ijk}A^{j}_{\mu}A^{k}_{\nu} \,,\\
F^{a}_{\mu\nu} & = & \partial_\mu C^{a}_{\nu}-\partial_\nu C^{a}_{\mu}+hf^{abc}C^{b}_{\mu}C^{c}_{\nu} \,,\\
D_{r}^\mu & = & \partial^\mu-\im h T_{r}^a C^{a\mu} \,.
%D_\mu(x) & = & \partial_\mu-\im g' \frac{Y}{2}B_\mu(x)-\im g T_{L}^m A^{m}_\mu(x)-\im g_s T_{c}^i G^{i}_\mu(x)-\im g_s T_{\F}^a C^{a}_\mu(x) \,.
\end{eqnarray}
where $C^{a}_{\mu}$ are the flavor gauge boson fields and $h$ is the flavor gauge coupling constant. The flavor generators $T_{r}^a$, $a=1,\dots,8$, have the form according to the flavor representation $r$ of the fermion fields $\psi_{Rr}$ and $\psi_{Lr}$ on which they act.

\section{Flavor representation setting}

\subsection{Flavor representation setting of known fermions}

There are more ways how to set three generations of known fermions into the flavor representations of $\SU{3}_\F$. The trio of chiral fields of a given electric charge and given chirality can be assigned into either triplet, $\mathbf{3}$, or anti-triplet, $\overline{\mathbf{3}}$, representation. We profit from this freedom to distinguish Dirac mass matrices of the charged fermions. The flavor transformation property of a Dirac mass matrix $m_\psi$ is given by the flavor transformation properties of chiral components of the fermion field, $\psi_{R}$ and $\psi_{L}$, which are in the representations $r_R$ and $r_L$,
\begin{equation}
m_\psi\propto\langle\bar{\psi}_{R}(r_R)\psi_{L}(r_L)\rangle\,:\ \ \ \overline{r}_R\times r_L \,.
\end{equation}
In general, there are four distinct mass matrices
\begin{equation}\label{m33nem33}
m^{\mathbf{3}\times\mathbf{3}}\neq m^{\overline{\mathbf{3}}\times\mathbf{3}}\neq
m^{\mathbf{3}\times\overline{\mathbf{3}}}\neq m^{\overline{\mathbf{3}}\times\overline{\mathbf{3}}} \,,
\end{equation}
which are however not completely independent as
\begin{eqnarray}\label{mass_matrices_relation}
m^{\overline{\mathbf{3}}\times\mathbf{3}} & = & \big[m^{\mathbf{3}\times\overline{\mathbf{3}}}\big]^\dag \,, \\
m^{\mathbf{3}\times\mathbf{3}} & = & \big[m^{\overline{\mathbf{3}}\times\overline{\mathbf{3}}}\big]^\dag \,.
\end{eqnarray}
It constrains our fermion flavor representation settings down to few cases listed in Tab.~\ref{FermionSetting} and their complex conjugates. The cases I and IV ascribe distinct role to the $u$-type quarks.

\begin{table}[t]
\begin{center}
\begin{tabular}{l|ccccc|c}
case & $q_L$ & $u_R$ & $d_R$ & $\ell_L$ & $e_R$ & $N$  \\
\hline
\hline
\vspace{-0.4cm}\\
$\vphantom{I^I}$I   & $\mathbf{3}$ & $\mathbf{3}$ & $\overline{\mathbf{3}}$ & $\overline{\mathbf{3}}$ & $\mathbf{3}$ & $3$  \\
II  & $\mathbf{3}$ & $\mathbf{3}$ & $\overline{\mathbf{3}}$ & $\overline{\mathbf{3}}$ & $\overline{\mathbf{3}}$ & $5$  \\
III & $\mathbf{3}$ & $\overline{\mathbf{3}}$ & $\mathbf{3}$ & $\overline{\mathbf{3}}$ & $\mathbf{3}$ & $3$  \\
IV  & $\mathbf{3}$ & $\overline{\mathbf{3}}$ & $\mathbf{3}$ & $\overline{\mathbf{3}}$ & $\overline{\mathbf{3}}$ & $5$ \\
\hline
\hline
\end{tabular}
\end{center}
\caption[Quark and lepton flavor settings within the flavor gauge model]{\small The flavor settings of electroweakly charged fermions. The number $N$ tells how many additional flavor triplets are necessary to cancel the flavor gauge anomaly. The notation is obvious: $q_L=(u_L,d_L)^\T$ and $\ell_L=(\nu_L,e_L)^\T$ are $\SU{2}_L$ doublets. Every symbol $u$, $d$, $\nu$ and $e$ denotes all three generations of a given fermion, i.e., $(u,c,t)$, $(d,s,b)$, $(\nu_e,\nu_\mu,\nu_\tau)$ and $e=(e,\mu,\tau)$, respectively. }
\label{FermionSetting}
\end{table}

The idea to distinguish the fermion mass matrices by their gauge representation setting was also pursued in the class of Extended-Technicolor models \cite{Appelquist:2003hn}.

It is not without interest that within one of the most beautiful models, the Georgi--Glashow $\SU{5}$ GUT model \cite{Georgi:1974gg}, the flavor representation setting of the case IV may arise quite naturally. All known electroweakly charged fermions are accommodated in two complex and therefore chiral $\SU{5}$ representations
\begin{equation}
\mathbf{5}_i=\beginm{c}d_{R1}\\d_{R2}\\d_{R3}\\(e_{L})^\C\\ (\nu_{L})^\C\endm_i \,,\ \ \
\mathbf{10}_i=\frac{1}{\sqrt2}
\beginm{ccccc}
0 & (u_{R}^3)^\C & -(u_{R}^2)^\C & -u_{L}^1 & -d_{L}^1 \\
-(u_{R}^3)^\C & 0 &  (u_{R}^1)^\C & -u_{L}^2 & -d_{L}^2 \\
(u_{R}^2)^\C & -(u_{R}^1)^\C & 0 & -u_{L}^3 & -d_{L}^3 \\
u_{L}^1 & u_{L}^2 & u_{L}^3 & 0 & -(e_{R})^\C \\
d_{L}^1 & d_{L}^2 & d_{L}^3 & (e_{R})^\C & 0
\endm_i \,.
\end{equation}
When gauging their flavor index $i$ in the most natural way, i.e., both being triplets, we obtain just the flavor representation setting of the case IV.

The charge conjugated fields are given by $\psi^{\C}=C\bar{\psi}^{\T}$ where $C=\im\gamma_0\gamma_2$. The case I and case II can be obtained in the flipped $\SU{5}$ model \cite{Barr:1982aa,Derendinger:1984aa}, which is obtained from the Georgi--Glashow model by interchanging $u_R\leftrightarrow d_R$, $e_R\rightarrow\nu_R$ and by setting the right-handed charged leptons $e_{Ri}$ into the $\SU{5}$ singlets.

\subsection{Need for right-handed neutrinos}

\subsubsection{Anomaly freedom}

The model would suffer from the flavor gauge anomaly without adding the proper number of new flavored fermions into the model. In order not to destroy the electroweak anomaly free balance, we are adding only the electroweak singlets. It is common to call them the \emph{right-handed neutrino} fields denoted by $\nu_R$. Let us also note that there is a possibility to add some balanced setting of new electroweakly charged fermions, like, for instance, whole new generation of fermions. This, however, we avoid completely as it would not fit in the scenario of gauging flavor index of available generations only.

The source of the gauge anomaly is the fermion loop triangle with gauge boson vertices. Contributions from all flavored fermions in each triangle type should cancel. Triangles involving only electroweak gauge bosons vanish according to the textbook analysis based on the key fact that $\sum_f Y_f=0$. Triangles involving one or two flavor gauge bosons vanish by the same arguments.

It remains to investigate triangles involving three flavor gluons. The flavor gauge anomaly is in general proportional to the triangle factor \cite{Peskin:1995ev}
\begin{eqnarray}\label{flavor_anomaly_coefficient}
\Delta_\F\equiv\sum_{\psi_r}\xi_\psi\Tr{T^{a}_r\{T^{b}_r,T^{c}_r\}}=\frac{1}{2}\sum_{\psi_r}\xi_\psi A(r) d_{abc} \,,
\end{eqnarray}
where $T^{a}_r$ are the flavor generators for a given fermion representation $r$. The coefficient $\xi_\psi=\pm1$ introduces the sign from chiral projectors $P_{R,L}=(1\pm\gamma_5)/2$ according to the chirality of fermions. $A(r)$ is the anomaly coefficient of the representation $r$ defined as
\begin{eqnarray}
\frac{1}{2}d^{abc}A(r) & = & \Tr{T^{a}_r\{T^{b}_r,T^{c}_r\}} \,. \label{anomalyC}
\end{eqnarray}
It is related to the cubic Casimir invariant $C_3(r)$
\begin{eqnarray}
\frac{5}{6}d(G)A(r) & = & 2 d(r)C_3(r) \,,
\end{eqnarray}
where
\begin{eqnarray}
d(r)C_3(r) & = & d^{abc}\Tr{T^{a}_r T^{b}_r T^{c}_r}\,, \label{CasimirI}
\end{eqnarray}
$d(r)$ is the dimension of the representation $r$ and $r=G$ is the adjoint representation.
For few lowest $\SU{3}$ representations, $A(r)$ and $C_3(r)$ are listed in the Tab.~\ref{table}.

The triangle factor $\Delta_\F$ \eqref{flavor_anomaly_coefficient} must vanish. The contribution from electroweakly charged fermions however does not vanish. In Tab.~\ref{FermionSetting} the number $N=3$ or $N=5$ indicates how many additional flavor triplets of right-handed neutrinos are needed in order to make the flavor gauge dynamics anomaly free.

Adding the triplets is not the only possibility. Specially balanced settings including higher representations, sextet, octet, decuplet, etc., lead to the anomaly free models as well. Notice that an additional pair of a complex multiplet and its conjugate (e.g., $\mathbf{3}+\overline{\mathbf{3}}$ or $\mathbf{6}+\overline{\mathbf{6}}$), as well as real representation multiplet (e.g., $\mathbf{8}$) do not contribute to the anomaly.

\begin{table}[t]
\begin{center}
\begin{tabular}{lc|cc|cc}
$r$ & $d(r)$ & $C(r)$ & $C_2(r)$ & $A(r)$ & $C_3(r)$ \\
\hline
\hline
\vspace{-0.4cm}\\
$\mathbf{3}(\overline{\mathbf{3}})$   & $3$      & $1/2$  & $4/3$  & $(-)1$  & $(-)10/9$ \\
$\mathbf{6}(\overline{\mathbf{6}})$   & $6$      & $5/2$  & $10/3$ & $(-)7$  & $(-)35/9$ \\
$\mathbf{8}$                          & $8$ & $3$    & $3$    & $0$     & $0$       \\
$\mathbf{10}(\overline{\mathbf{10}})$ & $10$     & $15/2$ & $6$    & $(-)27$ & $(-)9$ \\
\hline
\hline
\end{tabular}
\end{center}
\caption[List of important coefficients for $\SU{3}$ group ]{\small List of important coefficients for the lowest-dimensional representations of the group $\SU{3}$. }
\label{table}
\end{table}

\subsubsection{Asymptotic freedom}

On the other hand, we should not add too many right-handed neutrinos in order not to destroy the asymptotic freedom of the flavor gauge dynamics. The asymptotic freedom is given by the negative sign of the $\beta$-function near origin $h\rightarrow0^+$.

The two-loop $\beta$-function is given by \cite{Machacek:1983tz}
\begin{eqnarray}\label{beta2}
\beta(h) & = & \hspace{0.2cm}-\frac{h^3}{(4\pi)^2}\left[\frac{11}{3}C(8)-\frac{2}{3}N^{\mathrm{EW}}C(3)-\frac{2}{3}\sum_r N^{\nu_R}_r C(r)\right] \nonumber\\
& & \hspace{0.2cm}-\frac{h^5}{(4\pi)^4}\left[\frac{34}{3}C(8)^2-N^{\mathrm{EW}}\left(2C_2(3)+\frac{10}{3}C(8)\right)C(3)
\right. \nonumber\\
& & \hspace{0.2cm}\left.
-\sum_r N^{\nu_R}_r\left(2C_2(r)+\frac{10}{3}C(8)\right)C(r)\right] \,,
\end{eqnarray}
where $N^{EW}=15$ is the number triplets and anti-triplets of electroweakly charged fermions, and $N^{\nu_R}_r$ is the number of right-handed neutrino
multiplets of a given representation $r$. The coefficient $C(r)$ is defined as
\begin{eqnarray}
\delta^{ab}C(r) & = & \Tr{T^{a}_r T^{b}_r} \,
\end{eqnarray}
and it is related to the quadratic Casimir invariant $C_2(r)$
\begin{eqnarray}
d(r)C_2(r) & = & \Tr{T^{a}_r T^{a}_r} \,
\end{eqnarray}
through the relation
\begin{eqnarray}
d(r)C_2(r)  & = & d(G)C(r) \,.
\end{eqnarray}
These coefficients are listed in the table Tab.~\ref{table} for few lowest-dimensional $\SU{3}$ representations.

The sign of the $\beta$-function near the origin is given by the sign of the $h^3$ one-loop coefficient. Requiring a negative sign we obtain a constraint on the number of right-handed neutrinos through their contribution $\eta_{\nu_R}$ to the one-loop $\beta$-function
\begin{equation}\label{etaAF_inequality}
  \eta_{\nu_R}<9 \,,
\end{equation}
where
\begin{subequations}
\begin{eqnarray}
\eta_{\nu_R} & \equiv & \sum_r N^{\nu_R}_r C(r)  \label{etaAF}\\
                      & = & \frac{1}{2}N^{\nu_R}_3+\frac{5}{2}N^{\nu_R}_6+ 3N^{\nu_R}_8+\frac{15}{2}N^{\nu_R}_{10}+\ldots \,.
\end{eqnarray}
\end{subequations}
The inequality \eqref{etaAF_inequality} leaves us with multiplets of lower dimensionality.

\subsubsection{Asymptotically and anomaly free settings}

Now we can combine our requirements of the asymptotic and anomaly freedom of the model. From the point of view of this analysis, the cases I and III are equivalent and so are the cases II and IV, see Tab.~\ref{FermionSetting}. Therefore we will distinguish the cases only by the number $N=3$ or $N=5$.

If the right-handed neutrino setting would contain a decuplet or higher complex representation, we would need to compensate its high anomaly coefficient $A(r)\geq 27$ by too many other multiplets and make the one-loop $\beta$-function positive. Adding the higher real representation immediately makes the one-loop $\beta$-function positive. So we are left to combine only lower representations, $\mathbf{3}$, $\overline{\mathbf{3}}$, $\mathbf{6}$, $\overline{\mathbf{6}}$, and $\mathbf{8}$. The Tab.~\ref{AAfreeSettings} shows all possible asymptotically and anomaly free settings.

\begin{table*}[t]
\begin{center}
\begin{small}
\begin{tabular}{c|l|l}
$N$ &  \ settings of $\nu_R$ $\SU{3}_\F$ representations ($n,m,k\geq0$) & \qquad $\eta_{\nu_R}$  \\
\hline
\hline
\vspace{-0.4cm}\\
3 & $3\times\mathbf{3} \phantom{{}+12\times\overline{\mathbf{3}}}\quad + n\times(\mathbf{3}+\overline{\mathbf{3}})+m\times\mathbf{8}+k\times(\mathbf{6}+\overline{\mathbf{6}})$
& $3/2\phantom{1}\ +(n+3m+5k\leq7)$ \\
%\hline
5 & $5\times\mathbf{3} \phantom{{}+12\times\overline{\mathbf{3}}} \quad+n\times(\mathbf{3}+\overline{\mathbf{3}})+m\times\mathbf{8}+k\times(\mathbf{6}+\overline{\mathbf{6}})$
& $5/2\phantom{1}\ +(n+3m+5k\leq6)$ \\
%\hline
5 & $1\times\mathbf{6}+2\times\overline{\mathbf{3}}\phantom{1} \quad+n\times(\mathbf{3}+\overline{\mathbf{3}}) +m\times\mathbf{8}+k\times(\mathbf{6}+\overline{\mathbf{6}})$
& $7/2\phantom{1}\ +(n+3m+5k\leq5)$ \\
%\hline
3 & $1\times\mathbf{6}+4\times\overline{\mathbf{3}}\phantom{1}\quad+n\times(\mathbf{3}+\overline{\mathbf{3}})+m\times\mathbf{8}$
& $9/2\phantom{1}\ +(n+3m\leq4)$ \\
%\hline
3 & $1\times\overline{\mathbf{6}}+10\times\mathbf{3}\quad+n\times(\mathbf{3}+\overline{\mathbf{3}})$
& $15/2\ +(n\leq1)$          \\
%\hline
5 & $1\times\overline{\mathbf{6}}+12\times\mathbf{3}$ & $17/2$ \\
\hline
\hline
\end{tabular}
\end{small}
\end{center}
\caption[ List of all anomaly and asymptotically free settings of right-handed neutrino flavor representations ]{\small The list of all possible anomaly and asymptotically free settings of $\SU{3}_\F$ right-handed neutrino representations. Since the anomaly freedom is not affected by adding any number of pairs of conjugate representations or any number of real representations, the possibility of adding $n$ pairs of $\mathbf{3}$ and $\overline{\mathbf{3}}$, $m$ copies of $\mathbf{8}$ and $k$ pairs of $\mathbf{6}$ and $\overline{\mathbf{6}}$ is indicated in the second column. However, adding too many of such representations would spoil the asymptotic freedom, hence the third column shows for each setting the corresponding value of $\eta_{\nu_R}$, \eqref{etaAF}. The indicated inequalities correspond to the asymptotic freedom condition $\eta_{\nu_R}<9$, \eqref{etaAF_inequality}. Clearly, as the numbers $n$, $m$, $k$ are integers, there is consequently only a finite number of all possible settings.}
\label{AAfreeSettings}
\end{table*}

\subsubsection{Absence of the perturbative infrared fixed point}

The number of suitable versions of the model is further reduced by applying even more stringent limit coming from demand not to produce too small, i.e., sub-critical, pertubative infrared fixed point, say $\alpha^{*}_{\mathrm{F,\,IR}}<0.5$, where $\alpha_\mathrm{F}\equiv\frac{h^2}{4\pi}$. It would leave the system in the chirally symmetric phase and prevent the whole symmetry breaking mechanism from taking place.

\begin{figure}[t]
\begin{tabular}{c}
\begin{tabular}{cc}
\includegraphics[width=0.45\textwidth]{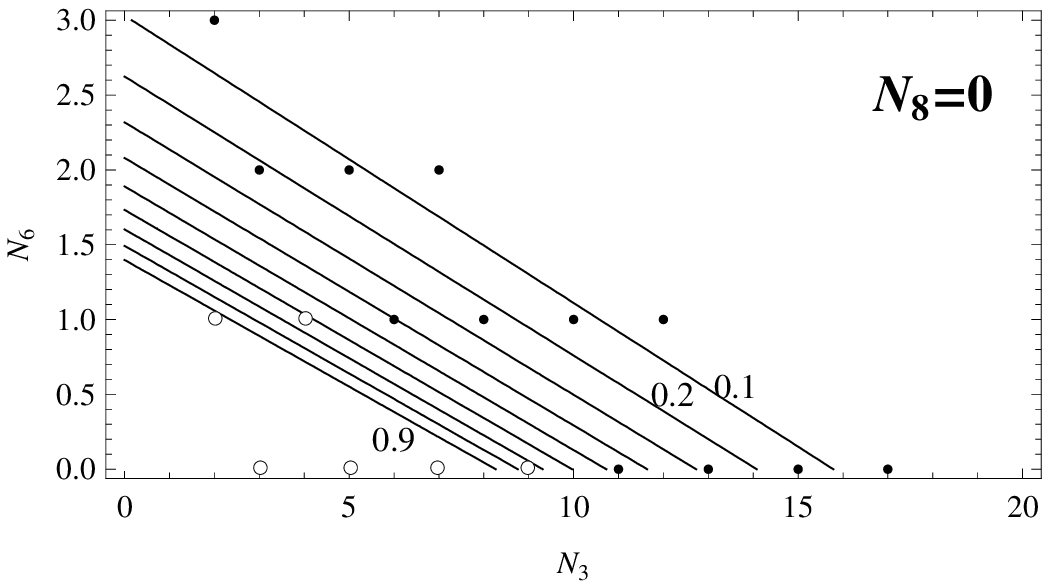} &
\includegraphics[width=0.45\textwidth]{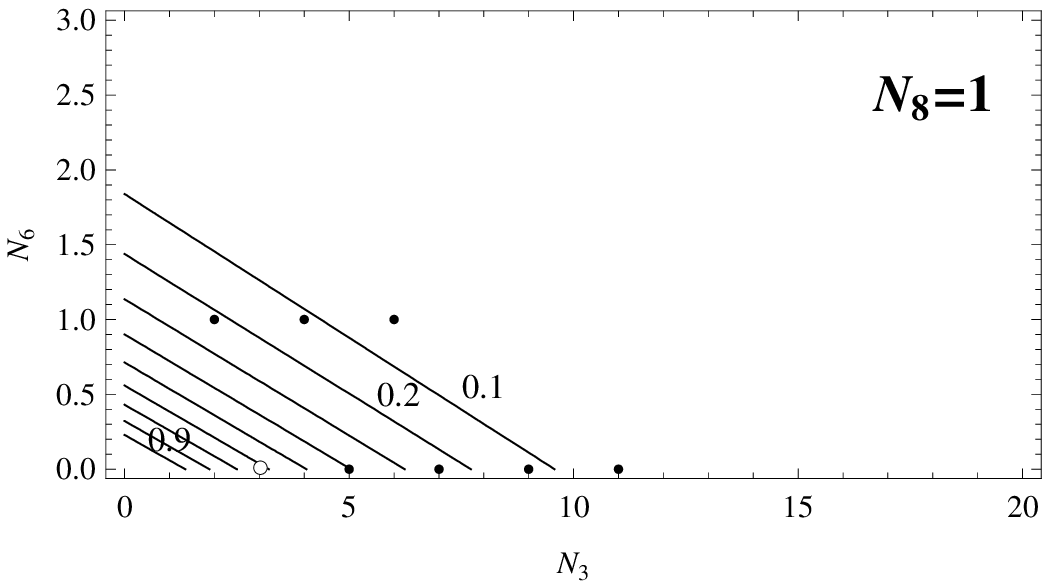}
\end{tabular} \\
\includegraphics[width=0.45\textwidth]{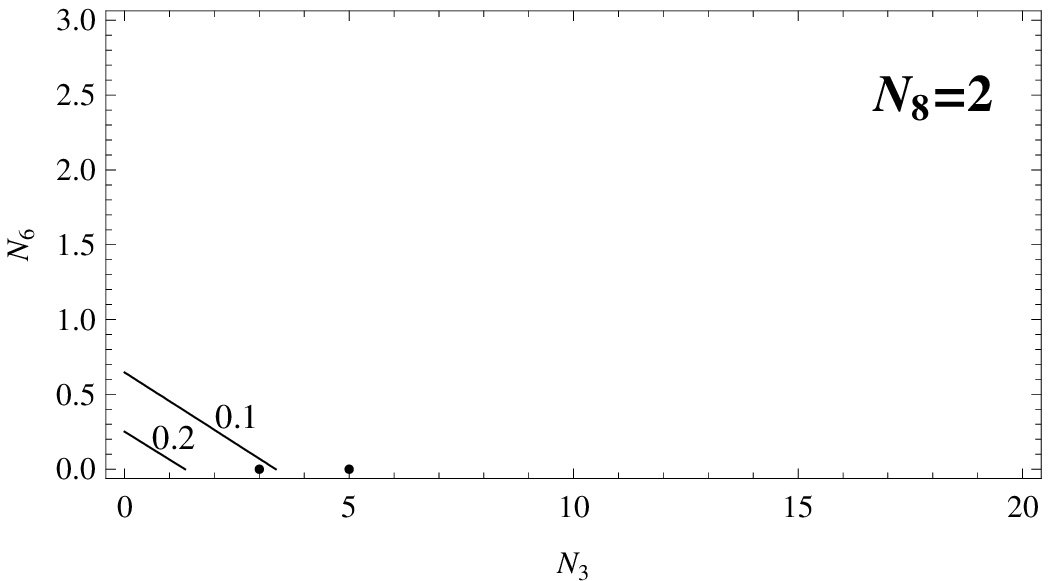}
\end{tabular}
\caption[Exclusion of model versions according to the perturbative infrared fixed point~$\alpha^{*}_{\mathrm{F,\,IR}}$]{The plots show various right-handed neutrino settings represented by dots in $(N_3,N_6)$-plane with $N_8=0$, $N_8=1$, and $N_8=2$. The contour lines represent given value of $\alpha^{*}_{\mathrm{F,\,IR}}$ calculated from the two-loop $\beta$-function \eqref{2loop_beta}. Black dots represent settings providing the values of $\alpha^{*}_{\mathrm{F,\,IR}}|_{2-\mathrm{loop}}$ smaller than $0.5$ which, therefore, can be understood as perturbative approximations of the infrared fixed points. White dots represent settings providing the values of $\alpha^{*}_{\mathrm{F,\,IR}}|_{2-\mathrm{loop}}$ larger than $0.5$ which. Therefore, the latter cannot be interpreted as the infrared fixed point of the theory, because the perturbative expansion of the $\beta$-function fails. While the former settings are disqualified by the smallness of their infrared fixed point, the latter are not.}
\label{IRfixedPoints}
\end{figure}

A zero of the two-loop $\beta$-function \eqref{beta2} gives an estimate of the perturbative infrared fixed point
\begin{equation}\label{2loop_beta}
\alpha^{*}_{\mathrm{F,\,IR}}=-4\pi\frac{-18+N^{\nu_R}_3+5N^{\nu_R}_6+6N^{\nu_R}_8}
                                       {-21+19N^{\nu_R}_3+125N^{\nu_R}_6+144N^{\nu_R}_8} \,.
\end{equation}
It gives a reliable estimate only if $\alpha^{*}_{\mathrm{F,\,IR}}$ comes out small. But that is sufficient to discard a number of versions. In Figs.~\ref{IRfixedPoints} we plot the value of $\alpha^{*}_{\mathrm{F,\,IR}}$ for various versions.

We choose the discriminating value of $\alpha^{*}_{\mathrm{F,\,IR}}$ being $0.5$ quite arbitrarily but motivated by QCD running coupling constant which is measured (still being in a perturbative regime) at the scale $1.7\,\mathrm{GeV}\gtrsim\Lambda_{\mathrm{QCD}}$ having the value $\alpha_\mathrm{s}(1.7\,\mathrm{GeV})\thickapprox0.35$ \cite{Nakamura:2010zzi}. Therefore we are left with only few versions of the model listed in Tab.~\ref{NUsettings}.

\begin{table}[t]
\begin{center}
\begin{tabular}{l|l|c|c}
\hline
\hline
    &       &        & approx. \\
$N$ & $\nu_R$ representation setting & chiral & vector-like \\
    &       &        & around $\Lambda_\mathrm{F}$ \\
\hline
3       & $3\times\mathbf{3}$ & \textbf{yes} & yes \\
        & $3\times\mathbf{3},\ 1\times(\mathbf{3},\overline{\mathbf{3}})$ & no & yes \\
        & $3\times\mathbf{3},\ 2\times(\mathbf{3},\overline{\mathbf{3}})$ & no & yes \\
        & $3\times\mathbf{3},\ 3\times(\mathbf{3},\overline{\mathbf{3}})$ & no & yes \\
        & $1\times\mathbf{6},\ 4\times\overline{\mathbf{3}}$ & \textbf{yes} & \textbf{no} \\
        & $1\times\mathbf{8},\ 3\times\mathbf{3}$ & no & \textbf{no} \\
\hline
5       & $5\times\mathbf{3}$ & \textbf{yes} & yes \\
        & $5\times\mathbf{3},\ 1\times(\mathbf{3},\overline{\mathbf{3}})$ & no & yes \\
        & $5\times\mathbf{3},\ 2\times(\mathbf{3},\overline{\mathbf{3}})$ & no & yes \\
        & $1\times\mathbf{6},\ 2\times\overline{\mathbf{3}}$ & \textbf{yes} & \textbf{no} \\
        & $1\times\mathbf{6},\ 2\times\mathbf{3},\ 1\times(\mathbf{3},\overline{\mathbf{3}})$ & no & \textbf{no} \\
\hline \hline
\end{tabular}
\end{center}
\caption[All viable versions of the flavor gauge model]{\small All viable versions of the flavor gauge model. }
\label{NUsettings}
\end{table}

\subsubsection{Chirality and non-vector-like nature}

The well defined versions of the flavor gauge model from Tab.~\ref{NUsettings} fall into various classes according to two criteria, their chirality and their approximate vector-like nature.

The versions containing right-handed neutrinos in real flavor representations $\mathbf{3}+\overline{\mathbf{3}}$ or $\mathbf{8}$ allow for the gauge invariant hard Majorana mass term. Therefore they are non-chiral. These versions of the flavor gauge model may be important when reproducing the realistic mass spectra mainly of neutrinos. On the other hand the presence of a bare parameter in the Lagrangian may indicate that such model is not fundamental. The origin of such mass parameter is not explained by the model. Rather, yet another dynamics operating at higher energy scale should be assumed to explain it. In this sense the chiral models appear to be more complete and more fundamental and therefore we prefer them.

From the high energy (around $\Lambda_\mathrm{F}$) perspective, the versions of the model that contain only $\mathbf{3}$ or $\overline{\mathbf{3}}$ are approximately vector-like with small non-vector-like perturbation given by the Standard Model gauge dynamics. This is due to the fact that right-handed anti-triplet Weyl spinor fields can be rewritten as charge conjugated left-handed triplet Weyl spinor fields and analogously for the left-handed fields. Rewritting all fermion fields as triplets we end up with manifestly vector-like setting having equal number of left- and right-handed fields due to the flavor gauge anomaly cancelation.

As an example we can take the minimal version, the case I with three right-handed neutrino triplets, and introduce new multi-component left- and right-handed Weyl spinors:
%\begin{eqnarray}\label{multi_spinor}
%\Psi_{L}^{(333)} & = & \Big(u^{1}_L \ u^{2}_L \ u^{3}_L \ d^{1}_L \ d^{2}_L \ d^{3}_L \ (d^{1}_R)^c \ (d^{2}_R)^c \ (d^{3}_R)^c \Big)^\T\,, \\
%\Psi_{R}^{(333)} & = & \Big(u^{1}_R \ u^{2}_R \ u^{3}_R \ (\nu_L)^c \ (e_L)^c \ e_R \ \nu^{\mathrm{i}}_R \ \nu^{\mathrm{ii}}_R \ \nu^{\mathrm{iii}}_R \Big)^\T\ .
%\end{eqnarray}
\begin{eqnarray}\label{multi_spinor}
\Psi_{L}^{(333)} = \beginm{c}u^{1}_L \\ u^{2}_L \\ u^{3}_L \\ d^{1}_L \\ d^{2}_L \\ d^{3}_L \\ (d^{1}_R)^c \\ (d^{2}_R)^c \\ (d^{3}_R)^c \endm\ \sim\ \mathbf{3}\,, \quad
\Psi_{R}^{(333)} = \beginm{c}u^{1}_R \\ u^{2}_R \\ u^{3}_R \\ (\nu_L)^c \\ (e_L)^c \\ e_R \\ \nu^{\mathrm{i}}_R \\ \nu^{\mathrm{ii}}_R \\ \nu^{\mathrm{iii}}_R \endm\ \sim\ \mathbf{3}\,.
\end{eqnarray}
The indices $1,\ 2,\ 3$ indicate color of quarks and the indices $\mathrm{i},\ \mathrm{ii},\ \mathrm{iii}$ label the right-handed neutrinos. Analogously, this field redefinition can be done in every version denoted in Tab.~\ref{NUsettings} as approximately vector-like. Omitting the Standard Model gauge dynamics the model Lagrangian \eqref{L_model} is rewritten as
\begin{eqnarray}\label{L_F_ala_QCD}
{\cal L}^{(333)}_\F & = & -\frac{1}{4}F_{\mu\nu}^a F^{\mu\nu a} +\bar\Psi^{(333)}_R\gamma^\mu(\im\partial_\mu+hT^{a}C^{a}_\mu)\Psi^{(333)}_R+\bar\Psi^{(333)}_L\gamma^\mu(\im\partial_\mu+hT^{a}C^{a}_\mu)\Psi^{(333)}_L \,,\quad\quad
\end{eqnarray}
where $T^{a}=\tfrac{1}{2}\lambda^{a}=T^{a}_\mathbf{3}$. The dynamics resembles that of QCD. As such it presumably prefers pairing in the $\mathbf{3}\times\overline{\mathbf{3}}$ channel forming the analogue of the chiral condensate \eqref{chiral_condensate}. It does not ensure the flavor symmetry breaking and without that it should rather confine. In this case the dynamical flavor symmetry breaking cannot be derived from the Lagrangian \eqref{L_F_ala_QCD} only. Without solving a complete system including the Standard Model gauge perturbations, the electric and color charge conserving and flavor symmetry breaking solution is only believed to be energetically more favorable than the flavor symmetry preserving one.

On the other hand, the versions of the model that contain right-handed neutrinos in higher representation $\mathbf{6}$, are essentially non-vector-like and prefer right-handed neutrino pairing in the Majorana channels $\mathbf{6}\times\overline{\mathbf{3}}$, or $\mathbf{6}\times\mathbf{6}$, that certainly break the flavor symmetry.

\begin{center}*\end{center}

The \emph{minimal} version with three right-handed neutrino triplets denoted by (333) was analyzed in the paper \cite{Benes:2011gi}. In this work we will pursue mainly the \emph{non-minimal} version, one of only two versions which are both non-vector-like and chiral. Its right-handed neutrino setting is
$(\mathbf{6},\overline{\mathbf{3}},\overline{\mathbf{3}},\overline{\mathbf{3}},\overline{\mathbf{3}})$, and we will denote it by (63333). We prefer it to the second version with $(\mathbf{6},\overline{\mathbf{3}},\overline{\mathbf{3}})$, because it is the version which gives a distinct role to the $u$-type quarks.

%Impossibility of tumbling \cite{Eichten:1982ef}

\section{Global symmetries}

Apart from the gauge symmetries, the complete classical Lagrangian possesses also a rich structure of global symmetries. We will study the symmetries only of the non-minimal version (63333).

The Lagrangian \eqref{L_model} with the Standard Model gauge dynamics turned off has a rather big chiral symmetry $G_{\chi}$. Turning the Standard Model gauge dynamics on, thanks to the specific alignment of the Standard Model gauge symmetry with respect to the flavor gauge symmetry alignment, the chiral symmetry is almost completely explicitly broken down to much smaller subgroup. Because the explicit breaking by the Standard Model gauge dynamics can be treated as a perturbation, the chiral symmetry $G_{\chi}$ may be relevant for classification of flavor bound states.

The remaining classically exact global symmetry has two subgroups corresponding to two sectors of the Lagrangian, the electroweakly charged sector possessing only four global $\U{1}$ symmetries, and the right-handed neutrino sector which possesses richer structure of the global symmetries. These global symmetries are symmetries of the Lagrangian, i.e. symmetries of the classical theory. Nevertheless quantum effects break them by axial anomalies. Some symmetries can survive being \emph{exact}, for some symmetries the breaking by anomaly is rather negligible and they can still be treated as good approximate symmetries, and some symmetries are not good symmetries at all because of the strong effect of their anomalies. In the following we recognize these global symmetries.

\subsection{Approximate chiral symmetry}

In the (63333) version of the model we have altogether $19$ flavor triplet or anti-triplet fermion fields out of which $15$ are electroweakly charged fermions, $u_{L}^i$, $d_{L}^i$, $e_{L}$, $\nu_{L}$, $u_{R}^i$, $d_{R}^i$ and $e_{R}$, and $4$ are right-handed neutrinos $\nu_{R}^s$. Rewriting all these fields into a triplet representation we can define the multi-component left- and right-handed Weyl spinors:
%\begin{eqnarray}\label{multi_spinor_63333}
%\Psi_{L}^{(63333)} & = & \Big(u^{1}_L \ u^{2}_L \ u^{3}_L \ d^{1}_L \ d^{2}_L \ d^{3}_L \ (d^{1}_R)^c \ (d^{2}_R)^c \ (d^{3}_R)^c \ (\nu^{\mathrm{i}}_R)^c \ (\nu^{\mathrm{ii}}_R)^c \ (\nu^{\mathrm{iii}}_R)^c \ (\nu^{\mathrm{iv}}_R)^c \Big)^\T\,, \\
%\Psi_{R}^{(63333)} & = & \Big(u^{1}_R \ u^{2}_R \ u^{3}_R \ (\nu_L)^c \ (e_L)^c \ e_R  \Big)^\T \,,
%\end{eqnarray}
\begin{eqnarray}\label{multi_spinor_63333}
\Psi_{L}^{(63333)}  =  \beginm{c}u^{1}_L \\ u^{2}_L \\ u^{3}_L \\ d^{1}_L \\ d^{2}_L \\ d^{3}_L \\ (d^{1}_R)^c \\ (d^{2}_R)^c \\ (d^{3}_R)^c \\ (\nu^{\mathrm{i}}_R)^c \\ (\nu^{\mathrm{ii}}_R)^c \\ (\nu^{\mathrm{iii}}_R)^c \\ (\nu^{\mathrm{iv}}_R)^c \endm\ \sim\ \mathbf{3}\,, \quad
\Psi_{R}^{(63333)}  =  \beginm{c}u^{1}_R \\ u^{2}_R \\ u^{3}_R \\ (\nu_L)^c \\ (e_L)^c \\ e_R \endm\ \sim\ \mathbf{3} \,,
\end{eqnarray}
where the indices $1,\,2,\,3$ indicate color of quarks and the indices $\mathrm{i},\,\mathrm{ii},\,\mathrm{iii},\,\mathrm{iv}$ label the right-handed neutrino triplets. Additionally we have the right-handed neutrino sextet $\nu_{R\mathbf{6}}$.

If we turn off the Standard Model gauge dynamics we have the Lagrangian \eqref{L_model} which can be rewritten in terms of \eqref{multi_spinor_63333} as
\begin{eqnarray}
{\cal L}^{(63333)}_\F & = & -\frac{1}{4}F_{\mu\nu}^a F^{\mu\nu a} +\bar{\nu}_{R\mathbf{6}}\gamma^\mu(\im\partial_\mu+hT^{a}_\mathbf{6}C^{a}_\mu)\nu_{R\mathbf{6}}\\
&& +\bar{\Psi}^{(63333)}_L\gamma^\mu(\im\partial_\mu+hT^{a}_\mathbf{3}C^{a}_\mu)\Psi^{(63333)}_L+\bar{\Psi}^{(63333)}_R\gamma^\mu(\im\partial_\mu+hT^{a}_\mathbf{3}C^{a}_\mu)\Psi^{(63333)}_R \nonumber
\end{eqnarray}
and at the classical level it possesses the chiral symmetry
\begin{equation}\label{chiral_symmetry_63333}
G_{\chi}=\U{13}_{L\mathbf{3}}\times\U{6}_{R\mathbf{3}}\times\U{1}_{R\mathbf{6}} \,,
\end{equation}
where $\U{13}_{\mathbf{3}L}$, $\U{6}_{\mathbf{3}R}$ and $\U{1}_{\mathbf{6}R}$ are the transformations of the left-handed flavor triplet multispinor $\Psi_{L}^{(63333)}$, the right-handed flavor triplet multispinor $\Psi_{R}^{(63333)}$ and the flavor sextet right-handed neutrino field $\nu_{R\mathbf{6}}$, respectively.
In the full model, this chiral symmetry is broken by the Standard Model gauge interactions. At least at high scales where the Standard Model gauge coupling constants are small, the chiral symmetry $G_{\chi}$ is the approximate symmetry.

\subsection{Global symmetry of electroweakly charged sector}

In the flavor gauge model the sector of electroweakly charged fermions is such that it does not possess any non-Abelian global symmetries. This is due to the specific alignment of four gauge symmetries.

The sector of electroweakly charged fermions possesses only some Abelian symmetries. In a free theory there are as many independent $\U{1}$ phases as there are individual complex fields. The electroweakly charged fermion sector is formed by $N_\F(N^{L}_\ell+N^{R}_\ell)+N_\mathrm{C} N_\F(N^{L}_q+N^{R}_q)=45$ fermion fields as in each of $N_\F=3$ families (flavors) there are $N^{L}_\ell=2$ left-handed and $N^{R}_\ell=1$ right-handed leptons, and $N^{L}_q=N^{R}_q=2$ left- and right-handed quarks of $N_\mathrm{C}=3$ colors. Non-Abelian gauge symmetries require the equality of phases of fields in the same multiplet.

When counting the number of phases in the Standard Model, gauge symmetries require to count color and weak isospin multiplets only once, i.e., $N_\mathrm{C}\rightarrow 1$ and $N^{L}_{\ell,q}\rightarrow1$ resulting in 15 fermion phases. Additionally the Higgs doublet brings 1 more phase and 11 independent constraints among the phases coming from the Yukawa interactions. In total there are $15+1-11=5$ independent phases. One of them is gauged being the weak hypercharge $Y$. Remaining four phases are baryon number $B$, and three individual lepton numbers $L_e$, $L_\mu$, and $L_\tau$.

In the flavor gauge model, additionally, the flavor index is gauged and we have to count each flavor triplet only once, i.e., $N_F\rightarrow1$. This results in 5 fermion phases, one for each independent multiplet, $q_{L}$, $\ell_{L}$, $u_{R}$, $d_{R}$ and $e_{R}$. One of the phases is again gauged as the weak hypercharge $Y$. For four other phases there can be chosen a basis formed by the baryon number $B$, the common lepton number $L$, the axial-baryon number $B_5$, and the axial-lepton number $L_5$,
\begin{equation}\label{EW_symmetry}
\U{1}_\mathrm{B}\times\U{1}_\mathrm{L}\times\U{1}_{\mathrm{B}_5}\times\U{1}_{\mathrm{L}_5} \,.
\end{equation}
The corresponding Noether currents are
\begin{subequations}\label{Abelian_currents}
\begin{eqnarray}
J^{\mu}_{\mathrm{L}} & = & \bar{\ell}_L\gamma^\mu \ell_L+\bar{e}_R\gamma^\mu e_R \,,\\
J^{\mu}_{\mathrm{L}_5} & = & -\bar{\ell}_L\gamma^\mu \ell_L+\bar{e}_R\gamma^\mu e_R \,,\\
J^{\mu}_{\mathrm{B}} & = & \frac{1}{3}(\bar{q}_L\gamma^\mu q_L+\bar{u}_R\gamma^\mu u_R+\bar{d}_R\gamma^\mu d_R) \,,\\
J^{\mu}_{\mathrm{B}_5} & = & \frac{1}{3}(-\bar{q}_L\gamma^\mu q_L+\bar{u}_R\gamma^\mu u_R+\bar{d}_R\gamma^\mu d_R) \,,
\end{eqnarray}
\end{subequations}
where the summations over the flavor, color and weak isospin indices are suppressed.

\subsection{Global symmetry of right-handed neutrino sector}

The sector of right-handed neutrinos is charged only by flavor. Therefore to recognize its symmetry it is enough to count the numbers $N^r$ of fields accommodated in a given type of the flavor representation $r$. In general, the symmetry reads
\begin{equation}
G_\mathrm{S}^{\mathrm{general}} \equiv \U{N^\mathbf{3}}\times\U{N^{\overline{\mathbf{3}}}}\times\U{N^\mathbf{6}}\times\U{N^{\overline{\mathbf{6}}}}\times\U{N^\mathbf{8}}\,
\end{equation}
and it is called the \emph{sterility symmetry}. For the (63333) version of the model, the sterility symmetry is
\begin{equation}\label{sterility_symmetry}
G_\mathrm{S}^{(63333)} = \U{1}_{\mathrm{S}_6}\times\U{1}_{\mathrm{S}_3}\times\SU{4}_\mathrm{S} \,.
\end{equation}
The corresponding Noether currents are%\footnote{ The peculiar values of $\U(1)_\S$ charges are chosen in order to normalize the corresponding anomalies in \eqref{Jdivergences}.  }
\begin{subequations}
\begin{eqnarray}
J^{\mu}_{\mathrm{S}_6} & = & \frac{1}{5}\bar{\nu}_{R\mathbf{6}}\gamma^\mu\nu_{R\mathbf{6}}=\frac{2}{5}\Tr\bar{\xi}_{R}\gamma^\mu\xi_{R} \,, \label{Abelian_sterility6_current}\\
J^{\mu}_{\mathrm{S}_3} & = & \frac{1}{4}\bar{\nu}_{R\overline{\mathbf{3}}}^s\gamma^\mu\nu_{R\overline{\mathbf{3}}}^s=\frac{1}{4}\bar{\zeta}_{R}^s\gamma^\mu\zeta_{R}^s \,, \label{Abelian_sterility3_current}\\
J^{\mu}_{\mathrm{S},\sigma} & = & \bar{\nu}_{R\overline{\mathbf{3}}}^s\left[S_\sigma\right]^{sr}\gamma^\mu\nu_{R\overline{\mathbf{3}}}^r=
\bar{\zeta}_{R}^s\left[S_\sigma\right]^{sr}\gamma^\mu\zeta_{R}^r \label{non_Abelian_sterility_current}
\,,
\end{eqnarray}
\end{subequations}
where the summation over the flavor index is suppressed. The indices $s,\,r=\mathrm{i},\dots,\mathrm{iv}$ run over four right-handed neutrino anti-triplets. Matrices $S_\sigma$, $\sigma=1,\dots,15$, are the $\SU{4}_\mathrm{S}$ generators. We denote sextet right-handed neutrinos as $\xi_R$ and anti-triplet right-handed neutrinos as $\zeta_R$:
\begin{subequations}\label{RHnu}
\begin{eqnarray}
\xi_{R} & \equiv & T^{\iota}_{\mathrm{(sym.)}}\nu^{\iota}_{R\mathbf{6}} \,,\\
\zeta^{s}_{R} & \equiv &
\nu_{R\overline{\mathbf{3}}}^s \,,
\end{eqnarray}
\end{subequations}
where $\iota=1,\dots,6$ is the flavor index. The six symmetric matrices $T^{\iota}_{\mathrm{(sym.)}}$ are given as the set of symmetric Gell-Mann matrices complemented by the unit matrix, i.e.,
\begin{equation}
T^{\iota}_{\mathrm{(sym.)}}=\Big\{\frac{1}{2\sqrt{3}}\openone,\,\frac{1}{2}\lambda_1,\,\frac{1}{2}\lambda_3,\,\frac{1}{2}\lambda_4,\,\frac{1}{2}\lambda_6,\,\frac{1}{2}\lambda_8\Big\} \,.
\end{equation}
The definitions \eqref{Abelian_sterility6_current} and \eqref{Abelian_sterility3_current} of the sextet and triplet sterility numbers to be $S_6=\tfrac{1}{5}$ and $S_3=\tfrac{1}{4}$ is chosen in order to normalize the anomaly coefficient in \eqref{Jdivergences}.
%\eqref{anomaly_coefficient}, which is proportional to $C(r)$, see Tab.~\ref{table}, to the same value for the sextet as for triplet. The choice $S_3=\tfrac{1}{4}$, similar reason applies to reflect the fact that there are four flavor triplets carrying the triplet sterility number.

\subsection{Global symmetries at quantum level}

\subsubsection{Axial anomaly in general}

At the classical level the Noether current $J^\mu(x)$ of an exact global symmetry satisfies the equation of continuity $\partial_\mu J^\mu(x)=0$. At the quantum level, the four-divergence of a Green's function containing the Noether current decides whether the symmetry is exact. When the four-divergence does not vanish at the quantum level despite the Noether current is conserved at the classical level, it is said that the symmetry is broken by the Adler--Bell--Jackiw \emph{axial anomaly} \cite{Adler:1969gk,Bell:1969ts}. The adjective `axial' reflects the fact that the anomaly appears whenever a quantum path integral measure is not invariant with respect to the axial transformation while the classical action is \cite{Fujikawa:1979ay}.

Properly applied regularization of the triangular loop integral contributing to the three-point function $T^{\mu\alpha\beta}_{ab}(x,y,z)=\langle J^\mu(x)A_{a}^\alpha(y)A_{b}^\beta(z)\rangle$ yields
\begin{equation}\label{4divT}
q_\mu T^{\mu\alpha\beta}_{ab}(q,k_1,k_2)=C_{ab}\frac{\im}{2\pi^2}\epsilon^{\alpha\beta\rho\sigma}k_{1\rho}k_{2\sigma}\,,
\end{equation}
where $k_{1\rho}$ and $k_{2\sigma}$ are outgoing momenta of the gauge fields $A_{a}^\alpha(y)$ and $A_{b}^\beta(z)$. The coefficient $C_{ab}$ is a symmetry group factor
\begin{equation}\label{anomaly_coefficient}
C_{ab} = g^2\Tr\big(\xi Q\{T_aT_b\}\big) \,,
\end{equation}
where $g$ is the coupling constant of the corresponding gauge dynamics and $T_a$ are generators of the gauge symmetry group normalized to $\Tr\{T_aT_b\}=\tfrac{1}{2}\delta_{ab}$. The coefficient $\xi=\pm\tfrac{1}{2}$ originates from the chiral projectors $P_{R,L}=(1\pm\gamma_5)/2$ of a given fermion forming the triangle loop. $Q$ is the matrix representation of the global symmetry charge given as $\int J^0(x)\d^3\vec{x}$. The trace is taken over all internal indices. The equation \eqref{4divT} can be translated into the equation for four-divergence of the Noether current
\begin{equation}
\partial_\mu J^\mu=C_{ab}\frac{1}{8\pi^2}F_{a\alpha\beta}\tilde{F}_{b}^{\alpha\beta} \,,
\end{equation}
where the field strength tensor $F_{a}^{\alpha\beta}$ corresponds to the gauge fields $A_{a}^\mu$, and its dual is defined as $\tilde{F}_{\alpha\beta}=\frac{1}{2}\epsilon_{\alpha\beta\rho\sigma}F^{\rho\sigma}$.

\subsubsection{Axial anomaly in the flavor gauge model}

We apply these general formulae to the flavor gauge model. The non-Abelian part of the sterility symmetry $J^{\mu}_{\mathrm{S},i}$ \eqref{non_Abelian_sterility_current} is anomaly free since its generators are traceless, thus we write
\begin{equation}
\partial_\mu J^{\mu}_{\mathrm{S},i} = 0 \,.
\end{equation}
On the other hand all Abelian currents \eqref{Abelian_currents}, \eqref{Abelian_sterility6_current} and \eqref{Abelian_sterility3_current} are anomalous and we can write:
\begin{equation}\label{Jdivergences}
\begin{array}{lcrrrr}
\partial_\mu J_{\mathrm{B}}^\mu & = & 3{\cal Y} & -3{\cal W} \,, & &  \vspace{5pt}\\
\partial_\mu J_{\mathrm{L}}^\mu & = & 3{\cal Y} & -3{\cal W} &  & -{\cal F} \,, \vspace{5pt}\\
\partial_\mu J_{\mathrm{B}_5}^\mu & = & \frac{11}{3}{\cal Y} & +3{\cal W} & +4{\cal G} & +4{\cal F} \,, \vspace{5pt}\\
\partial_\mu J_{\mathrm{L}_5}^\mu & = & 9{\cal Y} & +3{\cal W} &  & +3{\cal F} \,, \vspace{5pt}\\
\partial_\mu J_{\mathrm{S}_3}^\mu & = & & & & {\cal F} \,, \vspace{5pt}\\
\partial_\mu J_{\mathrm{S}_6}^\mu & = & & & & {\cal F} \,, \vspace{5pt}\\
\end{array}
\end{equation}
where
\begin{eqnarray}\label{anomaly_elements}
{\cal Y} & = & \frac{g_{1}^2}{8\pi^2}Y_{\mu\nu}\tilde{Y}^{\mu\nu} \,,\\
{\cal W} & = & \frac{g_{2}^2}{32\pi^2}W_{\mu\nu a}\tilde{W}^{\mu\nu}_a \,,\\
{\cal G} & = & \frac{g_{3}^2}{32\pi^2}G_{\mu\nu i}\tilde{G}^{\mu\nu}_i \,,\\
{\cal F} & = & \frac{h^2}{32\pi^2}F_{\mu\nu a}\tilde{F}^{\mu\nu}_a \,.
\end{eqnarray}
The first two equations \eqref{Jdivergences} for $J_{\mathrm{B}}^\mu$ and $J_{\mathrm{L}}^\mu$ agree with the textbook calculation which can be found for instance in \cite{Peccei:1998jv} considering different weak hypercharge convention.

Among the currents there are two linear combinations which correspond to the exact symmetries. These can be written as
\begin{subequations}\label{Lepton_number}
\begin{eqnarray}
J^{\mu}_{\mathrm{B}-\mathrm{L}-\mathrm{S}} & = & J^{\mu}_{\mathrm{B}}-J^{\mu}_{\mathrm{L}}-\left(a J^{\mu}_{\mathrm{S}_3}+(1-a)J^{\mu}_{\mathrm{S}_6}\right) \,, \\
J^{\mu}_{\mathrm{S}_3-\mathrm{S}_6} & = & J^{\mu}_{\mathrm{S}_3}-J^{\mu}_{\mathrm{S}_6} \,,
\end{eqnarray}
\end{subequations}
where the real coefficient $a$ is arbitrary. The $(B-L-S)$ symmetry is a direct analogue of the anomaly free $B-L$ symmetry in the Standard Model.

Further, one of the linear combinations, that which is broken dominantly by the QCD anomaly, is naturally identified with the current $J^{\mu}_\mathrm{P.-Q.}(x)$ of the Peccei--Quinn symmetry instrumental in solving the strong $CP$ problem \cite{Peccei:1977hh,Peccei:1977ur}.

\section{Flavor symmetry self-breaking by masses}

The flavor symmetry is not a property of the fermion mass spectrum, therefore it has to be dynamically broken. We do not introduce any other dynamics in order to provide the flavor symmetry breaking, but we assume that the flavor gauge dynamics self-breaks. The program of the self-breaking gauge theories started with the Abelian gauge symmetry \cite{Freundlich:1970kn,Jackiw:1973tr,Cornwall:1973ts} and its non-Abelian generalization was elaborated in \cite{Eichten:1974et,Smit:1974je}.

Because the flavor gauge model is formulated as a chiral gauge theory, at the Lagrangian level the masslessness of fermion fields is protected by the chiral gauge symmetries, the flavor and electroweak symmetries in particular. The masslessness of gauge fields is protected by the gauge nature of the symmetries. \emph{Massless fields can, however, excite massive particles}, if the protective symmetries are spontaneously broken.

\subsection{Fermions}
\label{ssec:fermions}

Massless fermion fields excite massive fermions if the ground state is not invariant under independent rotations of their left-handed and right-handed components \cite{Nambu:1961tp}. Operationally this manifests by nonzero chirality-changing parts $\Sigma_\psi(p^2)$ of the full propagators for $\psi=u,d,n,e$. For the charged fermions, the field $\psi$ is a Dirac field composed by the chiral components as
\begin{eqnarray}
\label{f=ude}
\psi &\equiv& \psi_L + \psi_R \quad\quad (\psi=u,d,e) \,.
\end{eqnarray}
For neutrinos, which are neutral and thus they can develop masses of both Dirac and Majorana types, it is more convenient to accommodate them in a single Nambu--Gorkov multispinor field
\begin{equation}\label{NGmultiplet}
n=\beginm{c} \nu_L+(\nu_{L})^\C \\
\nu_{R\overline{\mathbf{3}}}^\mathrm{i}+(\nu_{R\overline{\mathbf{3}}}^\mathrm{i})^\C \\ \vspace{-0.43cm}\\ \nu_{R\overline{\mathbf{3}}}^\mathrm{ii}+(\nu_{R\overline{\mathbf{3}}}^\mathrm{ii})^\C \\ \vspace{-0.43cm}\\
\nu_{R\overline{\mathbf{3}}}^\mathrm{iii}+(\nu_{R\overline{\mathbf{3}}}^\mathrm{iii})^\C \\ \vspace{-0.43cm}\\ \nu_{R\overline{\mathbf{3}}}^\mathrm{iv}+(\nu_{R\overline{\mathbf{3}}}^\mathrm{iv})^\C \\ \vspace{-0.43cm}\\ \nu_{R\mathbf{6}}+(\nu_{R\mathbf{6}})^\C   \endm
\,,
\end{equation}
where the flavor indices are suppressed.

The corresponding full propagators $S_\psi(p)$ are considered in the same general form as in \eqref{fermion_propagator}
\begin{eqnarray}
\label{sigma_ansatz}
S_\psi^{-1}(p) &=& \slashed{p}-\mathbf{\Sigma}_\psi(p^2) \,,
\end{eqnarray}
with
\begin{eqnarray}
\mathbf{\Sigma}_\psi(p^2) &=& \Sigma_\psi(p^2) P_L+\Sigma^{\dag}_\psi(p^2)P_R \,.
\end{eqnarray}
Notice, that we are again omitting the wave function renormalizations as in \eqref{inverse_fermion_propagator_Yukawa}. Both $S_\psi^{-1}(p)$ and $\Sigma_\psi(p^2)$ are now complex momentum-dependent matrices of the dimension $3\times3$ for charged fermions and $21\times21$ for neutrinos. Moreover, the neutrino matrix $\Sigma_n(p^2)$ is symmetric, in more detail, suppressing the momentum arguments,
\begin{equation}\label{NGselfenergy}
\Sigma_n=\beginm{c|c|c}
    L^{\mathbf{3}\times\mathbf{3}} & \ \ \ \ D^{\overline{\mathbf{3}}\times\mathbf{3}}_s \ \ \ \ &
                                                            D^{\mathbf{6}\times\mathbf{3}} \\ \hline
    D^{\mathbf{3}\times\overline{\mathbf{3}}}_r & R^{\overline{\mathbf{3}}\times\overline{\mathbf{3}}}_{rs} &
                    R^{\mathbf{6}\times\overline{\mathbf{3}}}_s \vphantom{\begin{array}{c}\ \\ \ \end{array}}\\ \hline && \vspace{-0.4cm}\\
    D^{\mathbf{3}\times\mathbf{6}} & R^{\overline{\mathbf{3}}\times\mathbf{6}}_r & R^{\mathbf{6}\times\mathbf{6}} \\ \endm \,.
\end{equation}
The diagonal blocks, $L^{\mathbf{3}\times\mathbf{3}}$, $R^{\overline{\mathbf{3}}\times\overline{\mathbf{3}}}$ and $R^{\mathbf{6}\times\mathbf{6}}$ are symmetric matrices, and $D^{\mathbf{3}\times\overline{\mathbf{3}}}_s=[D^{\overline{\mathbf{3}}\times\mathbf{3}}_s]^\T$, $D^{\mathbf{3}\times\mathbf{6}}=[D^{\mathbf{6}\times\mathbf{3}}]^\T$ and $R^{\overline{\mathbf{3}}\times\mathbf{6}}_s=[R^{\mathbf{6}\times\overline{\mathbf{3}}}_s]^\T$ where $s,\,r=\mathrm{i},\dots,\mathrm{iv}$.

The fermion mass spectrum is given by poles of the full propagator, i.e., by the solutions of the equation \eqref{m_from_Sigma}
\begin{eqnarray}
\det\big[p^2-\Sigma_\psi(p^2)\Sigma_\psi^{\dag}(p^2)\big] &=& 0 \,.
\label{poles_fermion}
\end{eqnarray}

Breaking of the flavor $\SU{3}_{\F}$ symmetry by the fermion self-energies can be written compactly as
\begin{eqnarray}
\mathbf{\Sigma}_\psi t_{\psi}^a - \bar t_{\psi}^a \mathbf{\Sigma}_\psi &\neq& 0 \,,
\end{eqnarray}
where the flavor generators in the bases \eqref{f=ude} and \eqref{NGmultiplet} are denoted as $t_{\psi}^a$ and they have the form
\begin{subequations}
\begin{eqnarray}
t_{\psi}^a &\equiv& T_{\psi_L}^a P_L + T_{\psi_R}^a P_R \quad\quad (\psi=u,d,e) \,, \\
t_{n}^a &\equiv& \beginm{ccc} T_{\mathbf{3}}^{a}P_R-[T_{\mathbf{3}}^{a}]^\mathrm{T}P_L & 0 & 0 \\
                            0 & \hspace{-0cm}\openone_{4\times4}\left(T_{\mathbf{3}}^{a}P_L-[T_{\mathbf{3}}^{a}]^\mathrm{T}P_R\right) & 0 \\
                            0 & 0 & \hspace{-0cm}T_{\mathbf{6}}^{a}P_R-[T_{\mathbf{6}}^{a}]^\mathrm{T}P_L \endm\,\hspace{0.5cm}
\end{eqnarray}
\end{subequations}
and where $\bar{t}_{\psi}^a \equiv \gamma_0 t_{\psi}^a \gamma_0$.

\begin{center}
\begin{figure}[t]
\begin{center}
\includegraphics[width=0.8\textwidth]{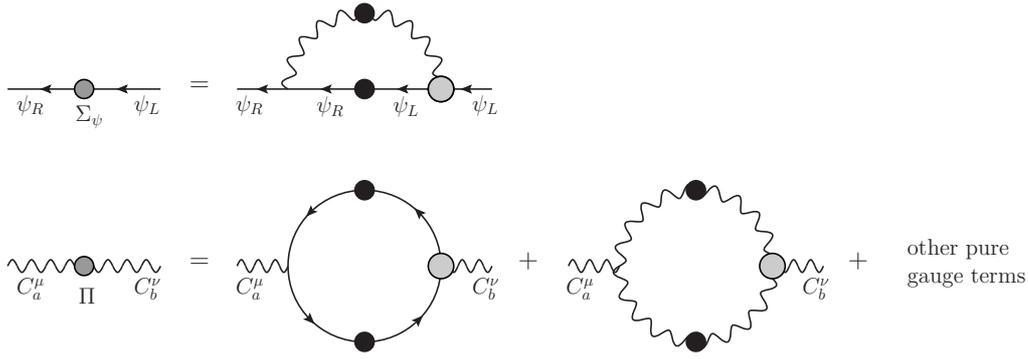}
\caption[Schwinger--Dyson equations within the flavor gauge model]{Mutual interactions of the flavor-gluons and the fermions, giving rise to the $\SU{3}_{\F}$ symmetry-breaking by fermion self-energies $\Sigma_\psi$ and flavor-gluon polarization tensor $\Pi_{ab}^{\mu\nu}$. Contributions of other pure gauge diagrams are omitted in the second line. The black and grey blobs on the lines stand for the full and 1PI propagators, respectively. The grey blobs on vertices stand for proper vertices. }
\label{fig:SDE}
\end{center}
\end{figure}
\end{center}

The exchange of the flavor gauge bosons $C_{a}^\mu$ between right-handed $\psi_R$ and left-handed $\psi_L$ chiral fermion components allows to draw a diagram for the fermion self-energy $\Sigma_\psi$, as it can be seen in the first equation in Fig.~\ref{fig:SDE}. Once the non-trivial fermion self-energies are generated then the flavor symmetry is spontaneously broken due to the chiral setting of fermion flavor representations.

\subsection{Flavor gauge bosons}

Massless gauge fields excite massive vector particles if their polarization tensor develops a Nambu--Goldstone pole as a result of the non-invariance of the ground state with respect to the global symmetry underlying the gauge one. This is called the Anderson--Higgs mechanism \cite{Anderson:1963pc,Higgs:1964pj,Guralnik:1964eu,Englert:1964et,Migdal:1965aa,'tHooft:1971rn} and we describe it in appendix~\ref{gauge_case}. The longitudinal polarization states of such massive vector particles emerge as the eaten Nambu--Goldstone bosons of the broken symmetry. Their massless propagators manifest as the massless pole in the polarization function $\Pi_{ab}(q^2)$
\begin{eqnarray}\label{massless_pole}
\Pi_{ab}(q^2) &=& -\frac{1}{q^2} M_{ab}^2(q^2) \,,
\label{PiIR}
\end{eqnarray}
defined from the transverse polarization tensor \eqref{pol_func}, where $M_{ab}^2(q^2)$ is a momentum-dependent symmetric $8\times8$ matrix, \emph{regular} at $q^2=0$, i.e., $M_{ab}^2(0)$ is a matrix residuum of the pole in \eqref{massless_pole}. Massiveness of the flavor gauge boson is then visible from their full propagator
\begin{eqnarray}
\Delta^{\mu\nu}_{ab}(q) &=&
-\frac{1}{q^2}
\bigg(g^{\mu\nu}-\frac{q^{\mu}q^{\nu}}{q^2}\bigg)\Big[\big(1+\Pi(q^2)\big)^{-1}\Big]_{ab}\quad
%\nonumber \\ &&
\label{Delta}
\end{eqnarray}
in the Landau gauge. Poles of this full propagator $\Delta^{\mu\nu}_{ab}(q)$ are given by the equation
\begin{eqnarray}
\det\big[q^2-M^2(q^2)\big] &=& 0 \,,
\label{M}
\end{eqnarray}
solutions of which define the flavor gauge boson mass spectrum. The flavor gauge boson polarization function $\Pi_{ab}(q^2)$ breaks the flavor $\SU{3}_{\F}$ symmetry once
\begin{eqnarray}
[T_a,\Pi] &\neq& 0 \,,
\end{eqnarray}
where $T_a$ are the $\SU{3}_{\F}$ generators in the adjoint representation and $\Pi$ is the polarization function with suppressed indices and momentum argument.

We emphasize that the presence of the flavor symmetry breaking fermion self-energies $\Sigma_\psi$ is instrumental in achieving the flavor symmetry breaking polarization tensor as a solution of the second equation in Fig.~\ref{fig:SDE}. Without the first term induced by fermions in non-vector-like flavor representation, the equation would be equivalent to that of pure gluo-dynamics whose solution is known to be symmetry preserving and confining.

\subsection{Flavor symmetry breaking scale}
The characteristic fermion flavor setting plays yet another important role. It makes the flavor gauge dynamics non-vector-like, what distinguishes it from QCD and makes it non-confining.

As well as the QCD, the flavor gauge dynamics is asymptotically free, i.e., the effective flavor charge in perturbative regime runs according to
\begin{equation}\label{AF}
\frac{\bar h^2(q^2)}{4\pi}=\frac{4\pi}{[6-\frac{2}{3}\eta_{\nu_R}]\ln{q^2/\Lambda^{2}_\mathrm{F}}}\,,
\end{equation}
where in the square parentheses there is the one-loop $\beta$-function coefficient given from \eqref{2loop_beta}. The parameter $\eta_{\nu_R}$ is the contribution from the right-handed neutrino sector \eqref{etaAF_inequality}. For the (63333) version its explicit value is given by
\begin{equation}
  \eta_{\nu_R}^{(63333)} = \frac{1}{2}N^{\nu_{R\mathbf{3}}}+\frac{5}{2}N^{\nu_{R\mathbf{6}}}=9/2 \,.
\end{equation}
%\begin{equation}\label{bF}
%b_\F=-(11-\frac{1}{3}N^\mathrm{EW}-\frac{2}{3}\eta_{\nu_R})\,.
%\end{equation}
Well above the scale of the flavor gauge dynamics, $\Lambda_\mathrm{F}$, the flavor gauge dynamics as well as the Standard Model gauge dynamics is weakly coupled. Decreasing the energy scale the effective flavor charge increases till it surpasses its critical value at the energy scale around $\Lambda_\mathrm{F}$. The fact that the flavor dynamics becomes strongly coupled causes the creation of the self-energies $\Sigma_\psi(p^2)$ and $\Pi_{ab}(q^2)$. Because of the non-vector-like nature of the flavor symmetry setting, these self-energies are flavor non-invariant and the flavor gauge symmetry becomes spontaneously broken. The flavor gauge bosons acquire masses by the Anderson--Higgs mechanism. Presumably the flavor gauge boson masses are of the same order of magnitude as the flavor scale $\Lambda_\mathrm{F}$.

The value of the scale $\Lambda_\F$ is from a theoretical point of view an arbitrary number. It is supposed to be fixed by a single measurement. From the phenomenological point of view however, it is the subject of several constraints.

The flavor gauge bosons have to be enormously heavy in order to suppress the processes with flavor changing neutral currents. Their masses and consequently the scale $\Lambda_\mathrm{F}$ have to be bigger than $10^6\,\mathrm{GeV}$ \cite{Eichten:1979ah}. In fact, the $\Lambda_\mathrm{F}$ has to be even much bigger. It is because the quark self-energies break spontaneously the Peccei--Quinn symmetry, naturally present in the model \eqref{Jdivergences}, and a composite axion arises. Therefore, in order to make the axion invisible, the quark self-energies have to be generated at the scales related to the axion scale $\Lambda_a$ lying in the so called axion window $10^{10}\,\mathrm{GeV}<\Lambda_a<10^{12}\,\mathrm{GeV}$ \cite{Raffelt:2006rj}. Therefore the flavor symmetry should be broken not lower than $\Lambda_a$, i.e., $\Lambda_\F\geq\Lambda_a$. Furthermore, within the model it is natural to connect the scale $\Lambda_\mathrm{F}$ with the seesaw scale $M_R$ leading to demand $\Lambda_\mathrm{F}>10^{13}\,\mathrm{GeV}$. Finally, the analysis of the top-quark and neutrino condensation, if taken seriously for the flavor gauge model, suggests the flavor scale to lie somewhere between the GUT and Planck scales, i.e., $\Lambda_\mathrm{F}\sim10^{17-18}\,\mathrm{GeV}$ \eqref{Lambda}.

Here we treat the scale $\Lambda_\mathrm{F}$ as a single number which tells us at which scale the flavor symmetry is broken. However we anticipate that the dynamics of the flavor gauge model generates whole sequence of scales out of which the highest one is $\Lambda_\mathrm{F}$.

\subsection{Flavor effective charge}

The flavor self-breaking scenario can be described by means of the flavor effective charge $\bar{h}_{ab}^2(q^2)$, which is in general $8\times8$ real-valued symmetric matrix. While for momenta greater than $\Lambda_\F$ the effective charge should be flavor symmetry preserving,
\begin{equation}
\bar{h}_{ab}^2(q^2)\ \stackrel{q^2>\Lambda_\F}{\longrightarrow}\ \bar{h}^2(q^2)\delta_{ab}\ \stackrel{q^2\rightarrow\infty}{\longrightarrow}\ 0 \,,
\end{equation}
below the $\Lambda_\F$ the effective charge should develop its flavor symmetry breaking matrix form and at even lower energies it should freeze at its infrared fixed point matrix $h_{ab}^*$
\begin{equation}\label{infrared_fixed_point_matrix}
\bar{h}_{ab}^2(q^2)\ \stackrel{q^2\rightarrow0}{\longrightarrow}\ h_{ab}^* \,.
\end{equation}
Especially the infrared behavior of the effective charge is important for generating masses. On the other hand, its ultraviolet behavior is important for damping of quantum fluctuations, in particular it determines how self-energies go to zero for asymptotically large momenta. While the ultraviolet part of the effective charge is fully determined by perturbation theory thanks to asymptotic freedom \eqref{AF}, the infrared part is completely terra incognita because it is the regime of strong coupling. That is why we do not even try to calculate the effective charge, we rather model it.

\subsection{Schwinger--Dyson equations for fermion self-energies}

The basic criterion for falsification of the flavor gauge model is its ability to reproduce complete fermion mass spectrum and all mixing matrices. It is an extremely ambitious task. Consider that we want to reproduce 9 masses of charged fermions, 4 parameters of the Cabibbo--Kobayashi--Maskawa (CKM) matrix, 2 neutrino squared mass  differences and 6 parameters of Pontecorvo--Maki--Nakagawa--Sakata (PMNS) matrix, only by a single parameter, the flavor gauge coupling constant, respectively by the fermion flavor representation setting. Unfortunately, to solve the fermion mass equations rigorously is a completely unfeasible program without resorting to rough simplifications and assumptions. Such solutions obtained from the over-simplified equations can hardly be compared with experiment. We believe however that at least some qualitative insights can be gained when analysing the simplified equations. We present partial results in appendices~\ref{sde} and \ref{separable_kernel}.

For fermion self-energies we consider the truncated Schwinger--Dyson equation
\begin{equation}\label{Sigma_SDE}
\Sigma(p^2) = -3\I \int\!\frac{\d^4k}{(2\pi)^4}\,
\bar h^2_{ab}\big((p-k)^2\big)\Delta_{0}((p-k)^2\big)\,
T_{R,a} \, \Sigma(k^2) \Big[k^2-\Sigma^{\dag}(k^2)\,\Sigma(k^2)\Big]^{-1} T_{L,b} \,,
\end{equation}
where we write the kernel renormalization-group invariantly in terms of the effective charge $\bar h^2_{ab}(q^2)$ and the bare flavor gauge boson propagator function $\Delta_{0}(q^2)$ \eqref{propagator_function} according to Pagels \cite{Pagels:1979ai}.

The effective charge $\bar h^2_{ab}(q^2)$ is in the flavor gauge model an $8\times8$ symmetric matrix function of momentum. The Schwinger--Dyson equation represents a matrix integral equation for complex matrix self-energies of fermions in a given flavor representation. For the charged fermions the self-energy is a $3\times3$ matrix, while for neutrinos the self-energy is a bigger matrix. In particular within the (63333) version it is a $21\times21$ symmetric matrix. The equations for various fermion representations are interconnected through the effective charge $\bar h^2_{ab}(q^2)$, for which we should write its own Schwinger--Dyson equation. At this stage however we do not do it. Instead we try to model $\bar h^2_{ab}(q^2)$ in order to reproduce the fermion mass spectrum. Our attempts are presented in appendices~\ref{sde} and \ref{separable_kernel}.

From the form of the Schwinger--Dyson equation without solving it, we can immediately extract a general relation between $d$-type and $e$-type self-energies. Because
\begin{subequations}\label{ed_relation} \begin{eqnarray}
& & T^{(d)}_{R,a}=T^{(e)}_{L,a}=\hphantom{-}T_{a} \,, \\
& & T^{(d)}_{L,a}=T^{(e)}_{R,a}=-T^{*}_{a} \,,
\end{eqnarray}
\end{subequations}
then the following relation holds,
%\begin{subequations}\label{d_e_relations}
\begin{eqnarray}
\Sigma_e &=& (\Sigma_d)^\dag \,.
\end{eqnarray}
%\end{subequations}
Notice that there is no analogous relation among $u$-type fermions and neutrinos.

\section{Physical view}
\label{Physical_view}

We will sketch the mass generation and the self-breaking of the flavor dynamics in terms of the technically simplest heuristic arguments:%, used in a similar context also in \cite{Nagoshi:1990wk}.
\begin{enumerate}
\item We imagine the generation of a massless pole in the flavor gauge boson polarization function \eqref{massless_pole} due to the strong flavor gauge dynamics by the effective mass term
    \begin{equation}\label{M_0}
        {\cal L}_{M_{0}}=M_{0}^2C_{a\mu}C_{a}^{\mu} \,,
    \end{equation}
    so far leaving the global $\SU{3}_\mathrm{F}$ unbroken. This consideration is not groundless. There are approaches to the QCD which demonstrate the dynamical generation of a single common gluon mass parameter as a result of strong coupling \cite{Cornwall:1973ts,Binosi:2009qm}.
\item Exchanges of massive flavor gauge bosons between fermion fields yield an effective four-fermion interaction
    \begin{eqnarray}\label{4fgfd}
        {\cal L}_{\psi^4} &=& -\frac{h^2}{M_{0}^2}j_a^{\mu}j_{a\mu} \,,
    \end{eqnarray}
    where $j_a^{\mu}$ is the flavor chiral current
    \begin{eqnarray}\label{flavor_current_63333}
        j_a^{\mu} &=& \bar{q}_L\gamma^\mu T_a q_L + \bar{u}_R\gamma^\mu T_a u_R - \bar{d}_R\gamma^\mu T^{*}_a d_R  \\
                  & & - \bar{\ell}_L\gamma^\mu T^{*}_a\ell_L + \bar{e}_R\gamma^\mu T_a e_R - \overline{\nu^{s}_{R\mathbf{3}}}\gamma^\mu T^{*}_a\nu^{s}_{R\mathbf{3}} + \overline{\nu_{R\mathbf{6}}}\gamma^\mu T^\mathbf{6}_a\nu_{R\mathbf{6}} \,. \nonumber
    \end{eqnarray}
\item Upon the Fierz rearrangements \eqref{SU3_Fierz} and \eqref{Dirac_Fierz} of the four-fermion interaction \eqref{4fgfd} we can identify the four-fermion interactions in the form $(\bar\psi_L\psi_R)(\bar\psi_L\psi_R)$ responsible for Dirac mass generation and in the form $(\overline{\psi^\C}\psi)(\overline{\psi}\psi^\C)$ responsible for Majorana mass generation.
\item The four-fermion dynamics \eqref{4fgfd} triggers a formation of various condensates $\langle\bar\psi_L\psi_R\rangle$ and $\langle\bar\psi\psi^\C\rangle$ in the attractive channels. They are solutions of their gap equations in analogy with \eqref{top_quark_gap_eq}. The non-trivial chirality changing condensates form massive poles in the fermion propagators. For illustration, we show here only one of the gap equations for a condensate of Dirac fermions,
    \begin{equation}\label{gapE}
        \langle\bar\psi_R\psi_L\rangle\ = \begin{array}{c}\includegraphics[width=0.25\textwidth]{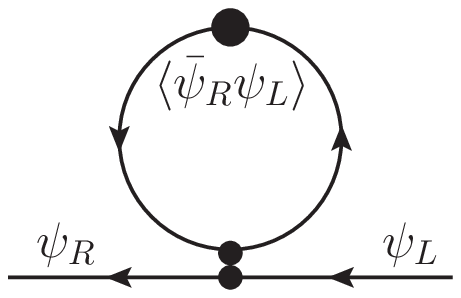}\end{array} \,.
    \end{equation}
\item Once some flavored condensate is formed, the flavor symmetry is broken. It produces a feedback to the flavor gauge boson masses in the form of splitting.
\end{enumerate}
Here we have formulated the process of the flavor symmetry breaking consecutively, but of course it happens at once in both fermion and flavor gauge boson sectors. Following this strategy, however, we can at least qualitatively learn at which scale the condensates are formed and what is the spectrum of bound states.

In a regime of very high energies ($>\Lambda_\F$) the system is completely symmetric, the flavor gauge bosons are massless. The strength of attraction among fermions, mediated by the massless flavor gauge bosons, can be estimated by the Most Attractive Channel (MAC) method \cite{Raby:1979my}. Even though it is a perturbative result we believe that its pattern is relevant also for the regime of strong coupling where the flavor gauge bosons become massive and the four-fermion interaction \eqref{4fgfd} is formed.

Decreasing the energy scale, the strength of the flavor gauge dynamics increases according to \eqref{AF}. The attractiveness of various channels depends on the flavor representations of participating fermions. Once the most attractive channel produces the flavor symmetry breaking at the energy scale $\Lambda_\F$, the MAC method looses its plausibility for the remaining pairing channels.

The attractiveness of different pairing channels
\begin{equation}
r_1\times r_2\rightarrow r_\mathrm{pair}
\end{equation}
is roughly measured by the quantity
\begin{equation}\label{AC}
\Delta C_2=C_2(r_1)+C_2(r_2)-C_2(r_\mathrm{pair}) \,,
\end{equation}
where $C_2(r)$ is the quadratic Casimir invariant for the representation $r$, see Tab.~\ref{table}. We write here some of the product representation decompositions for $\SU{3}$ (for details see, e.g., \cite{Cheng:2000ct} or \cite{Wesslen:2009aa}):
\begin{eqnarray}
\mathbf{3}\times\mathbf{3} & = & \mathbf{6}+\overline{\mathbf{3}} \,, \\
\mathbf{3}\times\overline{\mathbf{3}} & = & \mathbf{8}+\mathbf{1} \,, \\
\mathbf{3}\times\mathbf{6} & = & \mathbf{10}+\mathbf{8} \,, \\
\overline{\mathbf{3}}\times\mathbf{6} & = & \mathbf{15}+\mathbf{3} \,, \\
\mathbf{6}\times\mathbf{6} & = & \mathbf{15}+\mathbf{15}+\overline{\mathbf{6}} \,.
\end{eqnarray}

The strengths $\Delta C_2$ of attractive channels $(\mathrm{A.C.})_\psi$ among charged fermions $\psi=u,d,e$ are evaluated as
\begin{eqnarray}
(\mathrm{A.C.})_u=\overline{\mathbf{3}}\times\mathbf{3}\rightarrow\mathbf{1} \,,&\quad\quad&\Delta C_2=8/3 \,, \\
(\mathrm{A.C.})_d=\mathbf{3}\times\mathbf{3}\rightarrow\overline{\mathbf{3}} \,,&\quad\quad&\Delta C_2=4/3 \,, \\
(\mathrm{A.C.})_e=\overline{\mathbf{3}}\times\overline{\mathbf{3}}\rightarrow\mathbf{3} \,,&\quad\quad&\Delta C_2=4/3 \,.
\end{eqnarray}
The attractive channels $(\mathrm{A.C.})_n$, giving rise to different parts of the neutrino self-energy written in the Nambu--Gorkov formalism, are (compare with \eqref{NGselfenergy})
\begin{equation}
(\mathrm{A.C.})_n=\beginm{c|c|c}
    \mathbf{3}\times\mathbf{3}\rightarrow\overline{\mathbf{3}} & \ \ \ \ \overline{\mathbf{3}}\times\mathbf{3}\rightarrow\mathbf{1} \ \ \ \ &
                                                            \mathbf{6}\times\mathbf{3}\rightarrow\mathbf{8} \\ \hline
    \mathbf{3}\times\overline{\mathbf{3}}\rightarrow\mathbf{1} & \overline{\mathbf{3}}\times\overline{\mathbf{3}}\rightarrow\mathbf{3} &
                    \mathbf{6}\times\overline{\mathbf{3}}\rightarrow\mathbf{3} \vphantom{\begin{array}{c}\ \\ \ \end{array}}\\ \hline &&\vspace{-0.4cm}\\
    \mathbf{3}\times\mathbf{6}\rightarrow\mathbf{8} & \overline{\mathbf{3}}\times\mathbf{6}\rightarrow\mathbf{3} & \mathbf{6}\times\mathbf{6}\rightarrow\overline{\mathbf{6}} \endm \,.
\end{equation}
The corresponding strength $\Delta C_2$ \eqref{AC} of the attractiveness of the channels is
\begin{equation}
(\Delta C_2)=\beginm{c|c|c}
    4/3 & \ \ \ \ 8/3 \ \ \ \ & 5/3 \\ \hline
    8/3 & 4/3 & 10/3 \vphantom{\begin{array}{c}\ \\ \ \end{array}}\\ \hline &&\vspace{-0.43cm}\\
    5/3 & 10/3 & 10/3 \endm \,.
\end{equation}
The most attractive channels are those of the right-handed neutrino Majorana type, $\mathbf{6}\times\overline{\mathbf{3}}\rightarrow\mathbf{3}$ and $\mathbf{6}\times\mathbf{6}\rightarrow\overline{\mathbf{6}}$ with $\Delta C_2=10/3$. It naturally follows that, decreasing the energy scale, the right-handed neutrino pairing of Majorana type happens first. This fact brings nice features:
\begin{itemize}
\item It breaks the flavor symmetry providing no confinement.
\item It suggests the seesaw pattern of neutrino mass matrix.
\item The electroweak symmetry breaking is postponed to lower energies where electroweakly charged condensates are formed.
\end{itemize}

This can be put in contrast with the minimal version (333), where all fields are in triplets or anti-triplets. Thus it is approximately vector-like according to Tab.~\ref{NUsettings}. The most attractive channel is that of $u$-type quarks, $\mathbf{3}\times\overline{\mathbf{3}}\rightarrow\mathbf{1}$, with $\Delta C_2=8/3$. Therefore according to the MAC method, within the minimal version (333) the most attractive channel is a flavor singlet. It does not support our essential assumption that the flavor gauge dynamics self-breaks. Even if we assume that the QCD and electroweak dynamics are sufficiently relevant to trigger the non-vector-like nature of the model, it still remains difficult to justify the seesaw pattern of neutrino mass matrix as all of its blocks are governed by the same attractive channel $\mathbf{3}\times\mathbf{3}\rightarrow\overline{\mathbf{3}}$.

\subsection{Flavor symmetry breaking by the sterility condensation}

We can quantify the anti-sextet and the four triplet pairings by, so called, \emph{sterility condensates}
\begin{subequations}
\begin{eqnarray}\label{condensates}
\bra{0}\frac{1}{4}\epsilon^{ACE}\epsilon^{BDF}\overline{(\xi_{R}^{CD})^\C}\xi_{R}^{EF}\ket{0} & \propto
& \Lambda_{\mathrm{F}}^2\bra{0}\Phi_{6}^{AB}\ket{0} \,,\hspace{0.5cm} \\
\bra{0}\overline{(\xi_{R}^{AB})^\C}\zeta_{R}^{sB}\ket{0} & \propto
& \Lambda_{\mathrm{F}}^2\bra{0}\Phi_{3}^{s,A}\ket{0} \,,\hspace{0.5cm}
\end{eqnarray}
\end{subequations}
where we have introduced auxiliary scalar fields $\Phi_6$ and $\Phi^{s}_3$ of mass dimension one and the right-handed neutrino fields $\xi_{R}$ and $\zeta_{R}^{s}$ are defined in \eqref{RHnu}. The index $s=1,\dots,4$ is the $\SU{4}_\mathrm{S}$ sterility index. The indices, $A,B,C,\dots{}=1,2,3$, are the indices of the fundamental flavor representation, and $\epsilon^{ABC}$ is the totally anti-symmetric tensor. The auxiliary fields transform as a ($2$-index symmetric) anti-sextet $\overline{\mathbf{6}}$ and a triplet $\mathbf{3}$, respectively, under the flavor rotations ${\cal U}=\e^{\im\alpha^a T^a}$
\begin{subequations}
\begin{eqnarray}
\Phi_{6}' & = & {\cal U}^{\dag\mathrm{T}}\Phi_{6}{\cal U}^{\dag} \,,\\
\Phi_{3}^{s}\vphantom{\Phi}' & = & {\cal U}\Phi_{3}^{s}\,.
\end{eqnarray}
\end{subequations}
These flavor transformation properties follow from the flavor transformation properties of the elementary right-handed neutrino fields (for their definitions see \eqref{RHnu})
\begin{subequations}
\begin{eqnarray}
\xi_{R}' & = & {\cal U}\xi_{R}{\cal U}^{\mathrm{T}} \,,\\
{\zeta_{R}^{s}}' & = & {\cal U}^*\zeta_{R}^{s}\,,
\end{eqnarray}
\end{subequations}
and the fact that the totally anti-symmetric tensor $\epsilon^{ABC}$ is flavor invariant
\begin{equation}
{\cal U}^{AD}{\cal U}^{BE}{\cal U}^{CF}\epsilon^{DEF}=\epsilon^{ABC}\,.
\end{equation}
The quantum numbers $(\,S_3-S_6,\,S_3+S_6,\,\SU{4}_\mathrm{S}\,)$ of the scalar fields are
\begin{subequations}
\begin{eqnarray}
\Phi_{6}\ :\hspace{0.2cm} & & \left(\,-\frac{2}{5},\,+\frac{2}{5},\,\mathbf{1}\,\right) \,, \\
\Phi_{3}^{s}\ :\hspace{0.2cm} & & \left(\,+\frac{1}{20},\,+\frac{9}{20},\,\mathbf{4}\,\right) \,.
\end{eqnarray}
\end{subequations}

$\Phi_6$ and $\Phi_{3}^{s}$ are 18 complex scalar fields. They can be expressed in terms of twice as many real scalar fields out of which some are the Nambu--Goldstone fields of broken flavor and sterility symmetries:

\begin{eqnarray}\label{Phi6}
\Phi_6(x) & = & \e^{-2\im\alpha(x)}\e^{+2\im\beta(x)}\e^{-\im\theta^a(x) T^{a\mathrm{T}}}\beginm{ccc}\Delta_1(x) & 0 & 0 \\
                                                                0 & \Delta_2(x) & 0 \\
                                                                0 & 0 & \Delta_3(x) \endm\e^{-\im\theta^a(x) T^{a}} \,,
\end{eqnarray}
\begin{eqnarray}\label{Phi3}
 \Big\{\Phi^{1}_3(x),\Phi^{2}_3(x), \Phi^{3}_3(x), \Phi^{4}_3(x)^\T \Big\} & = & \\ && \hspace{-0.2\textwidth}\e^{-\frac{3}{4}\im\alpha(x)}\e^{+\frac{5}{4}\im\beta(x)}\e^{\im\gamma^\sigma(x)S^{\sigma}}\e^{\im\theta^a(x) T^{a}}
                    \left\{
                        \left(\begin{array}{c} 0 \\ 0 \\ 0 \\ 0                                         \end{array}\right) \,,\
                        \left(\hspace{-5pt}\begin{array}{c} 0 \\ 0 \\ 0 \\ \delta_2(x)                               \end{array}\hspace{-5pt}\right) \,,\
                        \left(\hspace{-5pt}\begin{array}{c} 0 \\ 0 \\ \delta_3(x) \\ \delta_1(x)                     \end{array}\hspace{-5pt}\right) \,,\
                        \left(\hspace{-5pt}\begin{array}{c} 0 \\ \delta_4(x) \\ \varepsilon_5(x) \\ \varepsilon_6(x) \end{array}\hspace{-5pt}\right)
                    \right\} \,. \nonumber
%\beginm{c}\Phi^{1}_3(x)^\T \\ \Phi^{2}_3(x)^\T \\ \Phi^{3}_3(x)^\T \\ \Phi^{4}_3(x)^\T \endm & = & \e^{-\frac{3}{4}\im\alpha(x)}\e^{+\frac{5}{4}\im\beta(x)}\e^{\im\gamma^i(x)s^{i}}\e^{\im\theta^a(x) T^{a}}\beginm{ccccc}
%                        \big(\hspace{-5pt} & 0 & 0 & 0 & \hspace{-5pt}\big) \\
%                        \big(\hspace{-5pt} & 0 & 0 & \delta_2(x) & \hspace{-5pt}\big) \\
%                        \big(\hspace{-5pt} & 0 & \delta_3(x) & \delta_1(x) & \hspace{-5pt}\big) \\
%                        \big(\hspace{-5pt} & \delta_4(x) & \varepsilon_5(x) & \varepsilon_6(x)   & \hspace{-5pt}\big)
%                    \endm \,.
\end{eqnarray}

The 25 Nambu--Goldstone bosons are (for majorons see section \ref{secV}):
\begin{itemize}
\item 8 of $\theta^a(x)$ corresponding to broken $\SU{3}_\mathrm{F}$:
    longitudinal polarizations of the flavor gauge bosons $C_{\mu}^a$
\item 15 of $\gamma^\sigma(x)$ corresponding to broken $\SU{4}_\mathrm{S}$:
    non-Abelian sterile majorons $\eta^\sigma$
\item 1 of $\alpha(x)$ corresponding to broken $S_3-S_6$:
    Abelian sterile majoron $\eta^0$
\item 1 of $\beta(x)$ corresponding to broken $S_3+S_6$:
    flavor axion $A$
\end{itemize}
The remaining components of $\Phi_6$ and $\Phi_{3}^{s}$ are 7 real and 2 complex scalars. They altogether can develop, in general, 9 $CP$-preserving vacuum expectation values $\phi_A$ and $\varphi_k$, and 2 $CP$-violating vacuum phases $\varsigma_k$.
\begin{itemize}
\item 3 real components of sextet field  $\Delta_A(x)\rightarrow\Delta_A(x)+\phi_A$, $A=1,2,3$
\item 4 real anti-triplet fields
    $\delta_k(x)\rightarrow\delta_k(x)+\varphi_k$, $k=1,2,3,4$
\item 2 complex anti-triplet fields  $\varepsilon_k(x)\rightarrow\varepsilon_k(x)+\varphi_k\e^{\im\varsigma_k}$, $k=5,6$
\end{itemize}

A general form of the condensate $\bra{0}\Phi_{6}\ket{0}$ is
\begin{equation}\label{cond6}
\bra{0}\Phi_{6}\ket{0}=\beginm{ccc}   \phi_1 & 0 & 0 \\
                                      0 & \phi_2 & 0 \\
                                      0 & 0 & \phi_3 \endm
\end{equation}
as follows from \eqref{Phi6}. A general form of the condensates $\bra{0}\Phi_{3}^{s}\ket{0}$ follows from \eqref{Phi3}. In general, the condensates $\bra{0}\Phi_{3}^{s}\ket{0}$ can be complex and they can have nontrivial alignment with respect to each other and also with respect to $\bra{0}\Phi_{6}\ket{0}$.

Not only for the sake of concreteness we choose here a special form of the triplet condensates
\begin{subequations}\label{cond3}
\begin{eqnarray}
\bra{0}\Phi_{3}^{s=1,2,3}\ket{0} & = & \ \ 0 \,, \\
\bra{0}\Phi_{3}^{s=4}\ket{0}\ \  & = & \beginm{ccc}\varphi_4 & \varphi_5 & \varphi_6 \endm^\T \,.
\end{eqnarray}
\end{subequations}
The motivation for this alignment is the fact that there is no dynamics which would distinguish or prefer among the condensates $\bra{0}\Phi_{3}^{s}\ket{0}$. Therefore, it seems natural to assume democratically that $\bra{0}\Phi_{3}^{s}\ket{0}=\bra{0}\Phi_{3}^{s'}\ket{0}$. The alignment \eqref{cond3} is nothing but the $\SU{4}_\mathrm{S}$ transform of this democratic alignment, i.e., physically equivalent. The main reason for this choice is that it leaves the $\SU{3}_\mathrm{S}$ sterility subgroup unbroken. This is necessary to protect the seesaw mechanism as we will discuss later in subsection~\ref{neutrino_pheno}. Without this special form the general condensates would break the sterility symmetry completely.

The condensates break the flavor symmetry completely while the electroweak symmetry breaking is postponed to the lower energies where the pairing of the electroweakly charged fermions occurs. The sextet sterility condensation is formally similar to the sextet color superconductivity \cite{Brauner:2003pj}.

\subsection{Masses from the sterility condensation}

The sterility condensation produces masses of all flavor gauge bosons. The masses can be estimated from the lowest order gauge invariant kinetic terms of the effective Lagrangian for the effective scalar fields
\begin{equation}\label{L_M_gauge}
{\cal L}_{M_\mathrm{gauge}}=(D^\mu\Phi_{3}^{s})^\dag D_\mu\Phi_{3}^{s}+\Tr(D^\mu\Phi_{6})^\dag D_\mu\Phi_{6} \,,
\end{equation}
where
\begin{subequations}
\begin{eqnarray}
D_\mu\Phi_{6} & = & \partial_\mu\Phi_{6}+\im h C^{a}_\mu(T^{a\mathrm{T}}\Phi_{6}+\Phi_{6}T^{a}) \,,\\
D_\mu\Phi_{3}^{s} & = & (\partial_\mu-\im h C^{a}_\mu T^a)\Phi_{3}^{s} \,.
\end{eqnarray}
\end{subequations}

In the effective Lagrangian ${\cal L}_{M_\mathrm{gauge}}$ \eqref{L_M_gauge} we substitute the effective scalar fields for their vacuum expectation value $\Phi\rightarrow\bra{0}\Phi\ket{0}$, and get the mass matrix for the gauge bosons
\begin{equation}\label{Mab}
M_{ab}^2=\big[M_{6}^2+M_{3}^2\big]_{ab} \,,
\end{equation}
where the mass matrices $M_{6}^2$ and $M_{3}^2$ with the specific form of the condensates \eqref{cond6} and \eqref{cond3} are
\begin{subequations}\label{Mgauge}
\begin{small}
\begin{eqnarray}
M_{6}^2 \hspace{-5pt} & = & \hspace{-5pt} h^2
\beginm{cccccccc}
(\phi_1+\phi_2)^2 \hspace{-10pt}& 0 & 0 & 0 & 0 & 0 & 0 & 0 \\
0 & \hspace{-10pt} (\phi_1-\phi_2)^2  \hspace{-10pt}& 0 & 0 & 0 & 0 & 0 & 0 \\
0 & 0 & \hspace{-10pt} 2(\phi_{1}^2+\phi_{2}^2)  \hspace{-10pt}& 0 & 0 & 0 & 0 & \hspace{-10pt} \frac{2}{\sqrt{3}}(\phi_{1}^2-\phi_{2}^2) \\
0 & 0 & 0 & \hspace{-10pt} (\phi_1+\phi_3)^2  \hspace{-10pt}& 0 & 0 & 0 & 0 \\
0 & 0 & 0 & 0 & \hspace{-10pt} (\phi_1-\phi_3)^2  \hspace{-10pt}& 0 & 0 & 0 \\
0 & 0 & 0 & 0 & 0 & \hspace{-10pt} (\phi_2+\phi_3)^2  \hspace{-10pt}& 0 & 0 \\
0 & 0 & 0 & 0 & 0 & 0 & \hspace{-10pt} (\phi_2-\phi_3)^2  \hspace{-10pt}& 0 \\
0 & 0 & \hspace{-10pt} \frac{2}{\sqrt{3}}(\phi_{1}^2-\phi_{2}^2) \hspace{-10pt} & 0 & 0 & 0 & 0 & \hspace{-10pt} \frac{2}{3}(\phi_{1}^2+\phi_{2}^2+4\phi_{3}^2)
\endm\,,\nonumber\\
&&\\
M_{3}^2 \hspace{-5pt} & = & \hspace{-5pt} \frac{h^2}{4}
\left(\hspace{-5pt}\begin{array}{cccccccc}
(\varphi_{4}^2+\varphi_{5}^2) \hspace{-13pt} & 0 & 0 & \varphi_{5}\varphi_{6} & 0 & \varphi_{4}\varphi_{6} & 0 & \frac{2}{\sqrt{3}}\varphi_{4}\varphi_{5} \\
0 & \hspace{-13pt} (\varphi_{4}^2+\varphi_{5}^2) \hspace{-13pt} & 0 & 0 & \varphi_{5}\varphi_{6} & 0 & -\varphi_{4}\varphi_{6} & 0 \\
0 & 0 & \hspace{-13pt} (\varphi_{4}^2+\varphi_{5}^2) \hspace{-13pt} & \varphi_{4}\varphi_{6} & 0 & -\varphi_{5}\varphi_{6} & 0 & \frac{1}{\sqrt{3}}(\varphi_{4}^2-\varphi_{5}^2) \\
\varphi_{5}\varphi_{6} & 0 & \varphi_{4}\varphi_{6} & \hspace{-13pt} (\varphi_{4}^2+\varphi_{6}^2) \hspace{-13pt} & 0 & \varphi_{4}\varphi_{5} & 0 & -\frac{1}{\sqrt{3}}\varphi_{4}\varphi_{6} \\
0 & \varphi_{5}\varphi_{6} & 0 & 0 & \hspace{-13pt} (\varphi_{4}^2+\varphi_{6}^2) \hspace{-13pt} & 0 & \varphi_{4}\varphi_{5} & 0 \\
\varphi_{4}\varphi_{6} & 0 & -\varphi_{5}\varphi_{6} & \varphi_{4}\varphi_{5} & 0 & \hspace{-13pt} (\varphi_{5}^2+\varphi_{6}^2) \hspace{-13pt} & 0 & -\frac{1}{\sqrt{3}}\varphi_{5}\varphi_{6} \\
0 & -\varphi_{4}\varphi_{6} & 0 & 0 & \varphi_{4}\varphi_{5} & 0 & \hspace{-13pt} (\varphi_{5}^2+\varphi_{6}^2) \hspace{-13pt} & 0 \\
\frac{2}{\sqrt{3}}\varphi_{4}\varphi_{5} & 0 &  \hspace{-13pt}\frac{1}{\sqrt{3}}(\varphi_{4}^2-\varphi_{5}^2) & -\frac{1}{\sqrt{3}}\varphi_{4}\varphi_{6} & 0 & -\frac{1}{\sqrt{3}}\varphi_{5}\varphi_{6} & 0 & \hspace{-13pt} \frac{1}{3}(\varphi_{4}^2+\varphi_{5}^2+4\varphi_{6}^2)
\end{array}\hspace{-5pt}\right)\,.\nonumber\\
&&
\end{eqnarray}
\end{small}
\end{subequations}

The sterility condensation produces also Majorana masses for right-handed neutrinos. The masses can be estimated from Yukawa terms of the effective Lagrangian for the effective scalar fields
\begin{equation}\label{L_M_sterile}
{\cal L}_{\mathrm{sterile}} = y_{36}\,\overline{(\zeta_{R}^{s})^\C}\xi_{R}\Phi_{3}^{s*}+ y_{6}\,\epsilon^{ACE}\epsilon^{BDF}\overline{(\xi_{R}^{AB})^\C}\xi_{R}^{CD}(\Phi_{6}^{EF})^\dag  +\hc \,,
\end{equation}
where the effective Yukawa coupling constants
\begin{equation}
y_{36}=\frac{4}{9}h^2 \nonumber \,,\
y_{6}=\frac{4}{9}h^2
\end{equation}
are obtained from the effective four-neutrino interaction $\sim(\overline{n^\C}\gamma_\mu t^an)(\bar{n}\gamma^\mu t^an^\C)$.

In the effective Lagrangian ${\cal L}_{\mathrm{sterile}}$ \eqref{L_M_sterile} we can substitute the scalars for their condensates \eqref{cond6} and \eqref{cond3}, and get the Majorana mass matrix for the right-handed neutrinos,
\begin{equation}\label{Msterile}
M_{R}=\frac{4}{9} h^2\beginm{c|c}
    0 & \begin{array}{c} 0 \\ 0 \\ 0 \\ \langle\Phi^{s=4}_3\rangle^\T \end{array}\\ \hline
    \begin{array}{cccc} 0 & 0 & 0 & \langle\Phi^{s=4}_3\rangle \end{array} & \langle\Phi_6\rangle \\ \endm \,.
\end{equation}
The condensates are rewritten in the $\nu_{R}$-basis \eqref{NGmultiplet}, instead of the matrix $\xi_{R}$-basis:
\begin{eqnarray}
\langle\Phi^{s=4}_3\rangle & = & \frac{1}{2}\beginm{ccc} \varphi_5 & \varphi_4 & 0 \\ \frac{\varphi_4}{\sqrt{3}} & \frac{\varphi_5}{\sqrt{3}} & -\frac{2\varphi_6}{\sqrt{3}} \\
\varphi_4 & -\varphi_5 & 0 \\ \varphi_6 & 0 & \varphi_4 \\ 2\varphi_4 & 2\varphi_5 & 2\varphi_6 \\ 0 & \varphi_6 & \varphi_5 \endm \,, \\
\langle\Phi_6\rangle & = & \frac{1}{2}\beginm{cccccc}
-\phi_3 & 0 & 0 & 0 & 0 & 0 \\
0 & \tfrac{\phi_3-2\phi_1-2\phi_2}{3} & \tfrac{\phi_1-\phi_2}{\sqrt{3}} & 0 & \tfrac{2\phi_3-\phi_1-\phi_2}{\sqrt{3}} & 0 \\
0 & \tfrac{\phi_1-\phi_2}{\sqrt{3}} & -\phi_3 & 0 & \phi_2-\phi_1 & 0 \\
0 & 0 & 0 & -\phi_2 & 0 & 0 \\
0 & \tfrac{2\phi_3-\phi_1-\phi_2}{\sqrt{3}} & \phi_2-\phi_1 & 0 & 4(\phi_1+\phi_2+\phi_3) & 0 \\
0 & 0 & 0 & 0 & 0 & -\phi_1 \endm \,.
\end{eqnarray}
The mass matrix $M_{R}$ \eqref{Msterile} has nine massless and nine massive eigenstates. Without the special choice of sterility condensates \eqref{cond6} and \eqref{cond3}, there would be only six massless eigenstates. The number of massless eigenstates is important for the success of the seesaw mechanism for three light active neutrinos. This will be discussed later in subsection~\ref{neutrino_pheno}.

\subsection{Electroweak symmetry breaking masses}\label{Dirac_masses}

Due to the sterility condensation of right-handed neutrinos in Majorana channels the flavor symmetry gets broken at the scale $\Lambda_\F$. The sterility condensation was introduced in the author's original work \cite{Smetana:2011tj}. In this subsection we set the direction how to extend the analysis to the condensation of charged fermions within the same approach of four-fermion interactions. %It was not published and it represents a subject of our future work.

Bellow the scale $\Lambda_\F$ the flavor gauge bosons are massive with the mass matrix $M_{ab}$ \eqref{Mab}. Masses of other fermions are generated at some lower scale due to their effective four-fermion interactions induced by the exchange of massive flavor gauge bosons. The Dirac masses are generated from interactions
\begin{subequations}\label{L4_Dirac}
\begin{eqnarray}
{\cal L}_{D,u} & = & -\left[h^2/M^{2}\right]_{ab}(\bar q_{L}\gamma_\mu T^{a}q_{L})(\bar u_{R}\gamma^\mu T^{b}u_{R}) \,,\\
{\cal L}_{D,d} & = & \hphantom{-}\left[h^2/M^{2}\right]_{ab}(\bar q_{L}\gamma_\mu T^{a}q_{L})(\bar d_{R}\gamma^\mu T^{*b}d_{R}) \,,\\
{\cal L}_{D,e} & = & \hphantom{-}\left[h^2/M^{2}\right]_{ab}(\bar \ell_{L}\gamma_\mu T^{*a}\ell_{L})(\bar e_{R}\gamma^\mu T^{b}e_{R}) \,,\\
{\cal L}_{D,\nu\zeta} & = & -\left[h^2/M^{2}\right]_{ab}
(\bar\ell_L\gamma_\mu T^{*a}\ell_L)(\bar\nu_{R\overline{\mathbf{3}}}^s\gamma^\mu T^{*b}\nu_{R\overline{\mathbf{3}}}^s) \,,\\
{\cal L}_{D,\nu\xi} & = & \hphantom{-}\left[h^2/M^{2}\right]_{ab}
(\bar\ell_L\gamma_\mu T^{*a}\ell_L)(\bar\nu_{R\mathbf{6}}\gamma^\mu T^{b}_6\nu_{R\mathbf{6}})
\end{eqnarray}
\end{subequations}
and Majorana masses are generated from interactions
\begin{subequations}\label{L4_Majorana}
\begin{eqnarray}
{\cal L}_{M,\zeta\xi} & = & \hphantom{-}\left[h^2/M^{2}\right]_{ab} (\bar\nu_{R\overline{\mathbf{3}}}^s\gamma^\mu T^{*a}\nu_{R\overline{\mathbf{3}}}^s)(\bar\nu_{R\mathbf{6}}\gamma^\mu T^{b}_6\nu_{R\mathbf{6}}) \,,\\
{\cal L}_{M,\zeta\zeta} & = & -\left[h^2/M^{2}\right]_{ab}
(\bar\nu_{R\overline{\mathbf{3}}}^s\gamma^\mu T^{*a}\nu_{R\overline{\mathbf{3}}}^s)(\bar\nu_{R\overline{\mathbf{3}}}^s\gamma^\mu T^{*b}\nu_{R\overline{\mathbf{3}}}^s) \,,\\
{\cal L}_{M,\nu\nu} & = & -\left[h^2/M^{2}\right]_{ab}(\bar\ell_L\gamma_\mu T^{*a}\ell_L)(\bar\ell_L\gamma_\mu T^{*b}\ell_L) \,.
\end{eqnarray}
\end{subequations}
The magnitude of these secondarily generated masses is assumed to come out critically scaled down to electroweak scale and lower. Also the magnitude of feedback corrections to masses of the right-handed neutrinos assisting on sterility condensation and of the flavor gauge bosons is assumed to come out of the same order, thus negligible. Therefore effectively below $\Lambda_\F$, the right-handed neutrinos act as having a hard Majorana mass term obtained from \eqref{L_M_sterile} and \eqref{Msterile} as
\begin{eqnarray}
{\cal L}_{M_R} & = & \overline{\nu_R}M_R\nu_{R}^\C + \hc
\end{eqnarray}
with all indices suppressed.

Further program towards electroweak symmetry breaking is standard and essentially it should follow the generalized analysis presented for the top-quark and neutrino condensation in section~\ref{top_and_nu_condensation}. As the complete analysis would be enormously baroque, we merely sketch it here.

The four-fermion interactions \eqref{L4_Dirac} and \eqref{L4_Majorana} underlie all the remaining fermion condensations and produce various composite bound states. We anticipate that out of the bound states there are weak isospin composite multiplets whose vacuum expectation values give rise to the masses of electroweakly charged fermions and which are therefore relevant for the electroweak symmetry breaking. This is the multitude of weak isospin doublets:
\begin{subequations}\label{H_composites}
\begin{eqnarray}
H^{(u)}_{AB} & \propto & \bar{u}_{R,A}q_{L,B} \,, \\
H^{(d)}_{AB} & \propto & \bar{d}_{R,A}q_{L,B} \,, \\
H^{(e)}_{AB} & \propto & \bar{e}_{R,A}\ell_{L,B} \,, \\
H^{(\nu3)s}_{AB} & \propto & \bar{\nu}^{s}_{R\overline{\mathbf{3}},A}\ell_{L,B} \,,\\
H^{(\nu6)}_{AB} & \propto & \bar{\nu}_{R\mathbf{6},A}\ell_{L,B} \,,
\end{eqnarray}
\end{subequations}
and a number of weak isospin triplets
\begin{eqnarray}
\Delta^{(\nu)}_{AB} & \propto & \bar{\ell}_{L,A}\ell^{\C}_{L,B} \,,
\end{eqnarray}
where $\ell^{\C}_{L}=\im\sigma_2C\bar{\ell}^{\T}_L$ and $C=\im\gamma_0\gamma_2$.
The composite bosons arise from bosonization of the Fierz rearranged interactions \eqref{L4_Dirac} and \eqref{L4_Majorana}. Below the condensation scale they become dynamical effective fields provided that the effective renormalizable Lagrangian, including their kinetic terms and Yukawa interactions, is generated. The parameters of the effective Lagrangian are to be evolved by means of renormalization group technique down to the electroweak scale with the initial condition given by the four fermion interactions.

The nonzero vacuum expectation values of their electrically neutral components, $v_{H}$ and $v_{\Delta}$, break spontaneously the electroweak symmetry and altogether they contribute to the electroweak scale $v$, proportionally to the magnitude of corresponding fermion masses. Using the observed spectrum of fermion masses there are practically only $\langle H^{(u)}\rangle$, $\langle H^{(\nu3)s}\rangle$ and $\langle H^{(\nu6)}\rangle$ which saturate $v$, provided the seesaw mechanism is governed by the scale $M_R\sim\Lambda_\F>10^{14}\,\mathrm{GeV}$. This is actually the motivation which led us to formulate the simplified top-quark and neutrino condensation model presented in chapter~\ref{top_and_nu_condensation}.

In principle, the complete bound state spectrum, including the Nambu--Goldstone modes eaten by gauge bosons and the effective sterility scalars, should occupy the coset representations of the approximate chiral symmetry \eqref{chiral_symmetry_63333} breaking $G_{\chi}\rightarrow\SU{3}_\mathrm{c}\times\U{1}_\mathrm{em}\times\U{1}_\mathrm{B}\times H_\mathrm{S}$ where $H_\mathrm{S}$ is an eventual unbroken subgroup of the sterility symmetry $G_\mathrm{S}$. Out of these bound states we will later study only those, whose existence is supported by a theorem, the Goldstone theorem. This will include Nambu--Goldstone states coupled to both exactly conserved and anomalous currents.

\subsection{Seesaw mechanism}\label{neutrino_pheno}

Let us discuss shortly the seesaw mechanism for neutrino masses. The sterility condensation within the non-minimal version (63333) naturally leads to the  huge Majorana mass of right-handed neutrinos \eqref{Msterile} $M_R\sim\Lambda_\F$. The Dirac neutrino masses are supposed to be as large as the electroweak scale, i.e.,
\begin{equation}\label{Hnu3s6}
\langle H^{(\nu3)s}\rangle\sim m_t\quad\mathrm{and}\quad\langle H^{(\nu6)}\rangle\sim m_t \,.
\end{equation}
Provided that the weak isospin triplet condensate is completely negligible $\langle\Delta^{(\nu)}\rangle\sim0$, the seesaw masses of the light active neutrinos, $m_{\nu_\mathrm{active}}\sim m_{t}^2/\Lambda_\F$, and of the sterile neutrinos, $m_{\nu_\mathrm{sterile}}\sim \Lambda_\F$, arise. In order to reproduce the light neutrino mass $m_{\nu_\mathrm{active}}\sim10^{-1}\,\mathrm{eV}$ while $m_{t}\sim10^{2}\,\mathrm{GeV}$, the flavor scale should be $\Lambda_\F\sim10^{14}\,\mathrm{GeV}$.

However, the system with the \emph{general} sterility condensation scheme is not directly able to accommodate all three light electroweak neutrinos. It is caused by the presence of six zero eigenvalues of general mass matrix $M_R$ \eqref{Msterile}. Instead of combining with super-heavy modes producing the seesaw spectrum, the three left-handed neutrino modes combine with three of the right-handed neutrino zero-modes to produce three pairs of quasi-degenerate modes of mass ($\sim m_t$). Those six modes in fact appear as three pseudo-Dirac active neutrinos in flagrant contradiction with observations.

There is a way out of this trouble, if some subgroup of the sterility symmetry $H_\mathrm{S}\subset G_\mathrm{S}$, which is able to prohibit the mixing of the left-handed neutrinos with the right-handed zero-modes, remains unbroken. The seesaw mechanism then acts only among the left-handed and super-heavy right-handed neutrino modes. We can say that we need the residual sterility symmetry $H_\mathrm{S}$ to protect the seesaw mechanism. The necessary residual sterility symmetry is achieved by imposing the \emph{special} sterility condensation scheme \eqref{cond3} which must be preserved also in the secondary condensation, i.e., for instance
\begin{equation}\label{Hnu3s=4}
\langle H^{(\nu3)s=1,2,3}\rangle=0\,,\quad\langle H^{(\nu3)s=4}\rangle\ne0 \,.
\end{equation}
The seesaw-mechanism-protecting symmetry is then $H_\mathrm{S}=\SU{3}_\mathrm{S}\subset\SU{4}_\mathrm{S}$ with a set of generators $S_\sigma$, with $\sigma=1,\dots,8$.

\section{Spectrum of majoron and axion-like particles}
\label{secV}

In this section we summarize our knowledge about the spectrum of (pseudo-)Nambu--Goldstone bosons. They arise from the spontaneous breaking of the global symmetries of the model triggered by the dynamical generation of fermion masses. Therefore the (pseudo-)Nambu--Goldstone bosons are composites of the fermions. Because some of the spontaneously broken currents have axial anomalies the corresponding bosons are massive.

The sterile bosons composed of neutrinos and their properties were analyzed in the author's original work \cite{Smetana:2011tj}. There we assumed just a single type of axial anomaly, induced by the flavor gauge dynamics. Such simplification  allowed us to use standard estimates for the pseudo-Nambu--Goldstone boson masses and their coupling properties \cite{Peccei:1998jv}.

Here we classify all of the (pseudo-)Nambu--Goldstone bosons as composites of all fermions, not only of neutrinos. This however represents a complication. The corresponding spontaneously broken currents have anomalies given by all four gauge dynamics, see \eqref{Jdivergences}. It is not a priori clear how to construct the mass matrix for the pseudo-Nambu--Goldstone bosons induced by such intricate structure of the anomalies, let alone the question how to identify appropriate mass eigenstates. Even though the rigourous treatment of this issue is the subject of our future work, we present here the conjectured spectrum of the pseudo-Nambu--Goldstone bosons. Under the assumption that the individual mass eigenstates correspond to individual anomalies, we estimate the masses by the standard methods \cite{Peccei:1998jv}.

\subsubsection{Spontaneously broken currents and their anomalies}

The global symmetries of the flavor gauge model either exact or anomalous, \eqref{EW_symmetry} and \eqref{sterility_symmetry},
\begin{equation}
\U{1}_\mathrm{B}\times\U{1}_\mathrm{L}\times\U{1}_{\mathrm{B}_5}\times\U{1}_{\mathrm{L}_5}
\times\U{1}_{\mathrm{S}_6}\times\U{1}_{\mathrm{S}_3}\times\SU{4}_\mathrm{S}
\end{equation}
are spontaneously broken by formation of self-energies, i.e., by nonzero vacuum expectation values of composite bosons \eqref{condensates} and \eqref{H_composites}
\begin{equation}
\langle H^{(u)}\rangle\,,\ \langle H^{(d)}\rangle\,,\ \langle H^{(e)}\rangle\,,\ \langle H^{(\nu3)s}\rangle\,,\ \langle H^{(\nu6)}\rangle\,,\ \langle\Phi^{s=4}_3\rangle\,,\ \langle\Phi_6\rangle\,,
\end{equation}
and other much smaller elements of the neutrino self-energy, down to
\begin{equation}
\U{1}_\mathrm{B}\times\SU{3}_\mathrm{S} \,.
\end{equation}
This gives rise to the spectrum of either true or pseudo Nambu--Goldstone bosons. The number of broken generators and so the number of Nambu--Goldstone states is 12. The complete analysis of the (pseudo-)Nambu--Goldstone boson spectrum is tedious and we do not undergo it here. We rather make several estimates according to which we introduce the terminology.

First of all, there are 6 true Nambu--Goldstone bosons which couple to the non-diagonal non-anomalous broken currents of the sterility coset $\SU{4}_\mathrm{S}/\SU{3}_\mathrm{S}$ with generators $S_\sigma$, $\sigma=9,\dots,14$, see \eqref{non_Abelian_sterility_current}. We call them \emph{non-Abelian sterile majorons} denoted by $\eta_\sigma$. They are exactly massless
\begin{equation}
m_{\eta_\sigma}=0 \,.
\end{equation}

The remaining 6 Nambu--Goldstone states couple to linear combinations of 7 diagonal currents
\begin{equation}
\mathcal{J}^{\mu}_A\ =\ \big(J^{\mu}_\mathrm{B},\ J^{\mu}_{\mathrm{L}},\ J^{\mu}_{\mathrm{B}_5},\ J^{\mu}_{\mathrm{L}_5},\ J^{\mu}_{\mathrm{S}_3},\ J^{\mu}_{\mathrm{S}_6},\ J^{\mu}_{\mathrm{S}_{\sigma=15}}\big)^\T \,.
\end{equation}
Some of them are anomalous what can be written in a closed form
\begin{equation}
\partial_\mu\mathcal{J}^{\mu}_A = \mathcal{A}_{AX}\omega_X \,,
\end{equation}
where $\omega_X=\big(\mathcal{Y},\mathcal{W},\mathcal{G},\mathcal{F}\big)^\T$ are the anomaly operators defined in \eqref{anomaly_elements}, and $\mathcal{A}_{AX}$ are the elements of the $7\times4$ matrix of anomaly coefficients see \eqref{Jdivergences}
\begin{equation}
\mathcal{A} = \beginm{cccc} 3 & -3 & 0 & 0 \\ 3 & -3 & 0 & -1 \\ \tfrac{11}{3} & 3 & 4 & 4 \\ 9 & 3 & 0 & 3 \\ 0 & 0 & 0 & 1 \\ 0 & 0 & 0 & 1 \\ 0 & 0 & 0 & 0 \\ \endm\,.
\end{equation}
By making linear combinations of the original currents $\mathcal{J}^\mu_A\rightarrow\mathcal{U}_{AB}\mathcal{J}^\mu_B$ we can bring the matrix of the anomaly coefficients into the form
\begin{equation}
\mathcal{U}\mathcal{A} = \beginm{cccc} 3 & -3 & 0 & 0 \\ 0 & 0 & 0 & 0 \\ 0 & 0 & 0 & 0 \\ 0 & 0 & 0 & 0 \\ x_1 & x_2 & x_3 & 1 \\ y_1 & y_2 & 1 & 0 \\ z_1 & 1 & 0 & 0 \\ \endm\,,
\end{equation}
where we left the first row for the unbroken baryon number, $\big[\mathcal{U}\cdot\mathcal{J}^\mu\big]_{A=1}=J^\mu_\mathrm{B}$. The baryon number is anomalous in the same way as in the Standard Model. It stays as a very good approximate symmetry. The baryon number transformation is able to eliminate the weak $\theta$-parameter \cite{Anselm:1993uj}.

\subsubsection{Conjectured spectrum of the (pseudo-)Nambu--Goldstone bosons}

Out of the rest of the matrix $\mathcal{U}\mathcal{A}$ we can make our conclusions for the particle spectrum:
\begin{itemize}
    \item The next three currents $\big[\mathcal{U}\cdot\mathcal{J}^\mu\big]_{A=2,3,4}$ are anomaly free and their generators $X_A$ are given as linear combinations
        \begin{equation}\label{XA_BLS}
         X_A = a_A(B-L-S_3)+b_A(S_3-S_6)+c_A S_{\sigma=15} \,.
        \end{equation}
        Clearly, the linear combination of lepton number and sterility numbers plays the role of the Standard Model lepton number. The exactly conserved, anomaly free charges $X_A$ correspond to the Standard Model $B-L$ charge. They are spontaneously broken by the full neutrino self-energy. This gives rise to three true Nambu--Goldstone bosons called \emph{standard majorons} and denoted as $J_A$. They are exactly massless,
        \begin{equation}
         m_J=0 \,.
        \end{equation}
    \item The last three currents $\big[\mathcal{U}\cdot\mathcal{J}^\mu\big]_{A=5,6,7}$ are anomalous and thus they are explicitly broken by instanton effects from the flavor gauge dynamics, from the QCD \cite{'tHooft:1976up} and from the weak isospin gauge dynamics \cite{Kuzmin:1985mm}. Via the anomaly breaking each dynamics introduces its typical mass scale suppressed by the corresponding topological factors. As a result the instantons of the flavor gauge dynamics provide the strongest symmetry breaking while the instantons of the weak isospin gauge dynamics provide the weakest symmetry breaking. This hierarchy is reflected by the triangular shape of the anomaly coefficient matrix $\big[\mathcal{U}\cdot\mathcal{A}\big]_{A=5,6,7}$. The first linear combination of the currents is broken dominantly by the flavor instantons, the second one is broken dominantly by the QCD instantons and the third one is broken by weak isospin instantons.

        On top of the explicit symmetry breaking, the corresponding symmetries are spontaneously broken by all of the fermion self-energies. Therefore three pseudo-Nambu--Goldstone bosons arise. We believe that their mass eigenstates reflect the hierarchical structure of the anomalies. These are the axion-like particles known from the literature:
        \begin{itemize}
            \item The lightest pseudo-Nambu--Goldstone boson is known as the Anselm--Uraltsev \emph{arion} denoted as $a^\prime$ \cite{Anselm:1981aw}. It is the `axion' of weak isospin gauge dynamics. Its tiny mass is estimated as \cite{Anselm:1990uy}
                \begin{eqnarray}\label{mass_arion}
                 m_{a^\prime} &\approx& M_W\left(\frac{8\pi^2}{g^2}\right)^2\e^{-\frac{4\pi^2}{g^2}}\approx 10^{-36}M_W \,,
                 \end{eqnarray}
                where we used the value \cite{Nakamura:2010zzi} of the weak coupling constant $g \doteq 0.653$. The tiny value of $m_{a^\prime}$ is given by both the smallness of the coupling constant $g$ and the screening effect due to the fact that the weak bosons are massive.
            \item The intermediate pseudo-Nambu--Goldstone boson couples to the linear combination of currents whose divergence is dominated by the QCD anomaly. The corresponding symmetry is the natural candidate for the Peccei--Quinn symmetry \cite{Peccei:1977hh,Peccei:1977ur}. Therefore we identify the intermediate pseudo-Nambu--Goldstone boson with the Weinberg--Wilczek \emph{axion} denoted by $a$ \cite{Weinberg:1977ma,Wilczek:1977pj}. Due to its compositeness caused by the flavor dynamics, the axion is invisible with mass \cite{Kim:1984pt}
                \begin{eqnarray}\label{mass_a}
                 m_{a} &\approx& \frac{m_\pi f_\pi}{\Lambda_{\F}}\approx(10^{-3}-10^{-7})\,\mathrm{eV} \,
                \end{eqnarray}
                for $\Lambda_{\F}\sim(10^{10}-10^{14})\,\mathrm{GeV}$.
                This mass is significantly higher than the mass of the arion. It is caused not only by the higher value of the coupling constant but also by the fact that the gluon fields stay massless, thus they do not screen the instanton effects.
            \item The heaviest pseudo-Nambu--Goldstone boson couples to the linear combination of currents whose divergence is dominated by the anomaly from the flavor gauge dynamics. We call it the \emph{flavor axion} and denote it by $A$. To estimate its mass we should combine aspects of both two previous cases. There is the screening effect due to the flavor gauge boson masses and the flavor gauge coupling is big. Therefore we should paraphrase the equation for arion \eqref{mass_arion},
                \begin{eqnarray}\label{mass_Arion}
                 m_{A} &\approx& \Lambda_\F\left(\frac{8\pi^2}{h^2}\right)^2\e^{-\frac{4\pi^2}{h^2}} \,.
                \end{eqnarray}
                The maximal value of the flavor axion mass $m_{A,\mathrm{max}}\approx16\Lambda_\F/\mathrm{e}^2$, where $\mathrm{e}$ is the Euler number, is reached for the value of the coupling constant $h=\sqrt{2}\pi$. By decreasing the value of the coupling constant the flavor axion mass falls down rapidly and already for $h=1$ it gets enormously suppressed $m_{A}\approx10^{-14}\Lambda_\F$.
        \end{itemize}
\end{itemize}
The complete and fully rigorous analysis of the (pseudo-)Nambu--Goldstone boson spectrum is the subject of our current and future research. It is clear that in order to link the appropriate linear combination of broken currents to a given mass eigenstate, we need to determine the right orthonormal basis of the (pseudo-)Nambu--Goldstone fields following the lines of \cite{Weinberg:1975gm}, but the instanton-induced mass matrix has to be taken into account additionally. Complete understanding of the issue requires also to clarify the origin of the mass formula for a screened axion-like particle \eqref{mass_arion}.

At this stage, rather for demonstrational purpose, we are going to discuss the coupling properties of one of the massless majorons $J$ and also of the flavor axion $A$.

\subsection{Coupling properties of the standard majoron}

In this subsection we study the coupling properties of the massless standard majoron $J$. It couples to $X=B-L-S$, the analogue of the Standard Model $B-L$. The charge $X$ is one of the spontaneously broken charges \eqref{XA_BLS}.

The charge $X$ is spontaneously broken only by neutrino self-energy, therefore the majoron $J$ interacts only with neutrinos. At the level of an effective Lagrangian, such interaction can be introduced generally by the effective Yukawa majoron-neutrino coupling term
\begin{subequations}\label{EffYukawa}
\begin{eqnarray}
{\cal L}_{\mathrm{eff,}Jnn} & = & y_{Jnn}\,(\bar n X n)\,J \,,
\end{eqnarray}
\end{subequations}
where $n$ is the neutrino Nambu--Gorkov multispinor \eqref{NGmultiplet}. At this level, the effective Yukawa coupling constant $y_{Jnn}$ is mere effective parameter. More precisely, it is the matrix in the Nambu--Gorkov space with the same dimension as the neutrino self-energy $\mathbf{\Sigma}_n(p)$ \eqref{NGselfenergy}. The Yukawa coupling constants can be related to the neutrino self-energy as discussed in section~\ref{PagelsStokar}.

For that purpose the Ward--Takahashi identity for proper vertices, described in appendix \ref{PagelsStokar}, leads to the Yukawa coupling
\begin{equation}\label{P_J_nu}
y_{Jnn}\simeq P_X(p,p)=\Lambda^{-1}_{X}(0)\big[\tilde{\mathbf{\Sigma}}_n X-\gamma_0 X\gamma_0\tilde{\mathbf{\Sigma}}\big] \,,
\end{equation}
where $P_X(p,p)$ is the Nambu--Goldstone pole vertex \eqref{P} for the standard majoron and $\Lambda_{X}(0)$ is the bilinear coupling of the majoron to the broken current $J^{\mu}_X(x)$. All fermions coupled to the current $J^{\mu}_X(x)$ contribute to the bilinear coupling $\Lambda_{X}(0)$ which can be approximated as a sum over all contributing fermions of one-loop expressions $I^{n}_X$ given by the integral \eqref{I_relation}. For the purpose of an order of magnitude estimate we can use $\Lambda_{X}(0)\approx \sum_n m_n$ where $m_n$ are masses of all neutrino states. Projecting only the active neutrinos $\nu$ out of the formula \eqref{P_J_nu} we end up with the estimate
\begin{equation}
y_{J\nu\nu}\sim \frac{m_\mathrm{light}}{\sum_n m_n} \,,
\end{equation}
the effective Yukawa coupling $y_{J\nu\nu}$ of standard majoron $J$ to active neutrinos $\nu$.

Now, in the minimal version (333), we could expect that masses of all neutrino eigenstates turn out to be of the same order, thus of the order of the active neutrino mass. That is why we can estimate the effective Yukawa coupling as
\begin{equation}
y_{J\nu\nu}^{(333)}\approx 10^{-1} \,.
\end{equation}
On the other hand, in the version (63333) where the seesaw mechanism is in work, the sum of all neutrino masses is dominated by masses of $N_\mathrm{heavy}$ super-heavy neutrinos, thus $\sum_n m_n\sim N_\mathrm{heavy}\Lambda_\F\approx10^{15}\,\mathrm{GeV}$.
That is why we can estimate the effective Yukawa coupling as
\begin{equation}
y_{J\nu\nu}^{(63333)}\approx 10^{-25} \,.
\end{equation}
That makes a qualitative difference between the two versions of the model. While the version (333) resembles more the triplet majoron models \cite{Gelmini:1980re} and it is in clash with the most recent limits on majoron-neutrino couplings \cite{Lessa:2007up}, the version (63333) resembles the singlet majoron models \cite{Chikashige:1980ui} and is by far safe from the experimental limits.

\subsection{Coupling properties of the flavor axion}

The flavor axion $A$ couples to the current with dominant flavor anomaly. Its dominant interactions are given by the anomalous coupling to the flavor gauge bosons
\begin{equation}\label{axionEWinteraction}
{\cal L}_{ACC}= \frac{h^2}{32\pi^2}\frac{A}{F_{A}}F_{\mu\nu a}\tilde{F}^{\mu\nu}_a \,,
\end{equation}
where $F_{A}$ is the flavor axion decay constant, and by the effective Yukawa coupling to super-heavy neutrinos
\begin{equation}\label{L_Hnn}
{\cal L}_{Ann}\sim \frac{m_n}{F_\mathrm{A}}A\bar{n}n \,.
\end{equation}
Because the Yukawa interactions are in general proportional to the fermion mass, their interactions with lighter fermions are enormously suppressed, thus the flavor axion is fairly invisible.

%The coupling of the flavor axion to the flavor anomaly has important consequences for the $CP$ properties of the flavor model. There is no reason why there should not be the $\theta$-term of flavor gauge dynamics in the effective Lagrangian. The $\theta$-parameter shifted by phase that makes the fermion masses real is eliminated by the Peccei--Quinn mechanism \cite{Peccei:1977hh,Peccei:1977ur}, where the flavor axion plays the role of the composite axion.

The flavor axion could decay into the heavy flavor gauge bosons due to the direct interaction \eqref{axionEWinteraction} induced by the flavor anomaly. The decay would be kinematically allowed if the flavor axion is heavier than twice the mass of $N_{C_\mathrm{light}}\leq8$ lightest flavor gauge bosons, $m_A<2M_C$. For the sake of rough estimate of the decay width, we omit the non-Abelian character of the flavor gauge bosons and also the differences of their masses using a common mass $M_C$. The matrix element ${\cal M}$ is given by the vertex \eqref{axionEWinteraction}
\begin{eqnarray}
\im{\cal M} & = & \begin{array}{c}\includegraphics[width=0.2\textwidth]{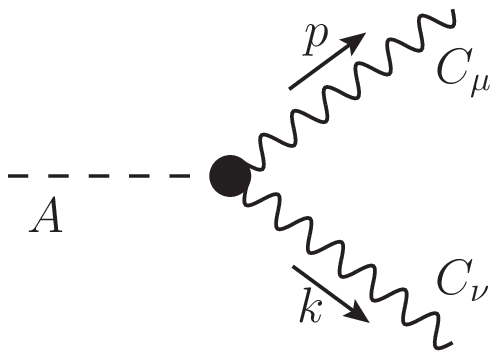}\end{array} \nonumber\\
    & = & \frac{h^2}{32\pi^2F_A}\varepsilon_{\mu}^*(p)\varepsilon_{\nu}^*(k)\epsilon^{\mu\nu\alpha\beta}p_\alpha k_\beta \,,
\end{eqnarray}
where $\varepsilon_{\mu}$ is the polarization vector of the flavor gauge boson $C_\mu$. The decay width then follows
\begin{equation}\label{DW_HM_CC}
\Gamma(A\rightarrow CC)=\frac{N_{C_\mathrm{light}}}{64\pi}\frac{h^4}{(32\pi^2)^2}\frac{m_{A}^3}{F_{A}^2}\left(1-\frac{4M_{C}^2}{m_{A}^2}\right)^{3/2} \,.
\end{equation}
After some order of magnitude assumptions $N_{C_\mathrm{light}} h^4\approx100$, and $m_A\sim M_C\sim F_A\sim \Lambda_\F$ we get an estimate
\begin{equation}
\Gamma(A\rightarrow CC)\approx 10^{-4}\Lambda_\F
\end{equation}
leading to very fast decay. If it is kinematically allowed, the decay to $N_\mathrm{heavy}$ super-heavy neutrinos of mass $m_{\mathrm{heavy}}\sim\Lambda_\F$ gives a contribution to the heavy sterile majoron decay width of comparable size with \eqref{DW_HM_CC}. From the effective vertex \eqref{L_Hnn} the decay width follows
\begin{equation}\label{DW_H_nn}
\Gamma(A\rightarrow nn)=\frac{N_\mathrm{heavy}}{8\pi}\frac{m_{\mathrm{heavy}}^2}{F_{A}^2} m_A\left(1-\frac{4m_\mathrm{heavy}^2}{m_{A}^2}\right)^{3/2}
\end{equation}
and a rough estimate is
\begin{equation}
\Gamma(A\rightarrow nn)\approx 10^{-1}\Lambda_\F \,.
\end{equation}

\section{Robustness of the flavor gauge model}

From the theoretical point of view, the flavor gauge model is consistently defined being built on the ground of the gauge principle. From the phenomenological point of view, the flavor gauge model is built to generate fermion masses, to explain their wild hierarchy and consequently to trigger the electroweak symmetry breaking.

On top of that, however, it has an unanticipated potential to address many of the particle physics issues. We have presented just few of them. The flavor gauge model predicts the specific sector of right-handed neutrinos, while not all of them turn out to be super-heavy. Some remain massless. The model provides the dynamical origin of the seesaw mechanism for the active neutrino masses. It predicts a number of composite majorons and composite axion-like particles. The composite boson spectrum includes the Weinberg--Wilczek axion connected with the solution of the strong $CP$ problem. All these sterile particles are potential candidates for being the dark matter constituents, see \cite{Nieuwenhuizen:2008pf,Kusenko:2009up,Bezrukov:2009th} for sterile neutrinos, \cite{Berezinsky:1993fm,Lattanzi:2008zz} for majorons and \cite{Holman:1982tb,Sikivie:2009fv,Hannestad:2010yi} for the axion. The sterility condensation can even generate the baryogenesis and drive the inflation of the Universe \cite{Barenboim:2008ds,Barenboim:2010nm}.

It requires intensive future work to make reliable predictions out of these indications.

\chapter{Conclusions}
%\pagenumbering{arabic}
%\input{P.tex}

\label{conclusion}

The topic of the thesis was to formulate the models of the electroweak symmetry breaking caused by the dynamically generated masses of quarks and leptons, to elaborate their basic features and to investigate their phenomenological consequences. We have presented two models, the model of strong Yukawa dynamics in chapter~\ref{part_strong_Yukawa_dynamics} and the flavor gauge model in chapter~\ref{gfd}. Prior to their exposure we have performed the basic analysis whether the main underlying idea, that only the known fermions can provide the electroweak symmetry breaking, is actually feasible. For that we elaborated the model of the top-quark and neutrino condensation in chapter~\ref{top_and_nu_condensation}. Because our approach towards the electroweak symmetry breaking is closely related to the renown models of dynamical electroweak symmetry breaking, the (Extended-)Technicolor models and the Top-quark Condensation models, we have presented their condensed overviews in chapter~\ref{dynamical_models} just after the introductory chapter~\ref{intro}.

\subsubsection{The top-quark and neutrino condensation model}

Within the top-quark and neutrino condensation model we have documented that the dynamical fermion mass generation has indeed a potential to trigger the electroweak symmetry breaking adequately. The basic ingredient is the seesaw mechanism for neutrinos with their sufficiently large Dirac masses. The Dirac masses then supplement the top-quark mass to saturate the electroweak scale. The model was designed as the two composite Higgs doublet model to be as simple as possible to still capture the main features of the electroweak symmetry breaking scenario. The model suggests that in order to reproduce the known mass spectrum, including the mass of the $125\GeV$ resonance, the number of right-handed neutrinos has to be rather large, $\mathcal{O}(100)$. This number may vary significantly depending on the particular realization of the underlying model. What should remain model independent is the prediction that the composite $125\GeV$ resonance should have a significant neutrino admixture. This feature brings about a suppression of its top-quark Yukawa coupling with respect to the Standard Model Higgs boson coupling. Additionally, the model predicts the existence of the spin-$0$ bound states with masses not far from the electroweak scale.

The top-quark and neutrino condensation scenario is interesting by itself even without referring to any of underlying fundamental models. To confront it more thoroughly with experimental data it is necessary to embed it within a more elaborated model which is able to address the mass spectrum and mixing of neutrinos. Ultimately, the improved model should be formulated in terms of a multitude of composite Higgs doublets, one for each participating Dirac mass. Apparently it offers a rich material for future work.

\subsubsection{The model of strong Yukawa dynamics}

The model is based on the strong Yukawa dynamics which acts among quarks and leptons to provide their masses dynamically. By elaborating the model we have gained an insight into the realm of non-perturbative calculations of fermion masses. Because of its non-gauge nature, we have managed to consistently formulate the Schwinger--Dyson equations for fermion self-energies. Next, we have used the approximate numerical methods to solve the equations and we have revealed the qualitative properties of the solutions. The resulting fermion self-energies exhibit a non-analytic dependence on the Yukawa coupling parameters. The non-analyticity provides the critical scaling of the self-energies which allows for reproducing strongly hierarchical mass spectra. In particular we have documented that the ratio of two fermion masses $m_1/m_2$ can be made arbitrarily small while keeping the Yukawa coupling constants of the same order $y_1/y_2\approx1$.

The ingenious mechanism, which underlies the generation of fermion self-energies, is the formation of the anomalous two-point Green's function $\langle\phi\phi\rangle$ of a complex scalar field $\phi(x)$. It is the mechanism for which the application beyond the electroweak symmetry breaking model building should be found, e.g., within some of non-relativistic condensed matter systems.

\subsubsection{The flavor gauge model}

The flavor gauge model is our most promising attempt to formulate a fundamental theory of fermion masses. First, it is based on the asymptotically free non-Abelian flavor gauge dynamics, which is therefore a UV complete dynamics. Second, in the infrared regime the flavor symmetry has to be broken. For that we have defined the flavor gauge dynamics as the chiral gauge theory, which does not allow any of bare masses in the Lagrangian. Therefore there is no need to invoke any other more fundamental theory which would have to explain such bare mass parameters. Third, the flavor gauge dynamics possesses just a single free parameter at the Lagrangian level. The complete mass spectrum of observed particles is expected to be parametrized by a handful of the effective low-energy parameters. Like in the QCD, they are at least in principle calculable from the first principles.

The aim of the model is indeed very ambitious. Correspondingly the elaboration of the model is a long-term program. We have started by identification of the favored version of the model, which is defined by particular non-minimal right-handed neutrino sector. It includes the right-handed neutrinos within the flavor sextet representation. We have ascribed the flavor symmetry breaking to the flavor sextet right-handed neutrino condensation in the Majorana channels. We have conjectured that the condensation of the right-handed neutrinos underlies their huge Majorana masses, the prerequisite for the seesaw pattern of the complete neutrino mass matrix. The seesaw mechanism with a huge right-handed neutrino Majorana masses is necessary not only for the explanation of the tininess of active neutrino masses, but also for the successful reproduction of the electroweak symmetry breaking and its scale, according to the conclusions of the top-quark and neutrino condensation model.

%Despite the fact how economic is the flavor gauge model, it is robust in addressing many particle physics phenomena from the lowest scales of neutrino physics up to the seesaw or GUT scales. To address these issues reliably requires to understand various subtleties of the quantum field theories which are often relevant exclusively for the chiral gauge theories. The flavor gauge model stimulated us to open many conceptual questions which are beyond the scope of this thesis, but they are clearly determining a course of our future research. Let us mention just few of the challenges:
%\begin{itemize}
%    \item the reliable formulation of a self-breaking of the chiral gauge theory,
%    \item the rigorous encountering of the effects of the momentum-dependence of self-energies on the observational quantities, mainly the determination of couplings of Nambu--Goldstone bosons to their constituent elementary fields,
%    \item the status of composite (pseudo-)Nambu--Goldstone bosons, mainly the determination of their masses and mass eigenstates in the presence of a general structure of axial anomalies,
%    \item the phenomenological impacts of topological non-trivialities of the chiral gauge dynamics.
%\end{itemize}

\subsubsection{Epilogue}

Our approach towards prominent issues of particle physics is led by our trust that the conservative quantum field theoretic methods are far from being exhausted. The currently accepted description of the elementary particle phenomena exhibits various tensions and challenges. One way to struggle with them is to replicate the operationally well known and established elements of quantum field theory in order to elaborate some more complicated construction under the auspices of a new exotic principle. This usually amounts to increase the number of free parameters.

The other way, which we have chosen, is to employ operationally not routine tools which are however introduced on the basis of economic, conservative and non-exotic principles. This usually amounts to just few free parameters. Both the prize and the price are high. On one hand, to explain all the observed phenomena by a reduced set of free parameters introduced by elegant principles is the theorist's dream. On the other hand, the ability to extract phenomenological outcomes can be paid only by new, fresh and astute ideas. We must conclude that the known principles simply transcend our current understanding of their realizations.

\appendix
\cleardoublepage
\addcontentsline{toc}{chapter}{Appendices}

\chapter{Approximate methods of solving Schwinger--Dyson equations}
\chaptermark{Approximate solving of Schwinger--Dyson eqns}
%\pagenumbering{arabic}
%\input{A.tex}

\label{approx_methods_for_SDE}

In this appendix we will explore various approximate methods of solving Schwinger--Dyson equations for fermion self-energy $\Sigma(p^2)$. In the types of models with which we deal in our work, the knowledge of fermion self-energies is completely essential, not only for reproducing the fermion mass spectrum. The fermion self-energies determine the masses of electroweak gauge bosons and the Yukawa couplings of various composite (pseudo-)Nambu--Goldstone bosons to their constituent fermions.

While elaborating our models we applied several methods to solve the Schwinger--Dyson equations. Even though none of them led to a completely satisfactory realistic solution, they provided us with at least qualitative insight into the complexity of the subject. In this appendix we summarize partial achievements which would otherwise not fit in the main text.

\subsubsection{The Schwinger--Dyson equations}

Generally, the Schwinger--Dyson equation has the form
\begin{equation}\label{main equation}
\Sigma(p^2)=\int\frac{\mathrm{d}^4k}{(2\pi)^4}F(k^2,\Sigma(k^2))G(k-p)\,.
\end{equation}
It is the integral equation for the unknown function $\Sigma(p^2)$, the fermion self-energy. In some cases the self-energy is a scalar complex or real valued function, like in the model of strong Yukawa dynamics described in chapter~\ref{part_strong_Yukawa_dynamics}. In other cases the self-energy is a general complex matrix function carrying information not only about the mass spectrum of fermions, but also about their mixing, like in the model of gauge flavor dynamics described in chapter~\ref{gfd}.

The function $F(k^2,\Sigma(k^2))$ is a chirality-changing part of the full fermion propagator \eqref{fermion_propagator}. The function $G(k-p)$ is a \emph{kernel} which is given by the dynamics responsible for the formation of the self-energy. Usually, the kernel is formed by a product of two vertices and a propagator of some intermediate field which provides an attraction among fermions. In this thesis, the intermediate field is either the scalar field in the model of strong Yukawa dynamics, or the flavor gauge boson in the model of flavor gauge dynamics. In the literature, wide variety of kernels can be found. For instance, in the work \cite{Wetterich:2006ii} the kernel is given by the exchange of a chiral tensor field.

\section{Usual approaches}

\subsubsection{Numerical-iterative method}

The Schwinger--Dyson equation \eqref{main equation} is an integral equation. Because the unknown function is under the integral, the integration cannot be performed analytically.

One can, however, try to solve the equation by means of a numerical-iterative method. The right-hand side of the Schwinger--Dyson equation is the functional of the self-energy, $\mathcal{F}[\Sigma]$. In terms of this functional the Schwinger--Dyson equation can be rewritten, suppressing the momentum arguments, as
\begin{equation}
\Sigma=\mathcal{F}[\Sigma]\,.
\end{equation}
If we plug some function $\Sigma_{0}(p^2)$ other than the exact solution into the functional, then $\Sigma_{0}\ne\Sigma_{1}=\mathcal{F}[\Sigma_{0}]$. We proceed by plugging the result of the integration $\Sigma_{1}(p^2)$ into the functional again and calculating $\Sigma_{2}=\mathcal{F}[\Sigma_{1}]$. We \emph{iterate} this step,
\begin{equation}
\Sigma_{I}=\mathcal{F}[\Sigma_{I-1}]\,,
\end{equation}
in order to minimize the difference between $\Sigma_I(p^2)$ and $\Sigma_{I-1}(p^2)$. If $|\Sigma_I(p^2)-\Sigma_{I-1}(p^2)|\rightarrow0$ then $\Sigma_I(p^2)$ converges to the exact solution.

For evaluating the functional a simple \emph{numerical integration} is used. After discretizing the integration momentum $k$, the integral can be approximated by a sum over the discrete values of $k$. In the following we denote the result of the numerical-iterative method as $\Sigma_{\mathrm{NI}}(p^2)$.

%Usually, we start by assuming the zeroth iteration of self-energy to be constant function, $\Sigma_0(p^2)=\mathrm{const}$

The complexity of the equations with the Minkowski metrics is given by poles of the integrand. The presence of the poles in the integrand may invalidate the numerical integration. Therefore one usually resorts to simplifications in order to avoid these poles, or at least some of them.

\subsubsection{Euclidean approximation}

Usually, the Euclidean approximation \cite{Roberts:1988yz} of the Schwinger--Dyson equation \eqref{main equation} is considered. This amounts to perform the Wick rotation within the integral, $k^0\rightarrow\im k^0$, and to make the same substitution for the outer momentum, $p^0\rightarrow\im p^0$. The fermion propagator is then substituted for its Euclidean version. Considering only the real-valued and non-matrix self-energy, we can write
\begin{equation}\label{F_Euclid}
F(k^2,\Sigma^2(k^2))=\frac{\Sigma(k^2)}{k^2+\Sigma^2(k^2)} \,.
\end{equation}

\subsubsection{Equal-time approximation}

In some cases the Schwinger--Dyson equation is subject of the Equal-time approximation \cite{LeYaouanc:1983iy,Adler:1984ri} where it is assumed that the dependence of the kernel $G(k-p)$ on the $k^0$ and $p^0$ is irrelevant. Consequently the self-energy $\Sigma(p^2)$ does not depend on $p^0$ either. The integration over remaining $k^0$ in $F(k^2,\Sigma(k^2))$ lowers the dimension of the integral and from the fermion propagator it produces
\begin{equation}
F(k^2,\Sigma^2(k^2))=\frac{\Sigma(k^2)}{\sqrt{k^2+\Sigma^2(k^2)}} \,.
\end{equation}
%Even after these simplifications the Schwinger--Dyson equations are difficult to solve. Therefore further approximations are used. One of the approaches is to adopt various Ans\"{a}tze on different parts of the equations. In this section we will describe a general method, called the \emph{trial method}, where we consider the Ansatz for the fermion self-energy.

\section{Analytical approach in Minkowski space}

Usually, the Wick rotation is the first step when solving the Schwinger--Dyson equations. One should be however aware of the fact that it may not be always an innocent act. Once the integrand contains some poles or branch cuts in the first and/or third quadrant of the complex plane, then the Wick rotation generates additional contributions to the integral. To evaluate these contributions it is necessary to know the poles or branch cuts. For that the exact solution has to be known. For practical calculations we omit this complication by neglecting these contributions, that is why we call the Wick rotation the approximation.

To appreciate the complexity of the issue of solving the Schwinger--Dyson equation in the full extent, i.e., in the Minkowski space, we present here the analytical expression calculated in detail in \cite{Bicudo:2003fd} for the integral\footnote{In the subsequent expressions we skip the infinitesimal $\epsilon$ keeping in mind that it always stands together with the corresponding mass squared, $M^2-\im\epsilon$. }
\begin{equation}\label{int_Bicudo}
      I(p^2)= \im\int\frac{\d^4k}{(2\pi)^4}\frac{m}{k^2-m^2+\im\epsilon}\left[\frac{1/2}{(k-p)^2-M_{1}^2+\im\epsilon}-\frac{1/2}{(k-p)^2-M_{2}^2+\im\epsilon}\right] \,.
\end{equation}
The integral \eqref{int_Bicudo} coincides with the integral in the Schwinger--Dyson equation derived under simplification $\Sigma(p^2)= m^2$ and $\Pi_N(p)=\mu^2$ within the model of strong Yukawa dynamics and its equation \eqref{1SDE_Minkowski}, where $M_{1,2}^2=M_{N}^2\mp\mu^2$ with real mass parameters $m$ and $\mu$. We have used here the symmetry breaking scalar propagator in the form of difference of two `standard' propagators \eqref{difference_propagator}. We can understand the integral $I(p^2)$ as the first iteration of the solution of the Schwinger--Dyson equation \eqref{1SDE_Minkowski} calculated analytically as $\Sigma_1(p^2)=I(p^2)$ and $\Sigma_0(p^2)=m$.

\begin{figure}
  \centering
  \includegraphics[width=0.7\textwidth]{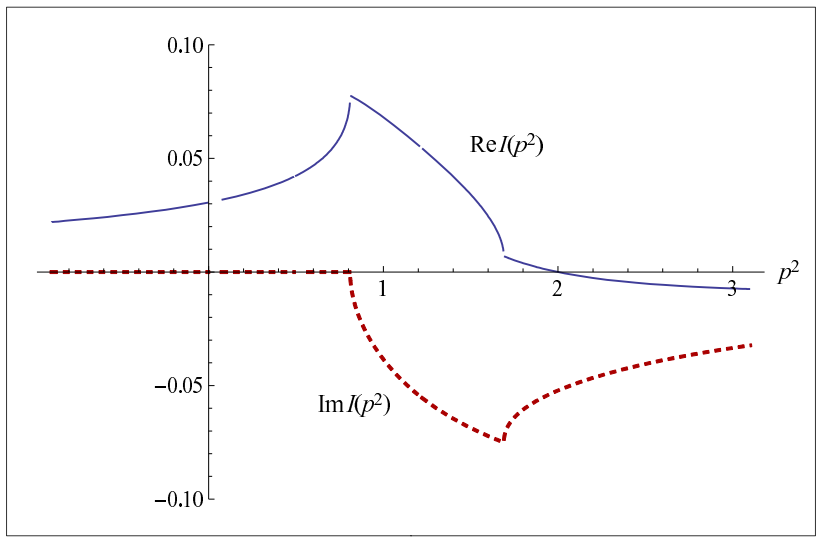}
  \caption[Real and imaginary parts of $I(p^2)$]{ The real and imaginary parts of $I(p^2)$ \eqref{bicudo} for $m=0.1$, $M_1=0.8$ and $M_2=1.2$. }\label{bicudo_fig}
\end{figure}

The analytic expression for the integral \eqref{int_Bicudo} reads
\begin{eqnarray}\label{bicudo}
I(p^2)&=&\sum_i \frac{\pi\xi_i m}{16p^2}\times\Biggl[ \left(p^2+M_i^2-m^2\right)\log\frac{m^2}{M_i^2} \\
&& +\left(p^2+M_i^2-m^2\right)\Delta(p^2,M_i,M_i)+\left(p^2-M_i^2+m^2\right)\Delta(p^2,M_i,m)\Biggr] \,, \nonumber
\label{analytical}
\end{eqnarray}
where the factors $\xi_1=+1$ and $\xi_2=-1$ introduce just signs of the `standard' propagators in \eqref{bicudo}, and
\begin{eqnarray}
 \Delta(p^2,M_i,X) & = & \frac{\rho(p^2,M_i)}{\sqrt{\rho^2(p^2,M_i)+X^2}} \Biggl[
 \log\im \left(1 - \frac{\sqrt{\rho^2(p^2,M_i)+X^2}}{\rho(p^2,M_i)} \right) \\
& & \hspace{3.5cm} -\log\im \left(1 + \frac{\sqrt{\rho^2(p^2,M_i)+X^2}}{\rho(p^2,M_i)} \right)
\Biggr]\,, \nonumber
\end{eqnarray}
where
\begin{equation}
  \rho(p^2,M_i)\equiv\sqrt{\frac{[p^2-(M_i+m)^2][p^2-(M_i-m)^2]}{4p^2}} \,.
\end{equation}
The real and imaginary parts of the integral $I(p^2)$ are plotted in Fig.~\ref{bicudo_fig}. The analytic expression of the imaginary part is
\begin{equation}\label{Im_I}
  \Im I(p^2) = -\sum_i \frac{\pi^2\xi_i m}{8p^2}\sqrt{[p^2-(M_i+m)^2][p^2-(M_i-m)^2]}\ \theta\big(p^2-(M_i+m)^2\big) \,,
\end{equation}
where $\theta(x)$ is the Heaviside step function.

We can see from \eqref{Im_I} that the integral $I(p^2)$ exhibits two thresholds at $p^2=(M_1+m)^2$ and $p^2=(M_2+m)^2$ which correspond to the fermion--boson production. These two thresholds are represented by two cusps in Fig.~\ref{bicudo_fig}. The analytic continuation of $I(p^2)$ into complex plane exhibits set of branch cuts which split apart from the cusps into the first quadrant of the complex plane.
It is conceivable that the exact solution for $\Sigma(p^2)$ shares the properties of $I(p^2)$. For the Wick rotation, this may be the complication as the integrand being a function of $\Sigma(p^2)$ shares the complex properties as well, including the branch cuts.

\section{Trial method}\label{trial_method}

The numerical-iterative method provides the way to find a solution which is a very good approximation of the exact solution of the Schwinger--Dyson equations. It is however time and numerical capacities demanding to investigate a dependence of the solution on some parameters. Therefore, in order to gain at least qualitative understanding of the behavior of the solution, we usually resort to more rough approximations. The goal of the trial method is to perform the integration analytically.

\subsubsection{Description of the trial method}

The trial method is simple. We \emph{try} to guess the type of a momentum dependence of the self-energy $\Sigma(p^2)$ in a way that the integration can be performed analytically. We introduce a \emph{trial} function $f(p^2)$ normalized as $f(0)=1$ as the Ansatz for the self-energy, $\Sigma(p^2)=m f(p^2)$. The dimensionful parameter $m$ represents the self-energy in the origin, $m=\Sigma(0)$. When we plug the Ansatz into the right hand side of the equation \eqref{main equation} we get
\begin{equation}
\int\frac{\mathrm{d}^4k}{(2\pi)^4}F(k^2,m f(k^2))G(k-p)=\varphi(m,p^2)
\end{equation}
with the function $\varphi(m,p^2)$ being the result of the integration. From the equation \eqref{main equation} we obtain the algebraic equation for the variable $m$
\begin{equation}\label{trial_equation}
  m f(p^2) = \varphi(m,p^2) \,. \\
\end{equation}

We can take the equation \eqref{trial_equation} as an infinite set of independent equations for the variable $m$, one equation for each value of $p^2$. Solutions of all the equations can be written as a function $\tilde{m}(p^2)$.

However for practical purpose, it is usually sufficient to calculate just $\tilde{m}(0)$, i.e., to use the equation \eqref{trial_equation} with $p=0$. The trial function, in general, can depend on some parameters $a_i$, so that $f(p^2)=f(p^2;a_i)$. In that case we need an additional equation for each parameter $a_i$. This can be achieved by taking the equation \eqref{trial_equation} for various values of $p_{i}^2\ne0$:
\begin{subequations}
\begin{eqnarray}
  m f(0;a_i) & = & \varphi(m,0;a_i) \,, \\
  m f(p_1^2;a_i) & = & \varphi(m,p_1^2;a_i) \,, \\
  m f(p_2^2;a_i) & = & \varphi(m,p_2^2;a_i) \,, \\
  & \dots & \nonumber
\end{eqnarray}
\end{subequations}

The choice of the trial function is crucially limited by the feasibility to calculate the integral in \eqref{main equation} analytically, or at least for a manageable numerical cost. The better Ansatz one chooses the more realistic result one obtains. This approach however disqualifies any systematic estimation of an error of the method. However, the end justifies the means.

\subsubsection{Examples}

As an example we will study a single equation derived in the model of strong Yukawa dynamics \eqref{1SDE_Euclidean} for a single non-matrix self-energy $\Sigma(p^2)$ within the Euclidean approximation. Therefore we have
\begin{equation}\label{FG_Yuk}
F(k^2,\Sigma(k^2))=\frac{\Sigma(k^2)}{k^2+\Sigma^2(k^2)}\quad\quad \mathrm{and} \quad\quad
G(k-p)=y^2\frac{\mu^2}{((k-p)^2+M^2)^2-\mu^4} \,,
\end{equation}
where $y$ is the real-valued coupling constant of the Yukawa dynamics and $M$ and $\mu$ are its mass parameters.

\begin{figure}[t]
\begin{tabular}{cc}
\includegraphics[width=0.45\textwidth]{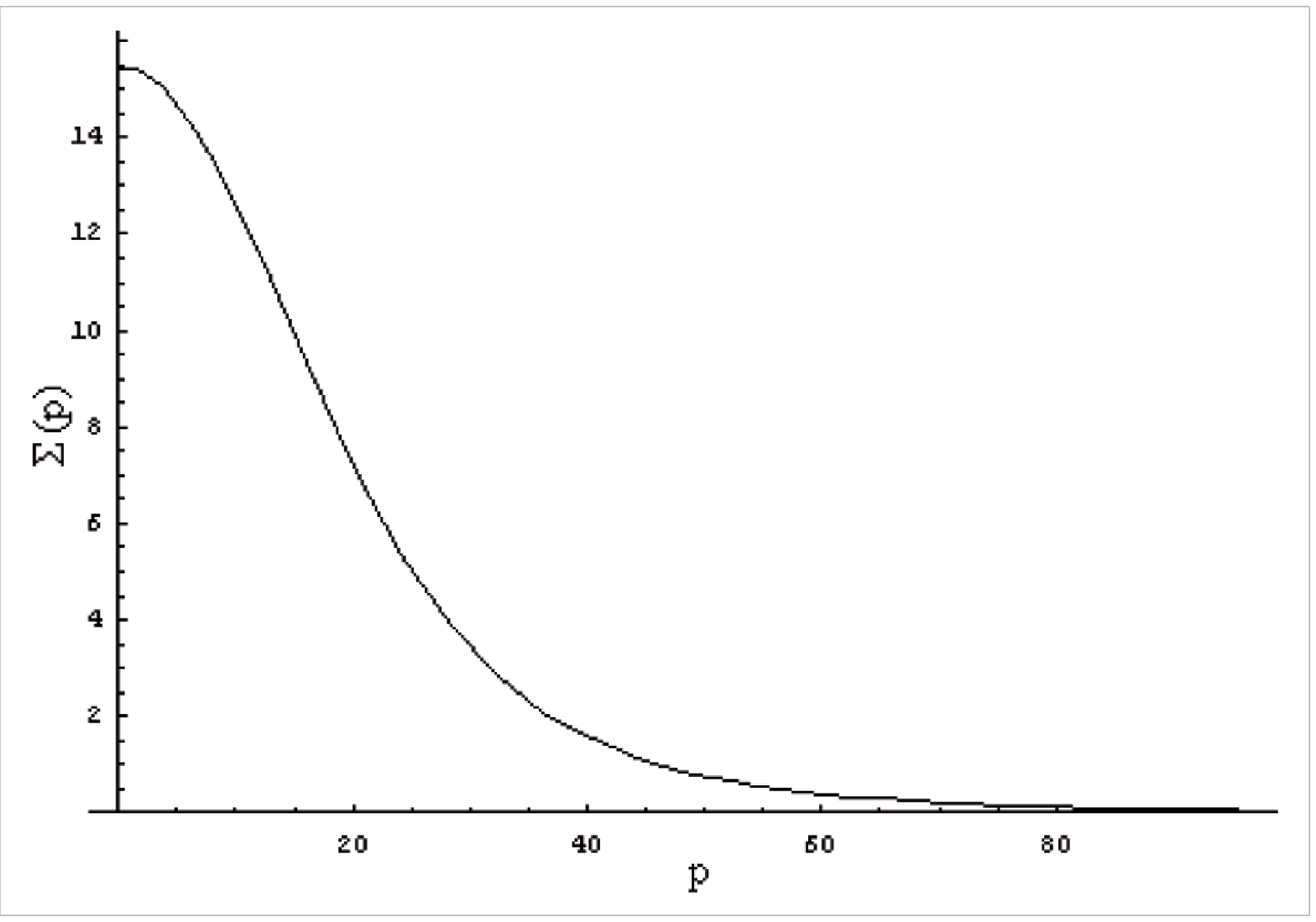} &
\includegraphics[width=0.45\textwidth]{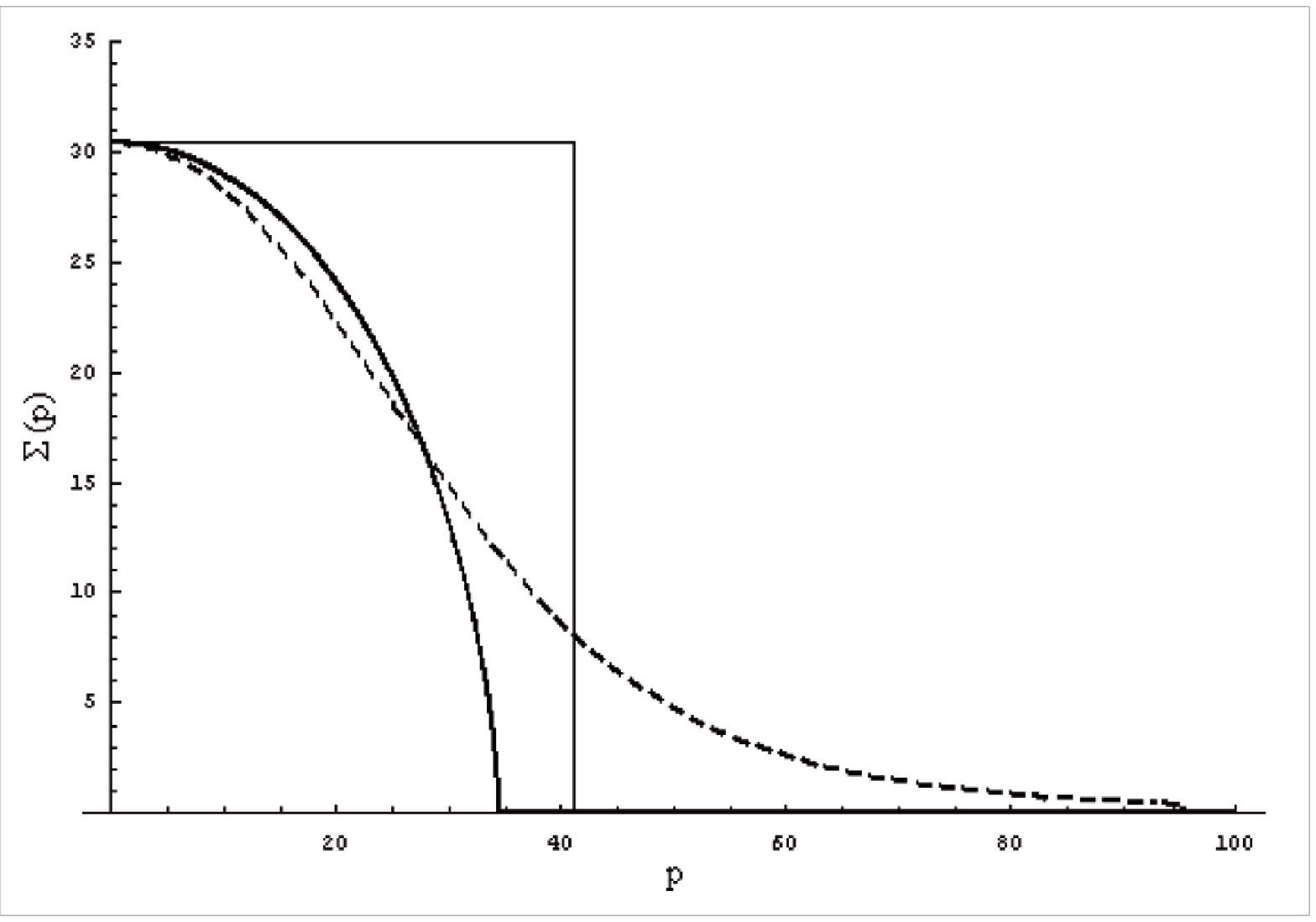} \\
a) & b)
\end{tabular}
\caption[Typical fermion self-energy obtained by numerical-iterative method, comparison of numerical-iterative and trial methods]{ a) The typical numerical-iterative solution $\Sigma_{\mathrm{NI}}(p^2)$ of the equation \eqref{main equation} with \eqref{FG_Yuk} for $y=128$, $M=10$, and $\mu=2$. b) A comparison of the step-function Ansatz (thin line), the solution of our method $\tilde{m}(p^2)$ (thick line), and the numerical-iterative solution (dashed line) for $y=170$, $M=10$, and $\mu=4$. In fact, we need to know the numerical-iterative solution to tune the cutoff $\Lambda=41.2$ so that our solution has a proper magnitude $\tilde{m}(0)=\Sigma_{\mathrm{NI}}(0)$. But even if we do not know how to tune the cutoff $\Lambda$, the method gives us a meaningful qualitative result. }
\label{fig:TS}
\label{fig:comparison}
\end{figure}

Under this setting the Schwinger--Dyson equation \eqref{main equation} can be solved easily by the numerical-iterative method. The resulting solution $\Sigma_{\mathrm{NI}}(p^2)$ is depicted in Fig.~\ref{fig:TS}a). We will compare our results with $\Sigma_{\mathrm{NI}}(p^2)$.

The simplest and rough trial function is the unit-step function
\begin{equation}
f(p^2)=\theta(\Lambda^2-p^2) \,.
\end{equation}
%We introduce the cutoff $\Lambda$ so that $\int f(p^2)\mathrm{d}p\approx \Lambda$.
The cutoff $\Lambda$ expresses the fact that the self-energy $\Sigma_{\mathrm{NI}}(p^2)$ decreases with momentum and the contributions from its tale above the cutoff into the integral is negligible. Inserting the Ansatz into the equation \eqref{trial_equation} we get
\begin{equation}
m\theta(\Lambda^2-p^2)=y^2\int_{k^2<\Lambda^2}\frac{\mathrm{d}^4k}{(2\pi)^4}F(k^2,m)G(k-p)=y^2 m\varphi(m,M,\mu,p^2,\Lambda^2) \,,
\end{equation}
from which we obtain transcendent but algebraic equation for the variable $\tilde{m}(p^2)$
\begin{eqnarray}
1=\varphi(\tilde{m},M,\mu,p^2,\Lambda^2) & \mathrm{for} & p^2<\Lambda^2\,, \\
\tilde{m}=0 & \mathrm{for} & p^2>\Lambda^2\,.
\end{eqnarray}
We have obtained the solution $\tilde{m}(p^2)$ depicted in Fig.~\ref{fig:comparison}b) (thick line). So the first goal is to gain better shape of the approximate solution function than the rough shape of the step-function.

\begin{figure}[t]
\begin{tabular}{cc}
\includegraphics[width=0.45\textwidth]{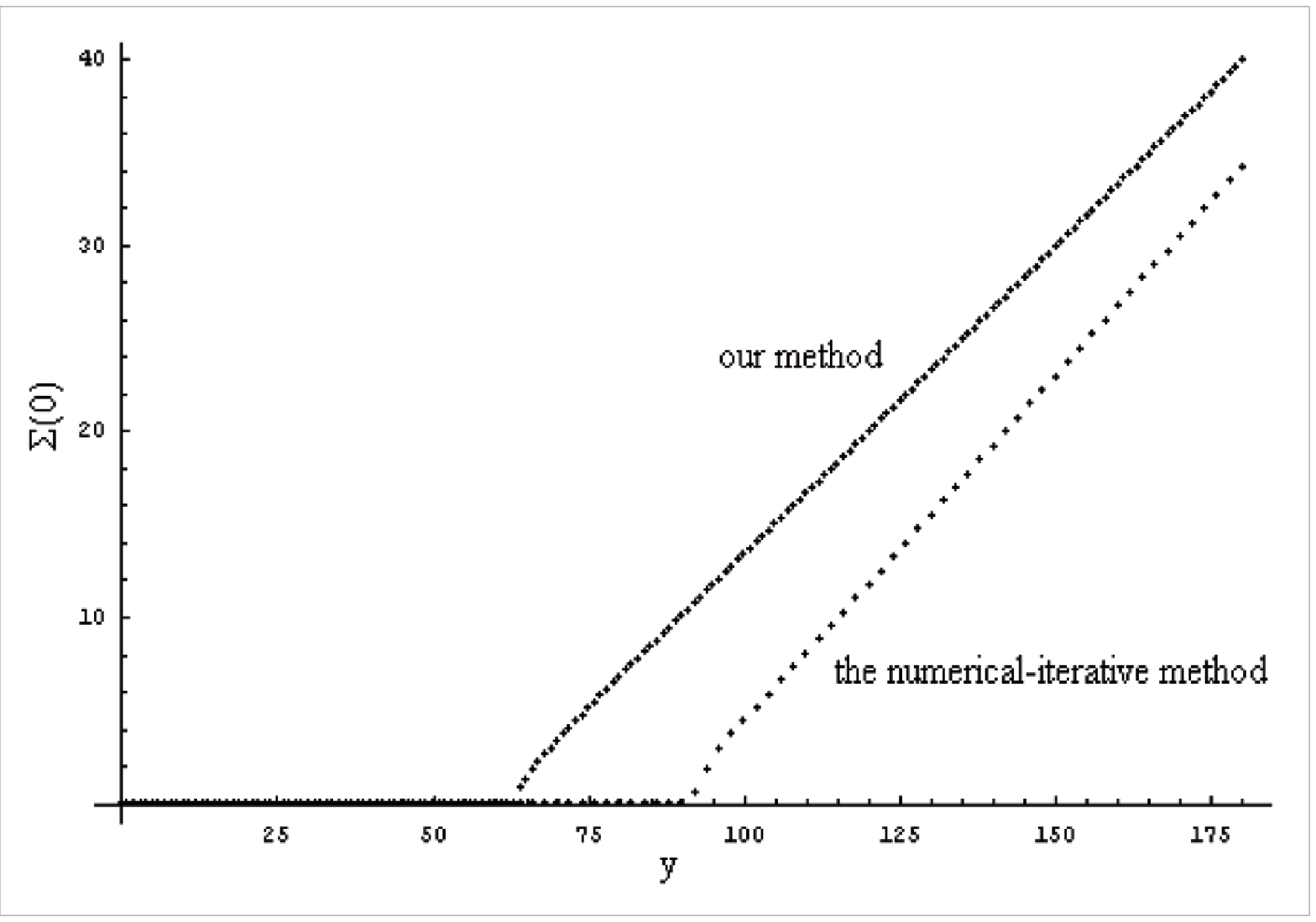} &
\includegraphics[width=0.45\textwidth]{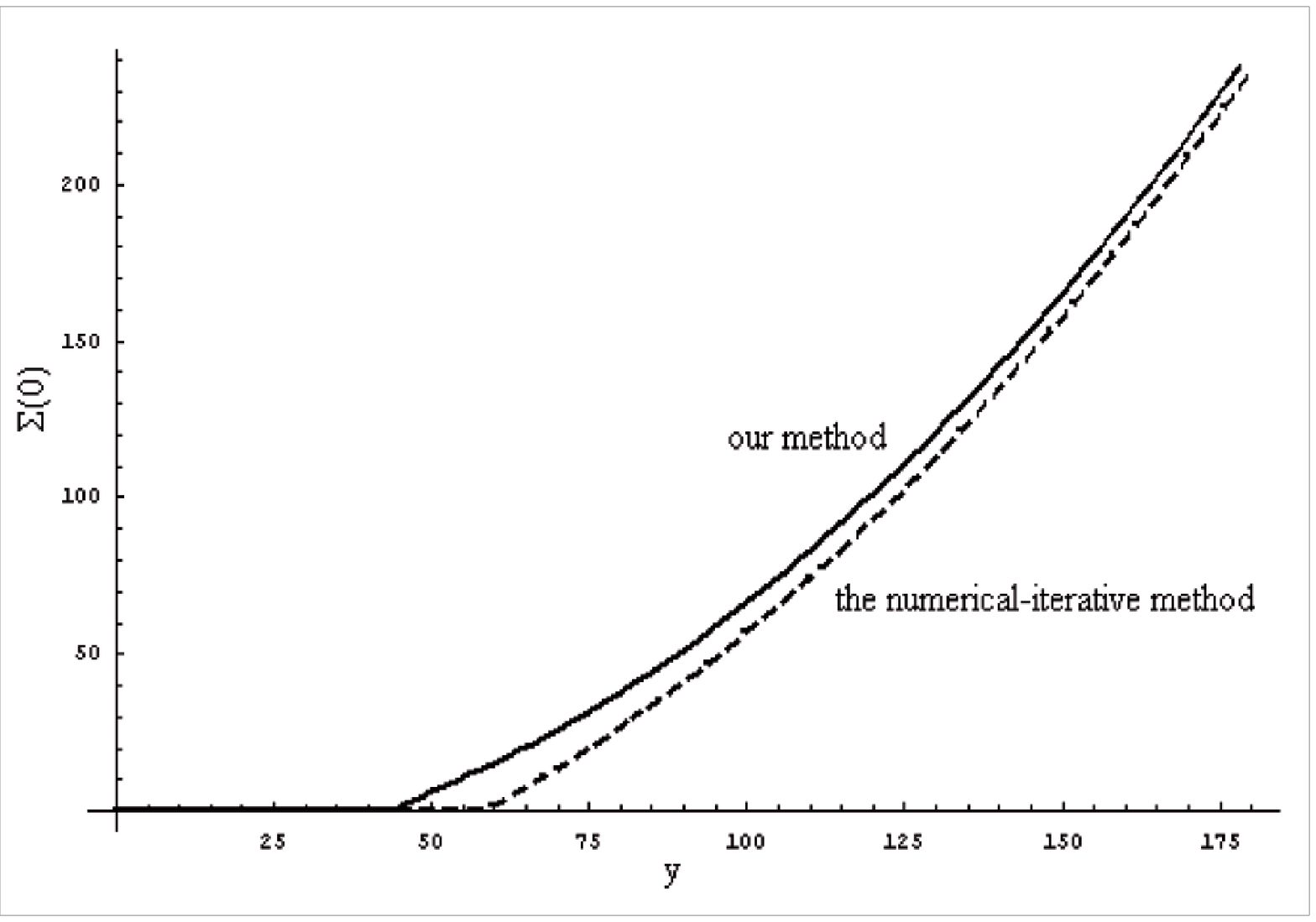} \\
a) & b)
\end{tabular}
\caption[ Comparison of $y$ dependance of $\Sigma(0)$ obtained by numerical-iterative and trial methods within Euclidean and Equal-time approximations ]{a) A comparison of the  $y$-evolutions of the magnitude of our solution $\tilde{m}(0)$ and the magnitude of the numerical-iterative solution $\Sigma_{\mathrm{NI}}(0)$ for $\Lambda=100$, $M=10$, and $\mu=4$, within the Euclidean approximation.
b) The $y$-evolutions of the magnitude of our solution $\tilde{m}(0)$ and the magnitude of the numerical-iterative solution $\Sigma_{\mathrm{NI}}(0)$ for $\Lambda=450$, $M=10$, and $\mu=4$, but now for the equations coming from the Equal-time approximation.  }
\label{fig:yevolutionET}
\label{fig:yevolution}
\end{figure}

Further we want to investigate the dependence of the magnitude of the solution on the coupling parameter $y$. The magnitude of the solution is given by $\tilde{m}(0)$. Therefore we use the equation for $p=0$. We compare the resulting dependence on the coupling constant $y$ with the dependence of the magnitude of the numerical-iterative solution given by $\Sigma_{\mathrm{NI}}(0)$ on $y$ in Fig.~\ref{fig:yevolution}a). From the plot it is possible to determine the critical value of the coupling constant $y$.

For further example, we apply the method on the same Schwinger--Dyson equation but now in the Equal-time approximation \cite{LeYaouanc:1983iy,Adler:1984ri}
\begin{equation}
\Sigma(p^2)=\pi y^2\int\frac{\mathrm{d}^3k}{(2\pi)^4}\frac{\Sigma(k^2)}{\sqrt{k^2+\Sigma^2(k^2)}}\frac{\mu}{[(k-p)^2+M^2]^2-\mu^2} \,.
\end{equation}
The result is given in Fig.~\ref{fig:yevolutionET}b).

In the presented examples, both the Euclidean and the Equal-time approximations avoid poles which are present in the original equations. Then the positive decreasing solutions exist. Therefore the step-function as an Ansatz seems to be reasonable. Clearly from Fig.~\ref{fig:yevolutionET} the trial method offers a qualitative agreement with the numerical-iterative method using much less effort.

\subsubsection{Combining the trial and the numerical-iterative methods}

The trial method can be combined with numerical-iterative method. One can assume that by changing the parameter $y$, the solution scales in some simple way. Then the numerical-iterative solution $\Sigma_{\mathrm{NI}}^{y_0}(p^2)$, obtained for a given single value of $y=y_0$ for the cost of the numerical method, can be used as the prototype profile function $f(p^2)$. For other values of $y$ one can use the Ansatz
\begin{equation}\label{Ansatz_SigmaNI}
\Sigma(p^2) = m\Sigma_{\mathrm{NI}}^{y_0}(p^2)/\Sigma_{\mathrm{NI}}^{y_0}(0) \,
\end{equation}
with single parameter $m$ and study its dependence on $y$. Obviously, it is not possible to calculate the integral analytically, it is necessary to calculate it numerically for each value of $y$. This is however still saving costs a lot, because within the numerical-iterative method the integral has to be numerically calculated many times for each value of $y$, not just once.

\section{Matrix Schwinger--Dyson equations}

\label{sde}

Under simplifying assumption, within the model of strong Yukawa interactions, the Schwinger--Dyson equations have the non-matrix form. On the other hand, within the flavor gauge model, the matrix structure of the Schwinger--Dyson equations is essential. Recall that by different flavor representation setting we intend to distinguish mass spectra of fermions of various electric charges \eqref{m33nem33}. In this section we will focus exclusively on the Schwinger--Dyson equations derived within the flavor gauge model \eqref{Sigma_SDE}. On top of that, we will consider the equations only for the charged fermions.

Upon the Wick rotation, the Schwinger--Dyson equation \eqref{Sigma_SDE} can be rewritten as
\begin{equation}\label{Sigma_SDE_Wick}
\Sigma(p^2) = 3\int\!\frac{\d^4k}{(2\pi)^4}\,\bar h^2_{ab}\big((p-k)^2\big)\Delta_{0}\big((p-k)^2\big)\,
T_{R,a} \, \Sigma(k^2) \Big[k^2+\Sigma^{\dag}(k^2)\,\Sigma(k^2)\Big]^{-1} T_{L,b} \,,
\end{equation}
where ${T_{R,a}}$ and ${T_{L,a}}$ are generators for either triplet or anti-triplet flavor representation of electrically charged fermions, therefore they are given by the Gell-Mann matrices either $\tfrac{1}{2}\lambda_a$ or $-\tfrac{1}{2}\lambda^{*}_a$. The effective charge $\bar{h}^2_{ab}(k^2)$ and the bare flavor gauge boson propagator function $\Delta_{0}(k^2)$ should be understood as already being Wick rotated.

First, we will discuss possibilities of achieving a solution which exhibits a critical scaling. Then, we will present an attempt to solve the Schwinger--Dyson equations by means of the approximation of a separable kernel.

\subsection{Critical scaling}

In the models of the type which we study in this thesis, a large hierarchy among the scale of new physics $M$ and the fermion masses $m$ is believed to arise dynamically. In this subsection we will demonstrate that reasonably simplified Schwinger--Dyson equations do indeed provide solutions with this property. They depend non-analytically on the coupling parameter $h$. Within various approximations we calculate the relation
\begin{equation}
  m=f(h^2-h_c^2)M \,,
\end{equation}
where $f(x)$ is the scaling function which is continuous and non-analytical in the origin. It is $f(x)=0$ for $x<0$ and increasing for $x\geq0$. By tuning $h$ close to its critical value $h_c$, arbitrarily small ratio $m/M$ can be achieved. This dependence of the solution $m$ on the coupling parameter is called the \emph{critical scaling}.

To demonstrate usefulness of the critical scaling, let us assume the Miransky scaling occurring in theories with the conformal phase transition \cite{Miransky:1996pd,Braun:2010qs}
\begin{eqnarray}\label{MirScaling}
m &=& M\e^{-4\pi/\sqrt{\Delta h^2}}\,,\quad\mathrm{where}\quad\Delta h^2\equiv h^2-h_{c}^2 \,.
\end{eqnarray}
With $M=10^3\, \rm TeV$ the neutrino mass $m_\nu \sim 1\,\mathrm{eV}$ is obtained for $\Delta h_{\nu}^2=\left(\tfrac{4\pi}{15\ln10}\right)^2\doteq 0.01$, and the mass of the top-quark $m_t\sim10^2\, \rm GeV$ is obtained for $\Delta h_{t}^2=\left(\tfrac{4\pi}{4\ln10}\right)^2\doteq 0.15$. Therefore the fermion mass hierarchy $\tfrac{m_t}{m_\nu}\sim10^{11}$ is achieved by effective coupling constants hierarchy $\Delta h_{t}^2/\Delta h_{\nu}^2\sim10$.

Within the flavor gauge model, the fermion mass generation is governed by the infrared behavior of the effective charge matrix, in particular, by the infrared fixed point matrix $h_{ab}^*$ \eqref{infrared_fixed_point_matrix}. In the Schwinger--Dyson equations \eqref{Sigma_SDE}, we have tried to model the infrared effective charge by various Ans\"{a}tze on the kernel $\bar h^2_{ab}(q^2)\Delta_{0}(q^2)$:
\begin{subequations}\label{pres_hbar}
\begin{eqnarray}
& & \mathrm{(A)}\quad \frac{\bar h^2_{ab}(q^2)}{q^2}=-\frac{h_{*}^2}{q^2}\left[M^2(q^2-M^2)^{-1}\right]_{ab} \,,\label{pole_pres_hbar}\\
& & \mathrm{(B)}\quad \frac{\bar h^2_{ab}(q^2)}{q^2-M_{0}^2}=-h_{*}^2\left[\big(M^2\big)^{-1}\right]_{ab} \,, \label{4f_pres_hbar} \\
& & \mathrm{(C)}\quad \mathrm{separable\ kernel} \,, \label{separable_pres_hbar}
\end{eqnarray}
\end{subequations}
where $h_*$ is a single infrared fixed point parameter and the matrix structure of the effective charge $\bar h^2_{ab}(q^2)$ is carried by the dimensionful matrix $M_{ab}$. For the Ansatz (B) we have encountered the flavor gauge boson mass generation \eqref{M_0} by including the corresponding massive pole at $q^2=M_{0}^2$ already into the bare propagator function $\Delta_{0}(q^2)=1/(q^2-M_{0}^2)$ according to \cite{Akiba:1985rr,Binosi:2009qm}. On the other hand, in the prescription for the Ansatz (A) we have let the bare propagator function to be $\Delta_{0}(q^2)=1/q^2$.

The Ansatz (B) is motivated by reproducing the four-fermion interaction dynamics and as such it requires regularizing the integral in the Schwinger--Dyson equation \eqref{Sigma_SDE} by a cutoff $\Lambda$. The cutoff $\Lambda$ has a physical interpretation. It is the energy above which the momentum dependent self-energy is truly negligible. The cutoff is however not necessary in the Ansatz (A) as the corresponding kernel makes the integral finite. The Ansatz (A) however exhibits a degenerate fixed point matrix $\bar h^2_{ab}(0)=h_{*}^2\delta_{ab}$ which will turn out to be an obstacle for reproducing the fermion mass hierarchy. On the contrary, the Ansatz (B) exhibits a general fixed point matrix $\bar h^2_{ab}(0)=h_{*}^2\left[\frac{M_{0}^2}{M^2}\right]_{ab}$. We study the separable kernel Ansatz (C) in a separate section \ref{separable_kernel}.

%We will see that each of Ans\"{a}tze exhibits some nice features but neither of them is fully satisfactory.

In this section we will sometimes work within the constant self-energy approximation of the Wick-rotated equation \eqref{Sigma_SDE_Wick} with $p=0$, i.e., considering the equation
\begin{equation}\label{Sigma_SDE_Wick_const}
\tilde{\Sigma} = \frac{3}{16\pi^2}\int_{0}^{\infty}\!k^2\d k^2\,\bar h^2_{ab}(k^2)\Delta_{0}(k^2)\,T_{R,a} \, \tilde{\Sigma}\Big[k^2+\tilde{\Sigma}^{\dag}\tilde{\Sigma}\Big]^{-1} T_{L,b} \,.
\end{equation}

\subsubsection{Degenerate case}

First we will illustrate the critical scaling on a non-realistic degenerate case when we take the effective charge in the form
\begin{equation}
\bar h_{ab}^2(q^2)=\bar h^2(q^2)\delta_{ab}
\end{equation}
in \eqref{Sigma_SDE_Wick}. This corresponds to $M_{ab}^2=M^2\delta_{ab}$ in \eqref{pole_pres_hbar} and \eqref{4f_pres_hbar}. It is interpreted as a degenerate mass spectrum of flavor gauge bosons. For this degenerate case the equation for the $u$-quarks exhibits the flavor \emph{singlet} solution
\begin{equation}\label{deg_gap_U}
\Sigma_u(p^2)=\beginm{ccc} 1 & 0 & 0 \\ 0 & 1 & 0 \\ 0 & 0 & 1 \\ \endm\sigma_u(p^2)\,,
\end{equation}
the equation for the $d$-quarks exhibits the flavor \emph{triplet} solution
\begin{equation}\label{deg_gap_D}
\Sigma_d(p^2)=\beginm{ccc} 0 & -\im & 0 \\ \im & 0 & 0 \\ 0 & 0 & 0 \\ \endm\sigma_d(p^2) \,.
\end{equation}
It follows from the identities
\begin{eqnarray}
\frac{1}{2}\lambda_a\beginm{ccc} 1 & 0 & 0 \\ 0 & 1 & 0 \\ 0 & 0 & 1 \\ \endm\frac{1}{2}\lambda_a & = & C_u\beginm{ccc} 1 & 0 & 0 \\ 0 & 1 & 0 \\ 0 & 0 & 1 \\ \endm \,, \\
\frac{1}{2}\lambda^{\T}_a\beginm{ccc} 0 & -\im & 0 \\ \im & 0 & 0 \\ 0 & 0 & 0 \\ \endm\frac{1}{2}\lambda_a & = & C_d\beginm{ccc} 0 & -\im & 0 \\ \im & 0 & 0 \\ 0 & 0 & 0 \\ \endm \,.
\end{eqnarray}
where $C_u=\frac{4}{3}$ and $C_d=\frac{2}{3}$. The self-energy functions $\sigma_\psi(p^2)$ for $\psi=u,d$ satisfy the gap equations
\begin{eqnarray}\label{gap_equation_degenerate}
\sigma_\psi(p^2) &=& \frac{3C_\psi}{16\pi^2}\int_{0}^{\infty}k^2\d k^2\bar h^2\big((p-k)^2\big)\Delta_{0}\big((p-k)^2\big)
\frac{\sigma_\psi(k^2)}{k^2+\sigma^{2}_\psi(k^2)} \,, \label{m}
\end{eqnarray}
To proceed we model the degenerate effective charge by means of the Ans\"{a}tze (A) and (B) in \eqref{pres_hbar} already Wick rotated.

(A) First, we make use of the Ansatz (A)
\begin{equation}\label{Ansatz_A}
\bar h^2(q^2)\Delta_{0}(q^2)=\frac{h_{*}^2}{q^2}\frac{M^2}{M^2+q^2} \,.
\end{equation}
%with $M^2$ being a mass parameter presumably of the same order of magnitude as the flavor gauge boson mass.
Next we adopt a simple momentum-dependent Ansatz for the fermion self-energy function $\sigma_\psi(p^2)$,
\begin{eqnarray}\label{abAnsatz}
\sigma_\psi(k^2)=\frac{(a+b)m_{\psi}^3}{ak^2+bm_{\psi}^2} \,,
\end{eqnarray}
with reasonably chosen parameters $a,b\geq0$. Notice that the Ansatz satisfies $\sigma_{\psi}(m_{\psi}^2)=m_{\psi}$, so that $m_{\psi}$ is directly the fermion mass. This is in contrast to the case in section~\ref{trial_method}, where we have taken $m=\Sigma(0)$ as the fermion mass.

If we set $a=0$ then $\sigma_{\psi}=m_{\psi}$ is a constant and the gap equations \eqref{m} with vanishing external momentum $p^2 = 0$ turn into the algebraic equations
\begin{eqnarray}\label{gap_eq_degenerate}
m_{\psi} &=& \frac{3C_\psi}{16\pi^2}\frac{m_{\psi}M^2}{M^2-m_{\psi}^2}\log{\frac{M^2}{m_{\psi}^2}} \,.
\end{eqnarray}
Within the approximation $\tfrac{m_{\psi}^2}{M^2}\ll1$ the solutions have the form \cite{Pagels:1979ai}
\begin{eqnarray}\label{expCritScaling}
m_u &=& M\exp[-2\pi^2/h_{*}^{2}] \,, \\
m_d &=& M\exp[-4\pi^2/h_{*}^{2}] \,,
\end{eqnarray}
exhibiting the appealing exponential critical scaling. Because the exponent is twice bigger for $d$-quarks than for $u$-quarks, the ratio of $m_d$ and $m_u$ scales exponentially as well
\begin{equation}
\frac{m_d}{m_u}=\exp[-2\pi^2/h_{*}^{2}]\,.
\end{equation}
The choice of the parameter $a=0$ leads to the exponential critical scaling \emph{\`{a} la} Miransky \eqref{MirScaling}, but it lacks the existence of a nonzero critical constant.

The nonzero critical constant occurs as a result of a UV damping of the self-energy encountered in the Ansatz \eqref{abAnsatz} once $a\ne0$. In order to work with simple analytic formulae we resort to simplification of using the momentum-dependent Ansatz only in the numerator of the integrand, while in the denominator we keep the self-energy function constant. Again we solve the equation only for vanishing external momentum, $p^2=0$. In the rest of this subsection we will consider the equation only for the $u$-quarks, because apart from the coefficient $C_\psi$ the equation for the $d$-quarks is completely analogous in the degenerate case. We can write the equation for a mass parameter $m_{u}$ as
\begin{subequations}\label{gap_equation_exp}
\begin{eqnarray}
m_{u}\frac{a+b}{b} &=& \frac{h_{*}^2}{4\pi^2}\int_{0}^{\infty}\d k^2\frac{M^2}{k^2+M^2}\frac{\frac{m_{u}^3(a+b)}{ak^2+bm_{u}^2}}{k^2+m_{u}^2} \\
  &=& \frac{h_{*}^2}{4\pi^2}\left[m_{u}\frac{a+b}{a-b}\frac{aM^2}{aM^2-bm_{u}^2}\ln\frac{a}{b}-\frac{m_{u}^3(a+b)}{aM^2-bm_{u}^2}\frac{M^2}{M^2-m_{u}^2}\ln\frac{M^2}{m_{u}^2}\right] \,.\quad\quad
\end{eqnarray}
\end{subequations}
For $m_{u}\ll M$ we approximate the gap equation as
\begin{eqnarray}\label{gap_eq_A}
m_{u} &=& \frac{h_{*}^2}{4\pi^2}m_{u}\left[\frac{b}{a-b}\ln\frac{a}{b}-\frac{b}{a}\frac{m_{u}^2}{M^2}\ln\frac{M^2}{m_{u}^2}\right] \,.
\end{eqnarray}
It can be rewritten as
\begin{eqnarray}\label{critScaling}
\frac{1}{h_{c}^2}-\frac{1}{h_{*}^2} &=& \frac{1}{4\pi^2}\frac{b}{a}\frac{m_{u}^2}{M^2}\ln\frac{M^2}{m_{u}^2} \,,
\end{eqnarray}
where
\begin{equation}\label{hcrit_A}
h_{c}^2=4\pi^2\frac{a-b}{b\ln\frac{a}{b}}
\end{equation}
is the critical coupling constant. The non-trivial solution exists only for $h_*>h_c$.

\begin{figure}
  \centering
  \includegraphics[width=1.0\textwidth]{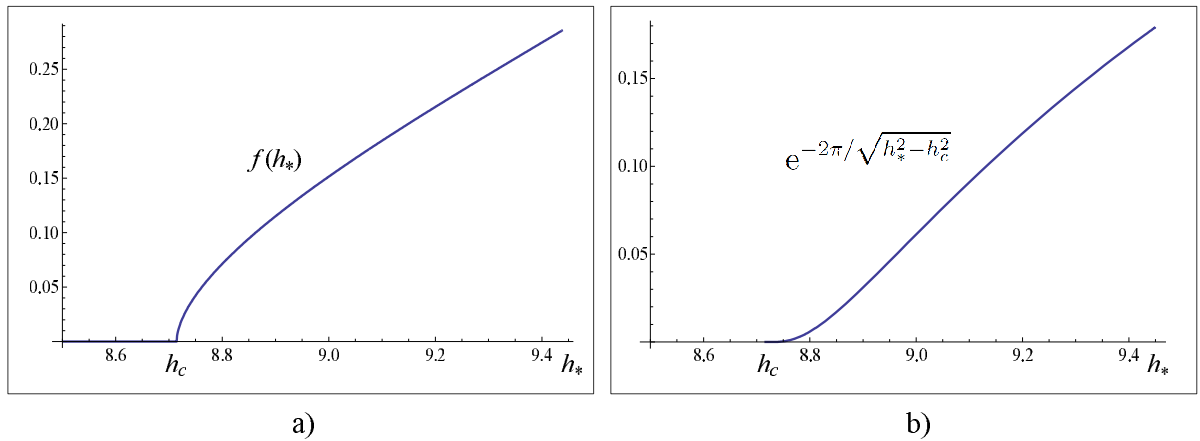}
  \caption[Square-root and Miransky critical scaling functions]{ Critical scaling of the fermion mass $m$ with respect to the scale $M$ given by $m=f(h_*)M$ around the critical coupling constant $h_c\doteq8.7$ obtained from \eqref{hcrit_A} for $a=2$ and $b=1$. a) The scaling function $f(h_*)$ calculated numerically from the gap equation \eqref{gap_equation_exp}. b) For comparison, the scaling function of the Miransky type \cite{Braun:2010qs} given by the formula \eqref{MirScaling} }\label{critical_scaling}
\end{figure}

The critical scaling following from \eqref{critScaling} allows for arbitrarily small fermion masses, $\tfrac{m}{M}\rightarrow0$, at the price of fine-tuning $h_*$ to be extremely close to $h_c$. The extreme fine-tuning is enforced by the square-root behaviour of the scaling function $f(h_*)$ near the critical point, as it can be seen in Fig.~\ref{critical_scaling}a). This fine-tuning is much weaker for the exponential critical scaling \eqref{expCritScaling} which, nevertheless, lacks the non-zero critical coupling constant. We believe that the ultimate critical scaling can exhibit combination of both behaviors, the exponential character and the non-zero critical scaling, like the Miransky scaling \eqref{MirScaling} does. For comparison we plot the Miransky scaling function in Fig.~\ref{critical_scaling}b). Both critical scalings obtained from \eqref{gap_eq_degenerate} and \eqref{critScaling} are the consequences of too rough approximations to exhibit both features simultaneously.

(B) For comparison we now use the Ansatz (B) in \eqref{4f_pres_hbar} already Wick-rotated for the degenerate effective charge as
\begin{equation}\label{Ansatz_B}
\bar h^2(q^2)\Delta_{0}(q^2)=\frac{h_{*}^2}{M^2} \,.
\end{equation}
This prescription requires the cutoff $\Lambda$ to regularize the gap equation. We use the simplest Ansatz for the self-energy function with $a=0$, because already the cutoff $\Lambda$ play the UV damping role in the gap equation. We write the gap equation again only for $m_u$ as for $m_d$ it is completely analogous,
\begin{subequations}
\begin{eqnarray}
m_{u} &=& \frac{h_{*}^2}{4\pi^2}\int_{0}^{\Lambda^2}\d k^2\frac{k^2}{M^2}\frac{m_{u}}{k^2+m_{u}^2} \\
&=& \frac{h_{*}^2}{4\pi^2}m_{u}\left[\frac{\Lambda^2}{M^2}-\frac{m_{u}^2}{M^2}\ln\frac{m_{u}^2}{\Lambda^2+m_{u}^2}\right] \,. \label{NJL_gap_U}
\end{eqnarray}
\end{subequations}
This leads to the equation which is equivalent to the equation \eqref{critScaling}
\begin{eqnarray}
\frac{1}{h_{c}^2}-\frac{1}{h_{*}^2} &=&  \frac{1}{4\pi^2}\frac{m_{u}^2}{M^2}\ln\frac{\Lambda^2}{m_{u}^2}
\end{eqnarray}
with the critical coupling constant
\begin{equation}\label{hcrit_B}
h_{c}^2=4\pi^2\frac{M^2}{\Lambda^2} \,.
\end{equation}
The critical coupling constant \eqref{hcrit_B} is substantially different from \eqref{hcrit_A}. It depends on $M$, while \eqref{hcrit_A} does not. In the more realistic non-degenerate case, this difference has big effect on achieving the fermion mass hierarchy.

\subsubsection{Non-degenerate case}

Our aim is to solve the Schwinger--Dyson equations \eqref{Sigma_SDE_Wick} in the more realistic case and obtain a non-degenerate fermion mass spectrum. In the simplest way, this can be achieved by assuming
\begin{equation}\label{non_degenerate}
M=\mathrm{Diag}\{M_1,M_2,M_3,M_4,M_5,M_6,M_7,M_8\} \,.
\end{equation}
This assumption defines the diagonal kernel, either \eqref{pole_pres_hbar} or \eqref{4f_pres_hbar}. We will work in the constant self-energy approximation using the equation \eqref{Sigma_SDE_Wick_const}. With the diagonal kernel it provides diagonal solutions in the form
\begin{equation}
\tilde{\Sigma}=\beginm{ccc} m_1 & 0 & 0 \\ 0 & m_2 & 0 \\ 0 & 0 & m_3 \\ \endm \,.
\end{equation}
For illustration we consider only the $u$-type fermions and for comparison we again use both Ans\"{a}tze \eqref{pole_pres_hbar} and \eqref{4f_pres_hbar}.

(A) First, we adopt the Ansatz \eqref{pole_pres_hbar}, i.e., in the Wick rotated equation \eqref{Sigma_SDE_Wick_const} we use
\begin{equation}\label{A_ansatz}
\bar h_{ab}^2(q^2)\Delta_{0}(q^2)=\frac{h_{*}^2}{q^2}\frac{M_{a}^2}{M_{a}^2+q^2}\delta_{ab} \,,
\end{equation}
where no summation over $a$ takes place.

As usual, we are interested in the solutions $m_i\ll M_a$. After neglecting the fermion masses $m_i$ in favor of $M_a$ we get the set of equations
\begin{subequations}\label{set_A}
\begin{eqnarray}
m_1 & = & \frac{3h_{*}^2}{64\pi^2}\left( m_2\ln{\frac{M_{12}^4}{m_{2}^4}}+m_3\ln{\frac{M_{45}^4}{m_{3}^4}}+\frac{2}{3}m_1\ln{\frac{M_{3}^3M_{8}}{m_{1}^4}}\right) \,,  \\
m_2 & = & \frac{3h_{*}^2}{64\pi^2}\left( m_1\ln{\frac{M_{12}^4}{m_{1}^4}}+m_3\ln{\frac{M_{67}^4}{m_{3}^4}}+\frac{2}{3}m_2\ln{\frac{M_{3}^3M_{8}}{m_{2}^4}}\right) \,, \\
m_3 & = & \frac{3h_{*}^2}{64\pi^2}\left( m_1\ln{\frac{M_{45}^4}{m_{1}^4}}+m_2\ln{\frac{M_{67}^4}{m_{2}^4}}+\frac{2}{3}m_3\ln{\frac{M_{8}^4}{m_{3}^4}}\right) \,,
\end{eqnarray}
\end{subequations}
where we have introduced $M_{12}=\sqrt{M_1M_2}$, $M_{45}=\sqrt{M_4M_5}$ and $M_{67}=\sqrt{M_6M_7}$. We can observe that the mass parameters $M_a$ stand indeed only under logarithms. It presents an obstacle in achieving the fermion mass hierarchy. We document it by the following example of a solution\footnote{We have fixed parameters $m_i$, $M_3$, $M_8$ and $M_{12}$, and numerically found the values of $M_{45}$, $M_{67}$ and $h_*$ }:
\begin{subequations}\label{result_A}
\begin{eqnarray}
&& \frac{M_8}{m_3}=10^{7}\,,\ \ \frac{m_1}{m_3}=-0.5\,,\ \ \frac{m_2}{m_3}=-0.2 \,,\ \  h_*=7.18 \,, \\
&& \frac{M_{12}}{M_8}=2000 \,,\ \ \frac{M_{3}}{M_8}=2000 \,,\ \ \frac{M_{45}}{M_8}= 0.017\,,\ \ \frac{M_{67}}{M_8}=0.371 \,.
\end{eqnarray}
\end{subequations}
In order to achieve the solution as much interesting as possible, we have enjoyed the freedom of choosing the signs of fermion masses. The signs are not observable as they can be always transformed away by the appropriate redefinition of the fermion fields.

This solution is however not satisfactory because even moderate hierarchy among fermion masses is paid by large hierarchy among $M_a$. Attempts to split further the fermion masses leads to a huge response in the ratios of $M_a$. Some of them quickly tend to be many orders of magnitude less than $M_8$. They even reach the order of $m_3$.

(B) Second, for completeness we present here the equations for the Ansatz (B) in \eqref{pres_hbar}, i.e., in the Wick rotated equation \eqref{Sigma_SDE_Wick_const} we use
\begin{equation}\label{B_ansatz}
\bar h_{ab}^2(q^2)\Delta_{0}(q^2)=\frac{h_{*}^2}{M_{a}^2}\delta_{ab} \,,
\end{equation}
where again no summation over $a$ takes place. This Ansatz however requires the introduction of the cutoff $\Lambda$ in the integrals. The equations for $m_i\ll M_a$ and $m_i\ll \Lambda$ read
\begin{subequations}\label{set_B}
\begin{eqnarray}
m_1 & = & \frac{3h_{*}^2}{64\pi^2}\left( m_2\frac{\Lambda^2}{N_{12}^2}F(m_2^2) + m_3\frac{\Lambda^2}{N_{45}^2}F(m_3^2) + m_1\left[\frac{\Lambda^2}{M_{3}^2}+\frac{\Lambda^2}{3M_{8}^2}\right]F(m_1^2)\right) \,,  \\
m_2 & = & \frac{3h_{*}^2}{64\pi^2}\left( m_1\frac{\Lambda^2}{N_{12}^2}F(m_1^2) + m_3\frac{\Lambda^2}{N_{67}^2}F(m_3^2) + m_2\left[\frac{\Lambda^2}{M_{3}^2}+\frac{\Lambda^2}{3M_{8}^2}\right]F(m_2^2)\right) \,,  \\
m_2 & = & \frac{3h_{*}^2}{64\pi^2}\left( m_1\frac{\Lambda^2}{N_{45}^2}F(m_1^2) + m_2\frac{\Lambda^2}{N_{67}^2}F(m_2^2) + m_3\left[\frac{4\Lambda^2}{3M_{8}^2}\right]F(m_3^2)\right) \,,
\end{eqnarray}
\end{subequations}
where
\begin{equation}
  F(m)\equiv1+\frac{m^2}{\Lambda^2}\ln\frac{\Lambda^2}{m^2} \,
\end{equation}
and
\begin{equation}
  \frac{1}{N_{ab}^2}\equiv \left[\frac{1}{M_{a}^2}+\frac{1}{M_{b}^2}\right] \,.
\end{equation}
Example of one solution of the equations is\footnote{We have fixed parameters $\Lambda$, $m_i$, $M_3$, $M_8$ and $N_{12}$, and numerically found the values of $N_{45}$, $N_{67}$ and $h_*$ }
\begin{subequations}\label{result_B}
\begin{eqnarray}
&& \frac{\Lambda}{M_8}=2\,,\ \ \frac{M_8}{m_3}=10^{8}\,,\ \ \frac{m_1}{m_3}=0.1\,,\ \ \frac{m_2}{m_3}=0.01 \,, \ \ h_*=6.26 \,, \\
&& \frac{N_{12}}{M_8}=6.55 \,,\ \ \frac{M_3}{M_8}=5 \,,\ \ \frac{N_{45}}{M_8}= 3.22\,,\ \ \frac{N_{67}}{M_8}=11.65 \,.
\end{eqnarray}
\end{subequations}
This solution exhibits the desired feature. The big hierarchy among fermion masses $m_3/m_2\sim10^2$ and the huge hierarchy $M_a/m_i\sim10^8$ is parametrized by the mass parameters $M_a$ exhibiting completely moderate hierarchy.

\subsubsection{Near-critical matrix solutions}

The difference between the results \eqref{result_A} and \eqref{result_B} of the two Ans\"{a}tze (A) and (B) can be understood from linearized gap equations.

According to \cite{Nagoshi:1995qw}, the leading texture of the fermion mass matrix is given by linearized gap equations. The idea is that being close to the critical point, the non-linear logarithmic terms on the right-hand sides of the gap equations are negligible with respect to the linear terms. As we approach to the critical point the approximation is getting better. Actually, being close to the critical point is needed in order to achieve the hierarchy $m/M\ll1$.

Already in the degenerate case, we can see the difference of the linear terms on the right-hand sides of the equations for the Ansatz (A) \eqref{gap_eq_A} and for the Ansatz (B) \eqref{NJL_gap_U}. For the Ansatz (B) the linear term depends on $M$, while for the Ansatz (A) does not. The same feature appears in the non-degenerate case for which we can write the linearized \emph{matrix} gap equations for the two Ans\"{a}tze:
\begin{eqnarray}
\mathrm{(A)}\ \ \ \ \ \ \tilde{\Sigma} & = & \frac{h_{*}^2}{h_{c}^2}\sum_a T_a \tilde{\Sigma} T_a \,, \\
\mathrm{(B)}\ \ \ \ \ \ \tilde{\Sigma} & = & \frac{h_{*}^2}{4\pi^2}\sum_a\frac{\Lambda^2}{M_{a}^2}T_a \tilde{\Sigma} T_a \,. \label{lin_B}
\end{eqnarray}
While the gap equation (A) does not allow any other solution than $\tilde{\Sigma}\propto\openone$, the solution of the gap equation (B) depends on $M_a$ and generally it is \emph{not} proportional to $\openone$.

The linearized gap equations do not determine the absolute magnitude of $\tilde{\Sigma}$. This is determined by the non-linear logarithmic terms. Also the departure from those linear solutions is determined by the non-linear logarithmic terms. For the Ansatz (A), the parameters $M_{a}^2$ appear under logarithms in \eqref{set_A}. In order to obtain little more interesting hierarchy among fermion masses the hierarchy of $M_{a}^2$ has to be wild. On the other hand in the case (B) the fermion mass hierarchy can be achieved already at the linearized level of the gap equation \eqref{lin_B} by corresponding choice of $M_{a}^2$ which does not need to be so wild.

\section{Approximation of separable kernel}
\label{separable_kernel}

The approximation of separable kernel is a powerful method which was used already to calculate consequences of the microscopic theory of superconductivity \cite{Bardeen:1957mv}. It is based on modeling the effective charge $\bar h^2\big((k-p)^2\big)$ in the kernel $G\big((k-p)^2\big)$ after the angular integration by a function with separated two participating momenta, $k(k)k(p)$. It allows to factor the dependence on the outer momentum $p$ out of the integral. Thus the Ansatz function $k(p)$ carries the momentum dependence of the fermion self-energy $\Sigma(p^2)\propto k(p)$. The approximation of a separable kernel is unique as it explicitly relates the fermion self-energy and the effective charge and models their momentum dependence by a single Ansatz.

The main aim of this section is to formulate the separable kernel approximation and to set a general formalism. At this level, we rather demonstrate the applicability of the method on the example of the flavor gauge model. Therefore we consider only the equations for $u$- and $d$-type fermions under the simplification of diagonal effective charge and diagonal self-energies. We derive the algebraic equations and present an example of a solution.

\subsubsection{The method}

The momentum dependence of the integrand in the Wick rotated equation \eqref{Sigma_SDE_Wick} takes place just by the combinations $k^2$ and $(p-k)^2=k^2+p^2-2pk\cos\theta$, where $\theta$ is the angle between two Euclidean vectors $k^\mu$ and $p^\mu$. Thus it does not depend on the remaining two angles over which the integration can be performed getting
\begin{equation}
\Sigma(p^2) =  \frac{3}{4\pi^2}\int_{0}^{\infty}\!k^3\d k^2\,K_{ab}(p,k)\,
T_{R,a} \, \Sigma(k^2) \Big[k^2+\Sigma^{\dag}(k^2)\,\Sigma(k^2)\Big]^{-1} T_{L,b} \,,
\end{equation}
where
\begin{equation}
  K_{ab}(p,k)\equiv\frac{1}{\pi}\int^{\pi}_0\bar h^2_{ab}(k^2+p^2-2pk\cos\theta)\Delta_0(k^2+p^2-2pk\cos\theta)\sin^2\theta\,\d\theta.
\end{equation}
The separable kernel approximation is given by the Ansatz
\begin{equation}\label{separable_eff_charge}
K_{ab}(p,k)=k_{ac}(p)k_{cb}(k) \,,
\end{equation}
leading to
\begin{equation}\label{saparable_SDE}
\Sigma(p^2) = \frac{3}{4\pi^2} T_{R,a}k_{ac}(p)\int^{\infty}_{0}\!k^3\d k\,\Sigma(k^2) \Big[k^2+\Sigma^{\dag}(k^2)\Sigma(k^2)\Big]^{-1} k_{cb}(k)T_{L,b} \,.
\end{equation}
Then the momentum-dependent self-energy is expressed in terms of the constant matrix $\mathcal{C}_{cb}$
\begin{equation}\label{Sigma_k_separable}
\Sigma(p^2) = T_{R,a}k_{ac}(p)\mathcal{C}_{cb}T_{L,b} \,,
\end{equation}
for which from \eqref{saparable_SDE} we have the gap equation
\begin{equation}\label{gap_C}
\mathcal{C}_{cb} = \frac{3}{4\pi^2} \int^{\infty}_{0}\!k^3\d k\,\Sigma(k^2) \Big[k^2+\Sigma^{\dag}(k^2)\Sigma(k^2)\Big]^{-1} k_{cb}(k) \,.
\end{equation}
Notice that each element $\mathcal{C}_{cb}$ is at the same time a matrix in the fermion flavor space.

At this point simplifications come. We assume the diagonal form of
\begin{eqnarray}
k_{ab}(p) &=& \delta_{ab}\kappa_b f(p) \,, \label{separable_k_f}
\end{eqnarray}
so that $\mathcal{C}_{cb}$ \eqref{gap_C} simplifies as
\begin{eqnarray}
\mathcal{C}_{cb} &=& C\delta_{cb}\kappa_b \,,  \label{separable_C}
\end{eqnarray}
where the matrix $C$ is the matrix in the fermion flavor space. By this simplification we parametrize the flavor symmetry breaking by 8 real parameters $\kappa_a$. It is motivated by the fact that once $k_{ac}(p)\mathcal{C}_{cb}$ in \eqref{Sigma_k_separable} is a diagonal matrix, the fermion self-energy comes out diagonal as well. The function $f(p)$ represents directly the momentum dependence of the fermion self-energy $\Sigma(p^2)$ through the relations \eqref{Sigma_k_separable} and \eqref{separable_k_f}. Hence we introduce the \emph{constant} matrix $\mathcal{N}$ carrying the matrix structure but not the momentum dependence of $\Sigma(p^2)$ as
\begin{eqnarray}\label{Sigma_Nf}
\Sigma(p^2) &=& \mathcal{N}f(p) \,.
\end{eqnarray}
Inserting \eqref{separable_k_f}, \eqref{separable_C} and \eqref{Sigma_Nf} into the equation \eqref{gap_C} we obtain the gap equation for the matrix $C$
\begin{equation}\label{C_intN}
C = \frac{3}{4\pi^2} \int^{\infty}_{0}\!k^3\d k\,\mathcal{N} \Big[\frac{k^2}{f(k)^2}+\mathcal{N}^{\dag}\mathcal{N}\Big]^{-1} \,,
\end{equation}
where
\begin{eqnarray}
\mathcal{N} &=& \sum_a T_{R,a}\kappa_{a}^2CT_{L,a} \,.
\end{eqnarray}
The matrix $\mathcal{N}$ can be diagonalized by the same bi-unitary transformation as $\Sigma(p^2)$ with the unitary matrices $\mathcal{U}$ and $\mathcal{V}$,
\begin{equation}
n=\mathcal{U}\mathcal{N}\mathcal{V}^\dag\equiv \beginm{ccc} n_1 & & \\ & \ddots &  \\ & & n_d \endm \,,
\end{equation}
where $d$ is the dimension of a given fermion multiplet. The gap equation for $n$ and the mixing matrices $\mathcal{U}$ and $\mathcal{V}$ then follows from \eqref{C_intN}
\begin{equation}\label{xxeqxx}
n = \sum_a \kappa_{a}^2\ \mathcal{U}T_{R,a}\mathcal{U}^\dag\ {\cal F}(n,\Lambda)\ \mathcal{V}T_{L,a}\mathcal{V}^\dag \,,
\end{equation}
where
\begin{equation}
{\cal F}(n,\Lambda)\equiv\beginm{ccc} F(n_1,\Lambda) & & \\ & \ddots &  \\ & & F(n_d,\Lambda) \endm \,.
\end{equation}
The function $F(n_i,\Lambda)$ is the result of a simple integral
\begin{equation}\label{fce_F_nL}
F(n_i,\Lambda) = \frac{3}{4\pi^2} n_i \int^{\Lambda}_{0}\!k^3\d k\,,
\Big[\frac{k^2}{f^2(k)}+n_{i}^2\Big]^{-1} \,
\end{equation}
where we have introduced $\Lambda$ to cutoff the integral in the case that the Ansatz function $f(p)$ is chosen to decrease too slowly. We want to stress, however, that within the asymptotically free theory the cutoff is not necessary as the asymptotically vanishing effective charge $\bar h(q^2)$ makes the integral in the Schwinger--Dyson equations \eqref{Sigma_SDE} perfectly finite. The eventual necessity of the cutoff $\Lambda$ is purely the consequence of the approximation.

For the sake of clarity, let us just briefly summarize the dimensionality of the introduced quantities, suppressing indices:
\begin{equation}
  k(p)\propto[\mathrm{mass}]^{-1}\,,\quad \mathcal{C},C\propto[\mathrm{mass}]^{2}\,,\quad \mathcal{N},n\propto[\mathrm{mass}]^2 \,,\quad \kappa\propto[\mathrm{mass}]^{0}\,,\quad f(p)\propto[\mathrm{mass}]^{-1}\,.
\end{equation}

The physical fermion masses are given by
\begin{equation}\label{n_m}
m_i=\xi_i n_i f(m_i) \,.
\end{equation}
The freedom to choose $\xi_i=\pm1$ follows from the fact that the signs of fermion masses are not observable. In the following we will make use of the property
\begin{equation}
F(\xi_i n_i,\Lambda) = \xi_i F(n_i,\Lambda) \,.
\end{equation}

\subsubsection{Application to the charged fermions}

In the flavor gauge model, all the charged fermions are in the triplet or anti-triplet flavor representation. Therefore $d=3$ and all corresponding matrices introduced above, like $\Sigma(p^2)$, $C$, $\mathcal{N}$, etc., are the $3\times3$ matrices. We adopt further simplification of diagonal self-energies, i.e.,
\begin{equation}
\mathcal{U}=\openone\,,\ \ \ \mathcal{V}=\openone \,.
\end{equation}
In the equation \eqref{xxeqxx}, instead of the numbers $n_i$ of arbitrary sign, we will use $\xi_i n_i$, where $n_i$ are positive from now on.

The equation \eqref{xxeqxx} is then rewritten for $u$-quarks as
\begin{equation}
4\beginm{c}\xi_{u1}n_{u1} \\ \xi_{u2}n_{u2} \\ \xi_{u3}n_{u3}\endm = \beginm{ccc} \kappa_{3}^2+\tfrac{1}{3}\kappa_{8}^2 & \mu_{12}^2 & \mu_{45}^2 \\ \mu_{12}^2 & \kappa_{3}^2+\tfrac{1}{3}\kappa_{8}^2 & \mu_{67}^2 \\ \mu_{45}^2 & \mu_{67}^2 & \tfrac{4}{3}\kappa_{8}^2 \endm\beginm{c}\xi_{ui}F_{ui} \\ \xi_{u2}F_{u2} \\ \xi_{u3}F_{u3} \endm \,
\end{equation}
and for $d$-quarks (and charged leptons) as
\begin{equation}
4\beginm{c}\xi_{d1}n_{d1} \\ \xi_{d2}n_{d2} \\ \xi_{d3}n_{d3}\endm = \beginm{ccc} -\kappa_{3}^2-\tfrac{1}{3}\kappa_{8}^2 & \nu_{12}^2 & \nu_{45}^2 \\ \nu_{12}^2 & -\kappa_{3}^2-\tfrac{1}{3}\kappa_{8}^2 & \nu_{67}^2 \\ \nu_{45}^2 & \nu_{67}^2 & -\tfrac{4}{3}\kappa_{8}^2 \endm\beginm{c}\xi_{di}F_{di} \\ \xi_{d2}F_{d2} \\ \xi_{d3}F_{d3} \endm \,
\end{equation}
%\begin{equation}
%4\xi_{di}n_{di} = \beginm{ccc} 0 & \nu_{12}^2 & \nu_{45}^2 \\ \nu_{12}^2 & 0 & \nu_{67}^2 \\ \nu_{45}^2 & \nu_{67}^2 & 0 \endm\xi_{di}F_{di}
%-\beginm{ccc} \kappa_{3}^2+\tfrac{1}{3}\kappa_{8}^2 & 0 & 0 \\ 0 & \kappa_{3}^2+\tfrac{1}{3}\kappa_{8}^2 & 0 \\ 0 & 0 & \tfrac{4}{3}\kappa_{8}^2 \endm\xi_{di}F_{di}\,,
%\end{equation}
where
\begin{eqnarray}
\mu_{ab}^2 &\equiv& \kappa_{a}^2+\kappa_{b}^2 \,, \\
\nu_{ab}^2 &\equiv& -\kappa_{a}^2+\kappa_{b}^2 \,
\end{eqnarray}
and
\begin{equation}
F_{fi}\equiv F(n_{fi},\Lambda) \,.
\end{equation}
We express the parameters $\mu_{ab}$ and $\nu_{ab}$, i.e., the parameters $\kappa_{a=1,2,4,5,6,7}$, in terms of $n_{fi}$, $\Lambda$, $\kappa_3$, $\kappa_8$ and the mass sign factors $\xi$'s:
\begin{equation}
\beginm{c}\mu_{12}^2 \\ \mu_{45}^2 \\ \mu_{67}^2 \endm =
\frac{1}{6}\beginm{ccc} \hspace{-1mm}\frac{\xi_{u1}\xi_{u2}}{F_{u1}F_{u2}}\hspace{-1mm} & 0 & 0 \\ 0 & \hspace{-1mm}\frac{\xi_{u1}\xi_{u3}}{F_{u1}F_{u3}}\hspace{-1mm} & 0 \\ 0 & 0 & \hspace{-1mm}\frac{\xi_{u2}\xi_{u3}}{F_{u2}F_{u3}}\hspace{-1mm} \endm
\beginm{ccc} F_{u1} & F_{u2} & \hspace{-1mm}-F_{u3}\hspace{-1mm} \\ F_{u1} & \hspace{-1mm}-F_{u2}\hspace{-1mm} & F_{u3} \\ \hspace{-1mm}-F_{u1}\hspace{-1mm} & F_{u2} & F_{u3} \endm
\beginm{c} \hspace{-1mm}12n_{u1}-(3\kappa_{3}^2+\kappa_{8}^2)F_{u1}\hspace{-1mm} \\ \hspace{-1mm}12n_{u2}-(3\kappa_{3}^2+\kappa_{8}^2)F_{u2}\hspace{-1mm} \\ 12n_{u3}-4\kappa_{8}^2F_{u3} \endm
\,,
\end{equation}
and
\begin{equation}
\beginm{c}\nu_{12}^2 \\ \nu_{45}^2 \\ \nu_{67}^2 \endm =
\frac{1}{6}\beginm{ccc} \hspace{-1mm}\frac{\xi_{d1}\xi_{d2}}{F_{d1}F_{d2}}\hspace{-1mm} & 0 & 0 \\ 0 & \hspace{-1mm}\frac{\xi_{d1}\xi_{d3}}{F_{d1}F_{d3}}\hspace{-1mm} & 0 \\ 0 & 0 & \hspace{-1mm}\frac{\xi_{d2}\xi_{d3}}{F_{d2}F_{d3}}\hspace{-1mm} \endm
\beginm{ccc} F_{d1} & F_{d2} & \hspace{-1mm}-F_{d3}\hspace{-1mm} \\ F_{d1} & \hspace{-1mm}-F_{d2}\hspace{-1mm} & F_{d3} \\ \hspace{-1mm}-F_{d1}\hspace{-1mm} & F_{d2} & F_{d3} \endm
\beginm{c} \hspace{-1mm}12n_{d1}+(3\kappa_{3}^2+\kappa_{8}^2)F_{d1}\hspace{-1mm} \\ \hspace{-1mm}12n_{d2}+(3\kappa_{3}^2+\kappa_{8}^2)F_{d2}\hspace{-1mm} \\ 12n_{d3}+4\kappa_{8}^2F_{d3} \endm
\,,
\end{equation}
where the minus signs in the last matrix on the right-hand side apply for $\mu_{ab}$ parameters and the plus signs apply for $\nu_{ab}$ parameters.

Thus we arrive to the expressions for $\mu_{ab}$ parameters
\begin{subequations}\label{eqs_mu}
\begin{eqnarray}
\mu_{12}^2 & = & \frac{\xi_{u1}\xi_{u2}}{6F_{u1}F_{u2}}\Big[
 F_{u1}\left(3n_{u1}-F_{u1}(3\kappa_{3}^2+\kappa_{8}^2)\right)+
 F_{u2}\left(3n_{u2}-F_{u2}(3\kappa_{3}^2+\kappa_{8}^2)\right) \\
 & & \hspace{0.5\textwidth} - F_{u3}\left(3n_{u3}-4F_{u3}\kappa_{8}^2\right)\Big] \,, \nonumber\\
\mu_{45}^2 & = & \frac{\xi_{u1}\xi_{u3}}{6F_{u1}F_{u3}}\Big[
 F_{u1}\left(3n_{u1}-F_{u1}(3\kappa_{3}^2+\kappa_{8}^2)\right)-
 F_{u2}\left(3n_{u2}-F_{u2}(3\kappa_{3}^2+\kappa_{8}^2)\right) \\
 & & \hspace{0.5\textwidth} + F_{u3}\left(3n_{u3}-4F_{u3}\kappa_{8}^2\right)\Big] \,, \nonumber\\
\mu_{67}^2 & = & \frac{\xi_{u2}\xi_{u3}}{6F_{u2}F_{u3}}\Big[-
 F_{u1}\left(3n_{u1}-F_{u1}(3\kappa_{3}^2+\kappa_{8}^2)\right)+
 F_{u2}\left(3n_{u2}-F_{u2}(3\kappa_{3}^2+\kappa_{8}^2)\right) \\
 & & \hspace{0.5\textwidth} + F_{u3}\left(3n_{u3}-4F_{u3}\kappa_{8}^2\right)\Big] \,, \nonumber
\end{eqnarray}
\end{subequations}
and for $\nu_{ab}$ parameters
\begin{subequations}\label{eqs_nu}
\begin{eqnarray}
\nu_{12}^2 & = & \frac{\xi_{d1}\xi_{d2}}{6F_{d1}F_{d2}}\Big[
 F_{d1}\left(3n_{d1}+F_{d1}(3\kappa_{3}^2+\kappa_{8}^2)\right)+
 F_{d2}\left(3n_{d2}+F_{d2}(3\kappa_{3}^2+\kappa_{8}^2)\right) \\
 & & \hspace{0.5\textwidth}- F_{d3}\left(3n_{d3}+4F_{d3}\kappa_{8}^2\right)\Big] \,, \nonumber\\
\nu_{45}^2 & = & \frac{\xi_{d1}\xi_{d3}}{6F_{d1}F_{d3}}\Big[
 F_{d1}\left(3n_{d1}+F_{d1}(3\kappa_{3}^2+\kappa_{8}^2)\right)-
 F_{d2}\left(3n_{d2}+F_{d2}(3\kappa_{3}^2+\kappa_{8}^2)\right) \\
 & & \hspace{0.5\textwidth} + F_{d3}\left(3n_{d3}+4F_{d3}\kappa_{8}^2\right)\Big] \,, \nonumber\\
\nu_{67}^2 & = & \frac{\xi_{d2}\xi_{d3}}{6F_{d2}F_{d3}}\Big[-
 F_{d1}\left(3n_{d1}+F_{d1}(3\kappa_{3}^2+\kappa_{8}^2)\right)+
 F_{d2}\left(3n_{d2}+F_{d2}(3\kappa_{3}^2+\kappa_{8}^2)\right) \\
 & & \hspace{0.5\textwidth} + F_{d3}\left(3n_{d3}+4F_{d3}\kappa_{8}^2\right)\Big] \,. \nonumber
\end{eqnarray}
\end{subequations}

\subsubsection{Numerical results}

In order to obtain results from the equations \eqref{eqs_mu} and \eqref{eqs_nu}, we need to specify the function $F_{fi}\equiv F(n_{fi},\Lambda)$. It is given by the integral \eqref{fce_F_nL}, where we have to adopt some Ansatz for the function $f(p)$ which through \eqref{Sigma_k_separable} and \eqref{separable_k_f} represents the momentum dependence of the self-energy $\Sigma(p^2)$.

%We will compare two Ans\"{a}tze:
%\begin{subequations}\label{Ansatze_separable}
%\begin{eqnarray}
%(I)&\quad\quad& f(p)=\frac{1}{p} \,, \\
%(II)&\quad\quad& f(p)=\frac{1}{p} \,,
%\end{eqnarray}
%\end{subequations}

To demonstrate how the method works, we will use very simple Ansatz
\begin{equation}\label{Ansatz_sep}
  f(p)=\frac{1}{p} \,.
\end{equation}
It yields the integral in \eqref{fce_F_nL} in a simple form
\begin{equation}\label{F_ansatz_I}
 F_{fi}\equiv F(n_{fi},\Lambda)=\frac{3}{8\pi^2}n_{fi}\,\ln\frac{\Lambda^2}{n_{fi}} \,.
\end{equation}
The parameters $n_{fi}$ are related to the fermion masses $m_{fi}$ through the Ansatz \eqref{Ansatz_sep} by the formula \eqref{n_m} leading to
\begin{equation}
  n_i = m_i^2 \,.
\end{equation}
For realistic fermion mass spectrum
\begin{eqnarray}\label{q_spectrum}
m_{u}=2\,\mathrm{MeV}\,, & m_{c}=1.3\,\mathrm{GeV}\,, & m_{t}=172\,\mathrm{GeV} \,, \nonumber\\
m_{d}=5\,\mathrm{MeV}\,, & m_{s}=104\,\mathrm{MeV}\,, & m_{b}=4.2\,\mathrm{GeV} \,,
\end{eqnarray}
and for setting of the parameters
\begin{equation}
  \Lambda=10^8\GeV\,,\quad\kappa_3=200\,,\quad\kappa_8=200\,,
\end{equation}
the example of a solution is
%\begin{subequations}
\begin{align}
  &\kappa_1=1.7\times10^4\,,&&\kappa_2=1.6\times10^4\,,  \\
  &\kappa_4=7.208\times10^6\,,&&\kappa_5=7.207\times10^6\,,  \\
  &\kappa_6=1.04052\times10^9\,,&&\kappa_7=1.04051\times10^9\,.
\end{align}
%\end{subequations}
Notice that the differences $|\kappa_a-\kappa_b|$ in a given row are of the same order of magnitude.

The parameters $\kappa_a$ through the definitions \eqref{separable_k_f} and \eqref{separable_eff_charge} parametrize the kernel $K(p,k)$. Therefore they parametrize the matrix infrared fixed point $h^{*}_{ab}$, which reflects the mass spectrum of the flavor gauge bosons. The huge hierarchy among the $\kappa_a$ parameters, $\frac{\kappa_7}{\kappa_8}\sim10^7$, is not satisfactory as it would mean the analogous hierarchy among flavor gauge boson masses. The hierarchy is actually even bigger than the fermion mass hierarchy $\frac{m_t}{m_u}\sim10^5$.

Partially, the hierarchy of the parameters $\kappa_a$ is an artefact of the Ansatz \eqref{Ansatz_sep}. With the constant Ansatz $f(p)=1/M$, which reproduces the effect of a four-fermion interaction, it is possible to achieve much more moderate hierarchy $\frac{\kappa_7}{\kappa_8}\sim10^3$. This response of the results on the choice of the Ansatz is similar to the situation discussed in the previous section, where we have compared the Ansatz (A) \eqref{A_ansatz} with the Ansatz (B) \eqref{B_ansatz}.

\chapter{The pole vertices for Nambu--Goldstone bosons}
%\pagenumbering{arabic}
\label{pole_vertices}

\label{NG_vertices}

In this appendix we will present formal approach towards calculating vertices for composite Nambu--Goldstone fields, in particular how they couple to their constituent fields. We will follow standard method \cite{Pagels:1979hd,Jackiw:1973tr,Cornwall:1973ts,Cornwall:1974vz} which is based on the Ward--Takahashi identities for the spontaneously broken currents.

In the first section \ref{PagelsStokar} we will derive approximate formula for the vertex function $P_a(p+q,p)$ in the case of spontaneously broken \emph{global} symmetry. We will focus only on the case when the symmetry is spontaneously broken by fermion self-energies $\Sigma(p^2)$. For simplicity, we will work in the approximation of constant self-energies. The key quantity is the bilinear coupling function $\Lambda_{ab}(q^2)$, which converts the Nambu--Goldstone bosons into the spontaneously broken currents $J^{\mu}_a(x)$. The bilinear coupling function $\Lambda_{ab}(q^2)$ is approximated by one-loop expression. The resulting formula agrees with the Pagels--Stokar formula \cite{Pagels:1979hd}.

In the second section \ref{gauge_case} we will consider the case when the spontaneously broken symmetry is gauged. In that case the corresponding gauge bosons acquire masses due to the presence of a massless pole in their polarization function $\Pi_{ab}(q^2)$. The massless pole is visualized as an exchange of the Nambu--Goldstone modes. We demonstrate that the masses of the gauge bosons are given by the bilinear coupling function $\Lambda_{ab}(q^2)$. We will first discuss these features rather generally, keeping in mind the context of the flavor gauge model and the flavor gauge symmetry self-breaking. Later we will correspondingly rephrase the electroweak symmetry breaking in the context of the Standard Model.

%In the last section \ref{flavor_symmetry_breaking}, we will discuss the flavor gauge symmetry self-breaking within the context of the flavor gauge model.

\section{Spontaneously broken global symmetry}

\label{PagelsStokar}

Once a continuous global symmetry is spontaneously broken, there is a corresponding massless Nambu--Goldstone boson field $\pi(x)$ coupled to the broken current $J^\mu(x)$ \cite{Goldstone:1961eq,Goldstone:1962es},
\begin{equation}
\bra{0}J^\mu(0)\ket{\pi(q)}=\im q^\nu F_\pi \,,
\end{equation}
where $F_\pi$ is the Nambu--Goldstone boson decay constant.

We assume that the symmetry is broken dynamically by formation of symmetry breaking self-energies for elementary fermion fields. Therefore the Nambu--Goldstone boson is a composite of the elementary fermion fields and as such it couples to them. In the original Lagrangian however there is no elementary field corresponding to the Nambu--Goldstone boson, let alone its interaction terms with elementary fields. At the level of the effective Lagrangian the Nambu--Goldstone field and its interaction terms can be introduced just by the symmetry arguments. Namely, because of the invariance under the Nambu--Goldstone field shift $\pi'(x)=\pi(x)+\alpha$, the most general interaction term of a single Nambu--Goldstone field $\pi(x)$ with two Dirac fermions $\psi_1(x)$ and $\psi_2(x)$ is given as a derivative coupling \cite{Peccei:1998jv}
\begin{eqnarray}
\mathcal{L}_\mathrm{NG}^\mathrm{fermion} & = & \im\frac{\partial_\mu\pi}{F_\pi}\bar{\psi}_1\big[a\gamma^\mu+b\gamma^\mu\gamma_5\big]\psi_2 + \hc  \nonumber\\
                                          & = & \frac{\pi}{F_\pi}\bar{\psi}_1\big[a(m_1-m_2)+b(m_1+m_2)\gamma_5\big]\psi_2 + \hc \,. \label{effective_NG_fermion_L}
\end{eqnarray}
where $m_1$ and $m_2$ are Dirac masses of the fermions $\psi_1$ and $\psi_2$, and $a$ and $b$ are the effective coupling parameters. The second row is obtained from the first row by using the fermion equations of motion. Notice that if the spontaneously broken symmetry mixes the fermion flavors then the corresponding Nambu--Goldstone boson does not need to be an eigenstate of parity.

In this section we will focus on a derivation of the coupling \eqref{effective_NG_fermion_L} from the first principles by means of a fundamental quantity, the fermion self-energy $\Sigma(p^2)$.

\subsubsection{Fermion self-energy}

We will treat both the charged fermion and the neutrino case on equal footing.

Let $J_{a,\psi}^{\mu}(x)$ be a current of the spontaneously broken symmetry acting on the fermion field multiplet $\psi(x)$, where $\psi=f$ for the case of charged fermions and $\psi=n$ for the case of neutrinos. For simplicity we assume the case of a simple symmetry group which is completely broken. The index $a$ labels the broken generators $T_{L}^{a}$ and $T_{R}^{a}$ which are, in general, different for left-handed and right-handed fermion fields. By the same index we will denote the set of Nambu--Goldstone fields $\pi_a(x)$.

The fermion field can be written in terms of its chiral components as
\begin{subequations}
\begin{eqnarray}
f & = & f_L+f_R \,, \\
n & = & \beginm{c} \nu_L+(\nu_{L})^\C \\ \nu_{R}+(\nu_{R})^\C \endm \,.
\end{eqnarray}
\end{subequations}
The neutrino field $n$ in the Nambu--Gorkov form satisfies $n=n^\C$. The spontaneously broken current is written for the two cases as
\begin{subequations}
\begin{eqnarray}
J_{a,\psi=f}^{\mu} & = & \bar{f}t_a\gamma^\mu f \,,\\
J_{a,\psi=n}^{\mu} & = & \frac{1}{2}\bar{n}t_a\gamma^\mu n \,,
\end{eqnarray}
\end{subequations}
with generators
\begin{subequations}\label{generators_ta}
\begin{eqnarray}
f:\ \ t^{a} & = & T_{L}^{a}P_L+T_{R}^{a}P_R \,, \label{Dirac_generator}\\
n:\ \ t^{a} & = & \beginm{cc} \left(T_{L}^{a}P_L-[T_{L}^{a}]^\T P_R\right) & 0 \\
                0 & \left(T_{R}^{a}P_R-[T_{R}^{a}]^\mathrm{T}P_L\right) \endm \,,
\end{eqnarray}
\end{subequations}
where $P_{L,R}$ are the chiral projectors. By \eqref{generators_ta} we encounter the possibility that the symmetry current is non-vector-like, i.e., the left- and right-handed fermions transform as different representations of the symmetry group with the corresponding generators $T_{L,R}^{a}$. %We keep the index $a$ for the sake of generality.

The full fermion propagator is
\begin{equation}\label{fermion_propagator}
S(p) \equiv \langle\psi\bar\psi\rangle = [\slashed{p}-\mathbf{\Sigma}(p^2)]^{-1} \,,
\end{equation}
where we omit the wave function renormalization. The symmetry breaking self-energy $\mathbf{\Sigma}(p^2)$ is in general a complex matrix which can be written as \cite{Benes:2009iz}
\begin{equation}\label{Sigma_obecne}
\mathbf{\Sigma}(p^2)=\Sigma(p^2)P_L+\Sigma^\dag(p^2)P_R \,.
\end{equation}
We can rewrite the propagator more explicitly in terms of $\Sigma(p^2)$, suppressing the momentum arguments
\begin{equation}
S(p) = (\slashed{p}+\Sigma^\dag)(p^2-\Sigma\Sigma^\dag)^{-1}P_L+(\slashed{p}+\Sigma)(p^2-\Sigma^\dag\Sigma)^{-1}P_R \,.
\end{equation}
The self-energy for neutrinos in the Nambu--Gorkov formalism is
\begin{equation}
\Sigma_n=\beginm{cc}
    \Sigma_L & \Sigma_D \\ \Sigma_{D}^\T & \Sigma_R \endm \,.
\end{equation}

\subsubsection{Fermion pole vertex}

When deriving the vertex of composite Nambu--Goldstone bosons with the constituent fermions we start with the Green's function which connects the broken current $J_{a,\psi}^{\alpha}(x)$ with fermion fields, i.e.,
\begin{eqnarray}
G^{\alpha}_a(x,y,z) & \equiv & \bra{0}TJ^{\alpha}_{a,\psi}(x)\psi(y)\bar{\psi}(z)\ket{0} \,,
\end{eqnarray}
which is rewritten in the momentum representation, and in terms of the proper vertex $\Gamma^{\alpha}_{a}$, as
\begin{eqnarray}
G^{\alpha}_{a}(p+q,p) & = & \im S(p+q)\Gamma^{\alpha}_{a}(p+q,p)\im S(p) \,.
\end{eqnarray}
The proper vertex satisfies the global Ward--Takahashi identity
\begin{eqnarray}\label{WTI_fermion}
q_\alpha\Gamma^{\alpha}_a(p+q,p) & = & S^{-1}(p+q)t_a-\bar t_a S^{-1}(p) \,,
\end{eqnarray}
where $\bar t_a \equiv \gamma_0 t_a \gamma_0$. Once the dynamics develops a symmetry breaking fermion self-energy then
\begin{eqnarray}\label{pseudo_commutator}
\mathbf{\Sigma}(p^2)t_a - \bar t_a \mathbf{\Sigma}(p^2) &\neq& 0
\end{eqnarray}
and the right-hand side of the Ward--Takahashi identity \eqref{WTI_fermion} does not vanish in the limit $q^\mu\rightarrow0$,
\begin{eqnarray}\label{WT_limit}
q_\alpha\Gamma^{\alpha}_a(p+q,p) &\stackrel{q\rightarrow0}{\ne}& 0 \,.
\end{eqnarray}

The Ward--Takahashi identity is the consequence of the symmetry of the Lagrangian and not of the symmetry of the ground state. Therefore it should remain valid also when the symmetry is spontaneously broken. From the non-vanishing limit \eqref{WT_limit} it follows that the proper vertex has to develop a pole, i.e., the leading ${\cal O}(q^{-1})$ part $\Gamma^{\alpha}_{a,\mathrm{lead.}}$ of the proper vertex
\begin{equation}
\Gamma^{\alpha}_a(p+q,p)=\Gamma^{\alpha}_{a,\mathrm{lead.}}(p+q,p)+{\cal O}(q^{0}) \,
\end{equation}
does not vanish and from \eqref{WTI_fermion}
\begin{equation}\label{Gamma_pole}
\Gamma^{\alpha}_{a,\mathrm{lead.}}(p+q,p)=-\frac{q^\alpha}{q^2}\left(\mathbf{\Sigma}(p^2)t_a-\bar t_a\mathbf{\Sigma}(p^2)\right) \,.
\end{equation}

The pole is interpreted in terms of the exchange of the massless Nambu--Goldstone bosons. Following this interpretation we pick up the Nambu--Goldstone part of the proper vertex
\begin{equation}
\Gamma^{\alpha}_{a}(p+q,p)=\Gamma^{\alpha}_{a,{\mathrm{NG}}}(p+q,p)+\ldots
\end{equation}
and approximate it by the expression
\begin{eqnarray}\label{Gamma_NG}
\Gamma^{\alpha}_{a,{\mathrm{NG}}}(p+q,p) & \equiv & \begin{array}{c}\includegraphics[width=0.3\textwidth]{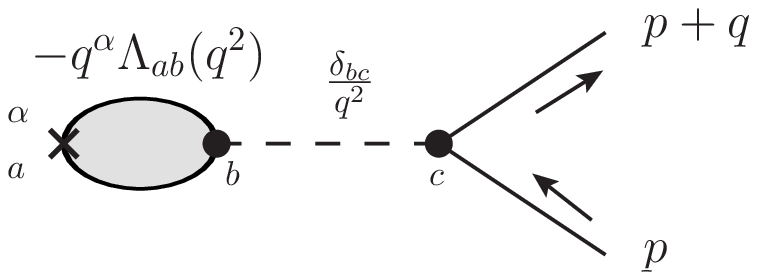}\end{array}  \nonumber\\
                   & \equiv & -\frac{q^\alpha}{q^2}\Lambda_{ab}(q^2)P_b(p+q,p) \,,
\end{eqnarray}
where the massless propagator $\tfrac{\delta_{bc}}{q^2}$ connects the function $-q^\mu \Lambda_{ab}(q^2)$ and the Nambu--Goldstone vertex $P_c(p+q,p)$ both \emph{regular} for $q=0$. Comparing the two expressions \eqref{Gamma_pole} and \eqref{Gamma_NG} for $q\rightarrow0$ we get the expression for the Nambu--Goldstone vertex for $q=0$
\begin{equation}\label{P}
P_a(p,p)=\Lambda^{-1}_{ab}(0)\big[\mathbf{\Sigma}(p^2)t_b-\bar t_b\mathbf{\Sigma}(p^2)\big] \,.
\end{equation}
We use here our simplifying assumption that the symmetry is simple and completely broken. In that case, $\Lambda_{ab}(0)$ is a square non-singular matrix and its inverse exists. The general case is operationally more involved but conceptually the same.

It is instructive to see the structure of the right-hand side of \eqref{P} for Dirac fermion fields, i.e., with the generators in the form of \eqref{Dirac_generator}. Using \eqref{Sigma_obecne} we get, suppressing the momentum arguments,
\begin{equation}
\mathbf{\Sigma}t^a - \bar t^a \mathbf{\Sigma}=\Sigma T^{a}_L P_L + \Sigma^\dag T^{a}_R P_R - T^{a}_R \Sigma P_L - T^{a}_L \Sigma^\dag P_R \,.
\end{equation}
For the \emph{axial} Abelian symmetry, when $T_L=-T_R=+\tfrac{1}{2}$, it reads
\begin{equation}
  \Sigma P_L - \Sigma^\dag P_R = \frac{1}{2}\big[(\Sigma - \Sigma^\dag)\openone-(\Sigma + \Sigma^\dag)\gamma_5\big] \,,
\end{equation}
what is of mixed parity due to the complex and matrix nature of $\Sigma$. Comparing it with \eqref{effective_NG_fermion_L} we can see that \eqref{P} is just what underlies the effective interaction term.

The calculation is completed by expressing the matrix $\Lambda_{ab}(0)$ in terms of $\Sigma(p^2)$. It can be approximated by a one-loop expressions $I_{ab}$ of all fields coupled to the broken current,
\begin{equation}\label{Lambda_bilinear}
\Lambda_{ab}(0)\approx\sum_{\mathrm{fields}}I^{\mathrm{field}}_{ab} \,.
\end{equation}
The fermion contribution $I^{\psi}_{ab}$ is expressed by the loop diagram
\begin{eqnarray}\label{I_def}
I^{\psi}_{ab} & = & \lim_{q\rightarrow0}\ \frac{q_\alpha}{q^2}\begin{array}{c}\includegraphics[width=0.15\textwidth]{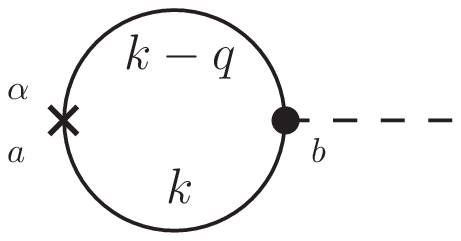}\end{array}  \nonumber\\
& = & -\im\lim_{q\rightarrow0}\int\hspace{-2mm}\frac{\d^4k}{(2\pi)^4}\Tr\left(\frac{\slashed{q}}{q^2}t_a S(k-q)P_b(k-q,k)S(k)\right) \,.
\end{eqnarray}

In order to be able to perform the limit in \eqref{I_def}, we assume the constant self-energy approximation and expand the $q$-dependent quantities up to ${\cal O}(q^1)$. The expansion of the fermion propagator is
\begin{equation}
\tilde{S}(k-q)=\tilde{S}(k)+\tilde{S}(k)\slashed q \tilde{S}(k)+{\cal O}(q^2) \,,
\end{equation}
and we \emph{assume} that the expansion for the Nambu--Goldstone vertex is
\begin{equation}
\tilde{P}_b(k-q,k)=\tilde{P}_b(k,k)+{\cal O}(q^2) \,,
\end{equation}
where the tilde means the constant self-energy approximation of the quantity. Finally, using the expression \eqref{P} for $\tilde{P}_b(k,k)$ we obtain the relation
\begin{eqnarray}\label{I_relation}
I^{\psi}_{ac}I^{\psi}_{bc} & = & -\im\lim_{q\rightarrow0}\int\hspace{-2mm}\frac{\d^4k}{(2\pi)^4}\Tr\left(\frac{\slashed{q}}{q^2}t_a\tilde{S}(k)\slashed q \tilde{S}(k)\big[\tilde{\mathbf{\Sigma}}t_b-\bar t_b\tilde{\mathbf{\Sigma}}\big]\tilde{S}(k)\right) \,.
\end{eqnarray}
This is a pretty general result. Its complexity is given by the complex-valued matrix structure of the fermion self-energy \eqref{Sigma_obecne}, for which it holds in general that $\tilde{\Sigma}\ne\tilde{\Sigma}^\dag$. Further, the complexity of the formula is carried by the general setting of the generators \eqref{generators_ta}.

To see how the formula works we will express it for a special case when the self-energy is real-valued scalar, i.e., $\tilde{\Sigma}=\tilde{\Sigma}^\dag$, for a single Dirac fermion. If the corresponding symmetry is \emph{vector-like}, i.e.,
\begin{equation}
T_{L}=\frac{1}{2}\quad\mathrm{and}\quad T_{R}=\frac{1}{2} \,,
\end{equation}
then
\begin{equation}
  \big[\tilde{\mathbf{\Sigma}}t-\bar t\tilde{\mathbf{\Sigma}}\big]\ =\ \tilde{\Sigma}\frac{1}{2}-\frac{1}{2}\tilde{\Sigma}\ =\ 0
\end{equation}
and the formula \eqref{I_relation} vanishes identically, $I^2=0$.

If the corresponding symmetry is \emph{axial}, i.e.,
\begin{equation}
T_{L}=\frac{1}{2}\quad\mathrm{and}\quad T_{R}=-\frac{1}{2} \,,
\end{equation}
then
\begin{equation}
  \big[\tilde{\mathbf{\Sigma}}t-\bar t\tilde{\mathbf{\Sigma}}\big]\ =\ -\tilde{\Sigma}\frac{1}{2}\gamma_5-\frac{1}{2}\gamma_5\tilde{\Sigma}\ =\ -\gamma_5\tilde{\Sigma} \,.
\end{equation}
The formula \eqref{I_relation} gets simplified to
\begin{eqnarray}
I^2 & = & -\im\int\hspace{-2mm}\frac{\d^4k}{(2\pi)^4}\Tr\left(\frac{\slashed{q}}{q^2}\big[-\tfrac{1}{2}\gamma_5\big]\frac{\slashed{k}+\tilde{\Sigma}}{k^2-\tilde{\Sigma}^2}\slashed{q}\frac{\slashed{k}+\tilde{\Sigma}}{k^2-\tilde{\Sigma}^2}\big[-\gamma_5\tilde{\Sigma}\big]\frac{\slashed{k}+\tilde{\Sigma}}{k^2-\tilde{\Sigma}^2}\right) \\
 & = & - 2\im\int\hspace{-2mm}\frac{\d^4k}{(2\pi)^4}\frac{\tilde{\Sigma}^2}{(k^2-\tilde{\Sigma}^2)^2} \label{simplified_PS}\,.
\end{eqnarray}
After the Wick rotation we obtain the formula
\begin{eqnarray}\label{simplified_Wick_PS}
I^2 & = &  2\int\frac{\d^4k}{(2\pi)^4}\frac{\tilde{\Sigma}^2}{(k^2+\tilde{\Sigma}^2)^2}=\frac{1}{8\pi^2}\int_{0}^{\Lambda^2} x\d x\frac{\tilde{\Sigma}^2}{(x+\tilde{\Sigma}^2)^2} \,,
\end{eqnarray}
where $\Lambda$ regularizes the integral compensating the fact that we have approximated $\Sigma(p^2)$ by a constant.

We use the formula \eqref{simplified_Wick_PS} even in the cases when the fermion self-energy is not a constant function, but it has a mild momentum dependence. We just replace $\tilde{\Sigma}$ by $\Sigma(p^2)$ in \eqref{simplified_Wick_PS}. In comparison with the original Pagels--Stokar formula \cite{Pagels:1979hd} or \cite{Benes:2012hz}, we are omitting the derivative term $\propto k^2\tfrac{\d}{\d k^2}\Sigma^{2}(k^2)$ in the integrand.

\section{Spontaneously broken gauge symmetry}

\label{gauge_case}

%In order to spontaneously break the gauge symmetry, it has to be \emph{chiral}.
Once the spontaneously broken symmetry is gauged, the observable outcomes changes significantly compared to the global symmetry case described in the previous section \ref{PagelsStokar}. The underlying principles and consequent formal apparatus however are the same.

Also in the gauge case the Nambu--Goldstone modes appear as a consequence of the Goldstone theorem \cite{Goldstone:1961eq,Goldstone:1962es}. Consequently there arises the bilinear coupling $-q^\mu \Lambda_{ab}(q^2)$ which converts the Nambu--Goldstone modes into the spontaneously broken currents. But now the spontaneously broken currents are gauged. Therefore the bilinear coupling, containing the corresponding gauge coupling constant, converts the Nambu--Goldstone modes into the gauge bosons. The resulting mixing of gauge boson fields with Nambu--Goldstone fields triggers the \emph{Schwinger mechanism} \cite{Schwinger:1962tp}, which is in the heart of the gauge boson mass generation, alias the \emph{Anderson--Higgs mechanism} \cite{Anderson:1963pc,Higgs:1964pj,Guralnik:1964eu,Englert:1964et,Migdal:1965aa,'tHooft:1971rn}.

%The chiral gauge theories do not allow any mass terms in the Lagrangian. They essentially deal with three types of massless fields whose existence is well motivated by field theoretical principles. These are a two-component Weyl spinor field being a fundamental representation of Lorentz group, a gauge vector field arising from a gauge principle that guarantees the renormalizability of a theory, and a Nambu--Goldstone field which is a consequence of the Goldstone theorem. All of them are massless fields, i.e., their Lagrangian misses mass terms. Their masslessness is always connected to the existence of a symmetry. While mass terms for Weyl fermions are forbidden by \emph{chiral} nature of the chiral gauge symmetry, mass terms for gauge bosons are forbidden also by the \emph{gauge} nature of the chiral gauge symmetry. Finally mass terms for Nambu--Goldstone fields are forbidden by the invariance with respect to the field shift in the direction of a vacuum degeneracy.
%
%It is of key importance for successful description of Nature, that massless fields can excite massive particles. It happens when their full propagators loose their massless pole and develop a massive pole.

%It is of key importance for successful description of Nature, that massless fields can excite massive particles.
In the rest of this appendix we discuss the gauge symmetry breaking. Finally, we apply these general considerations to the case of the electroweak gauge symmetry breaking and electroweak gauge boson mass generation within the Standard Model.

\subsubsection{Gauge invariance and massless gauge bosons}

The gauge invariance protects the masslessness of gauge boson \emph{fields}. Whether the masslessness is a property of the gauge boson \emph{particles} depends on dynamics and particular realization of the gauge symmetry.

A general full gauge boson propagator is
\begin{equation}
\Delta^{ab}_{\mu\nu}(q)=-\left[P_{\mu\nu}(q^2)\Delta^{ab}(q^2)+\xi\frac{q_\mu q_\nu}{(q^2+\im\epsilon)^2}\right]
\end{equation}
where $\xi^{ab}$ are gauge fixing parameters, the transverse projector is
\begin{equation}
P^{\mu\nu}(q)\equiv g^{\mu\nu}-\frac{q^\mu q^\nu}{q^2}\,,
\end{equation}
and the propagator function $\Delta^{ab}(q^2)$ is given by
\begin{equation}\label{propagator_function}
\Delta^{ab}(q^2)=\frac{1}{q^2+\im\epsilon}\left[(\openone+\Pi(q^2))^{-1}\right]^{ab} \ .
\end{equation}
The polarization tensor is the gauge boson proper self-energy defined as
\begin{equation}
\Pi^{ab}_{\mu\nu}(q)=[\Delta^{-1}(q)]^{ab}_{\mu\nu}-[\Delta^{-1}_0(q)]^{ab}_{\mu\nu} \,.
\end{equation}
Its transversality follows from the exact Ward identity
\begin{equation}
q_\mu\Pi_{ab}^{\mu\nu}(q)=0 \,,
\end{equation}
therefore it can be rewritten in terms of a Lorentz-scalar polarization function $\Pi^{ab}(q^2)$
\begin{equation}\label{pol_func}
\Pi^{ab}_{\mu\nu}(q)=-(g_{\mu\nu}q^2-q_\mu q_\nu)\Pi^{ab}(q^2) \,.
\end{equation}
The factor $\frac{1}{q^2+\im\epsilon}$ in the propagator \eqref{propagator_function} represents the massless pole which is responsible for the fact that the free gauge boson fields, i.e., when $\Pi^{ab}(q^2)=0$, excites massless particles. In the case of interacting gauge field, as far as the polarization function $\Pi^{ab}(q^2)$ is regular for $q^2=0$, like in the QED in four dimensions\footnote{ In the massless QED in four dimensions, $\Pi^{ab}(q^2)$ has a logarithmic singularity in origin. But it is not enough to overcome the $\frac{1}{q^2+\im\epsilon}$ and the photon still remains massless. }, the massless pole survives the quantum corrections. In this case the gauge invariance protects the masslessness of a gauge boson particles.

\subsubsection{Schwinger mechanism}

The dynamics can however generate a massless pole in the polarization function
\begin{equation}\label{pole_pol_func}
\Pi_{ab}(q^2)=-\frac{M^{2}_{ab}(q^2)}{q^2} \,,
\end{equation}
with $M^{2}_{ab}(q^2)$ regular in $q^2=0$. Then the massless pole in the propagator is canceled and instead the massive poles are generated as solutions of the equation
\begin{eqnarray}\label{mass_equation}
\det\big[q^2-M^2(q^2)\big] &=& 0 \,.
\end{eqnarray}
This is the Schwinger mechanism identified for the first time within the Schwinger's model of two-dimensional QED \cite{Schwinger:1962tp}. In the Schwinger's model the pole in the polarization function appears due to the axial anomaly which is the feature already of two-point functions in two dimensions, not of three-point functions like in four dimensions.

In four dimensions, the origin of the poles in the polarization function is different. There are approaches \cite{Ibanez:2012zk} which deal with a \emph{symmetry preserving pole}, $M^{2}_{ab}(0)=M^2\delta_{ab}$, which is developed purely on the non-perturbative basis because of the gauge dynamics itself gets strong. Standard approaches however assume the \emph{symmetry violating pole} $M^{2}_{ab}\equiv M^{2}_{ab}(0)\ne M^2\delta_{ab}$, which arises because of the presence of the Nambu--Goldstone modes due to the spontaneous gauge symmetry breaking. It is called the Anderson--Higgs mechanism \cite{Anderson:1963pc,Higgs:1964pj,Guralnik:1964eu,Englert:1964et,Migdal:1965aa,'tHooft:1971rn} and it is relevant for the electroweak and flavor gauge boson mass generation. We dedicate the rest of the section to describe it.

\subsection{Spontaneous gauge symmetry breaking}

\label{flavor_symmetry_breaking}

%Whenever a theory contains massless fields there is some symmetry connected to it. Whenever a lagrangian does not exhibit the corresponding symmetry there is no principle that would prevent the quantum corrections from generating mass operators which enter the lagrangian accompanied by mass parameters even if they were originally missing, take the Coleman--Weinberg model \cite{Coleman:1973jx} as an example.

%We find the most attractive and useful the combined scenario.
%The flavor gauge model and also some of the Extended-Technicolor models are built as strongly coupled asymptotically free chiral gauge theories without any elementary scalars. They provide an origin of the pole in the gauge boson polarization function, which is a combination of the two origins described above. When the dynamics gets strongly coupled it generates the massless pole in the analogy with the generation of the symmetry preserving pole. But because of the chiral setting of a symmetry the pole is symmetry violating. Such gauge dynamics is referred to as tumbling or \emph{self-breaking}. Apparently, because there are no elementary scalars in the theory, the pole corresponds to Nambu--Goldstone modes which are \emph{composites} of elementary fields, of the fermions and the gauge bosons themselves.

%The realization of this scenario is possible only if the symmetry setting of the gauge dynamics is non-vector-like, it has to be \emph{chiral}, for what the presence of fermions is instrumental.

Visualization of the massless poles of the composite Nambu--Goldstone modes is based on the analysis of the Slavnov--Taylor identities of underlying spontaneously broken gauge symmetries \cite{Eichten:1974et}. Their treatment is however very complicated, yet they provide the same argumentation as the Ward--Takahashi identities. There are even gauge fixing schemes within the Pinch technique, which consistently reduce the Slavnov--Taylor identities to the Ward--Takahashi identities \cite{Binosi:2009qm}. Therefore, in our considerations we use the Ward--Takahashi identities and follow the same reasoning described in the previous section \ref{PagelsStokar}.

In the gauge case, the three-point proper vertex $\Gamma^\mu_a(p+q,p)$ corresponds to a Green's function containing the gauge field with indices $\mu$ and $a$ instead of the global symmetry current $J^{\mu}_a$. For example, in the flavor gauge model, there are two types of three-point proper vertices relevant for the flavor gauge symmetry breaking. They are the flavor-gauge-boson-fermion proper vertex corresponding to the Green's function $\bra{0}TC^{\mu}_{a}(x)\psi(y)\bar{\psi}(z)\ket{0}$ and three-flavor-gauge-boson proper vertex corresponding to the Green's function $\bra{0}TC^{\mu}_{a}(x)C^{\alpha}_{b}(y)C^{\beta}_{c}(z)\ket{0}$.

If the propagators of the elementary fields develop the symmetry breaking parts then the right-hand sides of the Ward--Takahashi identities for three-point proper vertices $\Gamma^\mu_a(p+q,p)$ do not vanish in the limit $q\rightarrow0$
\begin{eqnarray}\label{non_zero_WTI}
q_\mu\Gamma^\mu_a(p+q,p) &\stackrel{q\rightarrow0}{\ne}& 0 \,.
\end{eqnarray}
It is in a direct analogy with the global symmetry case \eqref{WTI_fermion}.

\begin{figure}[t]
\begin{center}
\begin{tabular}{cc}
  \includegraphics[width=0.4\textwidth]{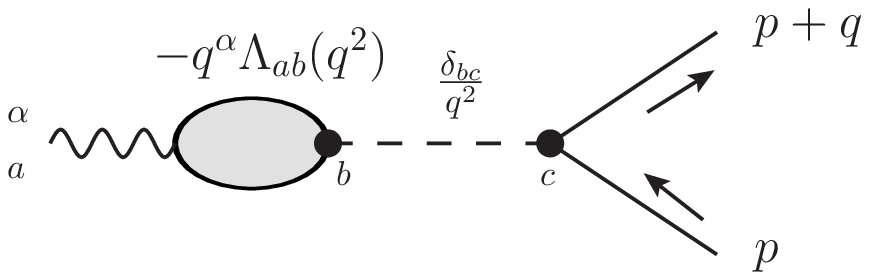} & \includegraphics[width=0.4\textwidth]{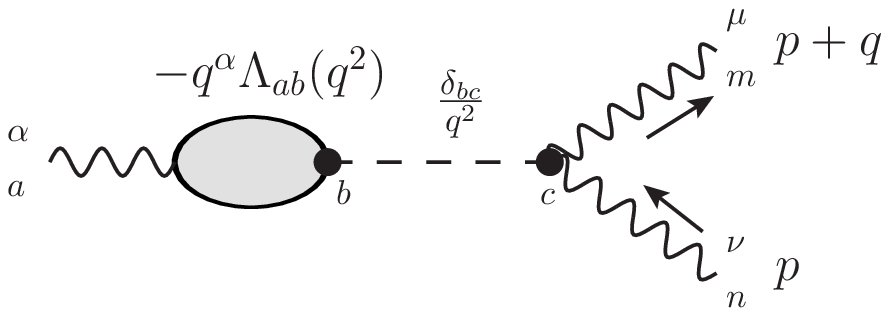} \\
  a) & b)
\end{tabular}
\caption[Nambu--Goldstone pole vertices ]{\small Diagrammatical expression of \eqref{eq_Gamma:rel:P}, i.e., the Nambu--Goldstone mode contribution to the proper vertex $\Gamma^\mu_a(p+q,p)$ for the flavor gauge model: a) the flavor-gauge-boson-fermion vertex and b) three-flavor-gauge-boson vertex. The exchange of the massless Nambu--Goldstone modes gives rise to the pole by means of the propagator $\tfrac{\delta_{bc}}{q^2}$. The Nambu--Goldstone modes couple to the fermions and flavor gauge bosons via the effective Nambu--Goldstone vertex $P_b(p+q,p)$. The bilinear coupling $-q^\mu\Lambda_{ab}(q^2)$ converts the flavor gauge bosons to the Nambu--Goldstone modes. }
\label{fig:pole}
\end{center}
\end{figure}

From the non-vanishing limit \eqref{non_zero_WTI} it follows that the corresponding proper vertex has to develop a pole. Physically, the pole of the form $q^\mu/q^2$ is interpretable as the Nambu--Goldstone field exchange, see Fig.~\ref{fig:pole}
\begin{eqnarray}\label{eq_Gamma:rel:P}
\Gamma^\mu_a(p+q,p)|_{\mathrm{NG}} &=& -q^\mu\Lambda_{ab}(q^2)\frac{1}{q^2}P_b(p+q,p) \,, \qquad
\end{eqnarray}
where $P_b(p+q,p)$ is a Nambu--Goldstone vertex function \emph{regular} in $q^\mu=0$. It just tells us that the Nambu--Goldstone modes are the composites of corresponding elementary fields. Once we know the Nambu--Goldstone vertex $P_b(p+q,p)$ we can in principle reconstruct the bilinear coupling $q^\mu\Lambda_{ab}(q^2)$ \emph{regular} in $q^2=0$, which converts the gauge bosons to the Nambu--Goldstone modes.

\begin{figure}[t]
\begin{center}
  \includegraphics[width=0.5\textwidth]{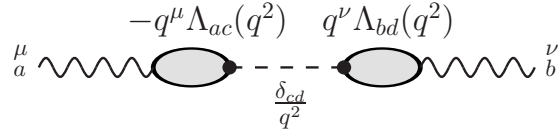}
\caption[Longitudinal part of pole part of polarization tensor]{ Diagrammatical expression of the longitudinal part of the pole part of the polarization tensor $\Pi^{\mu\nu}_{ab}(q)|_{\mathrm{long.}}$. }
\label{fig:pol_tens}
\end{center}
\end{figure}

Using the bilinear coupling we can reconstruct the longitudinal pole part of the polarization tensor as depicted in Fig.~\ref{fig:pol_tens}
\begin{eqnarray}
\Pi^{\mu\nu}_{ab}(q)|_{\mathrm{long.}} &\stackrel{q^2\rightarrow0}{=}&
-q^\mu\Lambda_{ac}(0)\,\frac{\delta_{cd}}{q^2}\,q^\nu\Lambda_{bd}(0)
\,.
\label{longPiNGpole}
\end{eqnarray}
There is however no doubt that the polarization tensor has to come out transverse. Therefore we can use the knowledge of the longitudinal part to write completely transverse pole part of the polarisation tensor,
\begin{eqnarray}
\Pi^{\mu\nu}_{ab}(q) &\stackrel{q^2\rightarrow0}{=}& \left(g^{\mu\nu}-\frac{q^\mu q^\nu}{q^2}\right)\Lambda_{ac}(0)\Lambda_{bc}(0)
\,.
\end{eqnarray}
Comparing this result with \eqref{pol_func} we can identify the pole polarization function
\begin{eqnarray}
\Pi_{ab}(q^2) &\stackrel{q^2\rightarrow0}{=}& -\frac{\Lambda_{ac}(0)\Lambda_{bc}(0)}{q^2}
\,,
\end{eqnarray}
where the residuum $\Lambda_{ac}(0)\Lambda_{bc}(0)$ gives us the gauge boson mass matrix \eqref{pole_pol_func}
\begin{eqnarray}
M^{2}_{ab} &=& \Lambda_{ac}(0)\Lambda_{bc}(0) \,.
\end{eqnarray}

If the gauge coupling is weak then the one-loop expression, \eqref{Lambda_bilinear}, is a good approximation for the bilinear coupling function $\Lambda_{ab}(q^2)$. The detailed analysis can be found in \cite{Benes:2012hz}. This approach, perturbative in the gauge coupling constant, however fails when the gauge dynamics is strongly coupled, i.e., it cannot be applied directly to the flavor gauge model. More sophisticated approach has to be developed. The Pinch technique \cite{Binosi:2009qm} seems to be a promising tool. It is however beyond the scope of this thesis, therefore we stay on the level of general considerations.

\subsection{The case of the Standard Model}
\label{EWSB_SM}

In this subsection, we reformulate the electroweak symmetry breaking by means of the visualization of Nambu--Goldstone modes described in the previous subsection~\ref{flavor_symmetry_breaking}. For that we work in the Landau gauge, $\xi=0$, in which the Nambu--Goldston propagators are massless.

\subsubsection{The electroweak symmetry breaking by the Higgs field vacuum expectation value}

The Lagrangian of the Higgs gauge sector of the Standard Model is
\begin{eqnarray}\label{Higgs_Lagrangian}
{\cal L}^\mathrm{SM}_{\mathrm{Higgs-gauge}} & = & (D_\mu\Phi)^\dag (D^\mu\Phi)-{\cal V}(\Phi^\dag\Phi) \,,
\end{eqnarray}
where the covariant derivative for the complex scalar Higgs field $\Phi(x)$, which is an $\SU{2}_L$ doublet with weak hypercharge the $Y_\Phi=+1$, is
\begin{equation}
D_\mu=\partial_\mu-\im g' \frac{1}{2}B_\mu-\im g \frac{1}{2}\sigma^a A^{a}_\mu \,,
\end{equation}
where $B_\mu$ and $A^{a}_\mu$ are the gauge boson fields corresponding to the weak hypercharge and the weak isospin, $g'$ and $g$ are their gauge coupling constants and $\sigma^a$ are Pauli matrices. The Higgs potential
\begin{equation}\label{Higgs_potential_app}
{\cal V}(\Phi^\dag\Phi)=-|\mu^2|\Phi^\dag\Phi+\lambda(\Phi^\dag\Phi)^2
\end{equation}
forces the Higgs field to condense into its classical field configuration $\Phi_{0}$ given by the minimum of ${\cal V}(\Phi^\dag\Phi)$, i.e.,
\begin{equation}\label{vacuum_condition_app}
\Phi_{0}^\dag\Phi_{0} \equiv \frac{v^2}{2}=\frac{|\mu^2|}{2\lambda} \,.
\end{equation}
Picking up one of the continuously degenerate non-trivial field configurations \eqref{vacuum_condition_app} as the ground state,
\begin{equation}\label{Phi_0}
\Phi_0\equiv\frac{1}{\sqrt{2}}\beginm{c} 0 \\ v \endm \,,
\end{equation}
leads to the spontaneous breaking of the symmetry of the Higgs potential \eqref{Higgs_potential_app}
\begin{equation}\label{SSB_O4_O3}
\OO{4}\sim\SU{2}_L\times\SU{2}_R \ \ \longrightarrow \ \ \OO{3}\sim\SU{2}_\mathrm{custodial} \,.
\end{equation}
The electroweak gauge symmetry subgroup gets spontaneously broken correspondingly $\SU{2}_L\times\U{1}_Y\rightarrow\U{1}_\mathrm{em}$. According to the Goldstone theorem, the symmetry breaking \eqref{SSB_O4_O3} gives rise to three Nambu--Goldstone boson fields $w^\pm(x)$ and $z(x)$. They are accommodated as three degrees of freedom in the elementary complex scalar field $\Phi(x)$,
\begin{equation}\label{Phi_shift}
\Phi(x)\equiv\beginm{c} -\im w^+(x) \\ \tfrac{1}{\sqrt{2}}\big[v+h(x)+\im z(x)\big] \endm \,.
\end{equation}

\subsubsection{The Nambu--Goldstone bosons in the Lagrangian}

Apart from other terms, the resulting Lagrangian contains the terms bilinear in the electroweak gauge boson fields, the mixing and kinetic terms of the gauge and Nambu--Goldstone boson fields
\begin{eqnarray}
{\cal L}^\mathrm{SM}_{\mathrm{Higgs-gauge}}  & = & \partial^\mu w^+\partial_\mu w^- + \frac{1}{2}\partial^\mu z\partial_\mu z + \frac{v^2}{8}\Big[\left(-g'B+g A_{3}\right)^2 +g^2 \left(A_{1}^2+A_{2}^2\right)\Big] \\
&   & - \frac{v}{2}\Big[\left(-g'B^\mu+g A_{3}^\mu\right)\partial_\mu z+\frac{g}{\sqrt{2}}\left(A_{1}^\mu+\im A_{2}^\mu\right)\partial_\mu w^++\frac{g}{\sqrt{2}}\left(A_{1}^\mu-\im A_{2}^\mu\right)\partial_\mu w^-\Big] + \dots \, \nonumber
\end{eqnarray}
The part of the Lagrangian bilinear in the gauge fields can be brought into its diagonal form by means of the gauge boson field redefinition
\begin{subequations}\label{Weinberg_rotation}
\begin{eqnarray}
W_{\mu}^\pm & = & \big(A^{1}_\mu\mp\im A^{2}_\mu\big)/\sqrt{2} \,,\\
Z_{\mu} & = & \cos\theta_WA^{3}_\mu-\sin\theta_WB_\mu \,,\\
A_{\mu} & = & \sin\theta_WA^{3}_\mu+\cos\theta_WB_\mu \,,
\end{eqnarray}
\end{subequations}
where $\tan\theta_W=g'/g$ defines the Weinberg angle $\theta_W$. We obtain the Higgs gauge Lagrangian \eqref{Higgs_Lagrangian} in terms of gauge boson mass eigenstates, $W$ and $Z$ bosons,
\begin{eqnarray}\label{WZ_NG_lagrangian}
{\cal L}^\mathrm{SM}_{\mathrm{Higgs-gauge}}  & = & \partial^\mu w^+\partial_\mu w^- + \frac{1}{2}\partial^\mu z\partial_\mu z + \frac{v^2}{4}g^2W^{+\mu}W^{-}_\mu+\frac{v^2}{8}(g^2+g^{\prime2})Z^{\mu}Z_\mu \\
& & - \frac{v}{2}\Big[\sqrt{g^2+g^{\prime2}}Z^\mu\partial_\mu z+gW^{-}_\mu\partial^\mu w^{+}+gW^{+}_\mu\partial^\mu w^{-}\Big] + \dots \, \nonumber
\end{eqnarray}
The mass and mixing terms can be treated as interaction terms. The Feynman rules for the mixing terms represent the tree-level expressions for the bilinear coupling function $\Lambda_{ab}(q^2)$ introduced in \eqref{eq_Gamma:rel:P}
\begin{eqnarray}
  \Lambda_{W}(q^2)|_\mathrm{tree} & = & g\frac{v}{2}\,, \\
  \Lambda_{Z}(q^2)|_\mathrm{tree} & = & \sqrt{g^2+g^{\prime2}}\frac{v}{2} \,.
\end{eqnarray}
The Lagrangian \eqref{WZ_NG_lagrangian} induces corrections to the free massless gauge boson propagator in the form of the polarization tensor.
At tree-level we obtain one-gauge-boson irreducible expression
\begin{eqnarray}\label{pic_polarization_tensor}
\Pi^{\mu\nu}_{W,\mathrm{tree}}(q) & = & \begin{array}{c}\includegraphics[width=0.45\textwidth]{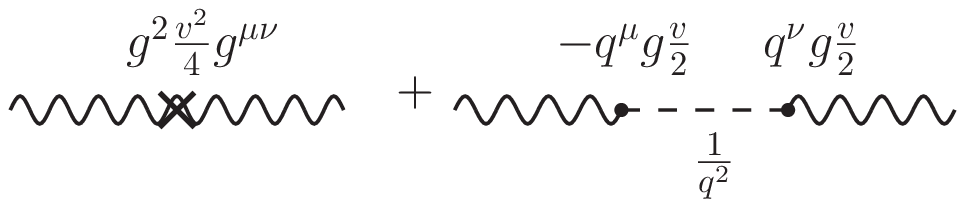}\end{array}
\end{eqnarray}
for polarization tensor of $W$ gauge boson. An analogous expression for $Z$ gauge boson is obtained under the replacement $g\rightarrow\sqrt{g^2+g^{\prime2}}$. These polarization tensors are manifestly transverse,
\begin{eqnarray}
\Pi^{\mu\nu}_{W,\mathrm{tree}}(q) & = & \left(g^{\mu\nu}-\frac{q^\mu q^\nu}{q^2}\right)\frac{g^2v^2}{4} \,, \\
\Pi^{\mu\nu}_{Z,\mathrm{tree}}(q) & = & \left(g^{\mu\nu}-\frac{q^\mu q^\nu}{q^2}\right)\frac{(g^2+g^{\prime2})v^2}{4} \,,
\end{eqnarray}
as they should be to satisfy the Ward identity $q_\mu\Pi^{\mu\nu}=0$. Note that at this leading tree-level, the polarization tensor for photon is zero as there are no corresponding terms analogous to \eqref{pic_polarization_tensor} in the Lagrangian \eqref{WZ_NG_lagrangian}, i.e.,
\begin{eqnarray}
\Pi^{\mu\nu}_{\gamma,\mathrm{tree}}(q) & = & 0 \,.
\end{eqnarray}
Extracting the polarization functions $\Pi_{W,Z}(q^2)$ from the polarization tensors $\Pi^{\mu\nu}_{W,Z}=-(q^2g^{\mu\nu}-q^\mu q^\nu)\Pi_{W,Z}(q^2)$, we can see that they develop massless poles due to the propagators of the Nambu--Goldstone fields $w^\pm$ and $z$ in \eqref{pic_polarization_tensor}
\begin{eqnarray}
\Pi_W(q^2) & = & -\frac{g^2v^2/4}{q^2} \,, \\
\Pi_Z(q^2) & = & -\frac{(g^2+g^{\prime2})v^2/4}{q^2} \,.
\end{eqnarray}
These poles are just what is needed in order to obtain massive propagators of vector bosons fully \emph{gauge invariantly}. The massless poles of the bare propagators are canceled by the massless Nambu--Goldstone poles of the polarization functions $\Pi_{W,Z}(q^2)$ and massive poles are generated at $q^2=M^2_{W,Z}$. Within the Landau gauge we can write the full propagators as
\begin{eqnarray}
\Delta^{\mu\nu}_{W,Z}(q) & = & -\left(g^{\mu\nu}-\frac{q^\mu q^\nu}{q^2}\right)\frac{1}{q^2(1+\Pi_{W,Z}(q^2))} = -\left(g^{\mu\nu}-\frac{q^\mu q^\nu}{q^2}\right)\frac{1}{q^2-M_{W,Z}^2} \,.
\end{eqnarray}
The resulting masses $M_{W,Z}^2$ are given as absolute values of residua of the massless Nambu--Goldstone poles in the polarization functions $\Pi_{W,Z}(q^2)$. We obtain
\begin{eqnarray}
M_{W} & = & \tfrac{1}{2}vg \,,\\
M_{Z} & = & \tfrac{1}{2}v\sqrt{g^2+g^{\prime2}} \,.
\end{eqnarray}
The photon propagator at this level stays bare and photon remains massless
\begin{eqnarray}
M_{\gamma} & = & 0 \,.
\end{eqnarray}

Qualitatively, all these features (the mass spectrum of the gauge bosons, transversality of the polarization tensor) are just results of underlying symmetry and its Nambu--Goldstone realization, and as such they hold to any loop order. Also the procedure can be followed in any gauge fixing scheme. Out of all possibilities, the unitary gauge is special because the mixing terms in \eqref{WZ_NG_lagrangian} and all other terms containing Nambu--Goldstone fields are just cancelled out from the Lagrangian \eqref{WZ_NG_lagrangian}. The gauge boson fields then occur simply as massive Proca fields. In all covariant gauges the Nambu--Goldstone fields stay in the Lagrangian but their propagators are gauge dependent, thus they never appear as asymptotic states. Instead the longitudinal components of gauge bosons become physical. This is the generally stated Anderson--Higgs mechanism usually rephrased by saying that the Nambu--Goldstone bosons are `eaten' by the gauge bosons in order to provide their masses.

\chapter{The Fierz identities}
%\pagenumbering{arabic}
%\input{C.tex}

\label{Fierz}

\subsubsection{Internal indices}

Have a general Hermitean $N\times N$ matrix $M$. We can decompose it in the basis of $N\times N$ unit matrix $\openone$ and set of $(N^2-1)$ generators of $\SU{N}$ in the fundamental representation $T_a$:
\begin{equation}
M=\openone m_0+T_a m_a \,.
\end{equation}
If we accept normalization of the generators
\begin{equation}
\Tr{T_aT_b}=\frac{1}{2}\delta_{ab} \,,
\end{equation}
then we have formulae for the coefficient of the decomposition
\begin{eqnarray}
m_0 & = & \frac{1}{N}\Tr{\openone M} \,, \\
m_a & = & 2\Tr{T_a M} \,.
\end{eqnarray}
The decomposition can be written element-wise
\begin{equation}
[M]_{ij}=\frac{1}{N}[\openone]_{ij}[M]_{kk}+2[T_a]_{ij}[T_a]_{kl}[M]_{lk} \,.
\end{equation}
To relate coefficients in front of the individual elements of matrix $M$, we rewrite the equation as
\begin{equation}
[\openone]_{ik}[\openone]_{lj}[M]_{kl}=\frac{1}{N}[\openone]_{ij}[\openone]_{lk}[M]_{kl}+2[T_a]_{ij}[T_a]_{lk}[M]_{kl} \,.
\end{equation}
This leads to the master equation from which Fierz identity for any two $N\times N$ Hermitean matrices can be derived
\begin{equation}
[\openone]_{ik}[\openone]_{lj}=\frac{1}{N}[\openone]_{ij}[\openone]_{lk}+2[T_a]_{ij}[T_a]_{lk} \,.
\end{equation}
The Fierz identity for two matrices $A$ and $B$ can be obtained by contracting the master equation with $[A]_{km}$ and $[B]_{nl}$ to get
\begin{eqnarray*}
[A]_{im}[B]_{nj} & = & \frac{1}{N}[\openone]_{ij}[BA]_{nm}+2[T_a]_{ij}[BT_aA]_{nm} \\
                 & = & \frac{1}{N^2}[\openone]_{ij}[\openone]_{nm}\Tr{BA}+4[T_a]_{ij}[T_b]_{nm}\Tr{T_bBT_aA} \\
                 & & +\frac{2}{N}[\openone]_{ij}[T_a]_{nm}\Tr{T_aBA}+\frac{2}{N}[T_a]_{ij}[\openone]_{nm}\Tr{BT_aA}
\end{eqnarray*}

\subsubsection{Application to $\SU{3}$}

For $\SU{3}$ generators in the basis of Gell-Mann matrices $T^a=\tfrac{1}{2}\lambda^a$ we have
\begin{subequations}\label{SU3_Fierz}
\begin{eqnarray}
{}[T_a]_{im}[T_a]_{nj} = -[T_a]_{im}[-T^{*}_a]_{jn} & = & \frac{4}{9}[\openone]_{ij}[\openone]_{nm}-\frac{1}{3}[T_a]_{ij}[T_a]_{nm} \,, \\
{}[T_a]_{im}[-T^{*}_a]_{nj} = -[T_a]_{im}[T_a]_{jn} & = & -\frac{1}{9}[\openone]_{ij}[\openone]_{nm}+\frac{4}{3}[T_A]_{ij}[T_A]_{nm}-\frac{2}{3}[T_S]_{ij}[T_S]_{nm} \,.\quad\quad
\end{eqnarray}
\end{subequations}
For the derivation of these results it is useful to know the following relations:
\begin{eqnarray}
\Tr{T_aT_bT_cT_d} & = & \frac{1}{12}(\delta_{ab}\delta_{cd}-\delta_{ac}\delta_{db}+\delta_{ad}\delta_{bc}) \\
& & +\frac{1}{8}(d_{abe}d_{cde}-d_{ace}d_{dbe}+d_{ade}d_{bce}) \nonumber\\
& & +\frac{1}{8}(d_{abe}f_{cde}-d_{ace}f_{dbe}+d_{ade}f_{bce}) \,, \nonumber\\
\Tr{T_aT_bT_cT_b} & = & -\frac{1}{12}\delta_{ac} \,, \\
\Tr{T_aT^{\T}_bT_cT_b} & = & \Tr{T_aT_bT_cT_d}\xi_{bd} \,,
\end{eqnarray}
where $\xi=\mathrm{Diag}(1,-1,1,1,-1,1,-1,1)$.
Other useful relations are:
\begin{eqnarray}
d_{abc}\delta_{bc} & = & 0 \,, \\
d_{abc}\xi_{bc} & = & 0 \,, \\
d_{aef}f_{bef} & = & 0 \,, \\
d_{aef}d_{cef} & = & \frac{5}{3}\delta_{ac} \,, \\
d_{abe}d_{cde}\xi_{bd} & = & \frac{1}{3}{\delta_{ac}+\xi_{ac}} \,.
\end{eqnarray}

\subsubsection{Dirac indices}

For completeness, from the reference \cite{Ripka:1997zb} we copy the Fierz identities for Dirac indices into both fermion-anti-fermion and fermion-fermion channels are
\begin{subequations}\label{Dirac_Fierz}
\begin{eqnarray}
\beginm{c} {}[\openone]_{ij}[\openone]_{kl} \\ {}[\gamma^\mu]_{ij}[\gamma_\mu]_{kl} \\ {}[\sigma^{\mu\nu}]_{ij}[\sigma_{\mu\nu}]_{kl} \\ {}[\gamma^\mu\gamma_5]_{ij}[\gamma_\mu\gamma_5]_{kl} \\ {}[\im\gamma_5]_{ij}[\im\gamma_5]_{kl} \endm & = & \beginm{ccccc}
\tfrac{1}{4} & \tfrac{1}{4} & \tfrac{1}{4} & -\tfrac{1}{4} & -\tfrac{1}{4} \\
1 & -\tfrac{1}{2} & 0 & -\tfrac{1}{2} & 1 \\
\tfrac{3}{2} & 0 & -\tfrac{1}{2} & 0 & -\tfrac{3}{2} \\
-1 & -\tfrac{1}{2} & 0 & -\tfrac{1}{2} & -1 \\
-\tfrac{1}{4} & \tfrac{1}{4} & -\tfrac{1}{4} & -\tfrac{1}{4} & \tfrac{1}{4}
\endm
\beginm{c} {}[\openone]_{il}[\openone]_{kj} \\ {}[\gamma^\mu]_{il}[\gamma_\mu]_{kj} \\ {}[\sigma^{\mu\nu}]_{il}[\sigma_{\mu\nu}]_{kj} \\ {}[\gamma^\mu\gamma_5]_{il}[\gamma_\mu\gamma_5]_{kj} \\ {}[\im\gamma_5]_{il}[\im\gamma_5]_{kj} \endm \,,\label{Dirac_Fierz_a} \\
\beginm{c} {}[\openone]_{ij}[\openone]_{kl} \\ {}[\gamma^\mu]_{ij}[\gamma_\mu]_{kl} \\ {}[\sigma^{\mu\nu}]_{ij}[\sigma_{\mu\nu}]_{kl} \\ {}[\gamma^\mu\gamma_5]_{ij}[\gamma_\mu\gamma_5]_{kl} \\ {}[\im\gamma_5]_{ij}[\im\gamma_5]_{kl} \endm & = & \beginm{ccccc}
-\tfrac{1}{4} & -\tfrac{1}{4} & -\tfrac{1}{4} & \tfrac{1}{4} & \tfrac{1}{4} \\
1 & -\tfrac{1}{2} & 0 & -\tfrac{1}{2} & 1 \\
\tfrac{3}{2} & 0 & -\tfrac{1}{2} & 0 & -\tfrac{3}{2} \\
1 & \tfrac{1}{2} & 0 & \tfrac{1}{2} & 1 \\
\tfrac{1}{4} & -\tfrac{1}{4} & \tfrac{1}{4} & \tfrac{1}{4} & -\tfrac{1}{4}
\endm
\beginm{c} {}[C]_{ik}[C]_{lj} \\ {}[\gamma^\mu C]_{ik}[C\gamma_\mu]_{lj} \\ {}[\sigma^{\mu\nu}C]_{ik}[C\sigma_{\mu\nu}]_{lj} \\ {}[\gamma^\mu\gamma_5C]_{ik}[C\gamma_\mu\gamma_5]_{lj} \\ {}[\im\gamma_5C]_{ik}[\im C\gamma_5]_{lj} \endm \,,\quad\quad\quad
\end{eqnarray}
\end{subequations}
%\begin{subequations}\label{Dirac_Fierz}
%\begin{eqnarray}
%{}[\gamma^\mu]_{ij}[\gamma_\mu]_{kl} & = & [\openone]_{il}[\openone]_{kj}+[\im\gamma_5]_{il}[\im\gamma_5]_{kj}-\frac{1}{2}[\gamma^\mu]_{il}[\gamma_\mu]_{kj}-\frac{1}{2}[\gamma^\mu\gamma_5]_{il}[\gamma_\mu\gamma_5]_{kj} \,,\label{Dirac_Fierz_a} \\
%{}[\gamma^\mu]_{ij}[\gamma_\mu]_{kl} & = &
%[C]_{ik}[C]_{lj}+[\im\gamma_5C]_{ik}[C\im\gamma_5]_{lj}-\frac{1}{2}[\gamma^\mu C]_{ik}[C\gamma_\mu]_{lj}-\frac{1}{2}[\gamma^\mu\gamma_5C]_{ik}[C\gamma_\mu\gamma_5]_{lj} \,,\quad\quad
%\end{eqnarray}
%\end{subequations}
where $\sigma_{\mu\nu}=\tfrac{\im}{2}[\gamma_\mu,\gamma_\nu]$ and $C=\im\gamma_0\gamma_2$ is the charge conjugation matrix. Its convention independent properties of $C$ are \cite{Benes:2012hz}
\begin{equation}
  C^\dag=C^{-1}\,,\quad C^\T=-C\,,\quad C^{-1}=-C\,.
\end{equation}

\bibliographystyle{JHEP}
\bibliography{references}

%
%\begin{thebibliography}{100}
%
%%\cleardoublepage
%\phantomsection
%\addcontentsline{toc}{chapter}{Bibliography}
%
%\newcommand{\bbTitle}[1]{\qm{#1}}
%
%%\newcommand{\bbArxiv}[2][]{\href{http://arxiv.org/abs/#2}{arXiv:#2 [#1]}}
%
%\newcommand{\bbArxivOld}[1]{\href{http://arxiv.org/abs/#1}{arXiv:#1}}
%\newcommand{\bbArxivNew}[2]{\href{http://arxiv.org/abs/#2}{arXiv:#2 [#1]}}
%\newcommand{\bbJournal}[1]{#1}
%\newcommand{\bbJournalDoi}[2]{\href{http://dx.doi.org/#2}{#1}}
%
%
%
%
%
%\end{thebibliography}

\end{document}